\documentclass[aps,onecolumn,prd,nofootinbib,superscriptaddress,preprintnumbers,longbibliography,eqsecnum,notitlepage,english,10pt]{revtex4-1}
\usepackage{amsfonts}
\usepackage{amsmath}
\usepackage{bm}
\usepackage{babel}
\usepackage{booktabs}
\usepackage{dsfont}
\usepackage[export]{adjustbox}
\usepackage{graphicx}
\usepackage{nicefrac}
\usepackage[dvipsnames]{xcolor}
\usepackage{braket}
\usepackage{mathtools}
\usepackage{multirow} 
\usepackage{tabularx}
\usepackage{orcidlink}
\usepackage{hyperref}
\hypersetup{
    colorlinks=true,
    urlcolor=blue,
    linkcolor = blue,
    urlcolor  = blue,
    citecolor = blue,
    anchorcolor = blue,
    pdftitle={ccc},
    pdfauthor={FMURDPB},
    pdfsubject={ccc}}
\usepackage{cleveref} 
    \crefformat{figure}{Fig.~#2#1#3}
    \crefformat{table}{Tab.~#2#1#3}
    \crefformat{equation}{Eq.~(#2#1#3)}
    \crefformat{section}{Sect.~#2#1#3}
\graphicspath{{./plots/}}
\renewcommand{\Re}{{\rm Re}}
\renewcommand{\Im}{{\rm Im}}
\newcommand{\Lagr}{\mathcal{L}}
\newcommand{\MeV}{{\rm ~MeV}}
\newcommand{\fm}{{\rm ~fm}}
\newcommand{\GeV}{{\rm ~GeV}}

\begin{document}

\title{
Universal parameters of the {\boldmath$\Lambda(1380)$}, the {\boldmath$\Lambda(1405)$} and their isospin partners from a combined analysis of Lattice QCD and experimental results
}
\author{Ferenc Pittler\orcidlink{0000-0003-4100-4289}}
\email{f.pittler@cyi.ac.cy}
\affiliation{Computation-based Science and Technology Research Center, The Cyprus Institute, Nicosia, Cyprus}
\author{Maxim~Mai\orcidlink{0000-0003-0119-0718}}
\email{maxim.mai@faculty.unibe.ch}
\affiliation{Albert Einstein Center for Fundamental Physics, Institute for Theoretical Physics, University of Bern, Sidlerstrasse 5, 3012 Bern, Switzerland}
\affiliation{Institute for Nuclear Studies and Department of Physics, The George Washington University, Washington, DC 20052, USA}
\author{Ulf-G.~Mei{\ss}ner\orcidlink{0000-0003-1254-442X}}
\email{meissner@hiskp.uni-bonn.de}
\affiliation{Helmholtz-Institut für Strahlen- und Kernphysik (Theorie) and Bethe Center for Theoretical Physics, Universität Bonn, 53115 Bonn, Germany}
\affiliation{Institute for Advanced Simulation (IAS-4), Forschungszentrum J\"ulich, D-52425 J\"ulich, Germany}
\affiliation{Peng Huanwu Collaborative 
Center for Research and Education, International Institute for Interdisciplinary and Frontiers, Beihang University, Beijing 
100191, China}
\author{{Ryan~F.~Ferguson}\orcidlink{0009-0004-5853-9671}}
\affiliation{School of Physics and Astronomy, University of Glasgow, Glasgow, G12 8QQ, United Kingdom}
\author{{Peter Hurck}\orcidlink{0000-0002-8473-1470}}
\affiliation{School of Physics and Astronomy, University of Glasgow, Glasgow, G12 8QQ, United Kingdom}
\author{{David G. Ireland}\orcidlink{0000-0001-7713-7011}}
\affiliation{School of Physics and Astronomy, University of Glasgow, Glasgow, G12 8QQ, United Kingdom}
\author{{Bryan McKinnon}\orcidlink{0000-0002-5550-0980}}
\affiliation{School of Physics and Astronomy, University of Glasgow, Glasgow, G12 8QQ, United Kingdom}

\begin{abstract}
We perform a global analysis of lattice and experimental data on negative-strangeness meson-baryon scattering using a large set of variations of the theoretical framework based on the Chiral Unitary Approach. For the former, the L{\"u}scher formalism is utilized taking into account all pertinent coupled-channel effects. Through this, systematic uncertainties related to data scarcity, potential ambiguities, and possible framework dependence are quantified for the first time. The implementation of information criteria and other statistical tools is discussed. As a final result we provide pole positions for isoscalar resonances at the physical and lattice points including statistical and systematic uncertainties. Predictions for the isovector states are also provided showing large uncertainties.
\end{abstract}

\maketitle
{\footnotesize\tableofcontents}

\section{Introduction and summary}

The hadron spectrum provides a manifestation of structure formation of the strong interaction, which remains a challenge to our understanding of the so successful Standard Model of particle physics. Primarily, due to advances in experimental techniques hundreds of new and predominantly excited hadrons have been discovered over the past century~\cite{ParticleDataGroup:2024cfk}. A partial ordering of the spectrum can be achieved through a simple quark model organizing mesons as quark-antiquark and baryons as three-quark states. However, this simple picture does not reflect the reality calling for a more comprehensive approach. For recent related reviews see Refs.~\cite{Thiel:2022xtb, Eichmann:2016yit, Doring:2025sgb, Mai:2022eur, Burkert:2025coj}. A prominent example of this kind is the negative-strangeness, isoscalar $\Lambda(1405)$ baryon which became a poster child of the two-pole structure~\cite{Oller:2000fj,Meissner:2020khl}. Currently, this is associated with two states, the $\Lambda(1405)$ and $\Lambda(1380)$. For a dedicated review including historical aspects see Ref.~\cite{Mai:2020ltx} as well as Refs.~\cite{Meissner:2020khl,Hyodo:2020czb} for a broader context.

Originally, the connection between the fundamental theory of the strong interaction, Quantum Chromodynamics (QCD), and the phenomenology of the two-pole structure of the $\Lambda(1405)$ was established using Chiral Perturbation Theory (CHPT) while extending the range of applicability through unitarization techniques. For details and connection to other approaches see Ref.~\cite{Mai:2020ltx}. Note further that the isovector pole also seen in Ref.~\cite{Oller:2000fj} (see also~\cite{Cieply:2016jby,Ikeda:2012au,Guo:2012vv,Lu:2022hwm}) has obtained much less attention but this issue will also be taken up here. Typically, the free parameters of this approach are fixed using the data available from experiments conducted over the last several decades. These parameters originate partly from the so-called low-energy constants, which encode the effects of integrated out heavy degrees of freedom of QCD but also reflect a certain degree of model dependence in the exact prescription of the unitarization procedure. This model dependence, along with ambiguities in the existing experimental input leads to  different predictions of the Chiral Unitary approaches (UCHPT) in regimes not covered by the experimental data. Illustrative examples of such systematic uncertainties are discussed in, e.g., Refs.~\cite{Cieply:2016jby, Mai:2020ltx}.

The problem can also be tackled from a different angle using Lattice QCD methodology~\cite{Menadue:2011pd, Engel:2012qp, Edwards:2012fx, Hall:2014uca, Hall:2014gqa}. In a most recent calculation, not too far away from the physical point (that is, quark masses that are only slightly larger than the physical ones) and using state-of-the-art methodology finite-volume spectra for the isoscalar channel of negative-strangeness meson-baryon interaction have been determined by the BaSc collaboration~\cite{BaryonScatteringBaSc:2023ori, BaryonScatteringBaSc:2023zvt}. While unphysical quark masses are used in this setup, CHPT underlying UCHPT allows one to extrapolate and connect these results to the physical point. Establishing this connection, along with the study of the model dependence within the UCHPT approaches constitutes the main goals of the present work. The central observations of our study can be summarized as follows:
\begin{itemize}
    \item UCHPT approaches based on the lowest order Chiral Lagrangian (called type M1 and M2 in what follows) do capture the main features of the interaction but fail to quantitatively describe the existing experimental data.
    \item UCHPT can accurately describe the lattice input. Still, the latter input alone does not seems sufficient for determining accurately both pole positions when model variations are taken into account.
    \item Within the most flexible models, the experimental input does lead to the pole structure similar to that of found in the existing literature. When extrapolating to the unphysical quark mass scenario (lattice point) the pole structure determined by the BaSc collaboration is confirmed. For the most flexible models we observe that ambiguities in the older cross-section data dominate the systematic uncertainties.
    \item Combined fits including all experimental and lattice inputs provide a very good description. Variations between different models are assessed using information criteria. Numerical results for the isoscalar poles, as well as predicted isovector poles are collected in \cref{tab:overall-poles}.
    \begin{table}[t]
        \caption{Numerical values for extracted pole positions (combined fit to lattice and experimental input) for the isoscalar and isovector case (in GeV). Different fits refer to variations of the UCHPT approach as explained in \cref{sec:UCHPT} (where
        M refers to the chosen interaction kernel and S to the method of regularization). The lattice point refers to quark mass setup used in Refs.~\cite{BaryonScatteringBaSc:2023ori, BaryonScatteringBaSc:2023zvt}.\label{tab:overall-poles}}
        \renewcommand{\arraystretch}{1.2}
        \begin{tabularx}{\linewidth}{|X|X|X|X|}
        \hline
        \multirow{8}{*}{$I=0$ $S=-1$}& Type & physical point            & lattice point\\
        \hline
        &\multirow{2}{*}{M3S1 $(F_{17})$} 
                        & $1.342^{+0.009}_{-0.010}-i\,0.028^{+0.013}_{-0.014}$ 
                        & $1.359_{-0.014}^{+0.010}-i\,0.100_{-0.050}^{+0.050}$ \\
                        && $1.432^{+0.002}_{-0.002}-i\,0.025^{+0.001}_{-0.001}$ 
                        & $1.460_{-0.005}^{+0.006}-i\,0.013_{-0.004}^{+0.005}$\\
        \cline{2-4}
        &\multirow{2}{*}{M3S2 $(F_{16})$}   
                        & $1.373_{-0.005}^{+0.002}-i\,0.082_{-0.009}^{+0.011}$ 
                        & $1.389_{-0.007}^{+0.006}$\\
                        && $1.423_{-0.001}^{+0.002}-i\,0.020_{-0.002}^{+0.001}$ 
                        & $1.466_{-0.004}^{+0.003}-i\,0.020_{-0.001}^{+0.001}$\\
        \cline{2-4}
        &\multirow{2}{*}{M3S3 $(F_{12})$}  
                        & $1.352_{-0.007}^{+0.004}-i\,0.091_{-0.006}^{+0.005}$ 
                        & $1.401_{-0.003}^{+0.003}$\\
                        && $1.420_{-0.001}^{+0.001}-i\,0.018_{-0.001}^{+0.001}$
                        & $1.478_{-0.004}^{+0.005}-i\,0.026_{-0.001}^{+0.001}$\\
        \hline
        \hline
        \multirow{4}{*}{$I=1$ $S=-1$}&\multirow{2}{*}{M3S1 $(F_{17})$}
                    & $1.351_{-0.015}^{+0.022}-i\,0.112_{-0.011}^{+0.010}$ 
                    & $1.333_{-0.024}^{+0.018}-i\,0.144_{-0.012}^{+0.008}$ \\
                    && $1.356_{-0.011}^{+0.005}-i\,0.035_{-0.014}^{+0.022}$
                    &\\
        \cline{2-4}
        &M3S2 $(F_{16})$   
                    & $1.430_{-0.040}^{+0.090}-i\,0.281_{-0.002}^{+0.170}$ 
                    & $1.430_{-0.060}^{+0.050}-i\,0.208_{-0.000}^{+0.090}$\\
        \cline{2-4}
        &M3S3 $(F_{12})$   
                    & $1.338_{-0.023}^{+0.021}-i\,0.215_{-0.004}^{+0.007}$ 
                    & $1.335_{-0.017}^{+0.025}-i\,0.143_{-0.005}^{+0.008}$\\
        \hline
        \end{tabularx}
    \end{table}
\end{itemize}

This paper is organized as follows. Details on the Lattice QCD input are provided in \cref{sec:LQCD} which also includes details of the finite-volume implementation. The experimental data are reviewed in \cref{sec:Experimental_data} discussing also their ambiguities. The Chiral Unitary approach is discussed in \cref{sec:UCHPT} including variation of the methodology reflecting systematic uncertainties. Additionally, potential complications due to 3-body on-shell states are evaluated in \cref{subsec:synthesis}. Fits and pertinent predictions are discussed in \cref{sec:fits} and \cref{sec:results}, respectively. Individual fit results are moved to the appendices for convenience.

\section{Lattice QCD input}
\label{sec:LQCD}

\subsection{Overview of the available quantities}

Recent advances in both theoretical frameworks  and computational techniques have enabled  Lattice QCD to make increasingly precise  predictions for the properties of strongly interacting  unstable particles (resonances) such as the  $\rho$, $\Delta$, and more recently, the  $\Lambda(1405)$. For recent reviews see Refs.~\cite{Morningstar:2025xmi, Mai:2022eur, Briceno:2017max}.  In lattice calculations one extracts information about  such states by computing the  discrete energy spectrum of multi-hadron  scattering states in a finite Euclidean  volume. This is achieved through Markov Chain Monte Carlo integration and analyzing the  exponential decay of correlation functions  constructed from suitably designed  interpolating operators. In his seminal work, Lüscher showed that the energy levels of two interacting hadrons in a finite spatial volume are shifted from their non-interacting values by an amount that is only power-law suppressed in the box size \cite{Luscher:1990ux,Luscher:1991cf}. 
These finite-volume energy shifts are directly  related to the infinite-volume scattering  phase-shift  \cite{Rummukainen:1995vs,Kim:2005gf}, and hence to the scattering amplitudes themselves. Subsequent works extended Lüscher's formalism to arbitrary spins and multiple coupled  channels~\cite{He:2005ey,Lage:2009zv,Briceno:2014oea, Bernard:2010fp, Gockeler:2012yj}. In the following subsection, we briefly summarize the  quantization condition used in the present  work.

\subsection{Implementation of the quantization condition}
\label{sec:quantization condition}

Experimental observables, such as cross sections and the finite-volume energy spectrum obtained from lattice simulations, can both be described using effective models in which the scattering amplitude is parameterized using a small number of input parameters. These parameters are constrained through a correlated $\chi^2$ fit to both types of data: the finite-volume lattice energy spectrum and experimental observables. 

Instead of parametrizing the unitary S-matrix directly, we work with the real, symmetric K-matrix, which is related to the S-matrix via
\begin{align}
    S=(1-iK)^{-1}(1+iK)\,.
\end{align}
Due to rotational invariance, the infinite dimensional K-matrix is diagonal in angular momentum space:
\begin{align}
    \langle J^{\prime}m_{J^{\prime}}\ell^{\prime}S^{\prime}a^{\prime}\vert K \vert Jm_J\ell Sa\rangle=\delta_{JJ^{\prime}}\delta_{m_Jm_J^{\prime}}K^{(J)}_{\ell^\prime S^{\prime}a^{\prime};\ell Sa}(s)\,,
\end{align}
where the states $\vert Jm_J\ell Sa\rangle$ are labeled by the total angular momentum $J$, its projection on the $z$-axis $m_J$, the total orbital angular momentum $\ell$ and spin angular momentum $S$ of the two particles, respectively, $a$ is the channel index and lastly $s$ is the Mandelstam variable, square of the total energy in the center-of-mass frame. In this section, we connect the parametrization of the K-matrix used in the quantization condition \cite{BaryonScatteringBaSc:2023ori} with the UCHPT parametrizations (e.g., Refs.~\cite{Bruns:2010sv,Mai:2012dt,Sadasivan:2022srs}) which provide what is often referred to as the Höhler's partial-wave amplitudes $f_{\ell\pm}(s)$~\cite{Hohler:1984ux,Chew:1957zz}, described later in \cref{subsec:unitarization} in terms of T-matrix. We also briefly summarize the quantization condition in multi-channel space applied in this work, closely following \cite{Morningstar:2017spu, Morningstar:2025xmi}.

\begin{figure}[t]
            \hspace{-12pt}              
    \begin{minipage}[t]{0.5\linewidth}
        \vspace{0pt}
        \includegraphics[height=6cm]{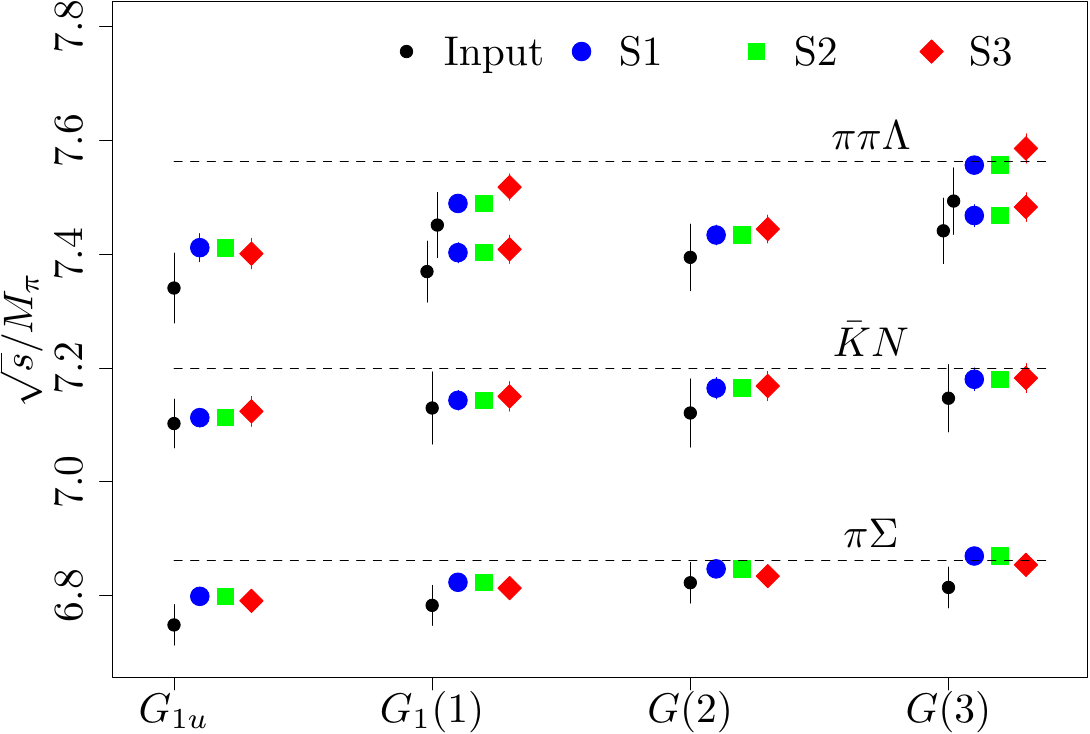}
    \end{minipage}%
    \begin{minipage}[t]{0.39\linewidth}
        \hspace{-12pt}
        \includegraphics[height=6cm]{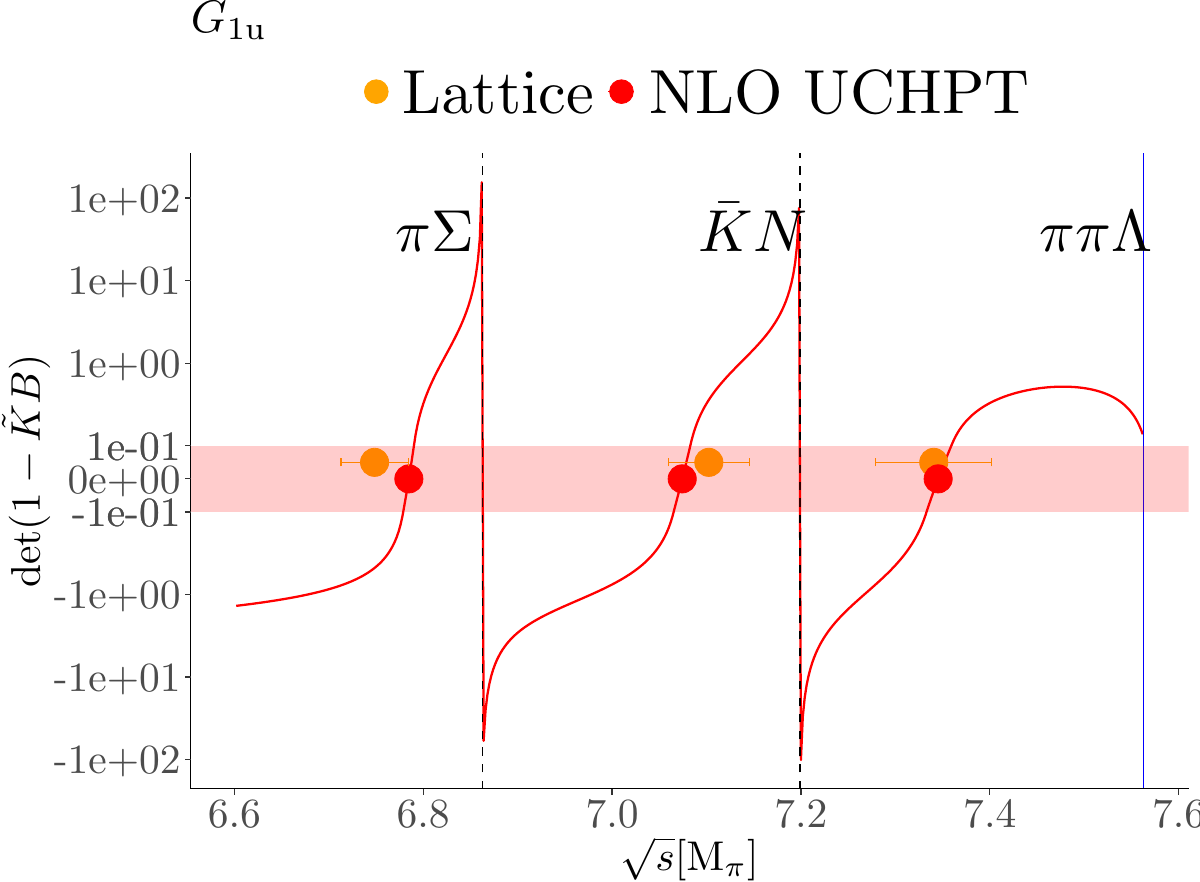}
    \end{minipage}
    \caption{
    Left: Summary of energy levels used as input in this work (black dots with error bars), together with our model estimates using different regularization schemes (S1, S2, S3). Right: Illustration of the quantization      condition as a function of the center-of-mass energy for the rest frame $G_{1u}$ irrep. Red band highlights  the area where the quantization condition is closely fulfilled, red circles indicate the prediction of UCHPT and orange circles with error bars shows the lattice results.}     
    \label{fig:spectrum_summary}
\end{figure}

To connect the two parametrizations, physical quantities are extracted from both meson-baryon scattering $MB\to MB$  Höhler partial-wave amplitudes and $\tilde{K}$~\cite{Morningstar:2017spu}. For example, using the former, the elastic scattering phase-shifts can be computed through a K-matrix like quantity as
\begin{align}
    f_{0+}(s)=\frac{1}{(\tilde K_E^{-1}(s)-ip_{\rm cm}(s))}\,
    \label{eq:f0plus_kmatrix}
    \Rightarrow \cot\delta=\frac{\Re f_{0+}}{\Im f_{0+}}=\frac{\tilde K_E^{-1}}{p_{\rm cm}}
    \Rightarrow p_{\rm cm}\cot\delta=\tilde K_E^{-1}\,.
\end{align}
Here $p_{\rm cm}$ denotes the magnitude of the three-momentum in the center of mass frame. In the same region, the phase-shifts can also be extracted using $\tilde K$, the matrix in the quantization condition for the finite-volume energy spectrum~\cite{Morningstar:2017spu}
\begin{align}
\label{eq:qc}
\mathrm{det}(1-\tilde{K}B^{\vec{P}})=0\,,
\end{align}
where $\vec{P}$ is the total momentum of the two-particle system, $\tilde{K}=\frac{2\pi}{Lp_{\mathrm{cm}}}K=\frac{2\pi}{L}\frac{1}{p_{\mathrm{cm}}\mathrm{cot}\delta}=\frac{2\pi}{L}\tilde{K}_E$, here  the first two equality follow from equations (16) and (18) in \cite{Morningstar:2017spu} and \eqref{eq:f0plus_kmatrix} is used in the last one.  The so-called box matrix  $B^{\vec{P}}$ does not depend on  interactions, it is a known, purely  kinematical matrix that depends on the  finite-volume. For real scattering  momenta, $B^{\vec{P}}$ is hermitian and diagonal in the channel-space 
\begin{align}
    \langle J^{\prime}m_{J^{\prime}}\ell^{\prime}S^{\prime}a^{\prime}\vert B^{\vec{p}}\vert Jm_J\ell Sa\rangle=-i\delta_{aa^{\prime}}    \delta_{SS^\prime}p_{\mathrm{cm},a}^{\ell+\ell^{\prime}+1}W^{\vec{P}a}_{\ell^{\prime}m_{\ell^\prime};\ell m_\ell}\langle J^{\prime}m_{J^{\prime}}\vert \ell^{\prime}m_{\ell^\prime},Sm_S\rangle\langle \ell m_\ell,Sm_S\vert Jm_J\rangle\,,
\end{align}
where $\langle j_1m_1j_2m_2\vert JM \rangle$ are the familiar Clebsch-Gordan coefficients and $W$ is defined in equation (6) in \cite{Morningstar:2017spu}. In practice, the infinite dimensional matrix in the determinant \eqref{eq:qc} is block-diagonalized by projecting onto the superposition of states that  transforms according to the irreducible representation (irrep) of the little group of $\vec{P}$, i.e., by performing a unitary basis transformation:
\begin{align}
    \vert \Lambda\lambda n J\ell Sa\rangle=\sum_{m_J}c_{m_{J}}^{J(-1)^{\ell};\Lambda\lambda n}\vert Jm_J\ell Sa\rangle\,,
\end{align}
where  $\Lambda$ is the irrep of the  little group of $\vec{P}$, $\lambda$ is the irrep row, $n$ is the occurrence of the particular irrep in the reducible  representation $\vert Jm_J\ell Sa\rangle$.  In each block, a truncation to $\ell \le  \ell_{\mathrm{max}}$ is imposed   to make the determinant condition  manageable; in the present work, we  consider only the $S$-wave,  $\ell_{\mathrm{max}}=0$.  We include energy levels from the rest  frame up to total momenta of three units  of lattice momenta \cite{BaryonScatteringBaSc:2023ori}.  The inter-channel interactions are  encoded in the dense $\tilde{K}_E$ matrix, while the box matrix remains  diagonal in the channel space. The matrix $\tilde{K}_E$ is a $10\times10$ matrix in the space of meson-baryon channels  with strangeness $S=-1$, as determined by the underlying $SU(3)$ symmetry
\begin{align}
    \mathcal{S}=\{K^-p, \bar K^0 n, \pi^0\Lambda, \pi^0\Sigma^0, \pi^+\Sigma^-, \pi^-\Sigma^+, \eta\Lambda, \eta \Sigma^0, K^+\Xi^-, K^0\Xi^0\}\,.
    \label{eq:ordering-channel}
\end{align}
For lattice energy spectrum analysis, we  further convert from the physical basis to the isospin basis and project onto channels with total isospin zero
\begin{align}
    \mathcal{S}^{I=0}=\lbrace \bar{K}N,\pi \Sigma, \eta\Lambda, K\Xi\rbrace\,.
    \label{eq:ordering-channel-LAT}
\end{align}
For phase-convention and explicit form of the projectors see, e.g., Ref.~\cite{Lu:2024ajt}. In predicting the finite-volume energy  spectrum we compute the box matrix for  the appropriate total momentum and irrep (a $4\times 4$ diagonal matrix), and  combine it with the corresponding  $4\times 4$ dense  $\tilde{K}_E$ to  evaluate the determinant in \eqref{eq:qc}.  An illustration of how the energy  spectrum constrains the scattering  amplitude is shown on the right part of \cref{fig:spectrum_summary} for the zero momentum case ($G_{1u}$ irrep).

In the spectrum we consider 14 energy levels, from all irreps dominated by $\ell =0$ lying   below the first relevant three-particle threshold  $\pi\pi\Lambda$. On left part of  \cref{fig:spectrum_summary} we show all the input energy levels together with our best  estimates using three different regularizations S1, S2 and S3, as discussed later in \cref{sec:UCHPT}.

\subsection{Details of lattice calculation}
\label{subsec:latticesimdetails}

The finite-volume lattice energy spectra have been generated using the D200 ensemble of the CLS collaboration \cite{Bruno:2014jqa}, which uses 2+1 flavor of non-perturbatively improved Wilson fermions and Lüscher-Weisz gauge action. The pion mass is slightly higher than the physical one, and the kaon mass is slightly lower than the physical one. We summarize the details of the ensemble relevant to the present work in \cref{tab:ensemble_param}. Correlation functions including two-hadron interpolating fields at the source / sink with different combinations of momenta  ($\pi(\vec{p}_1)\Sigma(\vec{p}_2),\bar{K}(\vec{p}_1)N(\vec{p}_2)$) up-to three lattice units of $\vec{p}_1+\vec{p}_2$ total momentum) were produced in order to determine all the energy levels in the isospin 0 sector below the lowest lying three-particle threshold ($\pi\pi\Lambda$). Correlation functions were evaluated using the stochastic Laplacian-Heaviside (sLapH) method \cite{HadronSpectrum:2009krc,Morningstar:2011ka}. The energy levels  were obtained using the ratio method, i.e., taking the ratio of properly diagonalized correlation matrices (Generalized eigenvalue problem principal correlators) with the single hadron correlators determining directly the relevant energy shift \cite{Cid-Mora:2023kuf}. The statistical errors are estimated via bootstrap resampling using 800 samples.

\begin{table}[h!]
\caption{Properties of the D200 ensemble including the masses of the light meson and baryon octet at the isospin symmetric point.\label{tab:ensemble_param}}
\begin{tabularx}{\linewidth}{|XX|XXX|XXXX|}
\hline
{$L$}[fm] & {$a$}[fm] & {$M_{\pi}$}[GeV] 
& {$M_{K}$}[GeV]
& {$M_{\eta}$}[GeV]
& {$m_{N}$}[GeV]
& {$m_{\Sigma}$}[GeV]
& {$m_{\Lambda}$}[GeV]
& {$m_{\Xi}$}[GeV]\\
\hline
4.05(4) & 
0.0633(7) &
0.2036(8) &
0.4864(5) &
0.5511    &
0.979(11) &
1.193(6)  &
1.132(4)  &
1.322(3) \\ 
\hline
\end{tabularx}
\end{table}

\section{Experimental input}
\label{sec:Experimental_data}

Below we review all experimental input included into the present study. To have a more transparent picture of the systematic uncertainties discussed in \cref{sec:UCHPT} we restrict ourselves only to data which are directly related to the meson-baryon scattering amplitudes in the energy region most relevant for the $\Lambda(1380)$ and $\Lambda(1405)$ states. Other input from, e.g., photon-induced reactions~\cite{Niiyama:2009zza, CLAS:2013rjt, CLAS:2013rxx, BGOOD:2021sog} is not included as it typically requires further parametrization of the reaction mechanism. For some studies of that type of data see Refs.~\cite{Anisovich:2020lec, Roca:2015tea, Mai:2014xna, Lutz:2004sg}.

\begin{itemize}
    \item {\bf Cross sections:} Most of the experimental data in the energy region relevant for a study of the first resonance region stem from the time not long after the initiation of the first large experimental programs on production of kaons in the 1950s. For this, mostly data from bubble chamber setups at CERN, LBNL, BNL or Bevatron were used until the mid 1980s. For a historical overview see Ref.~\cite{Mai:2020ltx}. The data has, therefore, quite large error bars and in certain cases there are systematic discrepancies between different data sets. Encouragingly, in the 2020s some progress occurred. As one of the most relevant recent developments in the field, the AMADEUS collaboration~\cite{Piscicchia:2022wmd} provided two new high-precision results based on data collected by the KLOE collaboration~\cite{Ambrosino:2004qx} on the $K^-p\to \pi\Sigma^0$ and $K^-p\to \pi\Lambda$ total cross sections. The impact of these data has been studied within a UCHPT model in Ref.~\cite{Sadasivan:2022srs}. In the energy range of interest, i.e., kaon momentum in the laboratory frame below $P_{\rm LAB}\le300\MeV$, there are 252 total cross section data~\cite{Ciborowski:1982et,Humphrey:1962zz, Sakitt:1965kh,Watson:1963zz,Piscicchia:2022wmd} (83, 47, 11, 11, 51, 49, corresponding to $K^-p\to K^-p$, $K^-p\to \bar K^0n$, $K^-p\to \pi^0 \Lambda$, $K^-p\to \pi^0 \Sigma^0$, $K^-p\to \pi^+ \Sigma^-$, $K^-p\to \pi^- \Sigma^+$ transitions)\footnote{The total cross section data in digitalized form can be accessed under \url{https://github.com/maxim-mai/Experimental-Data/tree/master/Lambda1405}.}. These data can be related to the partial-wave $f_{\ell\pm}$ derived form a given model. Neglecting higher partial waves, the explicit formula for the transition $\phi_\alpha B_\alpha \to \phi_\beta B_\beta$ reads
    \begin{align}
        \sigma_{\alpha\beta}=4\pi\frac{p_\beta(s)}{p_\alpha (s)}|f_{0+,\alpha\beta}(s)|^2\,,
    \label{eq:cross_section}
    \end{align}
    where $p_\alpha$ refers to $p_{\rm cm}$ in the meson baryon channel $\alpha$. For formulas including higher partial wave we refer the reader to Refs.~\cite{Sadasivan:2018jig,Hohler:1984ux}.

    \begin{figure}[hbt]
        \centering
        \includegraphics[width=1.0\linewidth]{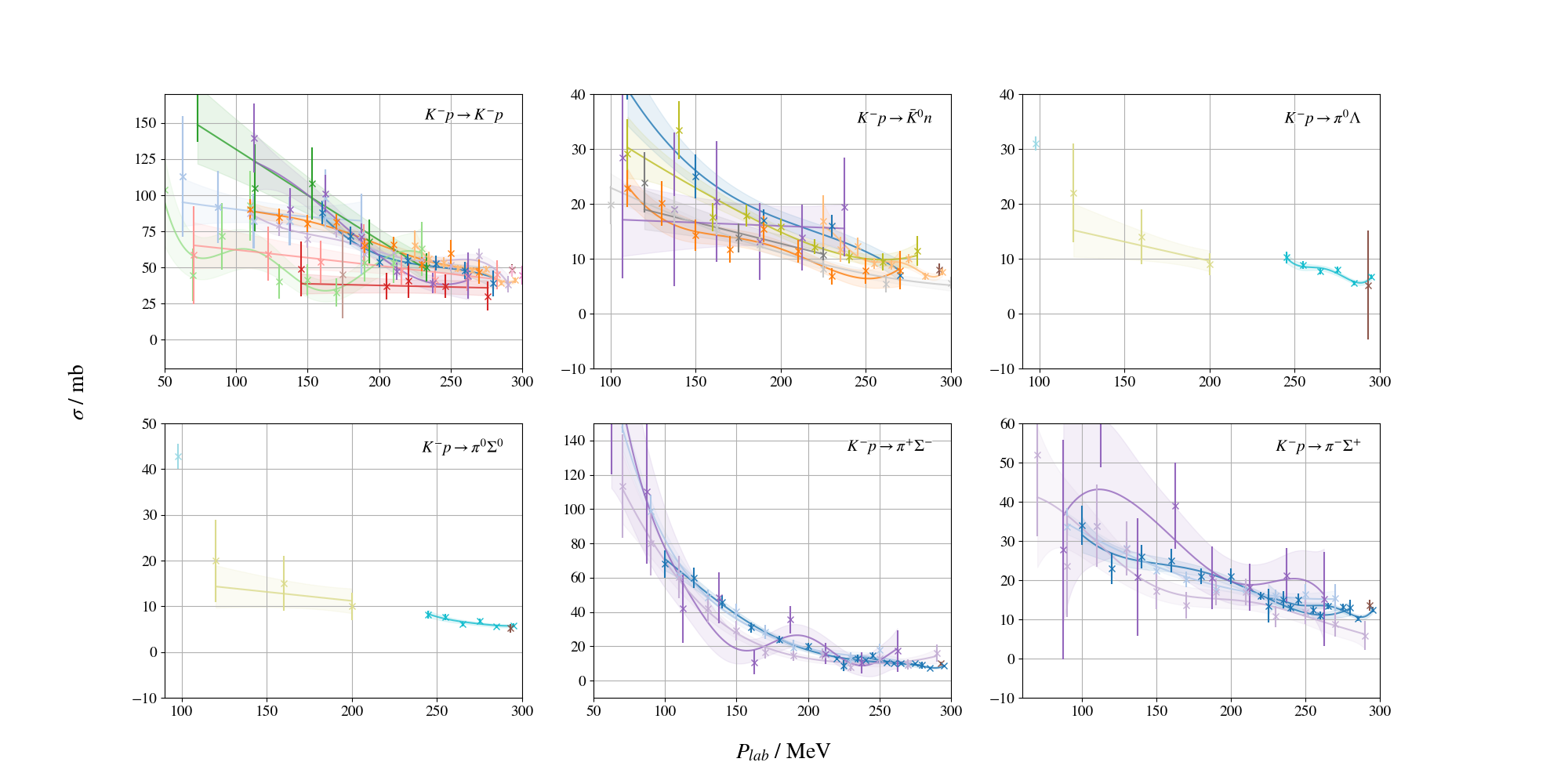}
        \caption{Total cross sections considered in this work. Different colors distinguish between various experiments~\cite{Ciborowski:1982et, Humphrey:1962zz, Sakitt:1965kh, Watson:1963zz, Piscicchia:2022wmd}. Shaded bands represent a Gaussian Process fit which is included to guide the eye \cite{ferguson_data-driven_2025}.}
        \label{fig:cross-section-data}
    \end{figure}
    By plotting the cross section data with different colours for separate experiments (along with a corresponding Gaussian Process fit if applicable), as we have in \cref{fig:cross-section-data}, it is readily seen that for several channels ($K^- p$, $\bar{K}^0 n$, $\pi^- \Sigma^+$), the data are not only widely varying at some energies, but are even inconsistent. This is likely due to the effect of different systematic uncertainties in each experiment. By using squares of residuals to determine a $\chi^2$ statistic, as is typically employed in analyses including this work, there is a chance that fits are drawn to the average of values, some of which are systematically wrong. 

    This issue is further illustrated in \cref{fig:cross-section-prob}, where the log probability surfaces for the data in \cref{fig:cross-section-data} are depicted. The procedure for generating these surfaces is described in \cite{ferguson_data-driven_2025}. Particularly in the case of the $K^- p \rightarrow \bar{K}^0 n$ channel, the multiple bands suggest inconsistent data. 

     Note also that we consider the scattering data for energies high enough so that the Coulomb effect in the charged channels can be neglected.
    \begin{figure}[hbt]
        \centering
        \includegraphics[width=1.0\linewidth]{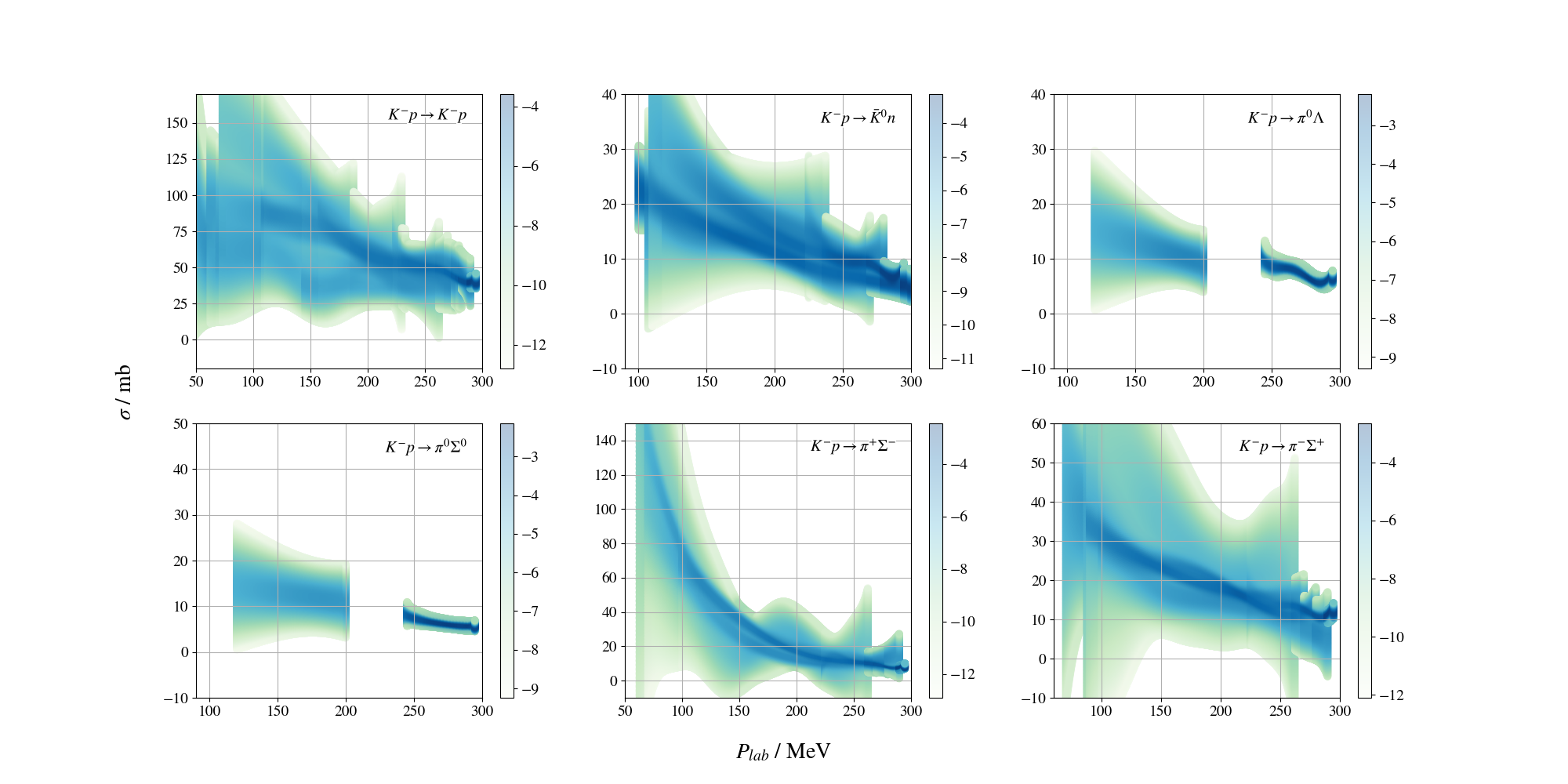}
        \caption{Log probability surfaces derived from the data illustrated in \cref{fig:cross-section-data}.}
        \label{fig:cross-section-prob}
    \end{figure}
    \item {\bf Threshold data:} At the $K^-p$ threshold, several ratios of the cross sections were measured some decades ago~\cite{Nowak:1978au,Tovee:1971ga}. Specifically, the values are $\gamma=2.38\pm0.04$, $R_c=0.664\pm0.011$ and $R_n=0.189\pm0.015$ which are related to the total cross section ratios as
    \begin{equation}
    \begin{aligned}[b]
    \gamma=\frac{\Gamma_{K^-p\rightarrow
     \pi^+\Sigma^-}}{\Gamma_{K^-p\rightarrow \pi^-\Sigma^+}}\,,
    \quad
    R_c=\frac{\Gamma_{K^-p\rightarrow
     {\rm charged~states}}}{\Gamma_{K^-p\rightarrow {\rm all~final~states}}}\,,
    \quad
    R_n=\frac{\Gamma_{K^-p\rightarrow\pi^0\Lambda}}{\Gamma_{K^-p\rightarrow
    \text{neutral states}}}\,\label{eq:threshold_ratios}.
    \end{aligned}
    \end{equation}
    Additionally, in a more recent kaonic hydrogen experiment at DAPHNE, the SIDDHARTA collaboration~\cite{Bazzi:2011zj} determined to a very high precision the energy shift and width due to strong interaction of the $K^-p$ system, i.e., $\Delta E=283\pm42$~eV, 
    $\Gamma/2=271\pm55$~eV. The complex energy shift in kaonic hydrogen is related to the $K^-p$ scattering length through the modified Deser formula~\cite{Meissner:2004jr} (and similarly
    for kaonic deuterium~\cite{Meissner:2006gx})
    \begin{equation}
    \begin{aligned}[b]
     \Delta E -i\Gamma/2=-2\alpha^3\mu^2_ca_{K^-p}
           \left(1-2a_{K^-p}\alpha\mu_c(\ln \alpha -1)\right)\,,
    \label{eq:Deser-type}
    \end{aligned}
    \end{equation}
    where $\alpha \simeq 1/137$ is the fine-structure constant, and $\mu_c$ is the reduced mass of the $K^-p$ system. For the discussion of higher-order corrections see Refs.~\cite{Cieply:2007nv,Shevchenko:2021swf}. We note that, recently, a new measurement of the kaonic deuteron system has been performed~\cite{Sgaramella:2024qdh} as well, which, however, is not part of this work. We wish to note that while older values ($\gamma, R_c,R_n$) offer very little constraint on the scattering amplitude, the SIDDHARTA results, indeed, do and should, thus, always be taken into account. A dedicated discussion can be found in Ref.~\cite{Mai:2020ltx}. 
    \item {\bf AMADEUS:} One of the latest experimental data points was taken by the AMADEUS collaboration~\cite{Piscicchia:2018rez}. Through the analysis of $K^-$ absorption processes on $^4$He the modulus of $|f_{0+}^{\pi^-\Lambda \to K^-n}(\sqrt{s}=1.4\GeV)|$ was extracted to high precision as $0.334\pm0.018\,\fm$. This is given in terms of Höhler's partial waves $f_{\ell\pm}$ as 
    \begin{align}
         f_{0+}^{\pi^-\Lambda \to K^-n}=\frac{1}{\sqrt{2}}\left(
         -f_{0+}^{\pi^0\Lambda \to K^-p}
         +f_{0+}^{\pi^0\Lambda \to \bar K^0n}\right)~.
    \end{align}
\end{itemize}

\section{Effective Field Theory and unitarized chiral perturbation approach}
\label{sec:UCHPT}

Lattice QCD provides a systematic way to access QCD Green's functions in the non-perturbative regime. In the intermediate steps, however, the methodology introduces by construction several approximations to the real-world (for a pedagogical introduction see Ref.~\cite{Gupta:1997nd}). Specifically, the calculation is performed with quarks occupying only intersections of typically cubic, and finite-volume lattice in Euclidean space-time (imaginary time). Additionally, a scale needs to be set defining the obtained quantities in physical units, since the computer algorithm does not have a notion of units. Finally, on a more practical side, quarks are often heavier than the physical ones for two reasons: 1) the physical pion appears in nature to be very light, that means very large lattice volumes are required to fit it in; 2) systems with heavier than physical pions have inelastic thresholds pushed to higher energies which effectively extends the range of applicability of the existing 2- and 3-body quantization conditions~\cite{Luscher:1990ux, Hansen:2014eka, Mai:2017bge}. 

For the present case, scale-setting, continuum extrapolation can be assumed to be  addressed in the provided finite-volume energy eigenvalues~\cite{BaryonScatteringBaSc:2023ori, BaryonScatteringBaSc:2023zvt}. A procedure for the finite-volume mapping including the related breakdown of rotational symmetry is discussed in~\cref{sec:quantization condition}. What remains is to establish a connection between the heavier-than physical pion mass results from the lattice with the experimental ones, often referred to as  {\it chiral extrapolation}. As the name suggests, the key point here is to use chiral symmetry to provide guidance on how hadron-hadron interactions behave with the changing pion mass. Specifically, we use Chiral Perturbation Theory (CHPT) extended to the meson-baryon sector~\cite{Gasser:1987rb, Bernard:1992qa, Tang:1996ca, Becher:1999he, Ellis:1997kc} to define a three-flavor meson-baryon interaction kernel at leading and next-to leading order. However, since the energy-regime of interest is large, non-perturbative effects are unavoidable. For the explicit calculation and breakdown of the convergence of the perturbative expansion see, e.g., Ref.~\cite{Mai:2009ce}. Therefore, the interaction kernel is iterated through a unitarization approach described in the following~\cref{subsec:unitarization}. This means that the extrapolation of the model-independent Lattice QCD results to the physical point (physical quark masses) is not accomplished through model-independent CHPT but a unitarized, somewhat model-dependent approach. This model dependence of the obtained results is one of the key questions we wish to discuss in this paper.

There is a plethora of approaches aiming to extend the range of applicability of the chiral series to the energy region of the $\Lambda(1405)$. An in-depth description of those including their differences and similarities is provided in the review~\cite{Mai:2020ltx}. To make the model dependence discussion more transparent, we rely here on one class of models described there, one which unites the simplicity of an algebraic formulation (vs., e.g., more sophisticated diagrammatic 4-dimensional integral equations of Ref.~\cite{Bruns:2010sv,Mai:2012dt}) with the phenomenological flexibility resulting in a wide range of applications~\cite{Kaiser:1996js, Oller:2000fj, Lutz:2001yb, Borasoy:2005ie, Ikeda:2012au}.

\subsection{Chiral Lagrangian and interaction kernel}

The general form of the chiral meson-baryon Lagrangian is written as an infinite series 
\begin{equation}
    \Lagr_{\phi B}=\Lagr^{(1)}_{\phi B}+\Lagr^{(2)}_{\phi B}+\Lagr^{(3)}_{\phi B}+\ldots\,
\end{equation}
of infinitely many terms ordered in Lagrangians with a fixed chiral order denoted above by the superscript. For the specific case of meson-baryon scattering, the leading-order Lagrangian contains three independent structures
\begin{align}
    \Lagr^{(1)}_{\phi B}&=\langle \bar{B} (i\gamma_\mu D^\mu-m_0)B\rangle
    +\frac{D}{2}\langle \bar{B}\gamma_\mu \gamma_5\{u^\mu,B\} \rangle 
    +\frac{F}{2}\langle \bar{B}\gamma_\mu \gamma_5[u^\mu,B] \rangle \,,
    \label{eq:LO-Lagrangian}
\end{align}
whereas the next-to-leading-order Lagrangian~\cite{Krause:1990xc} in its minimal form contains 14 independent structures~\cite{Frink:2005ru}
\begin{equation}
\begin{aligned}
    \Lagr^{(2)}_{\phi B}=
    &
    b_0 \langle\bar B B\rangle \langle\chi_+\rangle
    +b_{D} \langle\bar B \left\{\chi_+,B\right\}\rangle
    +b_{F} \langle\bar B \left[\chi_+,B\right]\rangle
    \\
    &\quad
    +b_{1} \langle\bar B  \left[u_\mu,\left[u^\mu,B\right]\right]\rangle
    +b_{2} \langle\bar B  \left[u_\mu,\left\{u^\mu,B\right\}\right]\rangle
    +b_3 \langle\bar B \left\{ u_\mu,\left\{ u^\mu,B\right\}\right\}\rangle
    +b_4 \langle\bar B  B\rangle \langle u_\mu u^\mu\rangle
    \\
    &\quad\quad
    +i\Big(
        b_{5}  \langle\bar B \sigma^{\mu\nu}\left[\left[u_\mu,u_\nu\right], B\right]\rangle
        +b_{6}  \langle\bar B \sigma^{\mu\nu}\left\{\left[u_\mu,u_\nu\right], B\right\}\rangle
        +b_7 \langle\bar B\sigma^{\mu\nu}u_\mu\rangle  \langle u_\nu B\rangle
    \Big)
    \\
    &\quad\quad\quad
    +\frac{i\,b_{8}}{2m_0}\Big( \langle\bar B \gamma^\mu\left[u_\mu,\left[u_\nu,\left[D^\nu, B\right]\right]\right]\rangle
    +\langle\bar B \gamma^\mu \big[D_\nu,\left[u^\nu,\left[u_\mu,B\right]\right]\big]\rangle\Big)
    \\
    &\quad\quad\quad\quad
    +\frac{i\,b_{9}}{2m_0}\Big( \langle\bar B \gamma^\mu\left[u_\mu,\left\{u_\nu,\left[D^\nu, B\right]\right\}\right]\rangle
    +\langle\bar B \gamma^\mu \big[D_\nu,\left\{u^\nu,\left[u_\mu,B\right]\right\}\big]\rangle\Big) 
    \\
    &\quad\quad\quad\quad\quad
    +\frac{i\,b_{10}}{2m_0}\Big( \langle\bar B \gamma^\mu\left\{ u_\mu,\left\{ u_\nu,\left[D^\nu,B\right]\right\}\right\}\rangle
    +\langle\bar B\gamma^\mu\left[D_\nu,\left\{ u^\nu,\left\{ u_\mu,B\right\}\right\}\right]\rangle\Big) \\
    &\quad\quad\quad\quad\quad\quad
    +\frac{i\,b_{11}}{2m_0}\Big( 2\langle\bar B \gamma^\mu \left[D_\nu,B\right]\rangle \langle u_\mu u^\nu\rangle
    +\langle\bar B \gamma^\mu B\rangle \langle\left[D_\nu,u_\mu\right]u^\nu + u_\mu \left[D_\nu,u^\nu\right]\rangle   \Big)\,,
\end{aligned}
\label{eq:NLO-Lagrangian}
\end{equation}
where $\langle\ldots\rangle$ denotes the trace in flavor space, $D_\mu B :=\partial_\mu B +\frac{1}{2}[[u^\dagger,\partial_\mu u],B]$, $m_0$ is the baryon octet mass in the chiral limit, and $D$, $F$ are the axial coupling constants. The next-to-leading order low-energy constants (LECs) $b_i$ split into the so-called symmetry breakers $b_{0,D,F}$ parameterizing the explicit chiral symmetry breaking through the non-vanishing quark masses and $\{b_i|i= 1,\ldots,11\}$ being referred to as dynamical LECs. All external currents except the scalar one are set to zero and 
\begin{align}
U =\exp\Bigl(i\frac{\phi}{F_0}\Bigr)\,,
\quad
u^2:=U\,,
\quad
u^\mu:=iu^{\dagger}\partial^\mu u +iu\partial^\mu u^{\dagger}\,,
\quad
\chi_\pm:=u^{\dagger}\chi u^{\dagger}\pm u\chi^{\dagger}u\,,
\quad
\chi:=2B_0\,\textrm{diag}(m_u,m_d,m_s)\,,
\end{align}
where $F_0$ and $B_0$ denote the pion decay constant and the constant related to the quark condensate in the chiral limit, respectively. The ground state octet mesons (Goldstone bosons of the theory) and baryons are included through
\begin{equation}
    \phi=\sqrt{2}
        \begin{pmatrix}
        \frac{\pi^0}{\sqrt{2}}+\frac{\eta}{\sqrt{6}} \!\!&\!\! \pi^+ \!\!&\!\! K^+ \\
        \pi^- \!\!&\!\! -\frac{\pi^0}{\sqrt{2}}+\frac{\eta}{\sqrt{6}} \!\!&\!\! K^0 \\
        K^- \!\!&\!\! \bar{K}^0 \!\!&\!\! -\frac{2}{\sqrt{6}}\eta
        \end{pmatrix}
    \,,
    \quad
    B=\begin{pmatrix}
        \frac{\Sigma^0}{\sqrt{2}}+\frac{\Lambda}{\sqrt{6}} & \Sigma^+ & p \\
        \Sigma^- & -\frac{\Sigma^0}{\sqrt{2}}+\frac{\Lambda}{\sqrt{6}}& n \\
        \Xi^-& {\Xi}^0 & -\frac{2}{\sqrt{6}}\Lambda
    \end{pmatrix}
    \,.
\label{eqn:fields}
\end{equation}

The above general Lagrangian defines all the Feynman diagrams as shown in Ref.~\cite{Mai:2009ce} calculating meson-baryon scattering in three-flavor CHPT. The completeness of momentum structures plays a crucial role, for instance, when constructing unitary gauge-invariant models for meson-photoproduction~\cite{Borasoy:2007ku, Ruic:2011wf, Mai:2012wy}, and warrants full accounting of all chiral logarithms. This may indeed become relevant when approaching the chiral limit as discussed and shown in Ref.~\cite{Bruns:2017gix}. It was shown, however, in Ref.~\cite{Mai:2012dt} that for antikaon-nucleon scattering in the physical region not too far from the two-body thresholds such effects are subdominant. An approach neglecting such effects was studied in Ref.~\cite{Mai:2014xna} and later including coupled-channel $S$- and $P$-waves in Ref.~\cite{Sadasivan:2022srs}.

Instead of this path, we proceed here with a closely related but computationally less expensive approach. It relies on calculating the chiral potential $V$ and iterating it to restore two-body unitarity. We use the normalization of Refs.~\cite{Borasoy:2005ie,Ikeda:2012au} and study three types of meson-baryon potentials
\begin{align}
    \text{M1}:~~~&V_{\alpha\beta}(s)=V_{\alpha\beta}^{\mathrm{WT}}(s)\,,\\
    \text{M2}:~~~&V_{\alpha\beta}(s)=V_{\alpha\beta}^{\mathrm{WT}}(s)+V_{\alpha\beta}^{\mathrm{BORN}s}+V_{\alpha\beta}^{\mathrm{BORN}u}(s)\,,\\
    \text{M3}:~~~&V_{\alpha\beta}(s)=V_{\alpha\beta}^{\mathrm{WT}}(s) +V_{\alpha\beta}^{\mathrm{BORN}s}+V_{\alpha\beta}^{\mathrm{BORN}u}(s)+V_{\alpha\beta}^{\rm{NLO}}(s)\,,
\label{eq:VNLO}
\end{align}
where $\alpha/\beta$ collect the  indices of the in/outgoing  meson-baryon states $\mathcal{S}$. Here, WT denotes the time-honored Weinberg-Tomozawa term and BORNx (x=$u$, $s$) the $s$- and $u$-channel Born terms, respectively.
Specifically, for the total strangeness $S=-1$, the relevant channels are listed in \eqref{eq:ordering-channel}.
All potentials except $V_{\alpha\beta}^{\rm{NLO}}(s)$ are obtained from the LO chiral Lagrangian~\cref{eq:LO-Lagrangian}, while the latter is deduced from the relevant part of the NLO Lagrangian
\begin{align}
    \nonumber
    \Lagr^{(2)}_{\phi B}=
    &
    b_0 \langle\bar B B\rangle \langle\chi_+\rangle
    +b_{D} \langle\bar B \left\{\chi_+,B\right\}\rangle
    +b_{F} \langle\bar B \left[\chi_+,B\right]\rangle
    \\
    &
    +d_1 \langle\bar B \left\{u_\mu,\left[u^\mu,B\right]\right\}\rangle
    +d_2 \langle\bar B \left[u_\mu,\left\{u^\mu,B\right\}\right]\rangle
    +d_3 \langle\bar B u_\mu\rangle \langle Bu^\mu\rangle
    +d_4 \langle\bar B B\rangle \langle u_\mu u^\mu\rangle
    \,.
    \label{eq:NLOsmall-Lagrangian}
\end{align}
It is notable that only the potentials in M2 and M3 include all terms at the given chiral order, namely leading and next-to-leading, respectively. Still, the M1 potential, the so-called Weinberg-Tomozawa term already captures the major aspects of the dynamics of the system correctly, but is at odds with the most recent threshold data~\cite{Bazzi:2011zj} even after the unitarization procedure, see, e.g., Ref.~\cite{Ikeda:2012au} or to foreshadow the results discussed later, see M1SxP fits in \cref{fig:fit-summary}.

Explicitly\footnote{These formulas are quite standard in the literature. However, a certain amount of typos and convention inconsistencies also became standard. To avoid this, explicit formulas are provided here.}, the above required potentials read as 
\begin{equation}
\begin{aligned}
    V_{(ai)(bj)}^{\rm WT}(s) &= 
        -\frac{\mathcal{N}_a\mathcal{N}_b}{8F_iF_j}C_{(ai)(bj)}^{\mathrm{WT}}  (2\sqrt{s}-m_a-m_b)\,,
    \\
    V_{(ai)(bj)}^{\mathrm{BORNs}}(s)&=\frac{\mathcal{N}_a \mathcal{N}_b}{12 F_iF_j} 
        \sum_{c=1}^{8} C_{(bj)(c)}^{[DF]} C_{(ai)(c)}^{[DF]} 
        \frac{1}{\sqrt{s + m_c}} \left( s - \sqrt{s} (m_a + m_b) + m_a m_b \right)\,,
    \\
    V_{(ai)(bj)}^{\mathrm{BORNu}}(s)&=
        -\frac{\mathcal{N}_a \mathcal{N}_b}{12 F_iF_j} 
        \sum_{c=1}^{8} C_{(ic)(b)}^{[DF]} C_{(jc)(a)}^{[DF]}
        \left[
        \sqrt{s} + m_c-
        \frac{(m_a + m_c)(m_b + m_c)}{2 \mathcal{N}_a^2 \mathcal{N}_b^2} \left(\sqrt{s}+m_a+m_b-m_c\right)
        \right.\\
    &\qquad\qquad\qquad
            +\left(
            \sqrt{s} + m_c-m_a-m_b
            - \frac{s + m_c^2 - M_i^2 - M_j^2 - 2 E_a E_b}{2 \mathcal{N}_a^2 \mathcal{N}_b^2}(\sqrt{s}+m_a+m_b-m_c)
            \right)\times\\
    &\qquad\qquad\qquad\qquad\qquad\qquad
            \left.
            \times
            \frac{(m_a + m_c)(m_b + m_c)}{4 p_{bj} p_{ai}}
            \ln 
            \left(
            \frac{s + m_c^2 - M_i^2 - M_j^2 - 2 E_a E_b - 2 p_{bj} p_{ai}}
                 {s + m_c^2 - M_i^2 - M_j^2 - 2 E_a E_b + 2 p_{bj} p_{ai}}
            \right)
        \right]\,,
    \\
    V_{(ai)(bj)}^{\rm NLO}(s) &= 
        \frac{\mathcal{N}_i\mathcal{N}_j}{F_iF_j}
        \left(
        C_{(ai)(bj)}^{[b_0b_Db_F]}-2C_{ij}^{[d_1d_2d_3d_4]}
        \left(
        E_iE_j+\frac{p_{bj}^2 p_{ai}^2}{3\mathcal{N}_a^2\mathcal{N}_b^2}
        \right)
        \right)\,.
\end{aligned}
\label{eq:Chiral-potentials}
\end{equation}
Here, we have explicitly written out the baryon/meson octet indices $\{a,b,c\}/\{i,j\}$ of the corresponding channel. Further, $s$ is the total energy squared, $\mathcal{N}_a=\sqrt{m_a+E_{a}}$; $E_{a}=\sqrt{m_a^2+p^2_{ai}}$; $E_{i}=\sqrt{M_i^2+p^2_{ai}}$; $p_{ai}=\sqrt{(s-(M_i+m_a)^2)(s-(M_i-m_a)^2)}/(2\sqrt{s})$. 
Meson and baryon masses are denoted by $M$ and $m$, respectively. The coefficient matrices $C^{\rm WT}$, $C^{[DF]}$, $C^{[b_0b_Db_F]}$, $C^{[d_1d_2d_3d_4]}$ are obtained from \cref{eq:NLOsmall-Lagrangian} but can also be obtained from the appendix of Ref.~\cite{Borasoy:2005ie}. The first matrix contains the LECs $F_{\pi,K,\eta}$, which appear explicitly in the denominator,  while the latter three matrices  include additional LECs, with the pertinent parameters put into in the superscript square-brackets, for convenience. Note that the leading order LECs $D,F$  are fixed and only the $b_i,d_i$ are to be determined by the fits. Additionally, $C^{[b_0b_Db_F]}$ depends explicitly on the quark masses, given in terms of the meson masses.

\subsection{Unitarization procedure and connection to observables}
\label{subsec:unitarization}

With the interaction kernel at hand a non-perturbative amplitude can be constructed. This typically involves some sort of resummation of an infinite set of diagrams. Usually guided by the S-matrix unitarity, typical methods  are the full four-dimensional Bethe-Salpeter, three-dimensional reduced Lippmann-Schwinger equations, $N/D$ or other dispersive tools. Using the above defined on-shell potentials projected to the $S$-wave~\cref{eq:Chiral-potentials}, the Bethe-Salpeter integral equation indeed reduces to an algebraic matrix equation (with respect to the channel space $\cal{S}$) 
\begin{equation}
\begin{aligned}
    T(s)&=-V(s)+T(s)G(s)V(s)=-V(s)-V(s)G(s)V(s)-V(s)G(s)V(s)G(s)V(s)-...\\
        &=-V(s)\frac{\mathds{1}}{1-V(s)G(s)}
        \,.
\end{aligned}
\label{eq:T-matrix}
\end{equation}
The infinite series on the right-hand side of the first line is written out to show the connection to an infinite set of loop diagrams. Clearly this set is still incomplete compared to all possible diagrams in CHPT to all orders. This is one of the sources of the model dependence acquired in this step. For further details, see the dedicated review~\cite{Mai:2020ltx}. 

The meson-baryon one-loop (channel $\alpha$) function is defined as
\begin{align}
    G_\alpha(s) = \int 
        \frac{d^4 l}{(2\pi)^4} \frac{i}{\left(l^2 - M_\alpha^2 + i\epsilon\right)\left((P -l)^2 - m_\alpha^2 + i\epsilon\right)}\,,
\label{eq:loop-function-integral}
\end{align}
which has an imaginary part $\Im\,G_\alpha(s)=-p_\alpha/(8\pi\sqrt{s})$. Therefore, T-matrix, indeed, automatically fulfills the partial-wave unitarity $\operatorname{Disc}T(s)=ip_\alpha/(4\pi\sqrt{s})|T(s)|^2$ for energy between $s=(m_\alpha+M_\alpha)^2$ and the next higher two-body threshold. For an introductory discussion of the S-matrix theory for hadron spectroscopy see the review~\cite{Mai:2025wjb}. This also allows one to relate the T-matrix to the K-matrix form used in \cref{sec:quantization condition}. In terms of the Höhler partial-wave~\cite{Chew:1957zz,Hohler:1984ux} $f_{0+}$ relevant for this study (c.f. \cref{eq:f0plus_kmatrix})
\begin{equation}
\begin{aligned}
    f_{0+}(s)=\frac{1}{8\pi\sqrt{s}}T(s)\,,
    \qquad
    f_{0+}(s)=\frac{1}{\tilde K_E^{-1}(s)-ip}\,. 
\end{aligned}
\label{eq:f0plus}
\end{equation}
Here, all quantities are matrices with respect to the channel space, i.e., $\tilde K_E\in \mathds{R}^{\cal{S} \times \cal{S}}$ and $p:=\operatorname{Diag}\{p_\alpha|\alpha\in\cal{S}\}$. The above nomenclature allows for straightforward relations to the observables. For example, the scattering length is simply given by $a=f_{0+}((m+M)^2)$ or total cross sections as shown in \cref{eq:cross_section}.

\subsection{Regularization schemes}

The four-dimensional meson-baryon loop integral $G$ is log-divergent and can be tamed in various ways using, e.g., a momentum cutoff or dimensional regularization. In the latter, and most frequently utilized form~\cite{Borasoy:2005ie, Lutz:2001yb,Ikeda:2012au, Oller:2000fj, Oller:2019opk,Feijoo:2021zau}, the loop function reads ($\alpha\in\cal{S}$)
\begin{align}
\label{eq:G_loop}
    G_\alpha(\sqrt{s}) =
    a_\alpha+
    \frac{1}{32\pi^2}\left(
        \log\left(\frac{m_\alpha^2}{\mu^2}\right)
        +\log\left(\frac{M_\alpha^2}{\mu^2}\right)
        -\frac{m_\alpha^2-M_\alpha^2}{s}\log\left(\frac{M_\alpha^2}{m_\alpha^2}\right)
        -2
        -\frac{8p_\alpha}{\sqrt{s}}\operatorname{arctanh}\left(\frac{2\sqrt{s} p_\alpha}{(m_\alpha+M_\alpha)^2-s}\right)
        \right)\,.
\end{align}
For the analysis of the experimental data, all masses are taken to their physical values while the regularization scale dependence is moved into the subtraction constants $a_\alpha$ channel-by-channel for a fixed scale $\mu$. Note that is equivalent of promoting the regularization scale $\mu$ to channel-by-channel $\mu_\alpha$~\cite{Mai:2012dt}. Since isospin breaking effects are far smaller than the available experimental precision, no distinction is made between subtraction constants in the same particle type leaving one with six free subtraction constants $\{a_{\bar KN},a_{\pi\Lambda}, a_{\pi\Sigma},a_{\eta\Lambda},a_{\eta\Sigma},a_{K\Xi}\}$ which are treated commonly as additional free parameters of the theory. Note that at the lattice point, isospin symmetry is exact and input is available for $I=0$ only. There, only four subtraction constants matter corresponding to the channels in \cref{eq:ordering-channel-LAT}.

Besides the choice of the resummation procedure and the choice of the interaction potential, there is yet another issue where a choice has to be made, namely the regularization procedure, which is also leading to systematic uncertainties in the UCHPT approach. In view of the recent experimental and more importantly Lattice QCD progress, the main phenomenological drawback of this is that by losing connection to the usual perturbative chiral expansion, the LECs cannot be compared easily between different approaches or to the perturbatively determined values. More importantly, chiral extrapolations of the amplitudes from unphysical (Lattice QCD) to physical quark masses will differ from one approach to another which was already observed in Ref.~\cite{Guo:2023wes}. Specifically, in contrast to the LECs, it is not clear how the subtraction constants depend on the quark masses since they absorb higher order terms.

\begin{figure}
    \centering
    \includegraphics[width=0.8\linewidth]{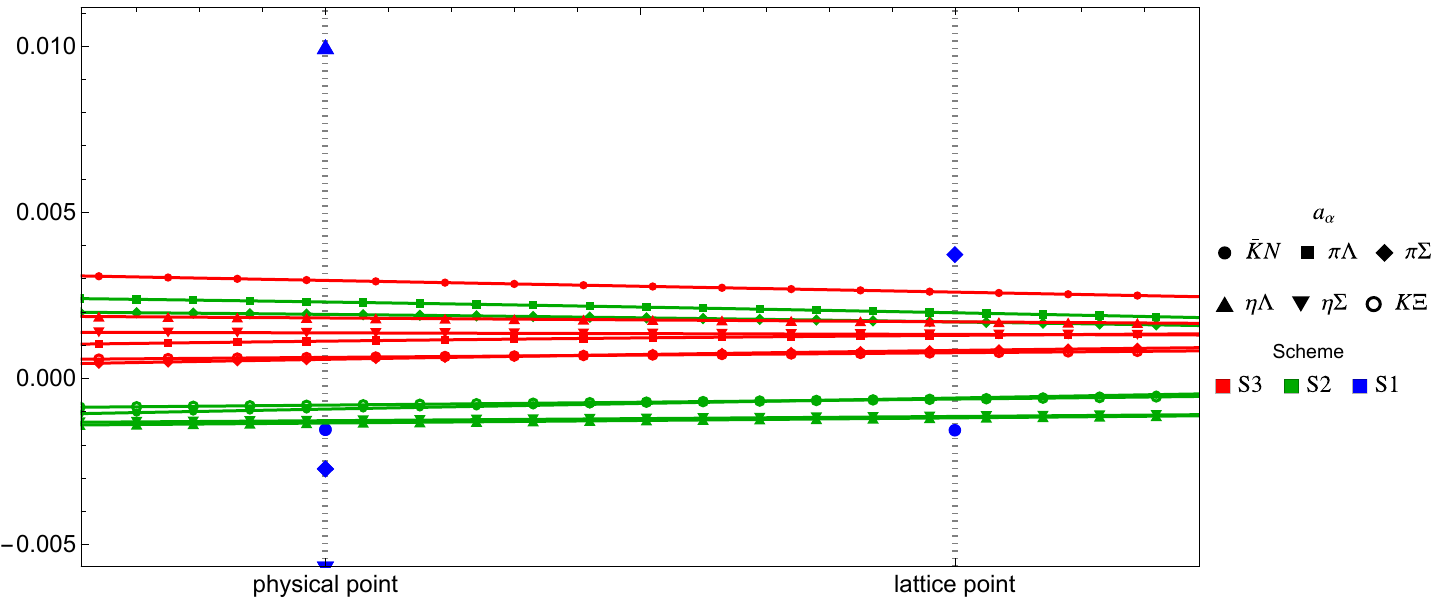}
    \caption{Comparison of subtraction constants obtained from global fits using different regularization schemes (S1, S2, S3) at different quark masses. Scheme S1 only provides values at the given quark-masses obtained from global fits $F_{17}$ (blue). S2 and S3-type subtraction constants are per construction pion mass dependent.}
    \label{fig:subtraction-constants}
\end{figure}
So far, removing the model dependence from these approaches entirely seems rather unrealistic. Thus, one cannot escape the imperative of quantifying how these above choices reflect themselves on observables or on obtained predictions, such as resonance pole positions or chiral extrapolations of the amplitudes. Therefore, in addition to the various choices of the interaction kernel we employ three types of regularization schemes widely used in the literature. 
\begin{itemize}
\item[S1]
    No assumption about the quark mass dependence of the subtraction constants. For each available quark-mass-setup and for fixed $\mu=1\GeV$, we fit a new set $\{ a_\alpha\}$ to the available data. Through this no extrapolation is possible, but the maximal possible freedom of the model is achieved. Note that seven LECs are still quark mass independent.
\item[S2]
    As proposed in Refs.~\cite{Lutz:2001yb, Hyodo:2008xr} one fixes the subtraction constants by demanding  
    \begin{align}
        T(\sqrt{s}=m)=V(\sqrt{s}=m)
        ~~~
        \Longleftrightarrow
        ~~~
        G(\sqrt{s}=m, a(\mu))=0\,.
    \end{align} 
    This was recently employed in Ref.~\cite{Guo:2023wes} and boils down to 
    \begin{align}
        a_\alpha=&
        -\frac{1}{32\pi^2}\left(
        \log\left(\frac{m_\alpha^2M_\alpha^2}{\mu^4}\right)
        -\frac{m_\alpha^2-M_\alpha^2}{s}\log\left(\frac{M_\alpha^2}{m_\alpha^2}\right)
            -2
            -\frac{8p_\alpha}{\sqrt{s}}
            \operatorname{arctanh}\left(\frac{2\sqrt{s} p_\alpha}{(m_\alpha+M_\alpha)^2-s}\right)
        \right)\Bigg|_{\sqrt{s}=m_p,\mu=m_\alpha}\,.
    \end{align} 
    Technically, this scheme is advantageous by reducing the number of free parameters, while connecting at the same time lattice point (unphysical quark masses) with the physical ones.
    \item[S3] Another scheme was proposed in Ref.~\cite{Oller:2019opk}. Similarly to S2 it sets a constraint on the loop function but at a different matching point. In particular, it is imposed that at the two-body threshold the  dimensionally and cutoff (hard cutoff $\Lambda$) regularized loop functions are identical. After matching the different expressions of the loop functions this boils down to 
    \begin{align}
        &a_\alpha=\frac{1}{16\pi^2}\left(
            1
            -\frac{2}{M_\alpha+m_\alpha}
            \left(
            m_\alpha\log\left(1+\sqrt{1+\frac{m_\alpha^2}{\Lambda^2}}\right)+
            M_\alpha\log\left(1+\sqrt{1+\frac{M_\alpha^2}{\Lambda^2}}\right)
            \right)
            +\log\left(\frac{\mu^2}{\Lambda^2}\right)
            \right)\,.
    \end{align}
    We fix again $\mu=1\GeV$, but fit a common channel-independent parameter $\Lambda$ for either lattice, experimental or both points simultaneously.    
\end{itemize}
In summary, the regularization scheme S1 makes no assumptions about the quark mass dependence of the subtraction constants but defines six free parameters ($a$'s) per quark-mass setup, and lacks predictive power outside of the fitted quark-mass regions. Schemes S2 (no free parameters) and S3 (one free parameter) on the other side make a certain assumption about the form of the loop integrals resulting in a higher predictive power also at other than fitted quark mass scenarios. We note that S1 scheme is similar/equivalent to fitting regularization scales~\cite{Cieply:2011nq,Bruns:2021krp} that can be matched to those used in dimensional regularization, see, e.g., Eq.~(2.13) in \cite{Bruns:2022sio}.

We will confront each scheme with the lattice and experimental data below. The resulting subtraction constants of the best fits to combined lattice and experimental data are collected as a function of the hadron masses extrapolated linearly ($m=x\,m_{\rm phys}+(1-x)\,m_{\rm latt}$ for any generic hadron mass $m$ and $x\in[0,1]$) between the physical and lattice point in \cref{fig:subtraction-constants}. We observe that S3 and S2 have similar order of magnitude of the determined constants, while most of the fitted constants (scheme S1) are, indeed, substantially larger. Among different variations of the scheme S1 fits (different starting points of the fits) we observe some clustering in a few cases but mostly the obtained values are very uncertain. This indicates that either S1 has too many/redundant parameters or the data is not allowing to fix them uniquely due to possible inconsistencies. For further details see \cref{subsec:fit-discussion}.

\subsection{Synthesis and evaluation of the UCHPT framework}
\label{subsec:synthesis}

The considered approach is based on CHPT in such a way that it matches CHPT amplitudes at the leading and next-to-leading order, when projecting to the $S$-wave. It captures part of the non-perturbative dynamics of the full CHPT to all orders but depends on the way how the truncation of higher orders is made. This results in a certain model dependence which is quantified in this work by varying
\begin{itemize}
    \item Truncation order of the interaction kernel: M1$[F_\pi,F_K,F_\eta]$, M2$[F_\pi,F_K,F_\eta,D,F]$, and\\ M3$[F_\pi,F_K,F_\eta,b_0,b_d,b_f,d_1,d_2,d_3,d_4]$. The free parameters are given in the square brackets, these are the quark-mass independent LECs. Note that the leading order LECs are kept fixed as $D=0.8$, $F=0.46$ and $F_{\pi,K,\eta}^{\rm phys}=\{92.4,110.0,118.8\}$~MeV and $F_{\pi,K,\eta}^{\rm latt}=\{93.2,108.2,121.1\}$~MeV.
    \item Regularization scheme: S1$[a_1,a_2,a_3,a_4,a_5,a_6]$, S2$[-]$, S3$[\Lambda]$. The free parameters are listed in the square brackets. Note that $\Lambda$ is quark-mass independent while  the $a$'s need to be fitted per quark-mass setup. At the unphysical point we have only four $[a_1,a_3,a_4,a_6]$ because only the isoscalar channel is available.
    \item Data: We will study if new lattice results~\cite{BaryonScatteringBaSc:2023ori, BaryonScatteringBaSc:2023zvt} (a) allow to fix the parameters of the models by themselves making reliable predictions for the physical point (b) are consistent with experimental data. In that, various combinations of Lattice QCD and/or experimental results will be used to fit the free parameters described before. 
\end{itemize}

Finally, we wish to discuss another yet mostly ignored limitation of the this and all current UCHPT approaches with respect to the intermediate three-body states. Specifically, in meson-baryon systems processes such as $\pi\Sigma\to\Lambda$ are allowed. Thus, an initial meson-baryon system can couple to meson-meson-baryon state in the intermediate step, e.g., $\pi\Sigma\to\pi\pi\Lambda \to\pi\Sigma$ etc.\,. Technically, this leads to new singularity structures~\cite{Mai:2025wjb,Mai:2020ltx} violating the  simple unitarity condition spelled out before \cref{eq:loop-function-integral}. 

\begin{figure}[t]
    \centering
    \begin{minipage}[t]{0.55\linewidth}
        \vspace{0pt}
        \includegraphics[height=6cm]{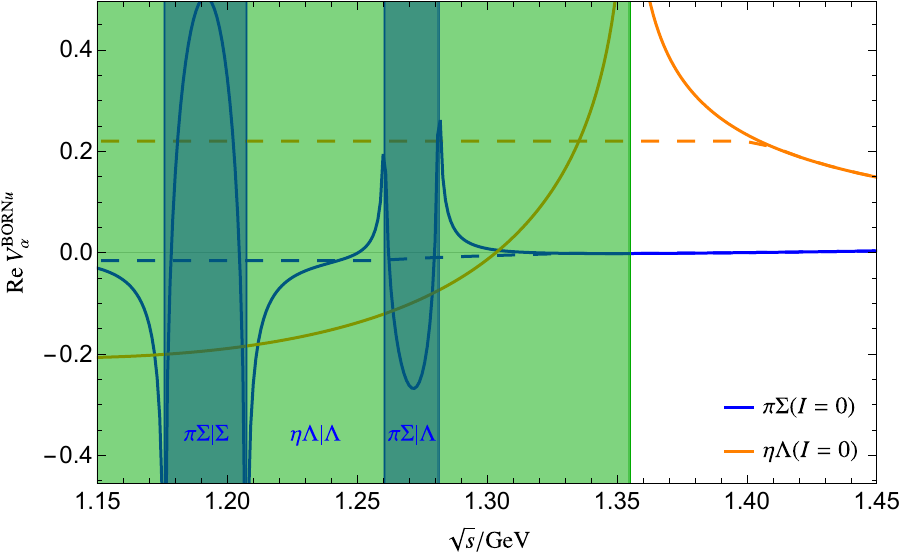}
    \end{minipage}%
    \hfill
    \begin{minipage}[t]{0.39\linewidth}
        \vspace{-12pt}
        \includegraphics[height=5.8cm]{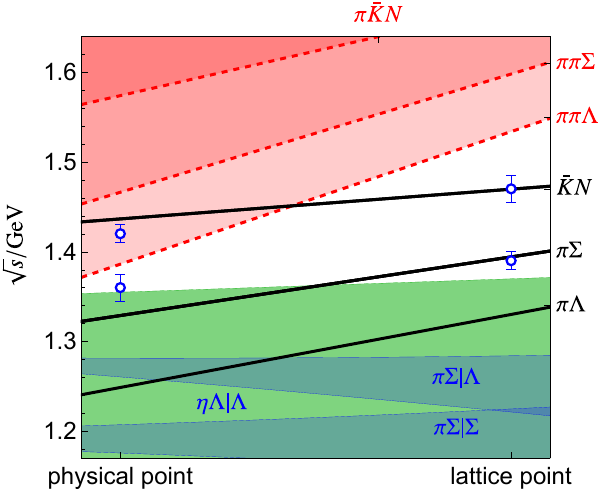}
    \end{minipage}
    \caption{Three-body related singularities for the physical and unphysical quark-mass setups. 
    Left: Singularities of the $u$-channel Born diagram~\cref{eq:Chiral-potentials} projected to $I=0$ for the initial/final states as specified in the legend for physical quark masses. Long-dashed lines represent the potential implemented in the UCHPT amplitude mitigating the appearance of $u$-channel left-hand cuts.
    Right: Relevant short left-hand cut associated with the $u$-channel exchange (green, blue areas), c.f. left figure. Black solid and red dashed lines denote the position of the right-hand cut branching points with respect to two- and three-body states, respectively. Energy region with no allowed three-body onshell states is the remaining white area. Blue dots with error bars denote the averaged result from the global analysis discussed in \cref{sec:results}.
    }
    \label{fig:3body-cuts}
\end{figure}

There are several types of new singularities. Most prominently, there is the \emph{right-hand cut}, occurring when the total energy is sufficient or higher than the sum of the masses of all three particles in question, e.g., $s\ge (2M_\pi+m_\Lambda)^2$. The positions of the branch points of these cuts are depicted as red-dashed lines in the right panel of \cref{fig:3body-cuts} including physical and unphysical (the one employed by the BaSc collaboration~\cite{BaryonScatteringBaSc:2023ori, BaryonScatteringBaSc:2023zvt}, $M_\pi\approx 200\MeV$) quark-mass scenarios. One observes clearly that the $\pi\pi\Lambda$ cut indeed is far above the estimated pole positions of $\Lambda(1405)$ when unphysical quark masses are employed. However, extrapolating down to the physical point this cut starts at lower energies and ultimately is just between the  estimated pole positions of the $\Lambda(1380)$ and the $\Lambda(1405)$. Thus, one cannot avoid the conclusion that the position of the latter state determined in the literature must carry a systematic, yet unknown uncertainty related to the neglected three-body states. Besides such phenomenological implications, this will play also a crucial role in future physical point Lattice QCD simulations. The development of pertinent tools, such as three-body scattering amplitudes and three-body quantization conditions has progressed strongly over the last few years~\cite{Mai:2021lwb} but will need another update in the future. Some steps in this direction are made recently for the similar case of the $\pi\pi N$ channel in relation to the Roper resonance, see Ref.~\cite{Severt:2022jtg} and references therein.

The exchange of a baryon in the $u$-channel also leads to the occurrence of the left-hand-cut or more specifically the baryon short left hand cut (sometimes called short baryon cut), for an in-depth discussion of such cuts see Ref.~\cite{Doring:2009yv}. Recently, studies of this type of singularities also became quite popular in the context of heavy meson scattering such as $DD^*$ or $BB^*$. The reason is that the left hand cut lies there (also and particular) for the unphysical pion mass scenarios on the lattice close to the two-body threshold, see, e.g.,  Refs.~\cite{Shrimal:2025ues, Vujmilovic:2024snz, Leskovec:2019ioa,Meinel:2022lzo,Collins:2024sfi,Du:2023hlu}. Thus, not accounting for the left-hand-cut leads to problems most apparent when dealing with the finite-volume Lattice QCD spectrum, see e.g., \cite{Meng:2023bmz,Dawid:2025wsn,Prelovsek:2025vbr}. For $S=-1$ the short left-hand cut occurs through, e.g., $\Lambda$ or $\Sigma$ exchange in the $\pi\Sigma\to\pi\Sigma$ transition, referred to as $\pi\Sigma|\Lambda$ and $\pi\Sigma|\Sigma$. Another example relevant for the isoscalar channel is $\eta\Lambda|\Lambda$. Indeed, such transitions are included in the UCHPT models (M2, M3) via the chiral potential $V^{{\rm BORN}u}$ from \cref{eq:Chiral-potentials}. For a general transition $\phi_1B_1\to\phi_2B_2$ including $u$-channel exchange of a baryon $B_x$, the limits of the singular region are given by
\begin{align}
    s=\frac{1}{2 m_x^2}\Bigg(&
    M_1^2 M_2^2 - M_2^2 m_2^2 + M_1^2 m_x^2 + M_2^2 m_x^2 + m_2^2 m_x^2 - m_x^4 + m_1^2 \left(-M_1^2 + m_2^2 + m_x^2\right) 
    \\
    &
    \pm\sqrt{\left(m_1^4 + \left(M_2^2 - m_x^2\right)^2 - 2 m_1^2 \left(M_2^2 + m_x^2\right)\right) \left(M_1^4 + \left(m_2^2 - m_x^2\right)^2 - 2 M_1^2 \left(m_2^2 + m_x^2\right)\right)}
    \Bigg)\,.
\end{align}
For $I=0$ transitions this is depicted in the left panel of \cref{fig:3body-cuts} by the shaded regions. There, one clearly sees that the chiral potential (the real part is plotted), indeed, has singularities in the shaded region. On the right, the same critical regions are depicted as a function of the quark mass extrapolated from the physical to the lattice point. We note that in the $\pi\Sigma$ channel this occurs at rather small energies, well below the region of interest for the $\Lambda(1380)$ and $\Lambda(1405)$. However, $\Lambda$ exchange in the $\eta\Lambda\to\eta\Lambda$ transition is only slightly below the $\Lambda(1380)$ bound state at the lattice point. Ultimately, this leads to a complex-valued K-matrix \cref{eq:f0plus} invalidating a simple application of Lüscher's quantization condition. Practically, in the current state of the art of the coupled channel UCHPT models, this is avoided by replacing the potential slightly above the critical region by a constant. The modified and used one-baryon exchange potential is depicted by the long-dashed line in the left panel of \cref{fig:3body-cuts}. So far, the coverage and precision of either experimental as well as lattice results is not sensible to these effects. However, in the long run we expect, that an extension of modern three-body formalisms like FVU~\cite{Mai:2017bge} or RFT~\cite{Hansen:2014eka} to such cases will be a better choice. These have been shown recently of being capable of dealing with strangeness channels~\cite{Alexandru:2020xqf,Dawid:2025doq,Dawid:2025zxc}, dealing with left-hand cuts~\cite{Dawid:2025wsn} extracting resonance pole positions for three-body states~\cite{Yan:2024gwp, Garofalo:2022pux, Sadasivan:2021emk, Mai:2021nul, Sadasivan:2020syi} from experimental and Lattice QCD spectra.

\section{Analysis}
\label{sec:fits}

\subsection{Pilot Study: finite-volume spectrum from UCHPT and experiment}
\label{subsec:pilotstudy}

\begin{figure}[t]
    \centering
    \includegraphics[width=0.9\linewidth]{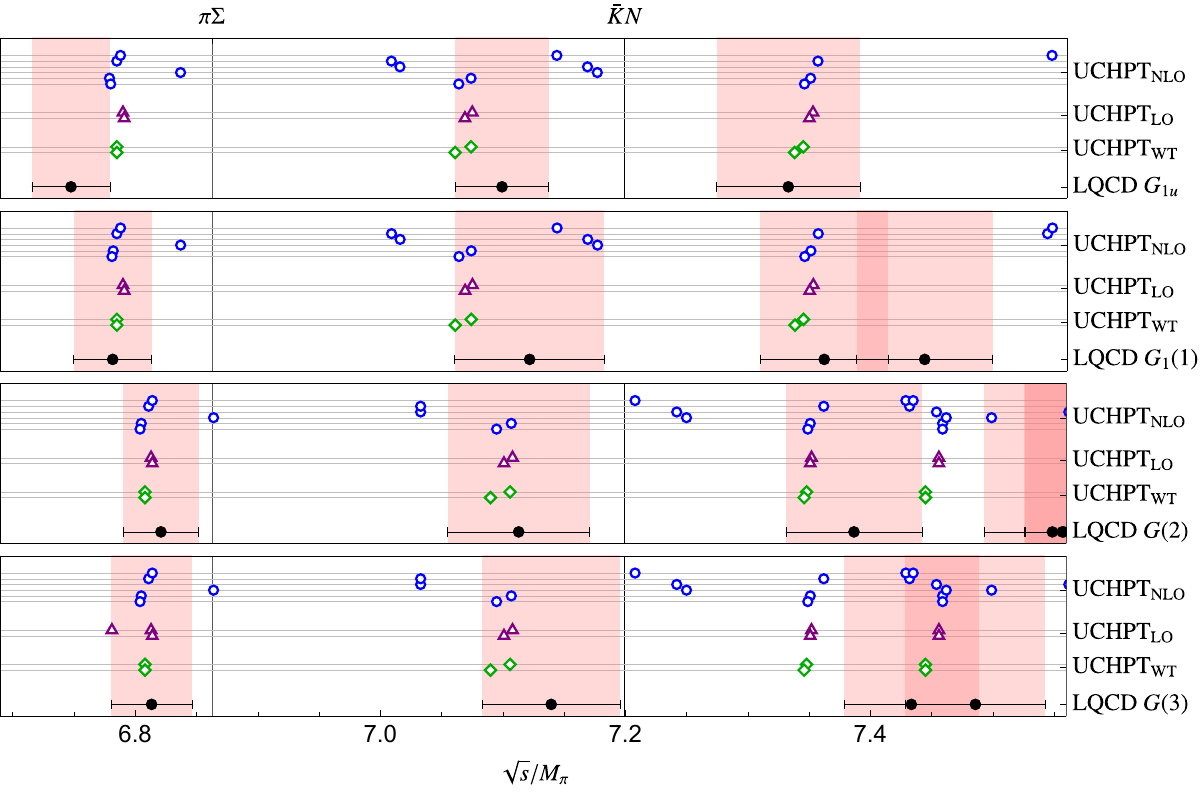}
    \caption{Fit-free prediction of the finite-volume spectrum relying on the contemporary WT (M1), LO (M2), and NLO type UCHPT models~\cite{Ikeda:2012au,Mai:2014xna,Sadasivan:2022srs}. Pink vertical bands show the Lattice QCD results~\cite{BaryonScatteringBaSc:2023ori, BaryonScatteringBaSc:2023zvt} for relevant irreps: $G_{1u}$, $G_1(1)$, $G(2)$, $G(3)$.}
    \label{fig:Prediction-fin-vol-spectrum}
\end{figure}

Before fitting the free parameters of the models, we wish to check if the finite-volume spectrum obtained from available models already matches the Lattice QCD spectrum~\cite{BaryonScatteringBaSc:2023ori, BaryonScatteringBaSc:2023zvt}. Specifically, we consider contemporary models of next-to-leading chiral order including diagrammatic or potential unitarization formalisms~\cite{Ikeda:2012au, Mai:2014xna, Sadasivan:2022srs}. These models, referred to as ${\rm UCHPT}_{\rm NLO}$ represent a fair spread (including M3 type models) of assumptions made in the derivation, sampling qualitatively possible model dependence, as discussed in \cref{sec:UCHPT}. We also include M1 (${\rm UCHPT}_{\rm WT}$) and M2 (${\rm UCHPT}_{\rm LO}$) type models available from Ref.~\cite{Ikeda:2012au}. All considered models rely on the scheme S1 with subtractions constants $a_\alpha$ assumed to be quark mass independent. All considered models describe at least the experimental data compiled in \cref{sec:Experimental_data} with similar quality ($\chi^2_{\rm dof}\approx 1$).

The finite-volume spectrum is predicted through the Lüscher formalism~\cite{Luscher:1990ux,Luscher:1991cf} implemented as discussed in \cref{sec:quantization condition}. The central values of the predicted energy eigenvalues in relevant irreps are depicted in \cref{fig:Prediction-fin-vol-spectrum} together with the Lattice QCD results. We observe that (ignoring correlations at the moment) for the most part all NLO but also WT and LO models agree with the provided spectrum. There are, however, stark exceptions to this observation like the second and third level of $G_{1u}$ where some NLO models predict different finite-volume spectrum. Further examples are the third level of the $G(3)$ irrep or the fourth lowest level of the $G(1)$ irrep which do not agree with either of the model predictions. It is noteworthy that the spread of the predictions for the NLO models is indeed expected since these models provide better description of the experimental data for the price of larger number of free parameters. Their predictions vary stronger than those of the WT or LO models.

Overall, we cannot escape the conclusion that new precise lattice results will provide an important constraint on the models.

\subsection{Fit details: loss function and degrees of freedom}
\label{subsec:AICBICChi2}

Stemming from quite different time periods and being results of vastly different experimental techniques, the quality of the experimental data summarized in \cref{sec:Experimental_data} is very different. More so if one considers combined analysis of Lattice QCD finite-volume spectra (\cref{sec:LQCD}) as adequate input for the UCHPT fits. Thus, a simple $\chi^2$ statistic as a measure of fit-quality is not sufficient and modifications are in order.

\begin{figure}[t]
\begin{minipage}[t]{0.49\linewidth}
\vspace{0pt}
\begin{tabular}{|l|l|l|c|r|r|r|}
\hline
Fit&UCHPT type&$N_{\rm data|exp.+lat.+m}$&$N_{\rm par.}$ & $\chi^2_{\rm dof}$&$AIC$&$BIC$\\
\hline
\hline
$F_{19}$&M1S1L&$0+14+0$&3&1.36&7.4&9.3\\
$F_{31}$&M1S2L&$0+14+0$&0&2.89&2.9&2.9\\
$F_{18}$&M1S3L&$0+14+0$&1&4.42&{\color{gray}6.4}&{\color{gray}7.1}\\
\hline
$F_{20}$& M2S1L&$0+14+0$ &3&1.42&7.4&9.3\\
$F_{32}$& M2S2L&$0+14+0$ &0&2.68&2.7&2.7\\
$F_{25}$& M2S3L&$0+14+0$ &1&3.54&5.5&6.2\\
\hline
$F_{01}$& M3S1L&$0+14+4$ &10&0.96 &21.0&29.9\\
$F_{15}$& M3S2L&$0+14+4$ &7 &0.90 &14.9&21.1\\
$F_{10}$& M3S3L&$0+14+4$ &8 &0.92 &16.9&24.0\\
\hline
\hline
$F_{21}$& M1S1P  &$258+0+0$&6&4.23 &{\color{gray}16.2}&{\color{gray}37.5} \\
$F_{28}$& M1S2P  &$258+0+0$&0&25.58&{\color{gray}25.6}&{\color{gray}25.6}\\
$F_{27}$& M1S3P  &$258+0+0$&1&30.28&{\color{gray}32.3}&{\color{gray}35.8}\\
\hline
$F_{22}$& M2S1P  &$258+0+0$&6& 8.87&{\color{gray}20.9}&{\color{gray}42.2}\\
$F_{29}$& M2S2P  &$258+0+0$&0&48.16&{\color{gray}48.2}&{\color{gray}48.2}\\
$F_{26}$& M2S3P  &$258+0+0$&1&18.69&{\color{gray}20.7}&{\color{gray}24.2}\\
\hline
$F_{30}$& M3S1P  &$258+0+0$&16&1.51&33.5&90.4\\
$F_{13}$& M3S2P  &$258+0+0$&7 &1.85&15.8&40.6\\
$F_{11}$& M3S3P  &$258+0+0$&8 &1.50&17.5&45.9\\
\hline
\hline
$F_{24}$& M1S3PL  &$258+14+0$&1&27.56&{\color{gray}29.6}&{\color{gray}33.2}\\
\hline
$F_{23}$& M2S3PL  &$258+14+0$&1&17.82&{\color{gray}19.8}&{\color{gray}23.4}\\
\hline
$F_{17}$& M3S1PL  &$258+14+4$&16&1.44 &33.4&91.4\\ 
$F_{16}$& M3S2PL  &$258+14+4$&7 &2.11 &16.1&41.5\\ 
$F_{12}$& M3S3PL  &$258+14+4$&8 &2.23 &18.2&47.2\\
\hline
\end{tabular}
\end{minipage}
~~~~
\begin{minipage}[t]{0.45\linewidth}
\vspace{0.2cm}
\includegraphics[height=12.1cm]{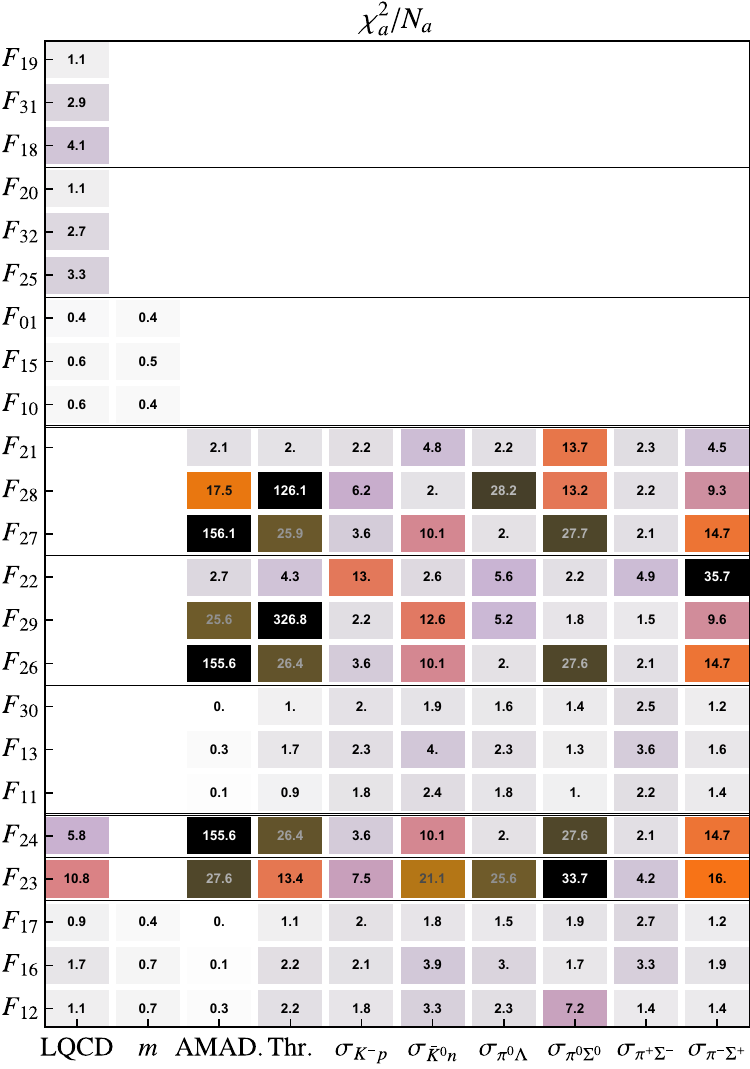}
\end{minipage}
\caption{Left: Summary of the fit results minimizing $\chi^2_{\rm dof}$ while using various model types, regularization schemes and data subsets. The last two columns show $AIC$ and $BIC$ values from \cref{eq:AIC-BIC-standard}, where results with large $\chi^2_{\rm dof}$ are grayed out. Right: Minimized $\chi^2$ from fits in table left hand side separated to the data type normalized by the number of data in that observable. Here "LQCD", "AMAD" and "Thr." refer to lattice, AMADEUS and threshold data discussed in \cref{sec:LQCD} and \cref{sec:Experimental_data}.}
\label{fig:fit-summary}
\end{figure}

As an example, consider one of the most modern experimental data inputs, the energy shift and width of the Kaonic hydrogen, in comparison with the oldest data available, the total cross sections for the $K^-p\to MB$. The former consists of only two quantities measured at 10-20\% accuracy at the $K^-p$ threshold. The latter data includes 252 data points at a few 10's\% statistical uncertainty depending on the channel and the kinematics which also carries considerable systematic uncertainties due to bin sizes and possible inconsistencies of the data. Using traditional $\chi^2$ definition aggregating both sources together would, thus, simply make the modern SIDDHARTA data entirely insignificant despite its superior quality. Lattice finite-volume spectra consist also of only 14 points and, thus, would appear as a insignificant contribution to the total $\chi^2$. In such cases of asymmetric data distribution among different observables, a weighted $\chi^2$ definition is more customary
\begin{align}
    \chi^2_{\rm dof}=
            \frac{\sum_a N_a}{A((\sum_a N_a)-N_{\rm par})}\chi^2_{\rm wt}
            \quad
            \text{with}~\chi^2_{\rm wt}=\sum_{a=1}^{A}\frac{\chi^2_a}{N_a}
            \quad
            \text{with}~\chi^2_a=\sum_{n=1,m=1}^{N_a}(f^a_n(\vec\aleph)-\hat f^a_n)\,
            [\hat{C}^{-1}_a]_{nm}
            \,
            (f^a_m(\vec\aleph)-\hat f^a_m)\,.
    \label{eq:chi2}
\end{align}
Here, we denote with $a/A$ the index/number of observables; $\vec \aleph$ is the  parameter vector of the model. Its size $N_{\rm par}$ can be read off from the \cref{fig:fit-summary}; $n/N_a$ is index/number of data in the data type $a$; $f^a/\hat f^a$ is model/datum result of the corresponding observable type $a$; $\hat C$ is the covariance matrix of the data. Note that in most cases, except Lattice QCD the latter is simply provided by a diagonal matrix of the form $[\hat{C}_a]_{nm}=\delta_{nm}({\hat{\sigma}}^a_n)^2$, where $\hat{\sigma}^a_n$ is the error for each 
datum $\hat{f}^a_n$. The total number of data is denoted by $N_{\rm data}=\sum_a N_a$.

The above $\chi^2_{\rm dof}$ is minimized in the fit with respect to $k$ parameters. 
However, given the various choices of interaction kernel (M1, M2, M3), and regularization scheme (S1, S2, S3), this analysis is both a \emph{parameter estimation} and \emph{model comparison} problem. Comparing models with different parameters should involve penalising models with more parameters.

A full model comparison would involve the estimation of Bayes factors via Markov Chain Monte Carlo which, given the number of parameters of interest, would be too computationally expensive with the current code. A crude model comparison may be effected using an \emph{information criterion} that attempts to balance goodness of fit against model complexity (see for example \cite{liddle_information_2007} for an example in an adjacent research area). Such criteria make use of the (log) maximum likelihood, with some adjustment for the number of parameters and size of data set. We use the Akaike Information Criterion ($AIC$) \cite{akaike_new_1974} and the Bayes Information Criterion \cite{schwarz_estimating_1978} ($BIC$), but we note that care must be taken in interpreting the numbers calculated for $AIC/BIC$ too seriously, and use them only as a guide for revealing gross features.

For a given maximum log-likelihood $\log{L}_{\mathrm{max}}$, the information criteria are defined as
\begin{align}
    AIC&=-2\log{L}_{\mathrm{max}}+2N_{\rm par}%
    \quad
    \text{and}
    \quad
    BIC=N_{\rm par}\log N_{\rm data}-2\log{L}_{\mathrm{max}}\,.
\label{eq:AIC-BIC-standard}
\end{align}
Comparing $AIC$/$BIC$ values for different models is only valid for fits to the same data points, but if this condition is true then the differences in $AIC$ or $BIC$ values between models can be related to the relative probability of a model being true. In our case, we use the $\chi^2_{\rm dof}$ as defined in~\cref{eq:chi2} as $-2\log L_{\mathrm{max}}$, and values are displayed in \cref{fig:fit-summary}. There, results for fits with too large $\chi^2_{\rm dof}$ are grayed out, furthermore, sensible comparison between different models can only be made for fits with equal number of data.

\subsection{Fit discussion}
\label{subsec:fit-discussion}
Following the description of the $\chi^2_{\mathrm{dof}}$ analysis, this section provides additional technical details of the fitting procedures used in this work. In the fits, subtraction constants and LECs will be constrained using two-hadron interacting spectrum and experimental quantities. In addition, the parameters, related to the explicit chiral symmetry breaking $b_0,b_D$ and $b_F$ can be constrained using the Lattice QCD estimation of single baryon masses provided by the BaSc collaboration \cite{BaryonScatteringBaSc:2023ori}.
The NLO CHPT formulas for the baryon masses reads as following \cite{Bernard:1995dp, Frink:2005ru}
\begin{align}
    m_N&=m_0-2\left(b_0+2b_F\right)M_\pi^2-4\left(b_0+b_D-b_F\right)M_K^2\,,\\
    m_\Lambda&=m_0-\frac{2}{3}\left(3b_0-2b_D\right)M_\pi^2-\frac{4}{3}\left(3b_0+4b_D-b_F\right)M_K^2\,,
    \\
   m_\Sigma&=m_0-2\left(b_0+2b_D\right)M_\pi^2-4b_0M_K^2\,, \\
   m_\Xi&=m_0-2\left(b_0-2b_F\right)M_\pi^2-4\left(b_0+b_D+b_F\right)M_K^2\,,
\end{align}
where the low-energy constants $b_0,b_D,b_F$ have been discussed before, and $m_0$ is the baryon octet mass in the chiral limit. In practice we avoid fitting $m_0$ directly with explicitly constructing the mass differences between the lattice and the physical point. In summary, we have 272 (258(experimental)+14(lattice finite-volume multihadron)+4(lattice finite-volume single baryon)) data points and construct a correlated $\chi^2_{\mathrm{dof}}$ as described in \cref{subsec:AICBICChi2}.

In the fits we are using Nelder-Mead minimization and check for absolute convergence by performing the fits using different  initial conditions. We are fitting models (M1 etc.) with increasing computational complexity using regularization S1, S2, S3 including lattice  and/or experimental data. In the end the best fits are selected using each regularization S1, S2, S3 using all of the available data. The pole positions are computed for each fit individually and can be found in the appendix. 

\begin{figure*}[t]
    \centering
    \includegraphics[width=0.3\linewidth]{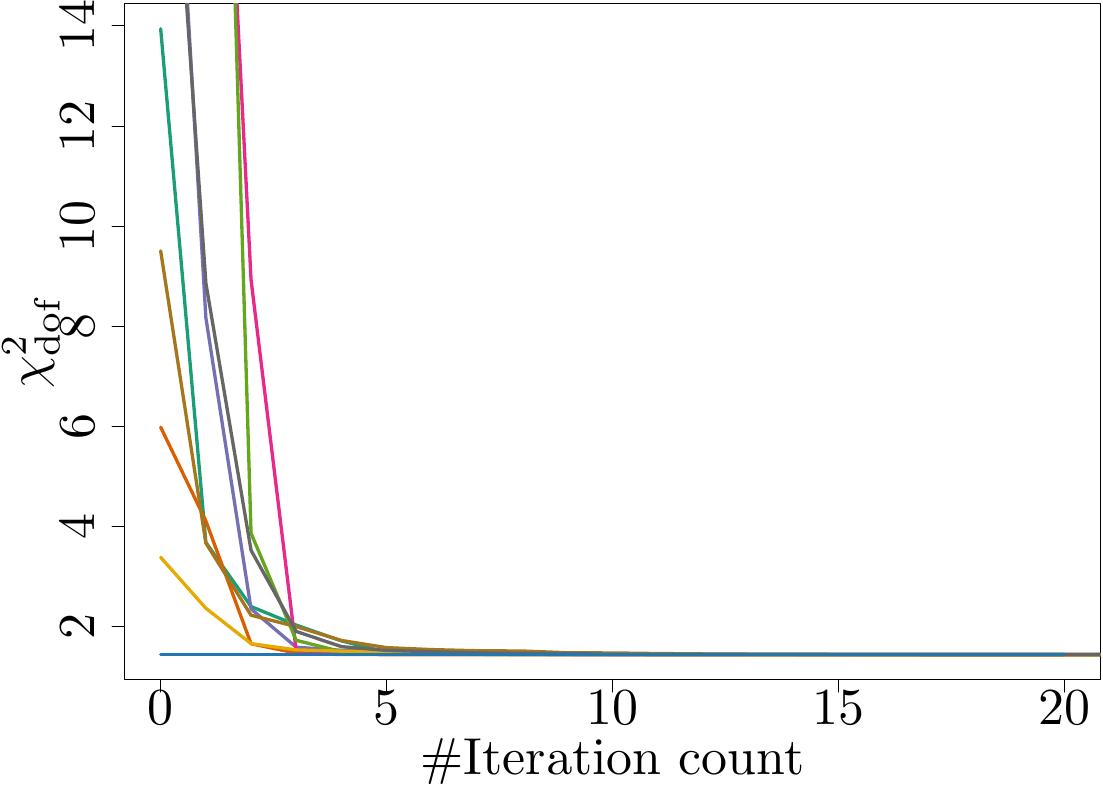}~~~~~
    \includegraphics[width=0.3\linewidth]{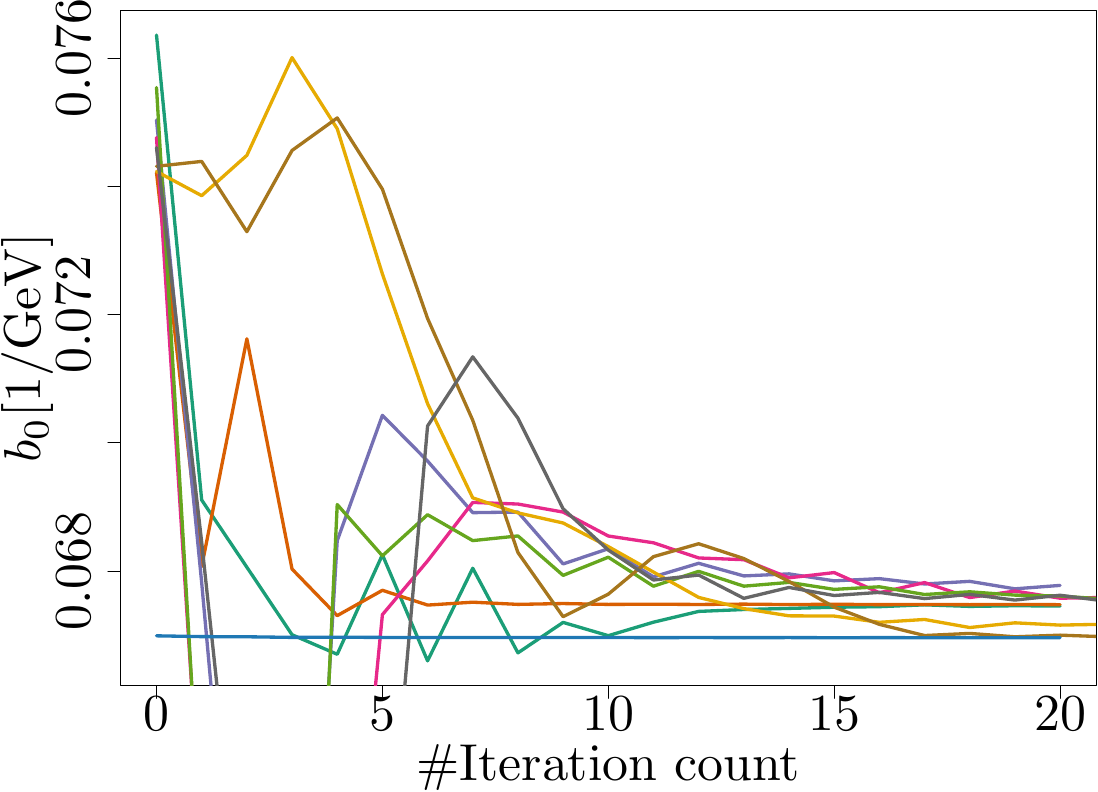}~~~~~
    \includegraphics[width=0.3\linewidth]{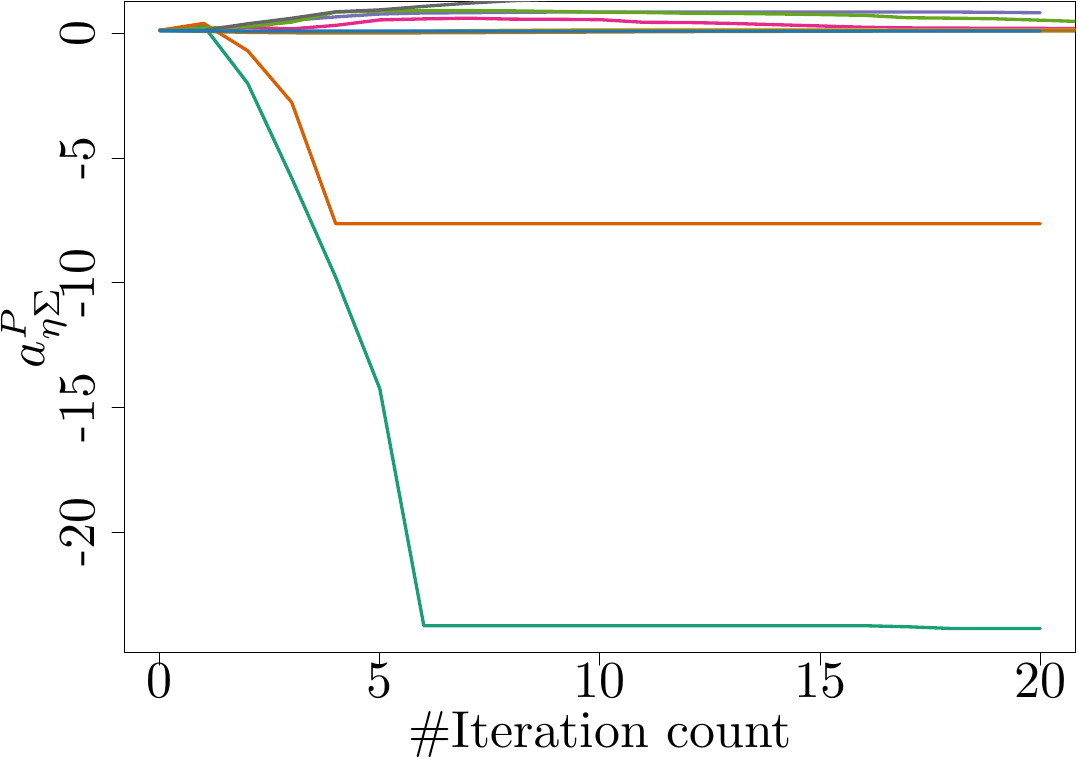}
    \caption{
    Convergence check of Fit 17 (M3S1PL) using nine different sets of initial parameters.
    Left:   Evolution of $\chi^2_\mathrm{dof}$ as a function of iteration number.
    Middle: Convergence behavior of the parameter $b_0$.
    Right:  Convergence behavior of the subtraction constant  $a_{\eta\Sigma}$ at the physical point.}
    \label{fig:S1behaviour}
 \end{figure*}

The same model parameters are constrained through the lattice and experimental data, although in the former case via the complex partial-wave amplitude $f_{0+}$ and in the latter case via the real $\vert f_{0+}\vert^2$.
Regularization $S1$ provides maximal freedom,   enabling a completely independent set of   subtraction constants at the physical and   lattice point. However, in the case of the  lattice data the projection to isospin zero   channel eliminates the dependence through  subtraction constants,   $a_{\pi\Lambda},a_{\eta\Sigma}$. In addition  during our numerical investigation we found out that the fit to the lattice data is insensitive to $a_{K\Xi}$.
  
To check the robustness of our fits, we select M3S1PL ($F_{17}$) and test its behavior under different random initial conditions. In this case, the minimization is performed in a 16-dimensional parameter space, and we examine whether and how the global minimum is consistently reached. The results are shown in \cref{fig:S1behaviour}. In the left panel, we show the convergence of $\chi^2_{\mathrm{dof}}$ as a function of iteration number. In all cases, $\chi^2_{\mathrm{dof}}$ converges to the same minimum value, indicating consistent fit quality across different initializations. In the middle panel, we present the convergence of the parameter $b_0$, which shows only a small spread in its final values. The right panel displays the convergence of the subtraction constants $a_{\eta\Sigma}$ at the physical point, where two outlier trajectories are observed. However, as seen in the left panel, these outliers still lead to approximately the same $\chi^2_{\mathrm{dof}}$ value, suggesting that the fit remains statistically valid. For the final fit selection, we choose the one where all subtraction constants are smaller than 0.05 in absolute value.

\begin{figure*}[t]
    \includegraphics[width=0.55\linewidth]{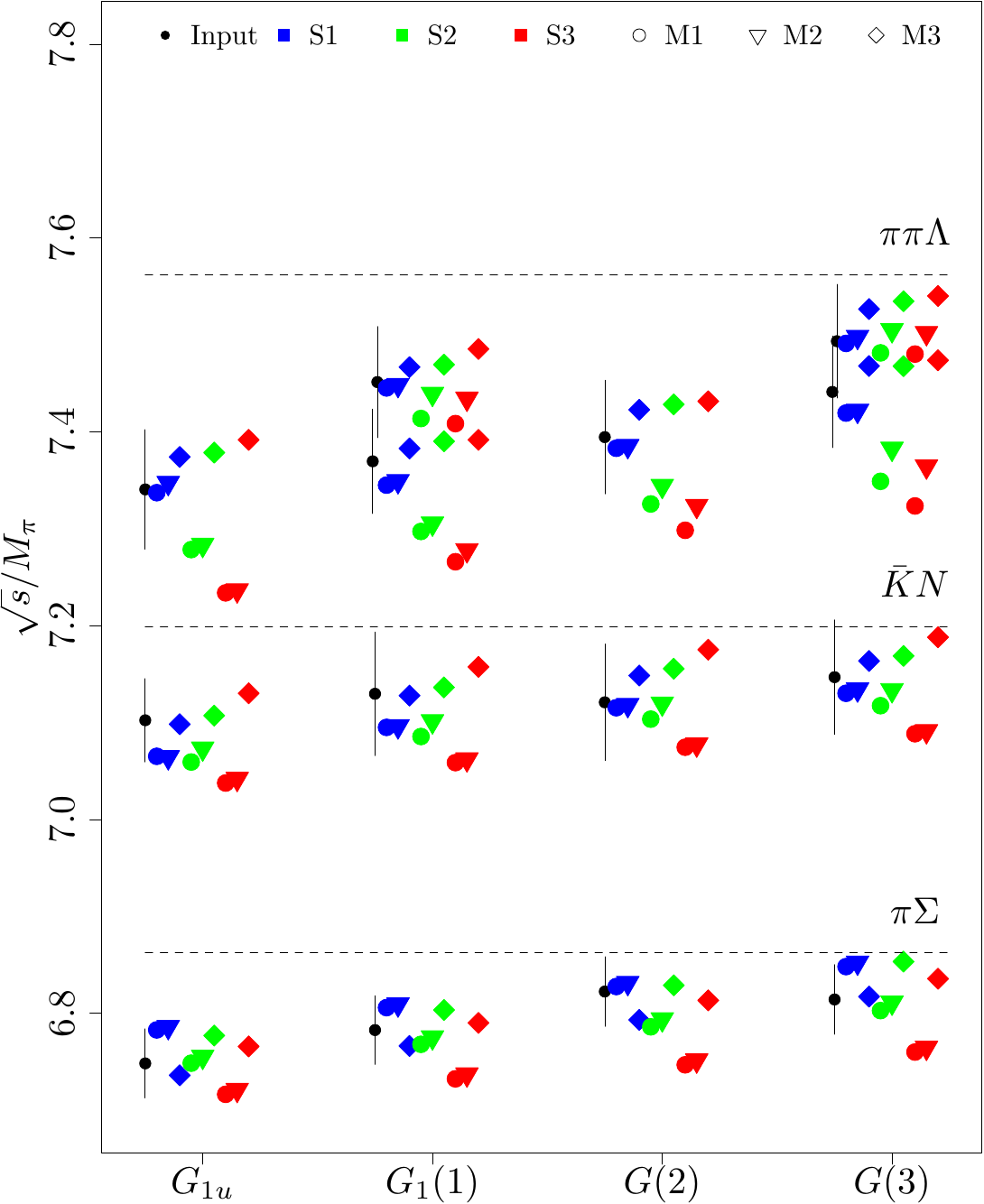}
    ~~~
    \includegraphics[width=0.4\linewidth]{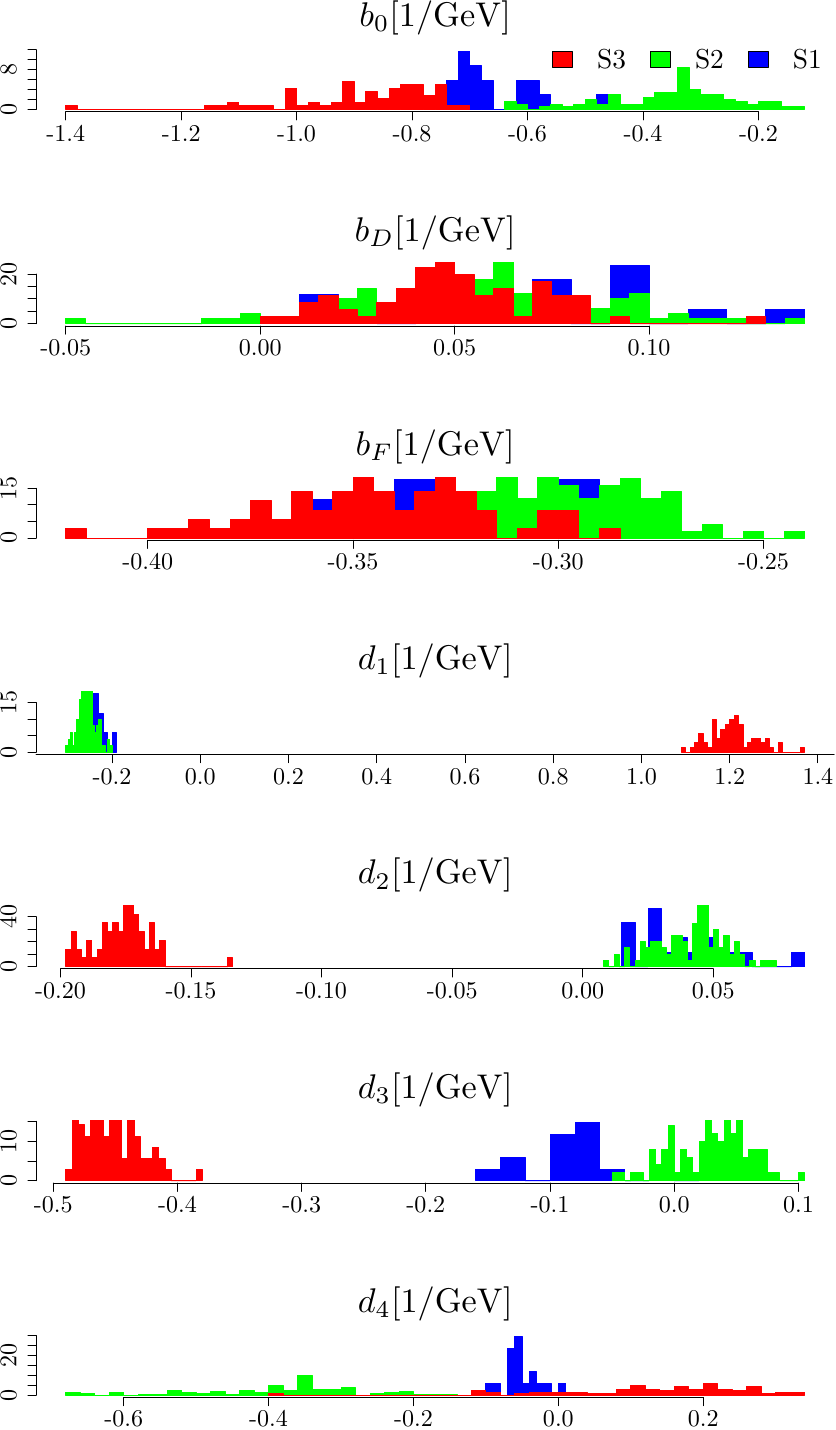}
    \caption{Left: Model (M1, M2, M3) predictions 
    for the lattice spectrum using different regularizations (S1, S2, S3). 
    Fits were done using only the lattice data. More detailed results including cross-correlations are provided for each fit in the Appendix. Individual $\chi^2$ values can be found in \cref{fig:fit-summary}.
    Right: Distribution of low-energy constants for the best combined fits for three different schemes.
    }
    \label{fig:spectrumfit_results1}
\end{figure*}

In the left panel of~\cref{fig:spectrumfit_results1} we show the fitted spectrum for all different available models and regularization schemes. We would like to point out that leading chiral order models M1 and M2 give very similar results, with the regularization S1 providing the best $\chi^2_\mathrm{dof}$ estimate. Note that in this case S3 has only one parameter, whereas the S2  prediction is parameter-free. On the right part of ~\cref{fig:spectrumfit_results1} we show the bootstrap samples distribution of the LECs using the three different regularizations fitting all the available data. Dynamical LECs ($d_1$,$d_2$,$d_3$) do agree in S1 and S2 schemes, but show large statistical and systematic uncertainties in $d_4$. Our results for the symmetry breaking parameters $(b_0,b_D,b_F)$ are compatible with those obtained in Ref.~\cite{Lutz:2024ubv} using the corresponding CLS ensembles, and also quite close to the values obtained in perturbative calculations~\cite{Mai:2009ce}.

\section{Results and discussion}
\label{sec:results}

After having determined the best parameters of each model with respect to the lattice and experimental input, we turn to the main goal of the paper, extracting transition amplitudes and their analytic structures. In the relevant energy region, the former are holomorphic functions except of meson-baryon right-hand cuts taken care of through the  unitarization procedure (\cref{subsec:unitarization}) and possible poles on the unphysical Riemann sheets associated with excited hadrons, see, e.g.,~\cite{Mai:2025wjb, Mai:2022eur, Burkert:2025coj, Willenbrock:2022smq}. 

In dealing with the 10 channel problem, there are $2^{10}$ Riemann sheets associated with the right-hand meson-baryon cuts. There are various ways how to label those. Most frequently~\cite{Cieply:2011fy,Cieply:2016jby} one denotes a Riemann sheet (full complex energy plane) by a sequence $[\pm\ldots\pm]$ referring to $\operatorname{sgn}(\Im(p_{\rm cm}(s)))$ in each two-body channel. For example, any experimental or lattice input is obtained on the real energy axis of the physical sheet $[+\ldots+]$. An unphysical sheet connected to the physical one between mass-ordered threshold $n-1$ and $n$ is denoted by $[\underset{1}{-}\ldots\underset{n}{-}+\ldots+]$. Riemann sheets with mixed order of $\pm$ are sometimes referred to as hidden sheets as they are connected to the physical real energy axis through a sequence of other sheets, see Fig. 3 in Ref.~\cite{Mai:2022eur}.

The pole positions of the $\Lambda(1405)$ and $\Lambda(1380)$ have been determined directly from lattice input using generic (EFT independent) tools~\cite{BaryonScatteringBaSc:2023ori,BaryonScatteringBaSc:2023zvt}, and from experimental results through UCHPT. While more can be said about the latter (see Fig. 7 from Ref.~\cite{Mai:2020ltx}) we use here as a reference PDG and BaSc values (in MeV)
\begin{align}
    \renewcommand{\arraystretch}{1.2}
    \begin{tabular}{cl|l}
    & physical point            & lattice point\\
    \cline{2-3}
    $\Lambda(1405)~~~$
    &$1417.7^{+6.1}_{-7.5}-i26.1^{+6.23}_{-8.2},\,1429^{+8}_{-7}-i12^{+2}_{-3},\,
    1434^{+2}_{-2}-i10^{+2}_{-1},\,
    1421^{+3}_{-2}-i19^{+8}_{-5}
    $
    &$1455^{+21}_{-21}-i12^{+6}_{-6}$    
    \\
    $\Lambda(1380)~~~$
    &$1325^{+15}_{-15}-i90^{+12}_{-18},\,
    1330^{+4}_{-5}-i56^{+17}_{-11},\,
    1388^{+9}_{-9}-i114^{+24}_{-25},\,
    1381^{+18}_{-6}-i81^{+19}_{-8}$
    &$1392^{+18}_{-18}$
    \end{tabular}
    \label{eq:ref-values}
\end{align}
All these poles are obtained on the Riemann sheet $[++----++++]$ in relation with the ordering provided in \cref{eq:ordering-channel}.

\begin{figure}[t]
    \centering
    \adjustbox{angle=90,lap=0pt}{~~~~~~~~~M3S1L $(F_{01})$}
    \includegraphics[height=4cm]{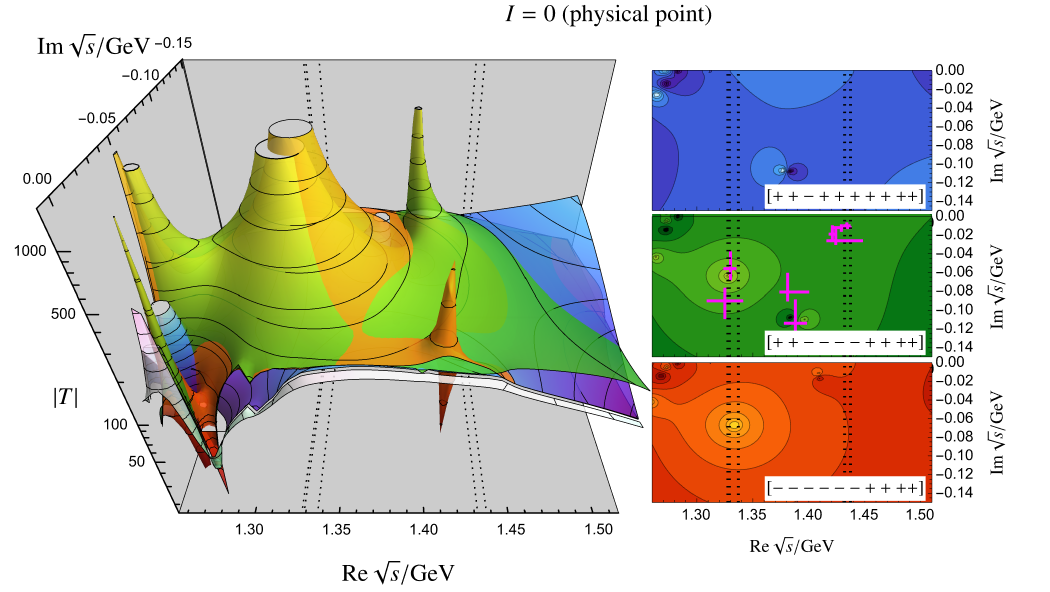}
    \includegraphics[height=4cm]{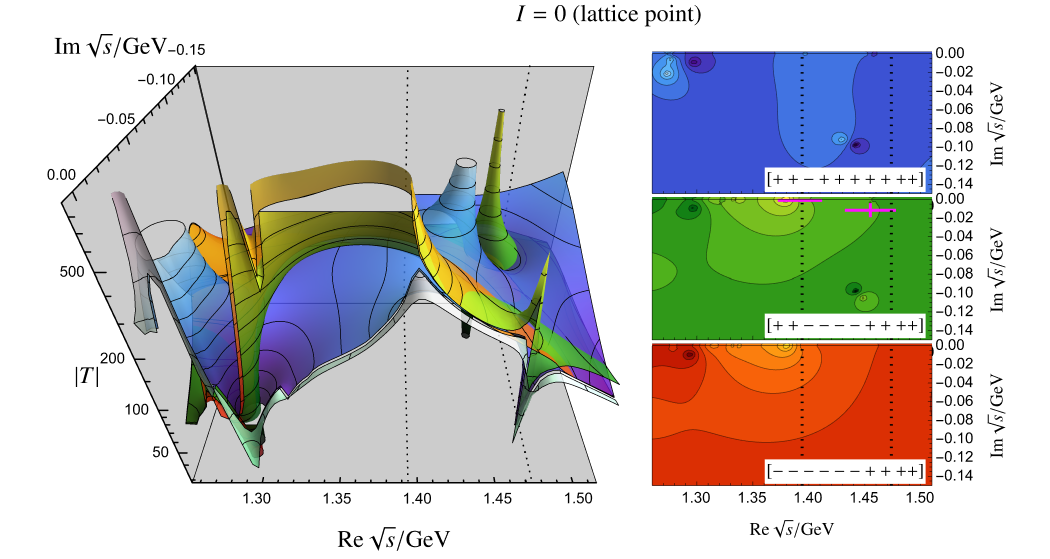}
    \\
    \adjustbox{angle=90,lap=0pt}{~~~~~~~~~M3S2L $(F_{15})$}
    \includegraphics[height=4cm]{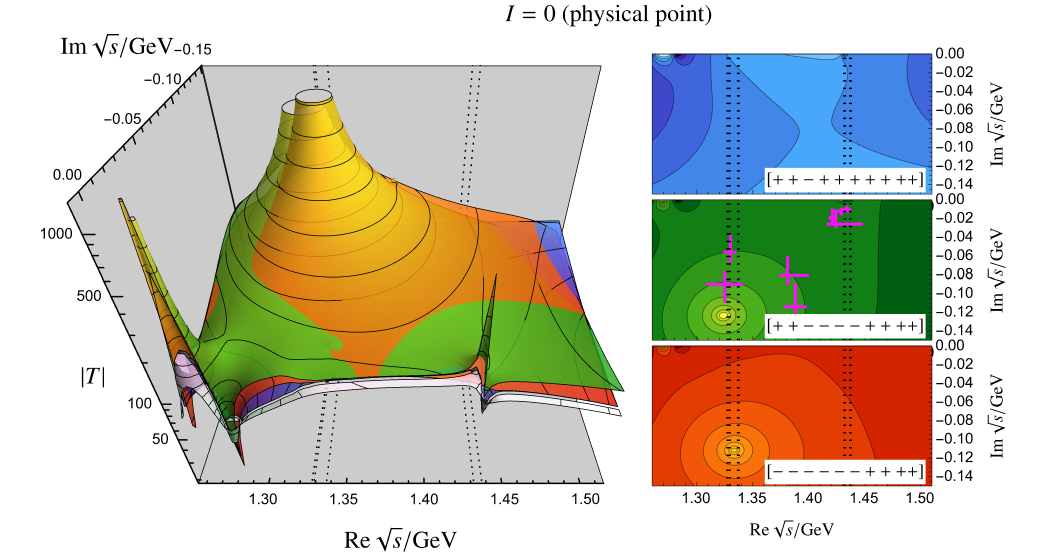}
    \includegraphics[height=4cm]{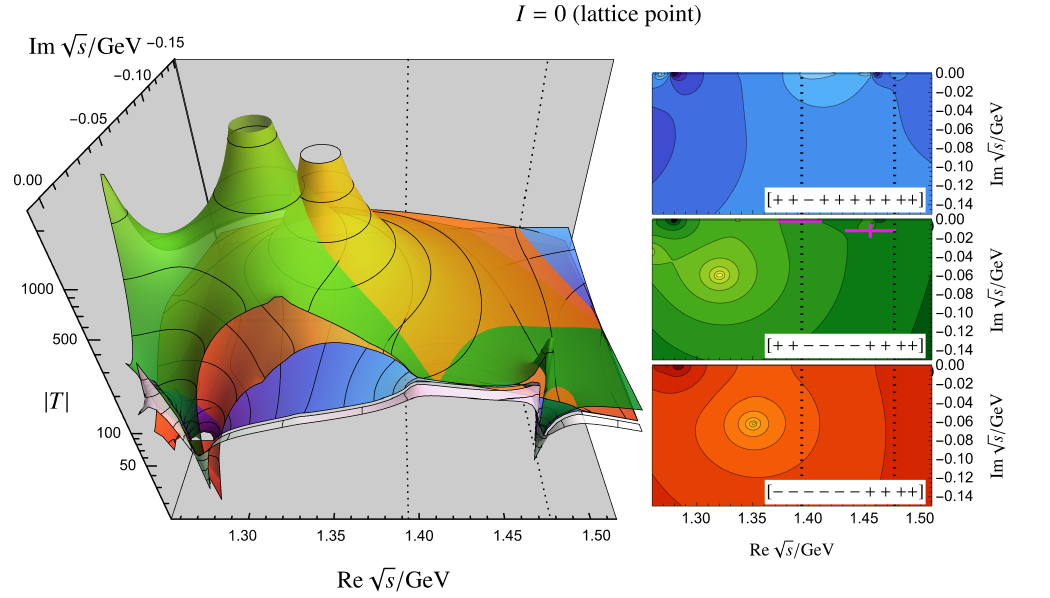}
    \\
    \adjustbox{angle=90,lap=0pt}{~~~~~~~~~M3S3L $(F_{10})$}
    \includegraphics[height=4cm]{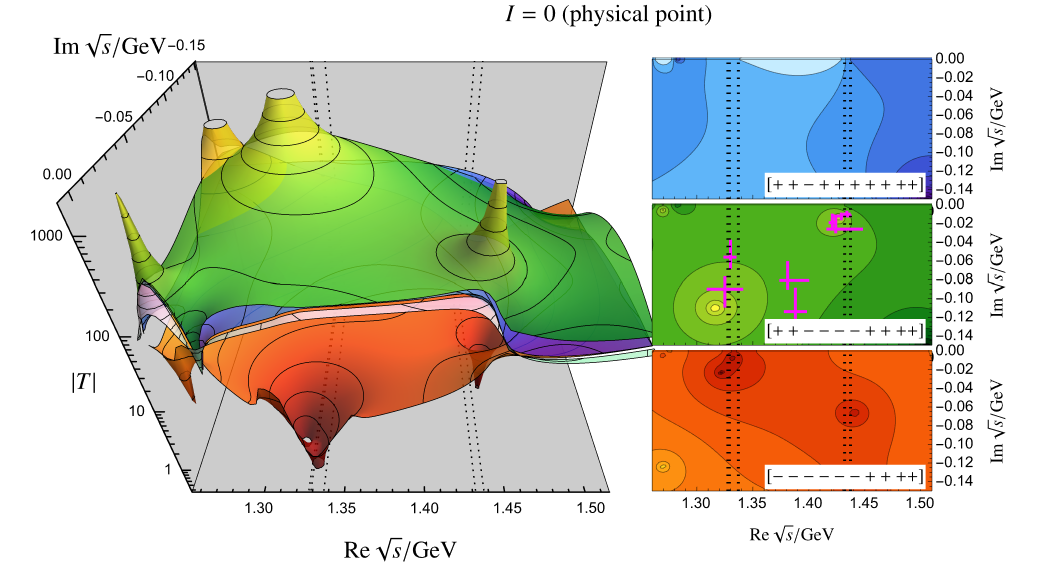}
    \includegraphics[height=4cm]{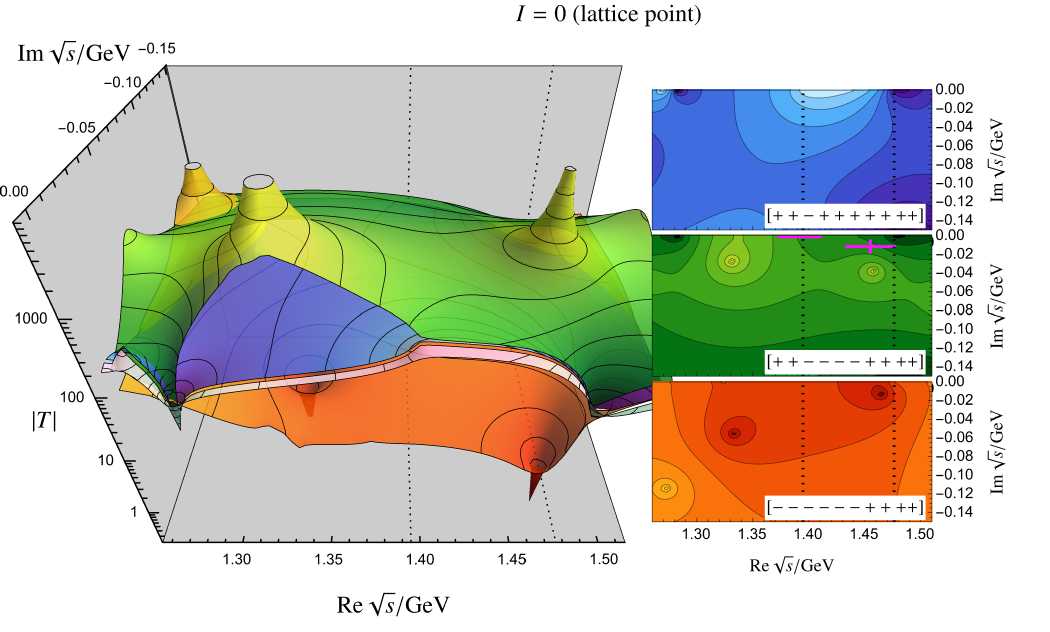}
    \caption{Isoscalar scattering amplitude on the next to the physical (light pink surface for $\Im\sqrt{s}>0$ in the 3D plot) unphysical Riemann sheets (color coded) of the solutions (M3Sx type)  fitted only to the Lattice QCD results. Extrapolation to the physical point is shown in the left column. More details of the fit and predictions thereof can be found in Appendix A.}
    \label{fig:RES-lat-to-phys}
\end{figure}

\subsection{From the lattice to the physical point}
\label{subsec:RES-latt-to-phys}

First, let us consider the case of Lattice QCD results being the only input to the UCHPT approach. Indeed, we have seen in \cref{subsec:pilotstudy} that the former is  indeed a non-trivial input. 

The M1-models $F_{19}$, $F_{31}$, $F_{18}$ in \cref{fig:fit-summary} provide an approximate description of the lattice results including only few free fit parameters as described in \cref{subsec:fit-discussion}. The S2 scheme provides such description quality even without any fits. Similarly, the M2-type ($F_{20}$, $F_{32}$, $F_{25}$) provide similar or slightly better fits to the data with no additional parameters. This shows that the exchange diagrams (Born terms) do matter in the description of the finite-volume spectrum. Regarding the isoscalar pole structure, at the lattice point we observe (see plots in the Appendix A) 
that all models provide the $\Lambda(1405)$ narrow pole and the $\Lambda(1380)$ broad pole. In a few but not all cases the latter becomes a virtual bound state as also obtained by the BaSc (K-matrix) analysis\footnote{An interesting data driven non-parametric approach based on Nevanlinna interpolation was proposed recently in Ref.~\cite{Salg:2025now}.}. More importantly the S1-type of models predict a resonance $\Lambda(1380)$ with non-negligible width having better $\chi^2_{\rm dof}$ than the other fits. Extrapolating to the physical point, while neglecting quark-mass dependence of the subtraction constants in the S1 as in \cref{subsec:pilotstudy}, we observe again a clear two-pole structure with the spread of poles due to variations of the model types reflecting the spread of the reference values. Regarding the isovector states, all models except of M1S1 do predict a state whose position, however, varies strongly with the model type.

Models of the M3-type ($F_{01}$, $F_{15}$, $F_{10}$ in \cref{fig:fit-summary}) have larger flexibility and do indeed provide an excellent description of the lattice input ($\chi^2_{\rm dof}\approx 1$), see \cref{fig:fit-summary}. On the most relevant sheet $[++----++++]$ we do again observe two poles both for the lattice point and after an extrapolation also at the physical point, see \cref{fig:RES-lat-to-phys}. However, all pole positions vary strongly between different models. For example, $F_{01}$ does agree with the reference values~\cref{eq:ref-values} on the $\Lambda(1380)$ pole position, but not on the corresponding $\Lambda(1405)$ value. Similar observations hold for the $F_{10}$ and $F_{15}$ fits. Interestingly, all chiral extrapolations to the physical point provide similar prediction for the $\Lambda(1380)$ which also qualitatively agrees with the reference values~\cref{eq:ref-values}. Position of the $\Lambda(1405)$ pole is on the other side not well predicted at the physical point.

We conclude that a combination of currently available Lattice QCD finite-volume spectra combined with the modern UCHPT approaches does indeed provide a proof for the existence of two states $\Lambda(1405)$ and $\Lambda(1380)$. However, it also seems that the pole positions are not yet fixed when taking into account systematic uncertainties of the UCHPT approaches. Information criteria from \cref{subsec:AICBICChi2} seem to prefer S2 and S3 type of fits due to the strong weight on the number of parameters.

\begin{figure}[tbhp]
    \centering
    \adjustbox{angle=90,lap=0pt}{~~~~~~~~~M3S2P $(F_{30})$}
    \includegraphics[height=4cm]{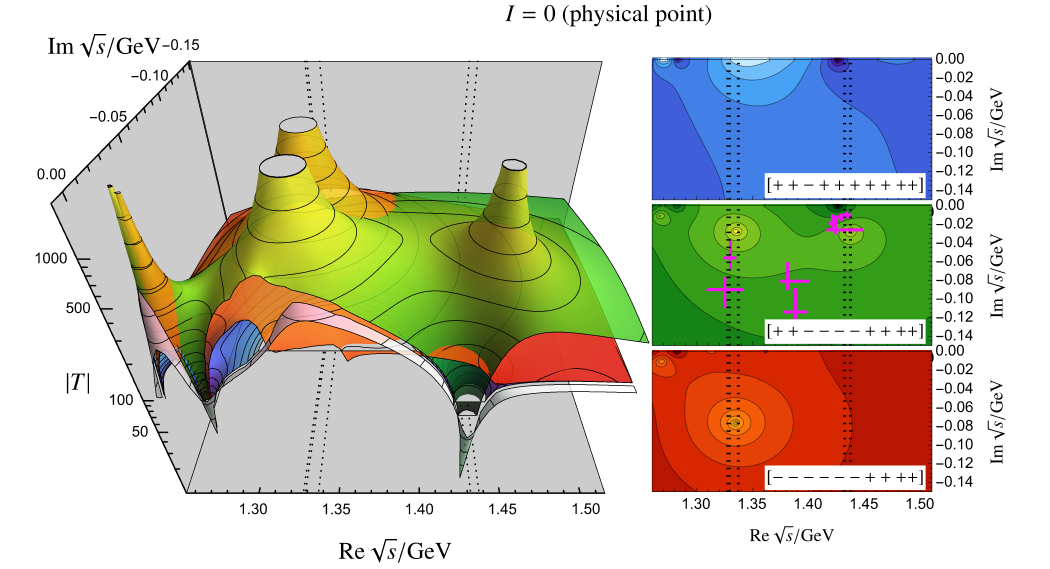}
    \includegraphics[height=4cm]{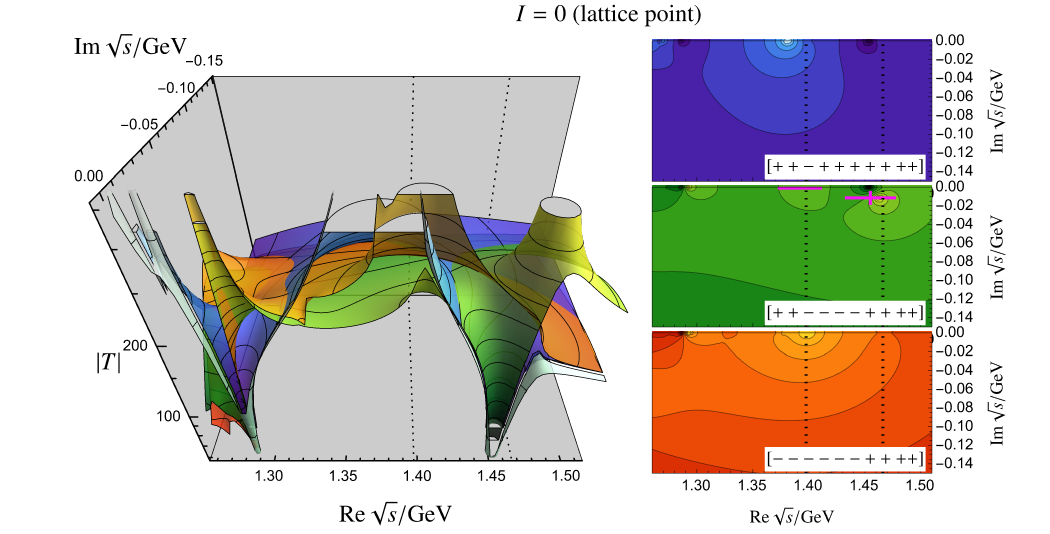}
    \\
    \adjustbox{angle=90,lap=0pt}{~~~~~~~~~M3S2P $(F_{13})$}
    \includegraphics[height=4cm]{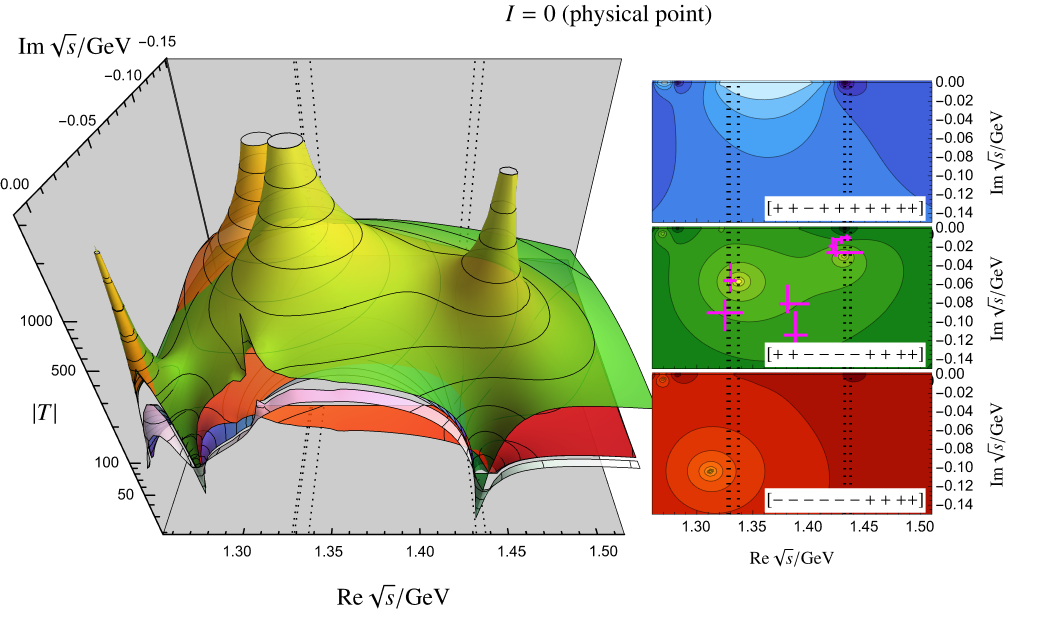}
    \includegraphics[height=4cm]{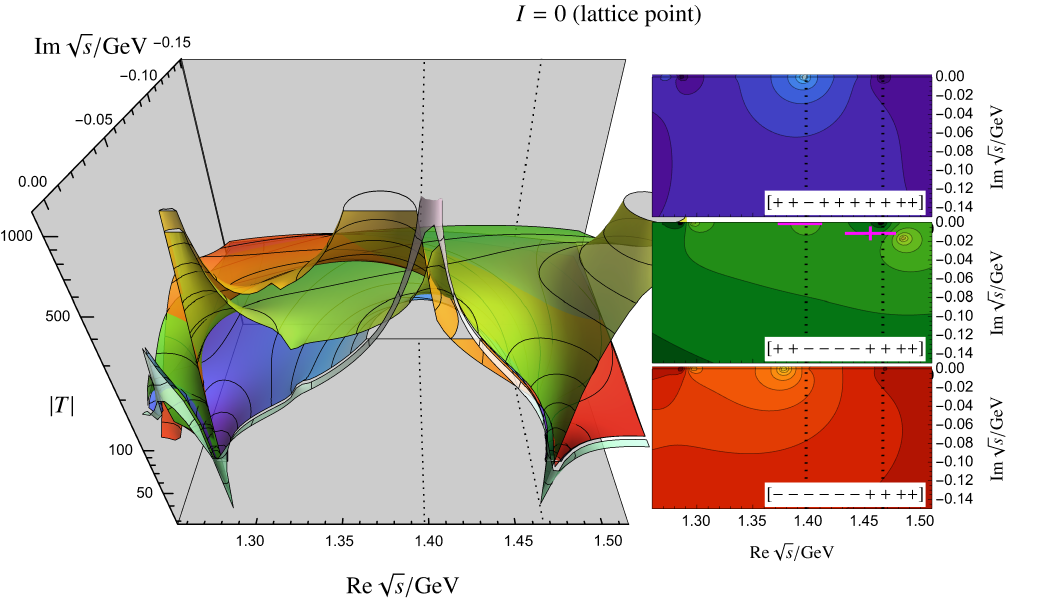}
    \\
    \adjustbox{angle=90,lap=0pt}{~~~~~~~~~M3S3P $(F_{11})$}
    \includegraphics[height=4cm]{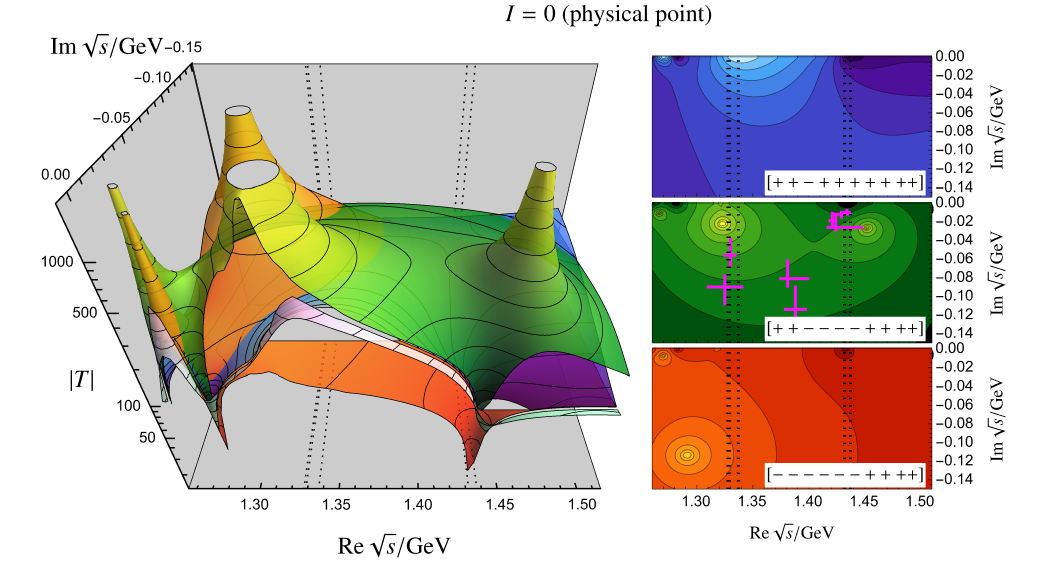}
    \includegraphics[height=4cm]{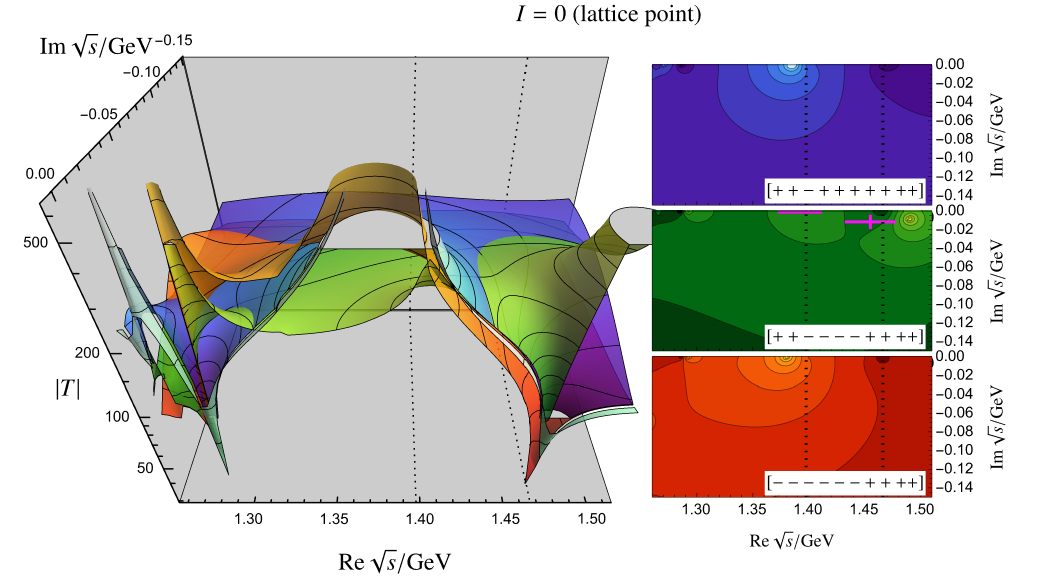}
    \caption{Isoscalar scattering amplitude on the next to the physical (light pink surface for $\Im\sqrt{s}>0$ in the 3D plot) unphysical Riemann sheets (color coded) of the solutions (M3Sx type)  fitted only to the experimental input. Extrapolation to the lattice point is shown in the right column. More details of the fit and predictions thereof can be found in Appendix B.}
    \label{fig:RES-phys-to-latt}
\end{figure}

\subsection{From the physical to the lattice point}
\label{subsec:RES-phys-to-lat}

Obviously, it is also possible to inverse the procedure of the last section, using only experimental data as input then predicting the pertinent lattice point results. One motivation behind this is to test the predictive power of the UCHPT approaches outside of the fitted quark-mass domain. 

The M1- and M2-type of models do not allow an adequate description of the experimental data as shown in \cref{fig:fit-summary}. A look on the separated contributions to the $\chi^2$ value with respect to different observable types reveals that this is mostly due to the threshold values including the so-important SIDDHARTA results~\cite{Bazzi:2011zj} and very recent AMADEUS data~\cite{Piscicchia:2018rez,Piscicchia:2022wmd}. The corresponding extracted isoscalar and isovector pole positions can be found in the Appendix B. Clearly, since the models are at odds with the data, the results scatter erratically and should not be over-interpreted.

The most flexible model type M3 ($F_{30}$, $F_{13}$, $F_{11}$) provides a reasonable description of all experimental data ($\chi^2_{\rm dof}\approx 1.5$) as shown in \cref{fig:fit-summary}. Again, S2 and S3 are favored due to the $AIC$ and $BIC$. At the physical point, we observe again the two-pole structure of the isoscalar states with well-fixed $\Lambda(1405)$ and less determined $\Lambda(1380)$ pole, see \cref{fig:RES-phys-to-latt}. At the lattice point all three solutions ($F_{30}$, $F_{13}$, $F_{11}$) provide a prediction of the pole structure which indeed overlaps with the BaSc determination~\cite{BaryonScatteringBaSc:2023ori} within $\approx2~\sigma$. Comparing this to the pertinent observation of the previous \cref{subsec:RES-latt-to-phys} it is reasonable to conclude that experimental data provides more strict constraints on the UCHPT approaches than the recent Lattice QCD results. In the isovector case poles are predicted in each model type for the lattice point. Their positions variate strongly with the chosen model type. At the physical point even less can be concluded with certainty. Indeed, this confirms the results of the previous meta-study~\cite{Cieply:2016jby} of various unitary models leading to vastly different predictions in the isovector case. Whether this can be mitigated through combined use of the lattice and experimental input is discussed in the next section.

\subsection{Combined analysis at the lattice and physical point}
\label{subsec:RES-phys-and-lat}

\begin{figure}[tbhp]
    \centering
    \adjustbox{angle=90,lap=0pt}{~~~~~~~~~M3S1PL $(F_{17})$}
    \includegraphics[width=0.42\linewidth]{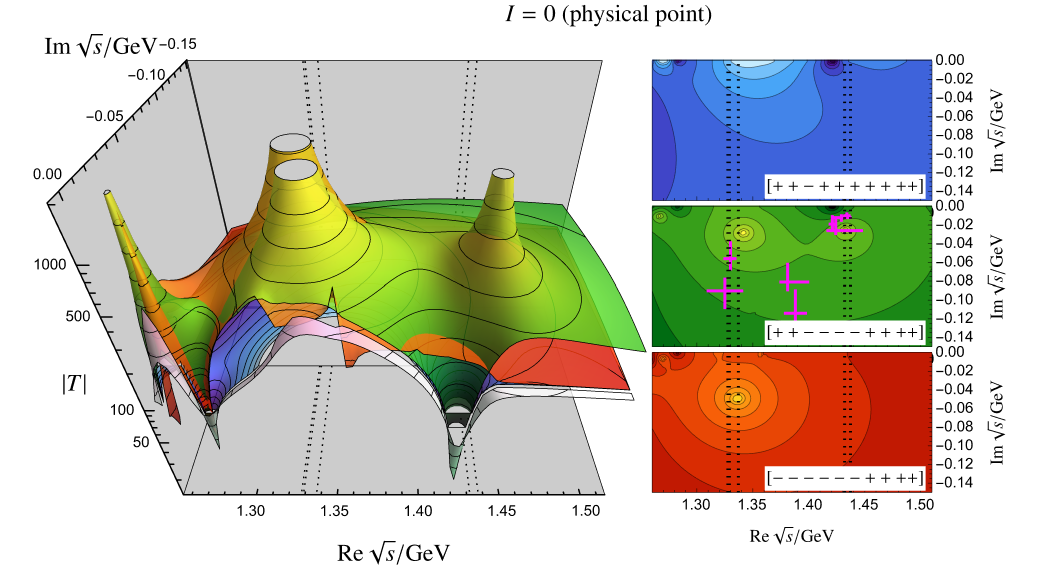}
    \includegraphics[width=0.42\linewidth]{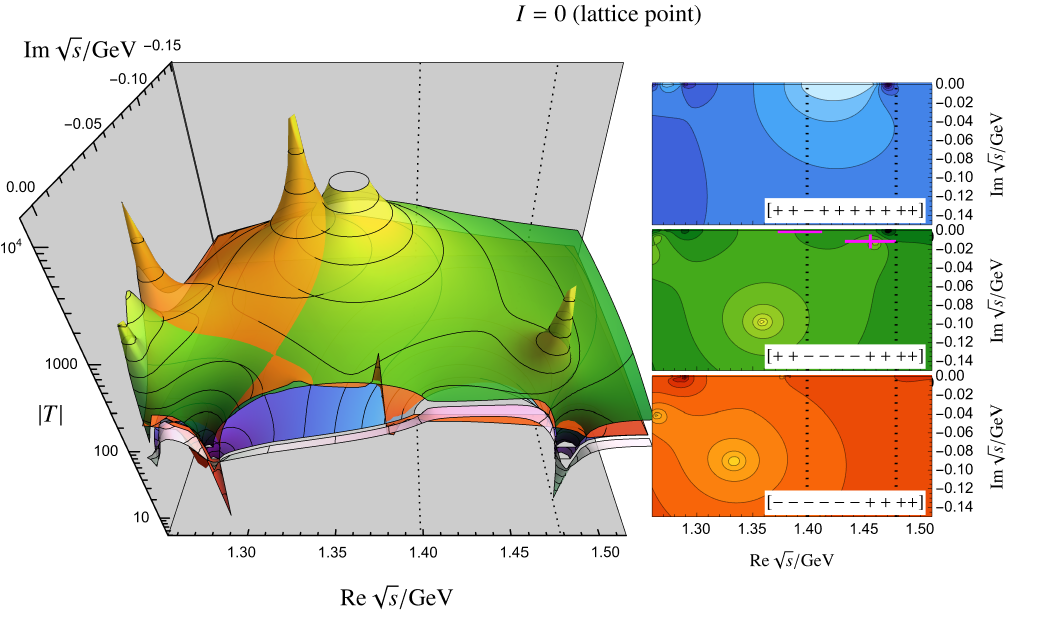}
    \\
    \adjustbox{angle=90,lap=0pt}{~~~~~~~~~M3S2PL $(F_{16})$}
    \includegraphics[width=0.42\linewidth]{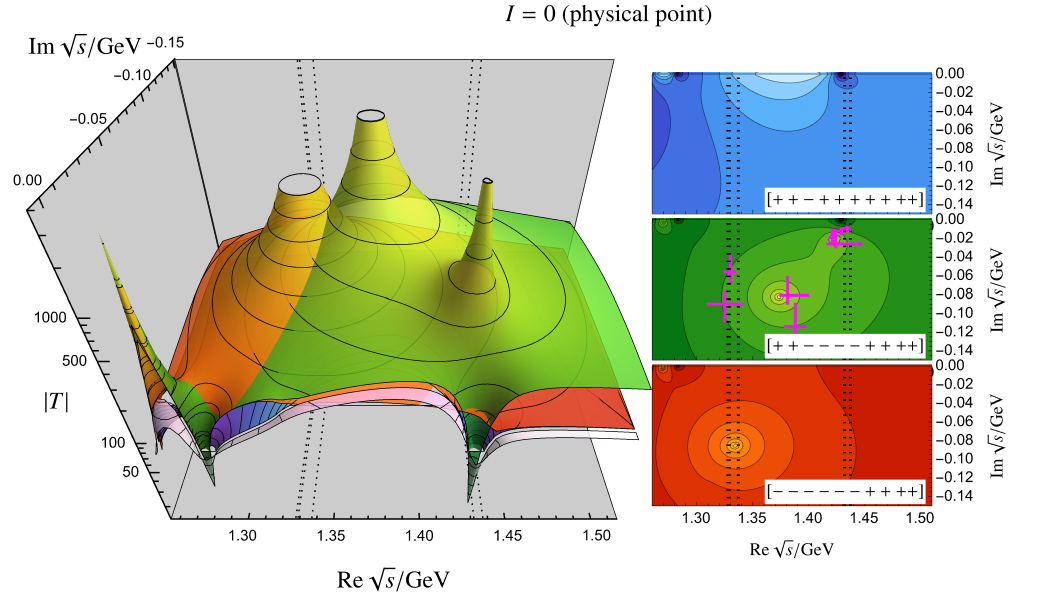}
    \includegraphics[width=0.42\linewidth]{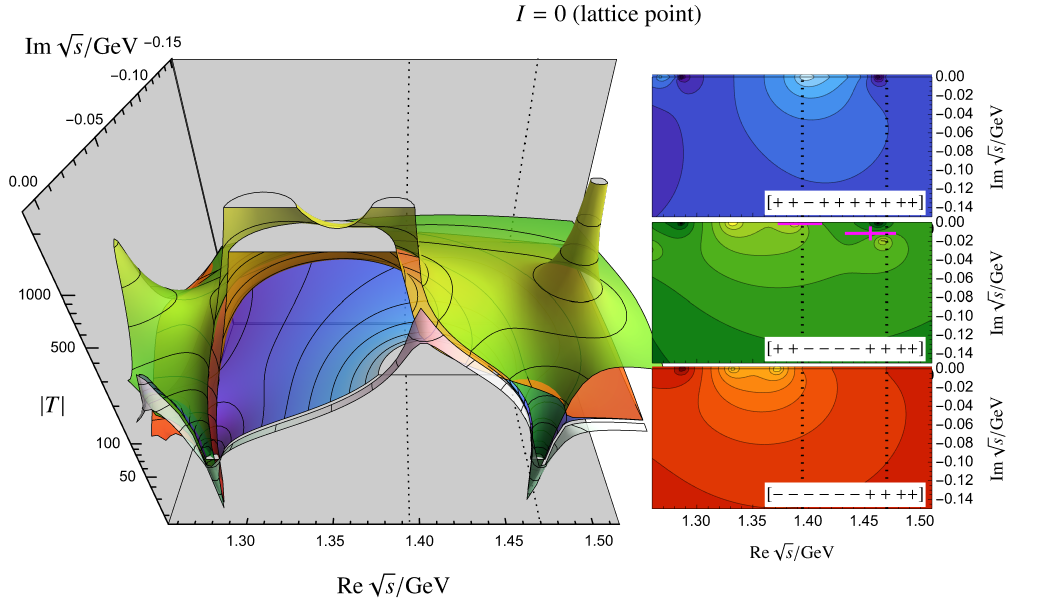}
    \\
    \adjustbox{angle=90,lap=0pt}{~~~~~~~~~M3S3PL $(F_{12})$}
    \includegraphics[width=0.42\linewidth]{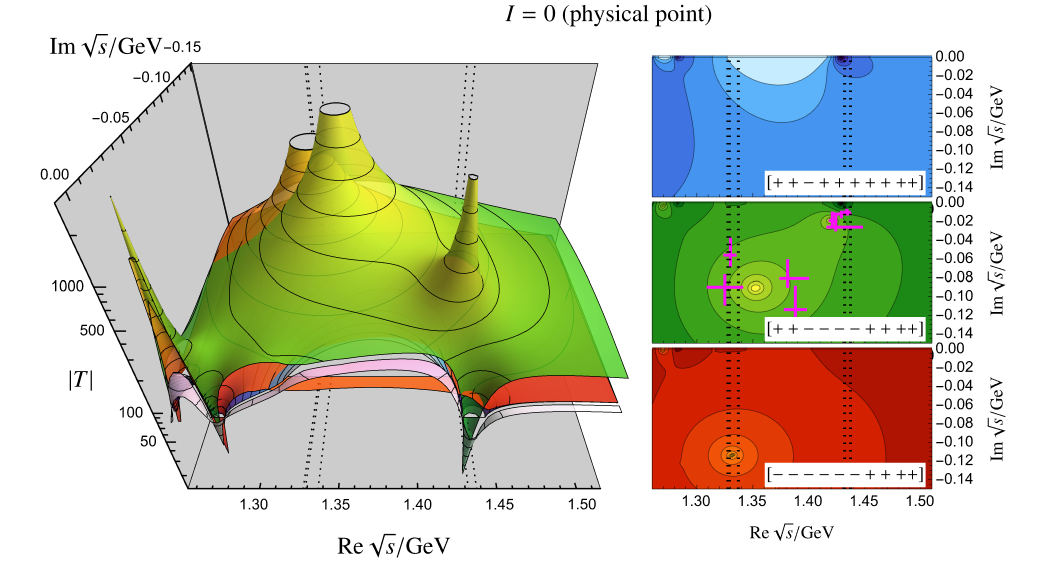}
    \includegraphics[width=0.42\linewidth]{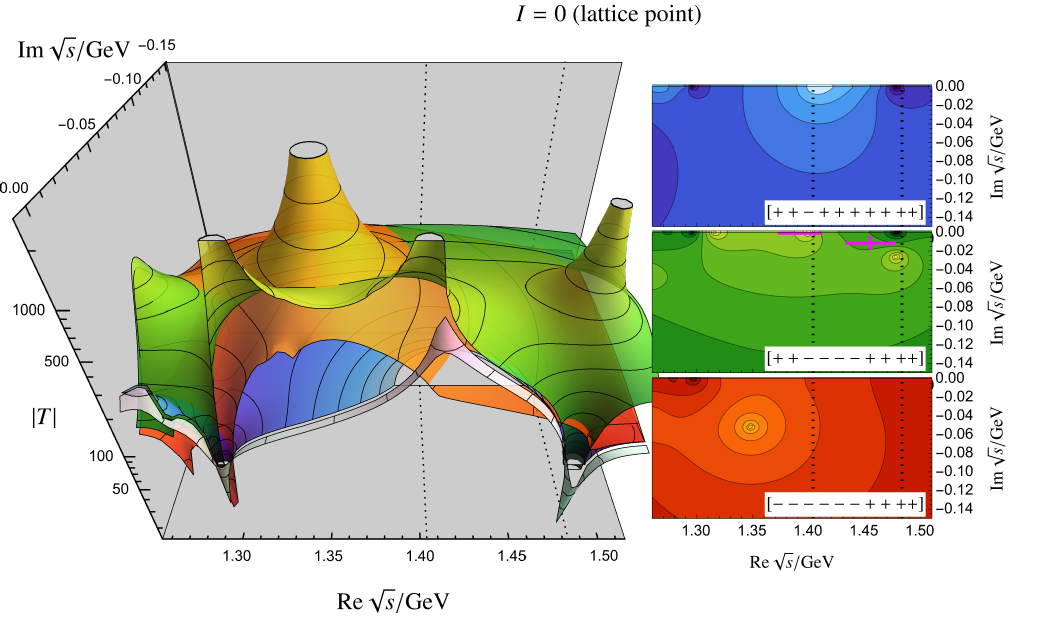}
    \caption{Isoscalar scattering amplitude on the next to the physical (light pink surface for $\Im\sqrt{s}>0$ in the 3D plot) unphysical Riemann sheets (color coded) of the M3Sx type approaches fitted to the lattice and experimental input. More details of the fit and predictions thereof can be found in the Appendix C.}
    \label{fig:RES-lat-and-phys}
\end{figure}

We have previously seen (\cref{subsec:RES-latt-to-phys} and \cref{subsec:RES-phys-to-lat}) that both lattice and experimental input can be successfully fit through the UCHPT model providing in some cases also sensible predictions outside of the fitted range. Still, a more detailed examination shows also that uncertainties are sizable. Thus, a combined fit to lattice and experimental input is performed in this work for the first time.

First, as shown in \cref{subsec:RES-phys-and-lat}, Weinberg-Tomozawa and leading-chiral order UCHPT models (M1 and M2) types are effectively ruled out by not being able to describe the (near)-threshold SIDDHARTA and AMADEUS data. For completeness the results of the combined fit can be found in \cref{fig:fit-summary}. Note that only S3 type are needed to be refit to the combined input, while S1 and S2 types decouple $\chi^2$ contributions from lattice and experiment when no NLO parameters ($b$'s) are used.

The results of the M3-type models for all three regularization schemes S1, S2, S3 are provided in the last three rows (see $F_{17}$, $F_{16}$, $F_{12}$) of \cref{fig:fit-summary}. Overall, a good $\chi^2_{\rm dof}$ is obtained with a relatively flat distribution of the individual contributions from different observables, see right panel of \cref{fig:fit-summary}. Biggest contributions come from cross section data, which again points to the systematic uncertainties within experimental data discussed in \cref{sec:Experimental_data}. Further details of the fit are provided in the Appendix C.

Figure \ref{fig:RES-lat-and-phys} shows the isoscalar pole positions on the relevant unphysical Riemann sheets for the lattice and physical point. All, except the lower pole ($\Lambda(1380)$) of M3S1 type at the lattice point, agree with the reference values (magenta in the figure). Notable is, however, that because of the unknown quark-mass dependence, the subtraction constants $a$ in the S1-type are fitted separately for lattice and physical point. One consequence of this is that the fit is too volatile, depending strongly on the starting values as discussed in \cref{subsec:fit-discussion}. Secondly, this fit is also disfavored in comparison to S2 and S3 by both information criteria despite smaller $\chi^2_{\rm dof}$.

\begin{figure*}[t]
    \includegraphics[width=0.99\linewidth]{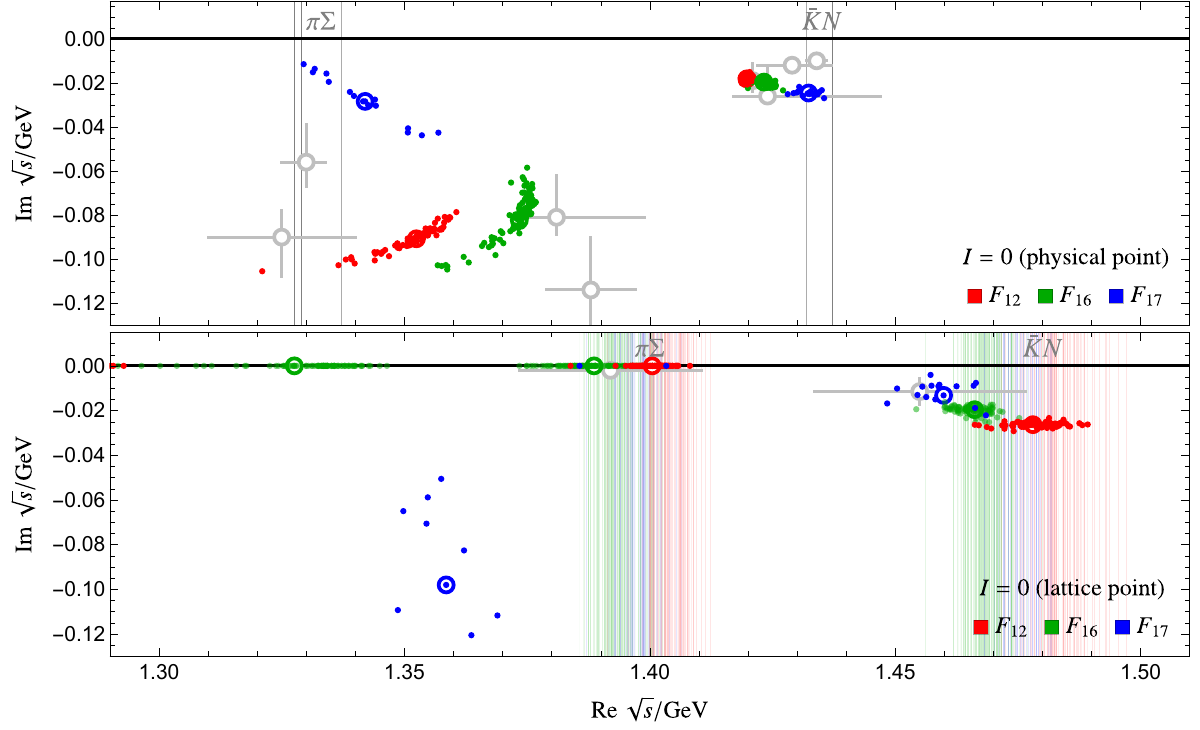}
    \caption{Pole positions for $I=0$ using best global models M3S1 (fit $F_{17}$), M3S2 (fit $F_{16}$), M3S3 (fit $F_{12}$). Pole positions are obtained on the $[++----++++]$ Riemann sheet for physical and unphysical (c.f., $M_\pi\approx 200\MeV$ etc.) quark masses. Circled pole positions are central fits compared to resampled fits as described in the main text. Vertical lines show positions of the two-body thresholds (resampled for the lattice point). Gray dots show the reference values from fits to either lattice or experimental input~\cite{BaryonScatteringBaSc:2023ori,Ikeda:2012au,Guo:2012vv,Mai:2014xna}.}
    \label{fig:BSpoleI0}
\end{figure*}

\begin{figure*}
    \centering
    \includegraphics[width=0.99\linewidth]{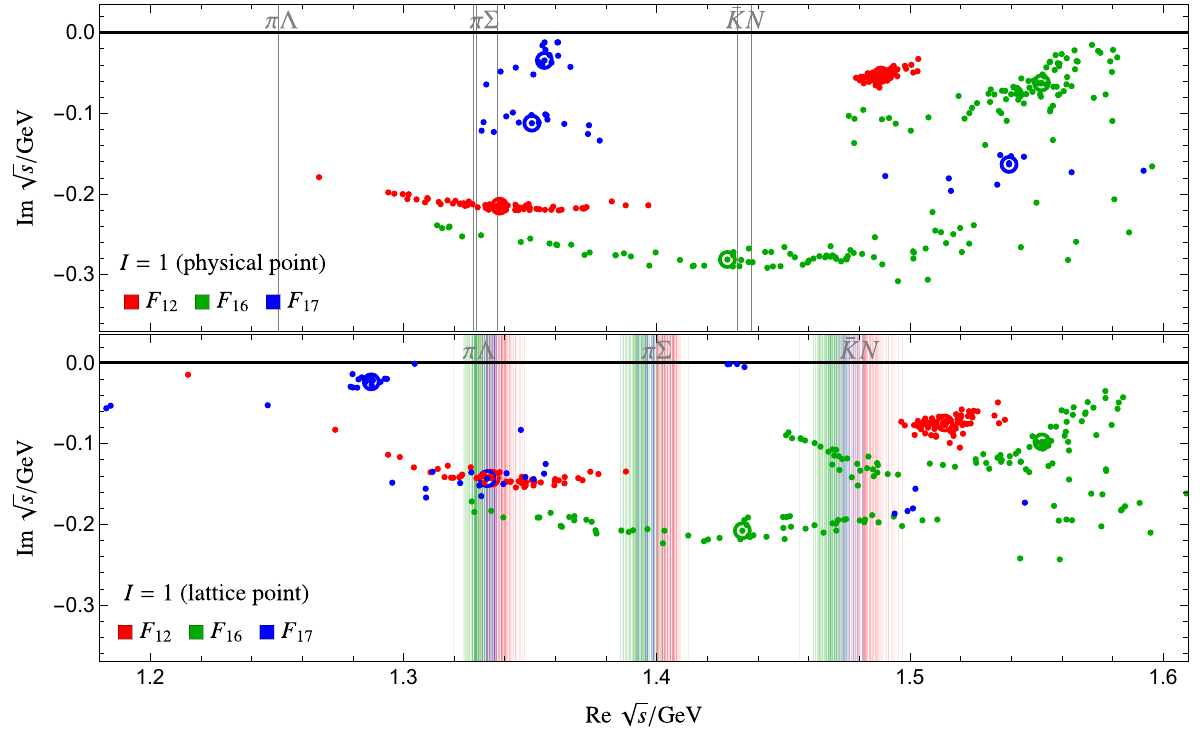}
    \caption{Pole positions for $I=1$ using best global models M3S1 (fit $F_{17}$), M3S2 (fit $F_{16}$), M3S3 (fit $F_{12}$). Pole positions are obtained on the $[++----++++]$ Riemann sheet for physical and unphysical (c.f., $M_\pi\approx 200\MeV$ etc.) quark masses. Circled pole positions are central fits compared to resampled fits as described in the main text. Vertical lines show the positions of the two-body thresholds (resampled for the lattice point).}
    \label{fig:BSpoleI1}
\end{figure*}

To further examine the uncertainty associated with the pole positions we resample the obtained fits by varying the input according to the provided (statistical) uncertainties. In the Lattice QCD case this is directly accomplished using provided bootstrap samples, whereas in the experimental case parametric bootstrap samples are generated by drawing synthetic datasets from uncorrelated, Gaussian-distributed data points, using the reported central values as the means and the quoted uncertainties as the standard deviations. The final result is provided in the summary plot \cref{fig:BSpoleI0} where systematic (model types) and statistical errors (re-sampling) are included and compared to the reference values from literature. Note that the latter were obtained through fits to either experimental or lattice input. The physical point result agrees nicely with the previous phenomenological fits also reflecting the large uncertainty of the $\Lambda(1380)$ pole positions. The $\Lambda(1405)$ is narrowed down to a very small region. At the lattice point, the position of the latter state supports the CHPT-independent determination of the lattice collaboration (BaSc~\cite{BaryonScatteringBaSc:2023ori, BaryonScatteringBaSc:2023zvt}) but tends to be slightly larger in real and imaginary part. In fits S2 ($F_{16}$) and S3 ($F_{12}$) the  $\Lambda(1380)$ is found for all bootstrap samples on the real axis just below the $\pi\Sigma$ threshold. Note that thresholds at the lattice point are also subject to resampling which is very important to keep track of. Individual thresholds are represented by the color-coded vertical lines. We also observe a second virtual bound state pole (c.f., $\sqrt{s}\approx 1.33$ for $F_{16}$) which is required due to analyticity. For a related discussion in the context of this and other excited hadrons see, e.g., Ref.~\cite{Mai:2019pqr, Guo:2018zss, Molina:2015uqp, Doring:2016bdr, Bruns:2024yga, Guo:2023wes,Zhuang:2024udv}. Numerical values are provided in~\cref{tab:overall-poles}. A critical observation is, however, that there is a non-negligible set of solution (S1-type $F_{17}$) predicting a $\Lambda(1380)$ resonance pole away from the real axis. We have checked explicitly that the poles are smoothly varying when moving along a linearized trajectory between the lattice and physical point. While disfavored by the $AIC$ or $BIC$ the existence of such solutions draws at least a shadow of a doubt that the pole positions of the $\Lambda(1380)$ state are resolved through the currently available (lattice and experimental) input. Of course the existence of both poles is undisputed by this and seems to be now solidified by the combination of UCHPT, lattice and experimental inputs.

As a final observation we also provide predictions of the pole positions for the isovector case including systematic and statistical uncertainties. The result is depicted in \cref{fig:BSpoleI1} referring again to the Riemann sheet $[++----++++]$ connected to the physical real axis between $\pi\Sigma$ and $\bar K N$ thresholds. At the physical point we observe for all fit types a broad state with a width of around $\Gamma\approx 200-400\MeV$ and mass above $1300\MeV$. Solution $F_{17}$ provides a second state with a lower width which is possibly a sign for an over-fit. Other states far above $\bar K N$ threshold also exist but their influence on the observables at real energies are expected to be negligible. At the lattice point the poles mostly do not move much except of the narrow $F_{17}$ pole. Presumably, this is simply due to the large width of the found states, which therefore have little effect on the real energy-axis where the input either from experiment or lattice is provided at. Numerical values are provided in \cref{tab:overall-poles}. We conclude that existence of the isovector, negative-strangeness excited baryon state is very likely but because of its large width its position is currently very hard to resolve. Turning this argument around, this means that lattice result in this sector are highly desired.

\begin{acknowledgments}
We thank BaSc collaboration for providing details of the lattice calculation, and Peter Bruns, Fernando Romero-López, Miguel Salg, Juan Fernandez de la Garza, Feng-Kun Guo and Christoph Hanhart for entertaining and useful discussions.
The work of MM was funded through the Heisenberg Programme by the Deutsche Forschungsgemeinschaft (DFG, German Research Foundation) – 532635001.
The work of UGM was supported in part by the CAS President's International Fellowship Initiative (PIFI) (Grant No.~2025PD0022)
by the MKW NRW under the funding code NW21-024-A, and
by the Deutsche Forschungsgemeinschaft (DFG, German Research Foundation) as part of the CRC 1639 NuMeriQS – project no. 511713970.
The work of FP was supported by the projects PulseQCD, DeNuTra, MuonHVP(EXCELLENCE/0524/0269,EXCELLENCE/524/0455, EXCELLENCE/524/0017) co-financed by the European Regional Development Fund and the Republic
of Cyprus through the Research and Innovation Foundation.
RFF, PH, DGI and BM were supported by the United Kingdom’s
Science and Technology Facilities Council (STFC) from
Grant No. ST/Y000315/1.
\end{acknowledgments}

\bibliography{BIB.bib}

\clearpage
\renewcommand{\subsection}[1]{\par\medskip\noindent\textbf{#1}\par\medskip}
\renewcommand{\section}[1]{\centering\par\medskip\noindent\textbf{#1}\par\medskip}

\section{Appendix A: Detailed fit results. Fits to the lattice input.}
\addcontentsline{toc}{section}{Appendix A: Detailed fit results. Fits to the lattice input.}
\label{app:fit-results-lattice}

\subsection{M1S1L ($F_{19}$)}

\begin{figure*}[h]
\begin{minipage}[t!]{0.23\textwidth}
\centering
 \caption{Subtraction constants at the lattice point for M1S1L ($F_{19}$).}
    \label{tab:app-19}
\begin{tabular}{|c|c|}
\hline
$\chi^2_{\rm dof}$ & 1.36 \\
\hline
$a_{\bar{K}N}$ & 2.108217e-03 \\
$a_{\pi\Lambda}$ & -1.079700e-01 \\
$a_{\pi\Sigma}$ & 2.108217e-03 \\
$a_{\eta\Lambda}$ & 2.939451e-04 \\
$a_{\eta\Sigma}$ & 2.163700e-01 \\
$a_{K\Xi}$ & 3.948000e-02 \\
\hline
\end{tabular}
\end{minipage}
\hfill
\begin{minipage}[t!]{0.7\textwidth}
    \vspace{-6pt} 
    \includegraphics[width=\linewidth]{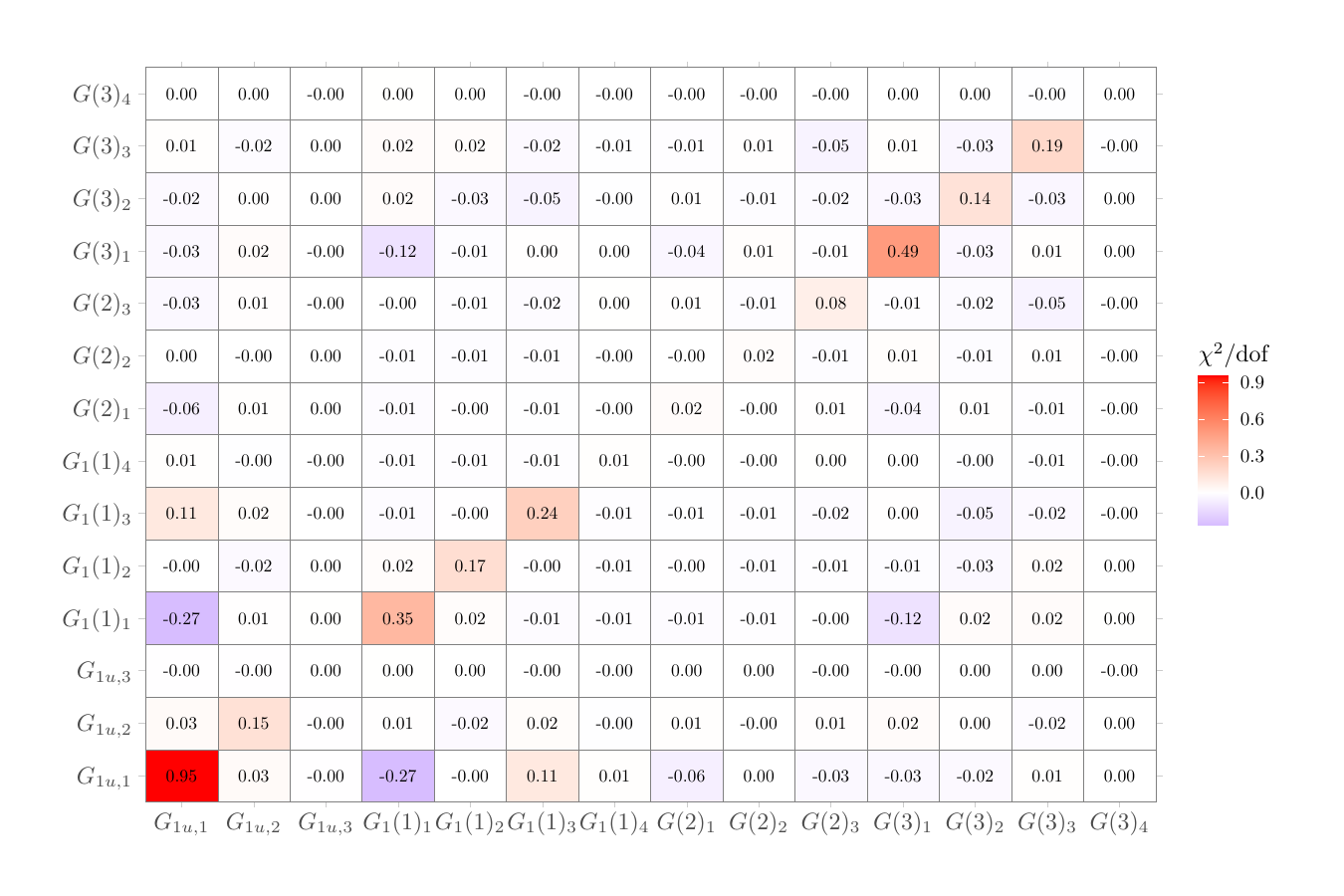}
    \vspace{-26pt}
    \caption{Heat map of correlated $\chi^2_{{\rm dof},ij}$, highlighting the relative impact of each energy level on the total fit quality.}
    \label{fig:partial-chi2_M1S1L}
\end{minipage}
\end{figure*}
\vspace{-6pt}


\begin{figure*}[h]
    \includegraphics[height=4cm]{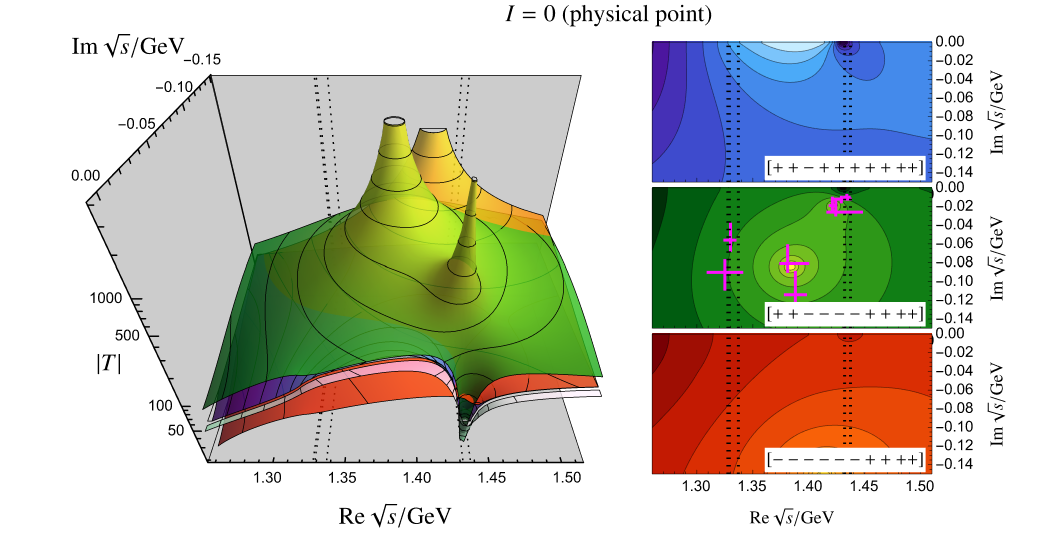}
    \includegraphics[height=4cm]{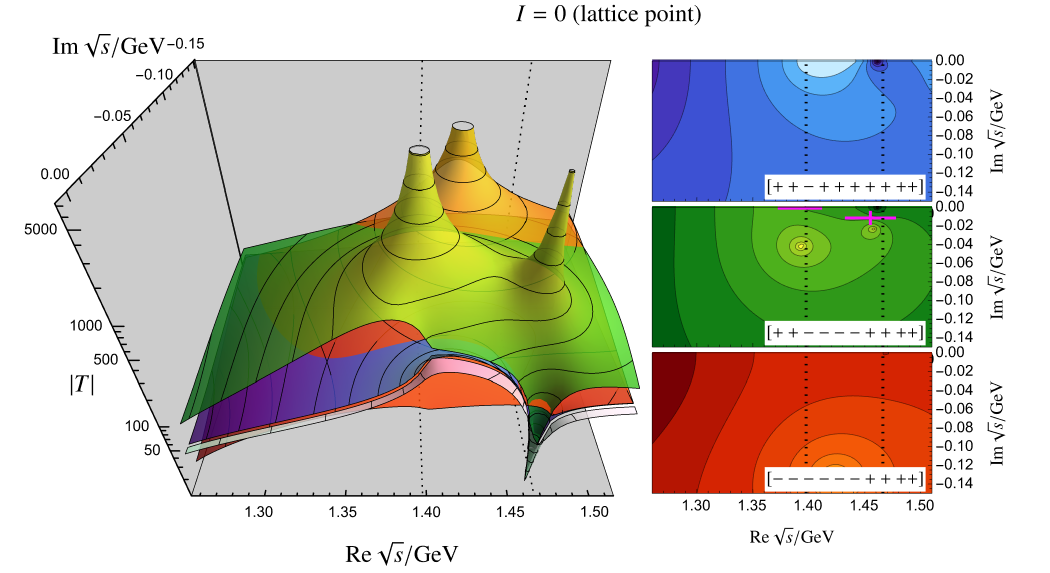}\\
    \includegraphics[height=4cm]{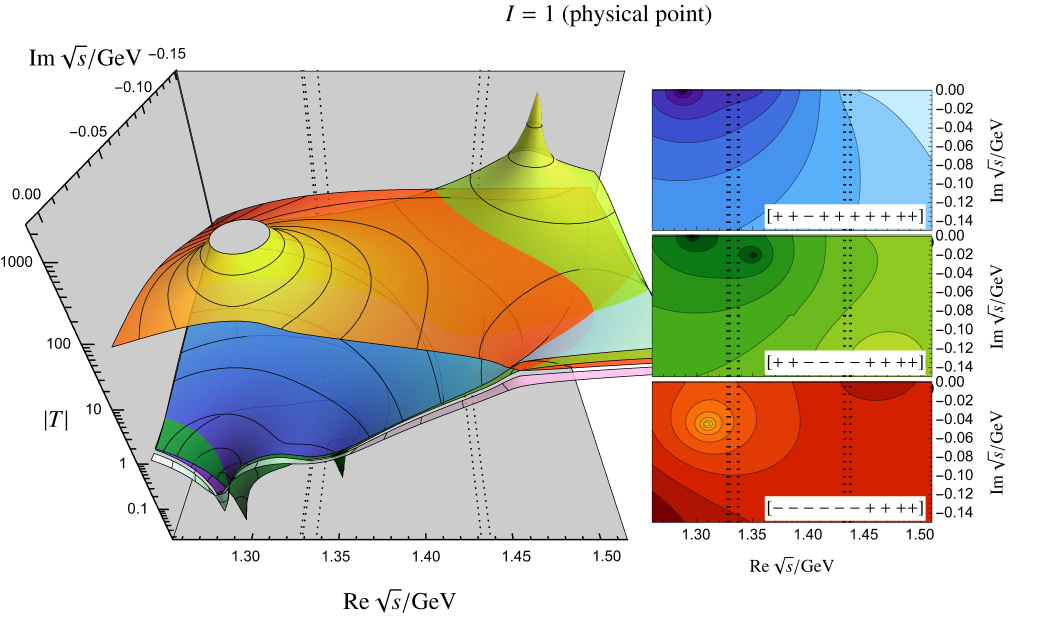}
    \includegraphics[height=4cm]{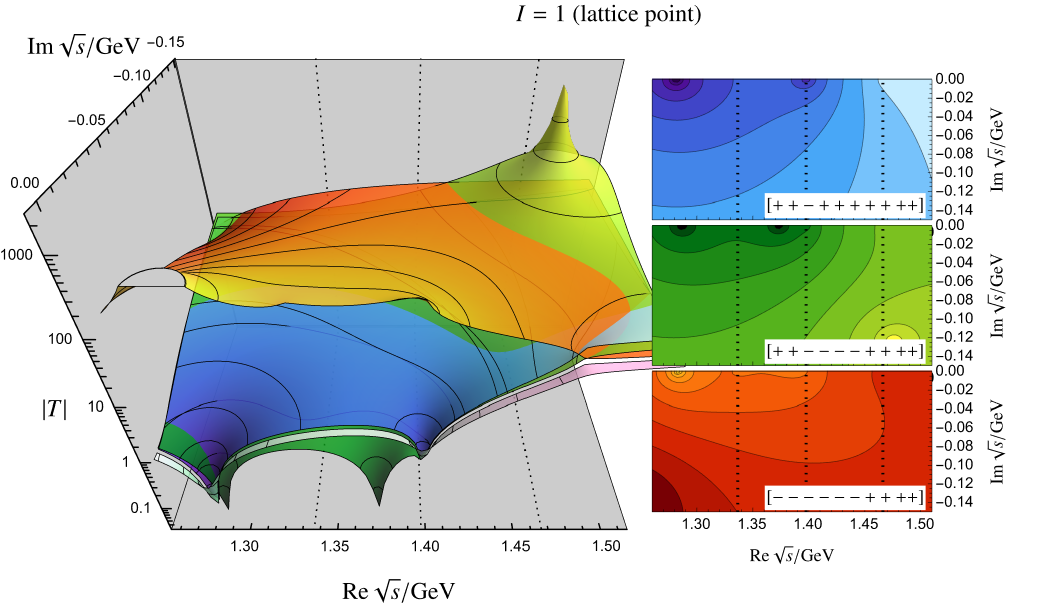}
    \caption{Isoscalar and isovector projected absolute value of the $\pi\Sigma\to \pi\Sigma$ scattering amplitude on the unphysical second Riemann sheets. Nomenclature as in the main text. Magenta crosses represent literature values from Refs.~\cite{BaryonScatteringBaSc:2023ori, BaryonScatteringBaSc:2023zvt} for the lattice point and Refs.~\cite{Ikeda:2012au,Guo:2012vv,Mai:2014xna}.}
    \label{fig:app-19}
\end{figure*}

\clearpage
\subsection{M1S2L ($F_{31}$)}

\begin{figure*}[h]
\begin{minipage}[t!]{0.23\textwidth}
    \vspace{-9pt} 
\centering
\caption{Total $\chi^2_\mathrm{dof}$ for the parameter-free M1S2L ($F_{31}$).}
    \label{tab:details_31}
\begin{tabular}{|c|c|}
\hline
$\chi^2_{\rm dof}$ & 2.89 \\
\hline
\end{tabular}
\end{minipage}
\hfill
\begin{minipage}[t!]{0.7\textwidth}
    \vspace{-6pt} 
    \includegraphics[width=\linewidth]{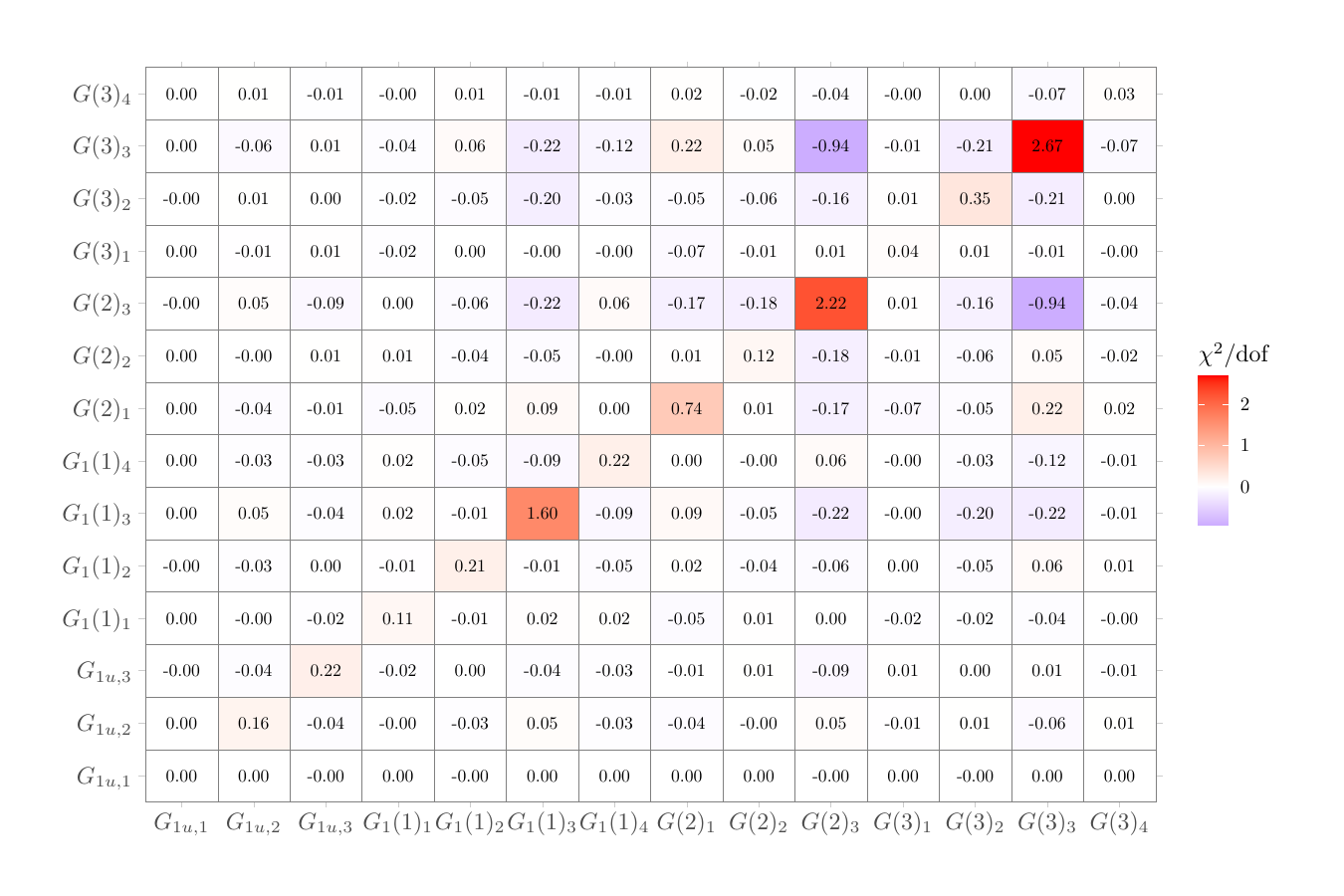}
    \vspace{-26pt}
    \caption{Heat map of correlated $\chi^2_{{\rm dof},ij}$, highlighting the relative impact of each energy level on the total fit quality.}
    \label{fig:partial-chi2_M1S2L}
\end{minipage}
\end{figure*}

\begin{figure*}[h]
    \includegraphics[height=4cm]{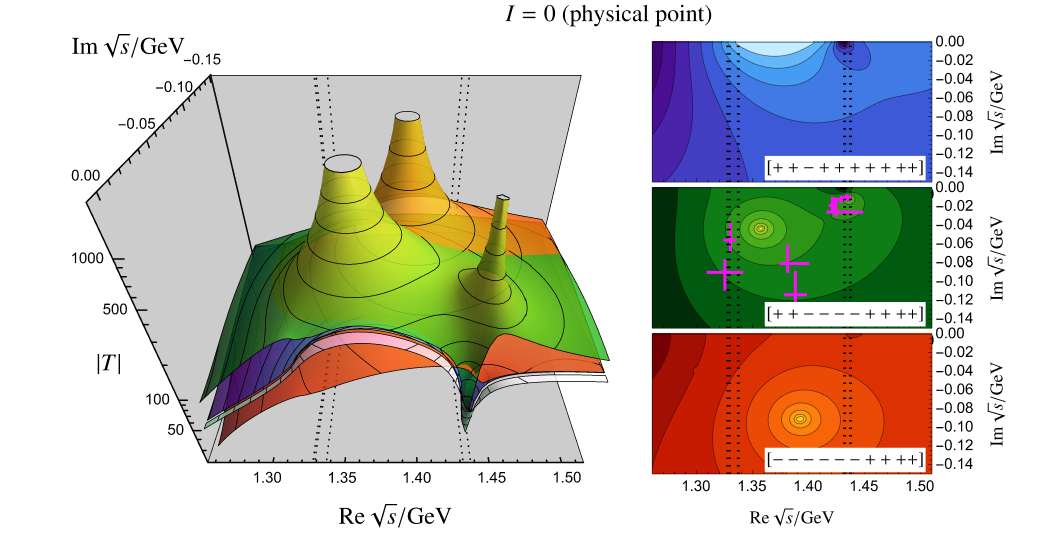}
    \includegraphics[height=4cm]{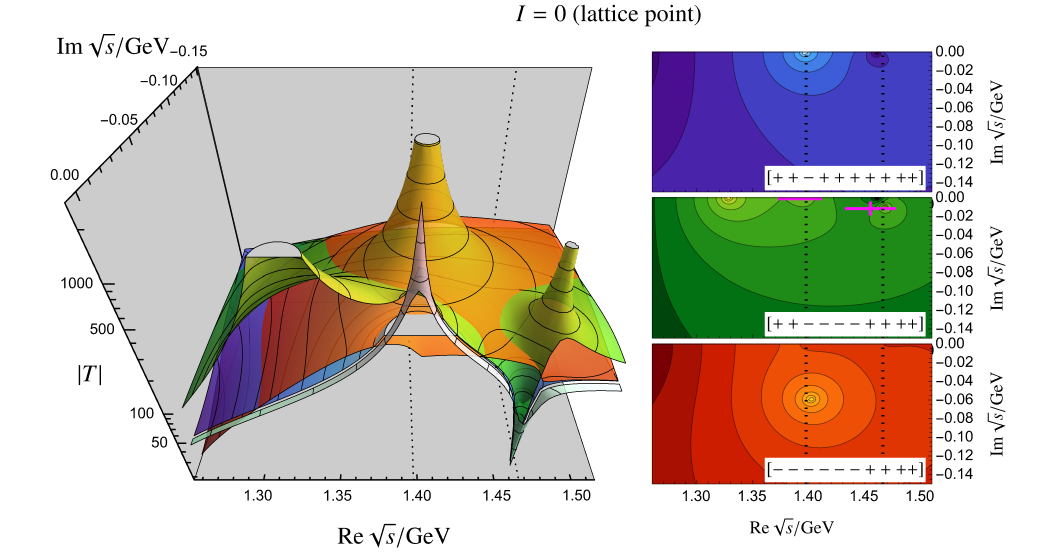}\\
    \includegraphics[height=4cm]{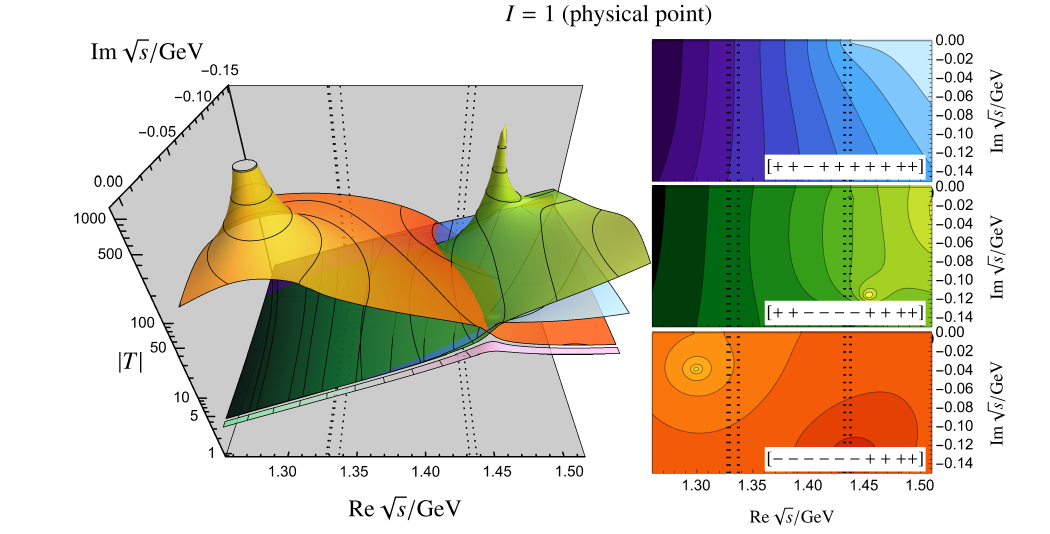}
    \includegraphics[height=4cm]{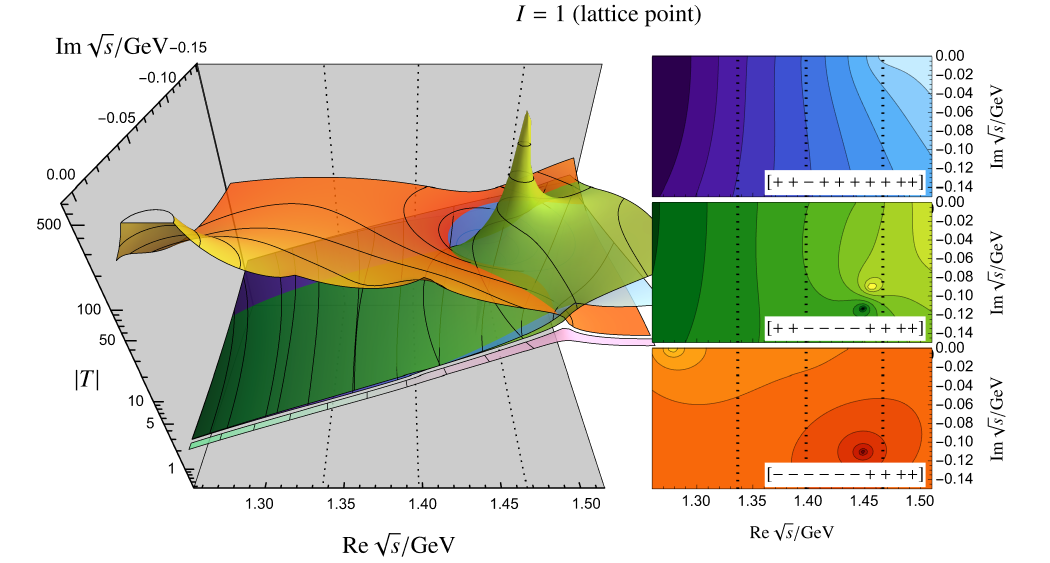}
    \caption{Isoscalar and isovector projected absolute value of the $\pi\Sigma\to \pi\Sigma$ scattering amplitude on the unphysical second Riemann sheets. Nomenclature as in the main text. Magenta crosses represent literature values from Refs.~\cite{BaryonScatteringBaSc:2023ori, BaryonScatteringBaSc:2023zvt} for the lattice point and Refs.~\cite{Ikeda:2012au,Guo:2012vv,Mai:2014xna} for the physical point.}
    \label{fig:app-31}
\end{figure*}

\clearpage
\subsection{M1S3L ($F_{18}$)}

\begin{figure*}[h]
\begin{minipage}[t!]{0.23\textwidth}
    \vspace{-9pt} 
\centering
\caption{Total $\chi^2_\mathrm{dof}$ and the $\Lambda$ parameter
for M1S3L ($F_{18}$).}
    \label{tab:details_18}
\begin{tabular}{|c|c|}
\hline
$\chi^2_{\rm dof}$ & 4.42 \\
\hline
$\Lambda[\mathrm{GeV}]$ & 0.6892541 \\
\hline
\end{tabular}
\end{minipage}
\hfill
\begin{minipage}[t!]{0.7\textwidth}
    \vspace{-6pt} 
    \includegraphics[width=\linewidth]{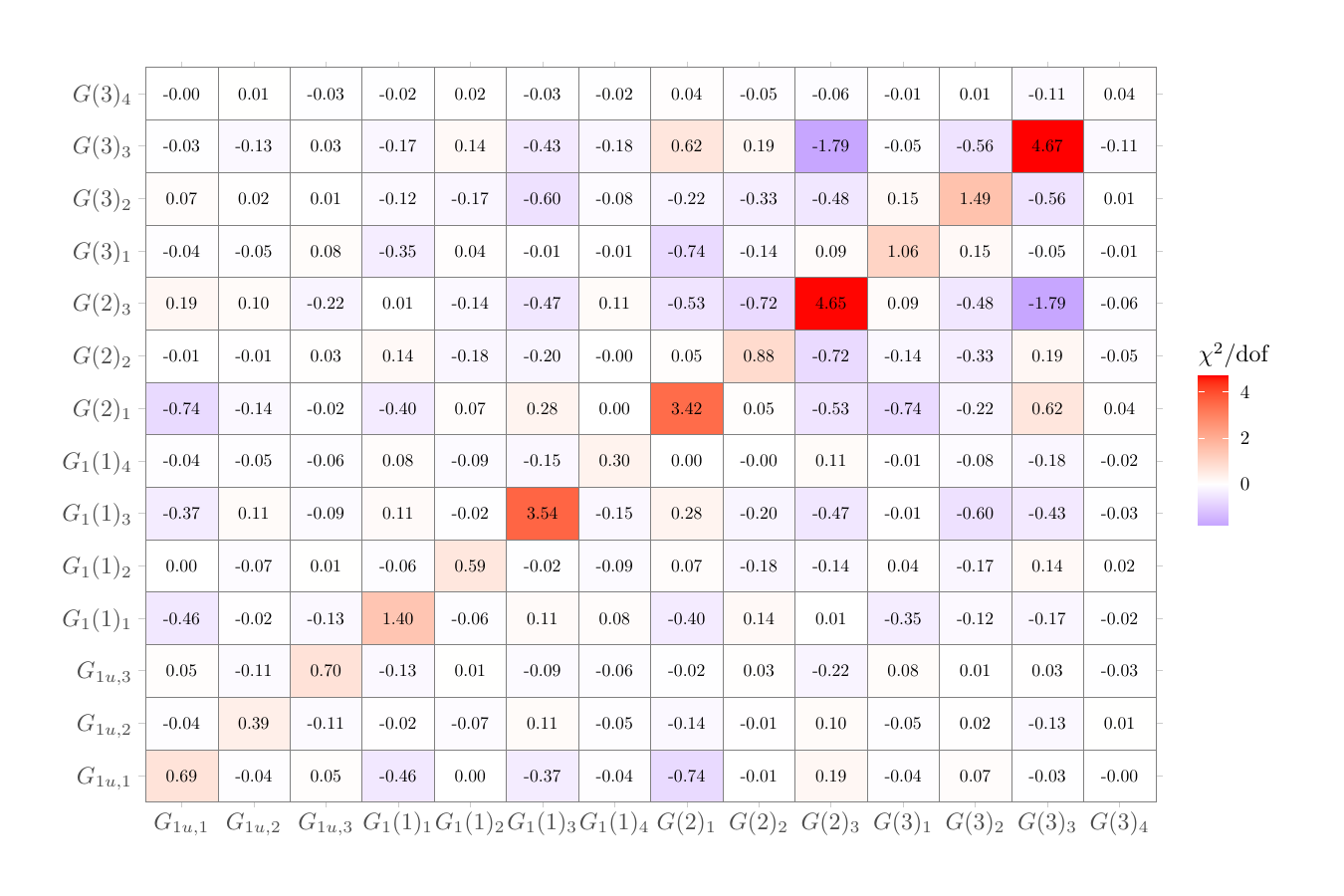}
    \vspace{-26pt}
    \caption{Heat map of correlated $\chi^2_{{\rm dof},ij}$, highlighting the relative impact of each energy level on the total fit quality.}
    \label{fig:partial-chi2_M1S3L}
\end{minipage}
\end{figure*}

\begin{figure*}[h]
    \includegraphics[height=4cm]{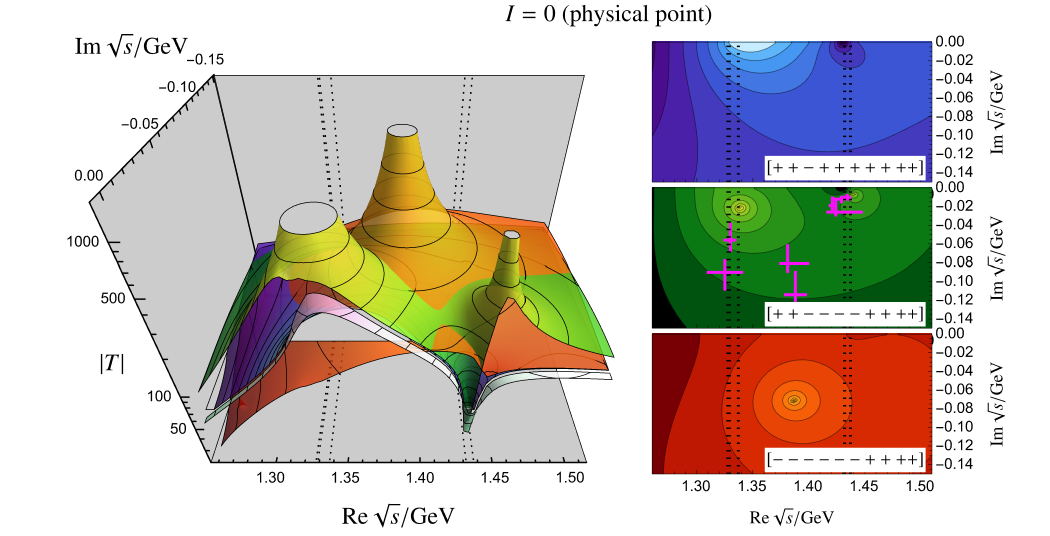}
    \includegraphics[height=4cm]{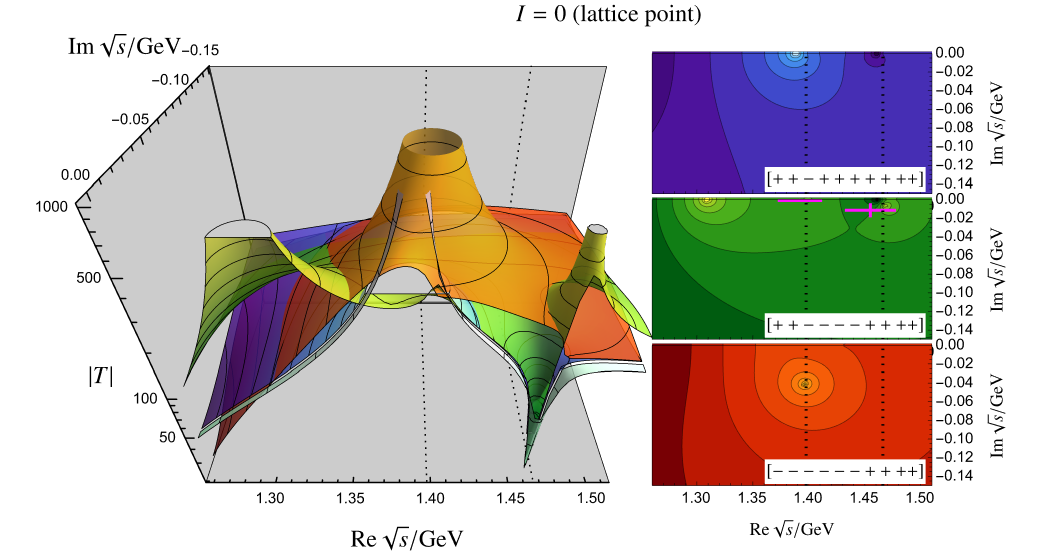}\\
    \includegraphics[height=4cm]{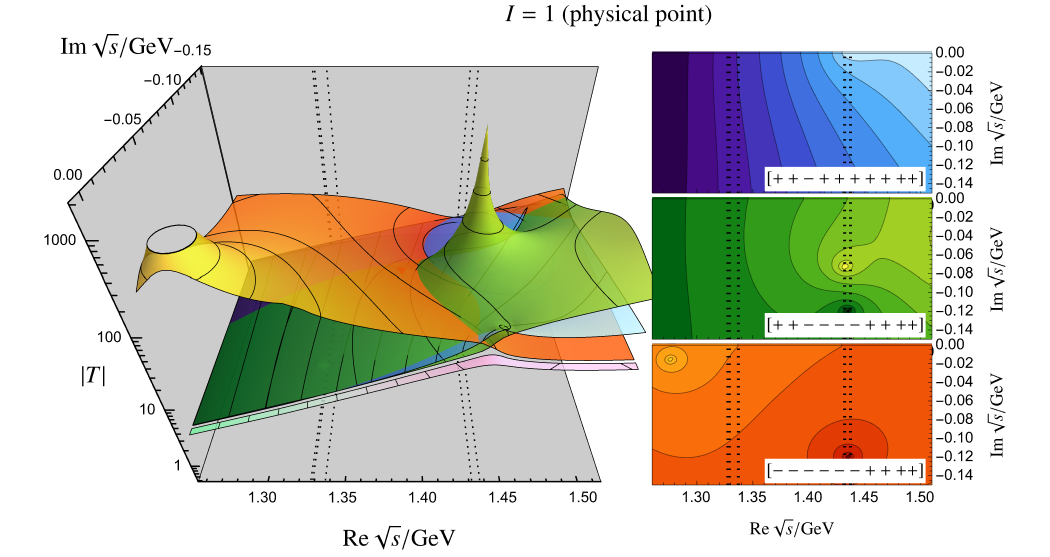}
    \includegraphics[height=4cm]{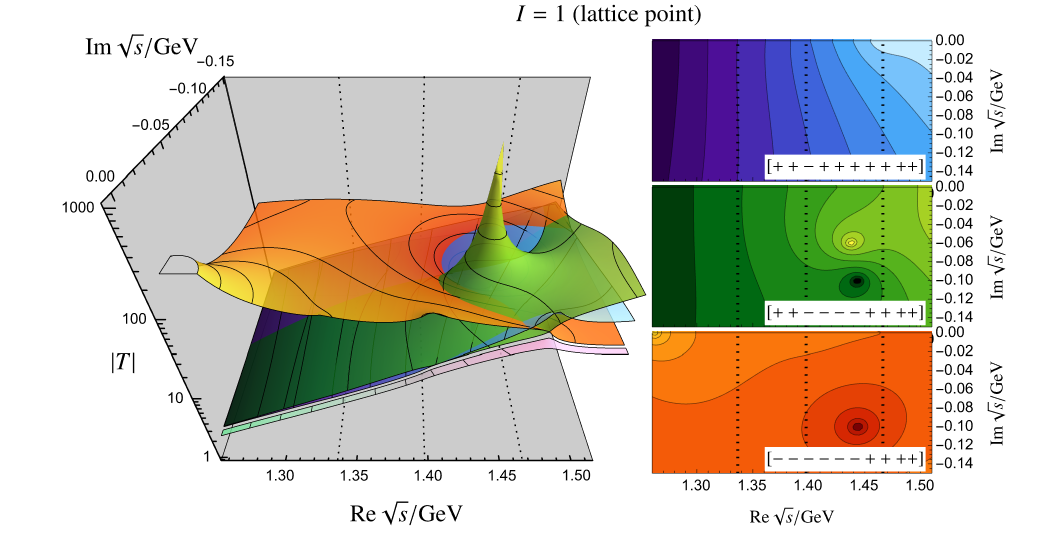}
    \caption{Isoscalar and isovector projected absolute value of the $\pi\Sigma\to \pi\Sigma$ scattering amplitude on the unphysical second Riemann sheets. Nomenclature as in the main text. Magenta crosses represent literature values from Refs.~\cite{BaryonScatteringBaSc:2023ori, BaryonScatteringBaSc:2023zvt} for the lattice point and Refs.~\cite{Ikeda:2012au,Guo:2012vv,Mai:2014xna} for the physical point.}
    \label{fig:app-20}
\end{figure*}

\clearpage
\subsection{M2S1L ($F_{20}$)}

\begin{figure*}[h]
\begin{minipage}[t!]{0.23\textwidth}
    \vspace{0pt} 
\centering
\caption{Subtraction constants at the lattice point for the M2S1L ($F_{20}$).}
\label{tab:details_20}
\begin{tabular}{|c|c|}
\hline
$\chi^2_{\rm dof}$ & 1.42 \\
\hline
$a_{\bar{K}N}$ & 2.249408e-03 \\
$a_{\pi\Lambda}$ & -1.079700e-01 \\
$a_{\pi\Sigma}$ & 2.249408e-03 \\
$a_{\eta\Lambda}$ & -5.364455e-03 \\
$a_{\eta\Sigma}$ & 2.163700e-01 \\
$a_{K\Xi}$ & 3.948000e-02 \\
\hline
\end{tabular}
\end{minipage}
\hfill
\begin{minipage}[t!]{0.7\textwidth}
    \vspace{-6pt} 
    \includegraphics[width=\linewidth]{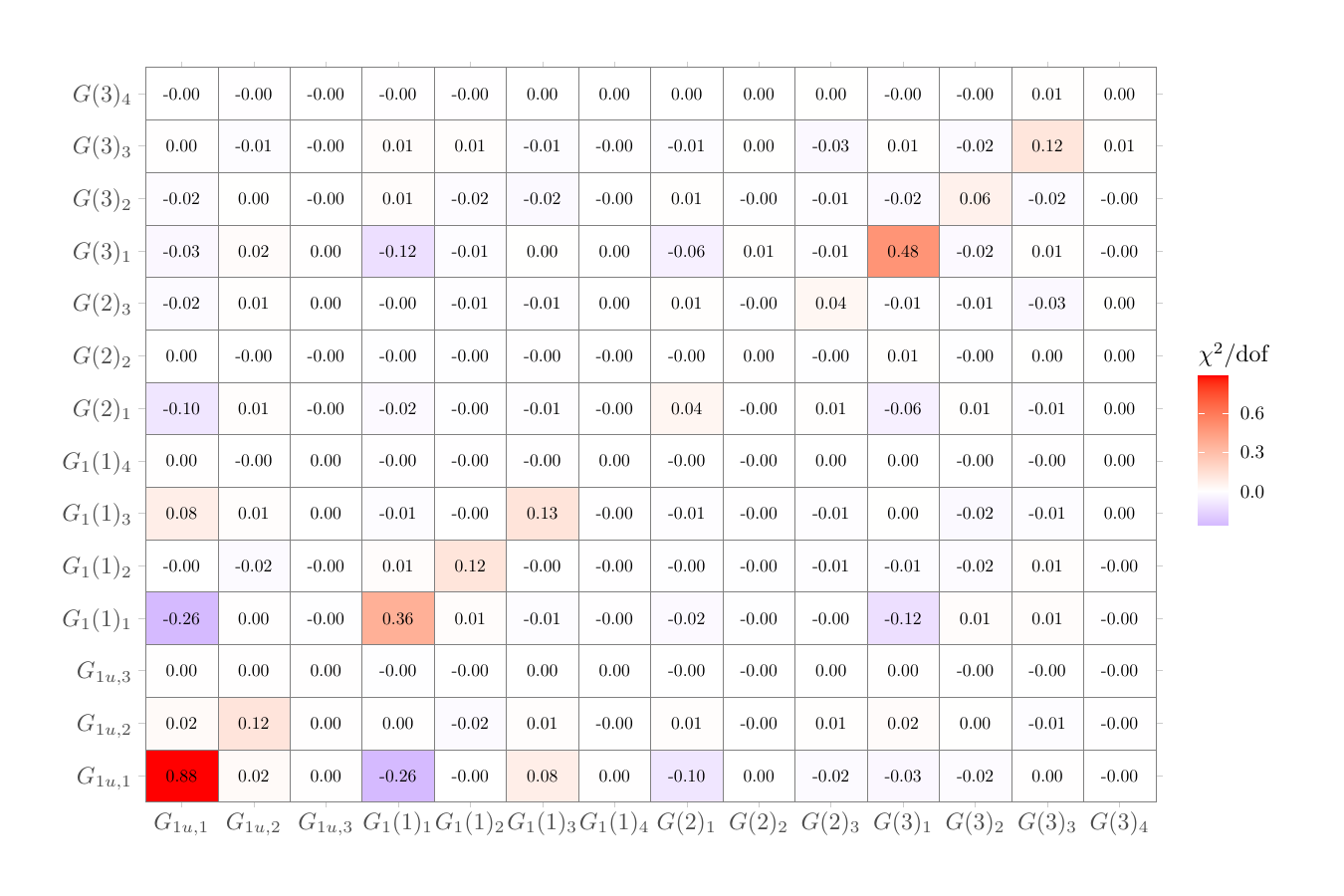}
    \vspace{-26pt}
    \caption{Heat map of correlated $\chi^2_{{\rm dof},ij}$, highlighting the relative impact of each energy level on the total fit quality.}
    \label{fig:partial-chi2_M2S1L}
\end{minipage}
\end{figure*}

\begin{figure*}[h]
    \includegraphics[height=4cm]{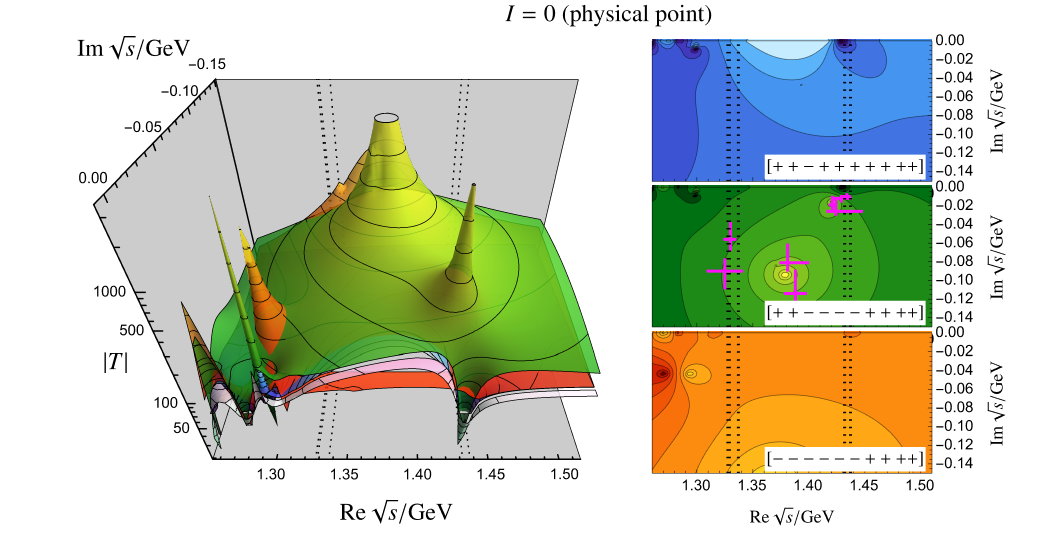}
    \includegraphics[height=4cm]{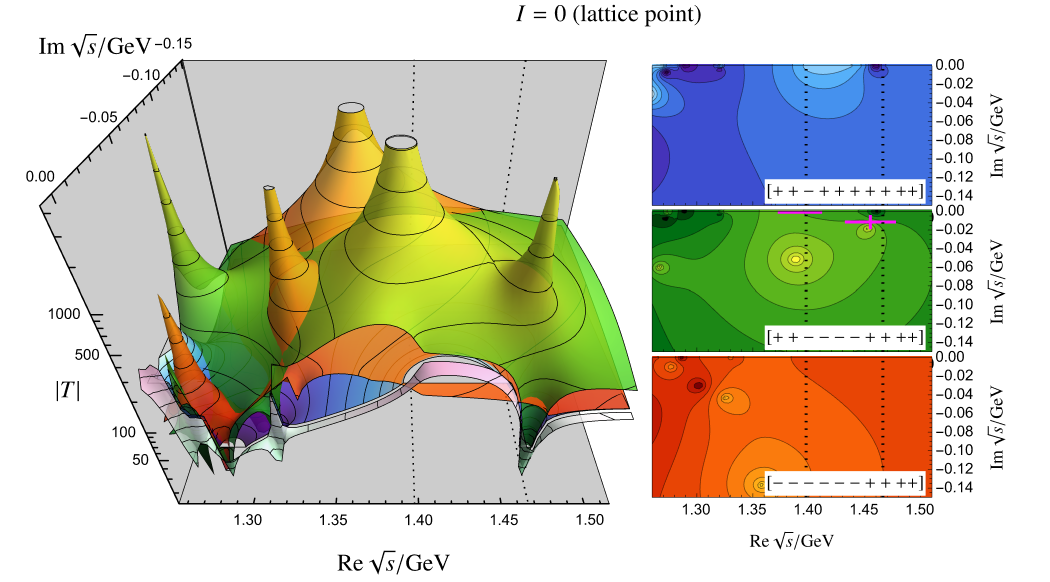}\\
    \includegraphics[height=4cm]{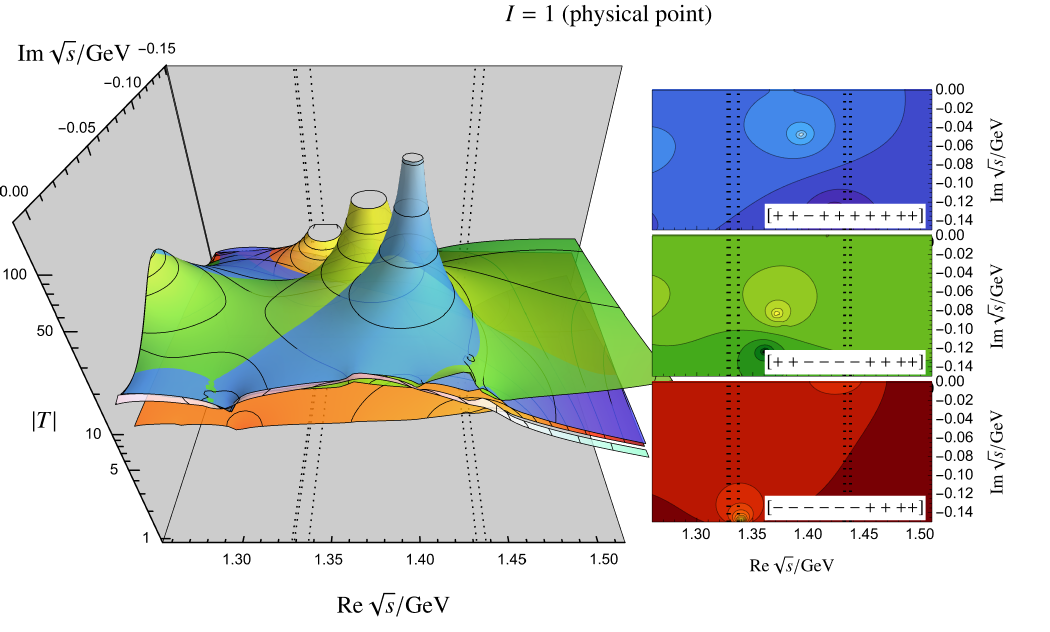}
    \includegraphics[height=4cm]{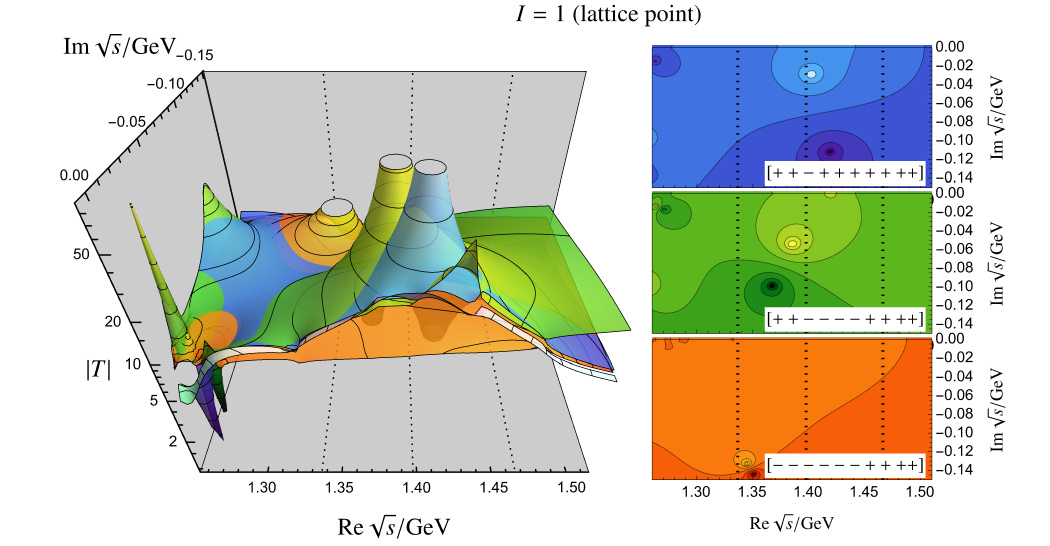}
    \caption{Isoscalar and isovector projected absolute value of the $\pi\Sigma\to \pi\Sigma$ scattering amplitude on the unphysical second Riemann sheets. Nomenclature as in the main text. Magenta crosses represent literature values from Refs.~\cite{BaryonScatteringBaSc:2023ori, BaryonScatteringBaSc:2023zvt} for the lattice point and Refs.~\cite{Ikeda:2012au,Guo:2012vv,Mai:2014xna} for the physical point.}
    \label{fig:app-32}
\end{figure*}

\clearpage
\subsection{M2S2L ($F_{32}$)}

\begin{figure*}[h]
\begin{minipage}[t!]{0.23\textwidth}
    \vspace{0pt} 
\centering
\caption{Total $\chi^2_\mathrm{dof}$ for the parameter-free
 M2S2L ($F_{32}$).}
\label{tab:details_32}
\begin{tabular}{|c|c|}
\hline
$\chi^2_{\rm dof}$ & 2.68 \\
\hline
\end{tabular}
\end{minipage}
\hfill
\begin{minipage}[t!]{0.7\textwidth}
    \vspace{-6pt} 
    \includegraphics[width=\linewidth]{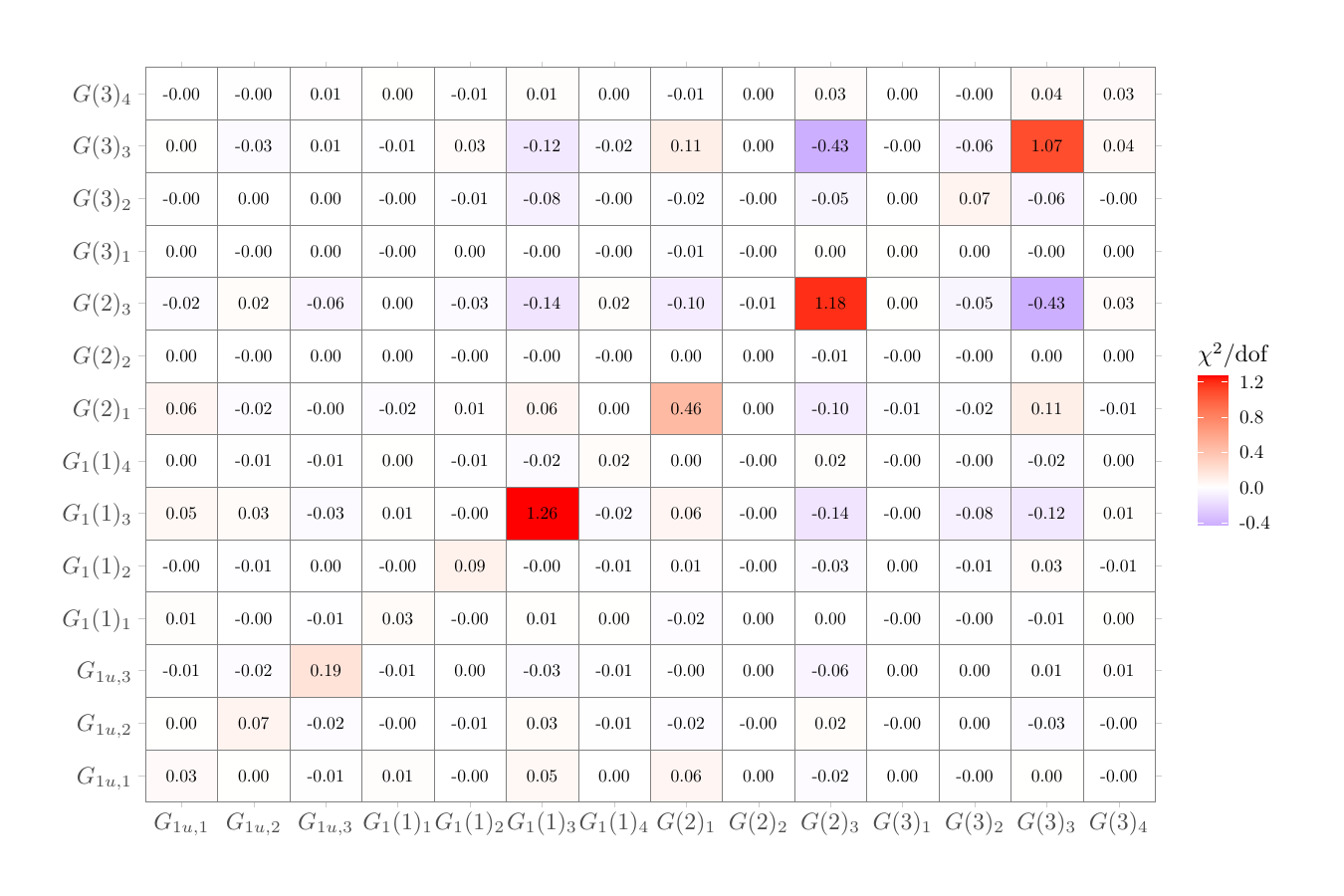}
    \vspace{-26pt}
    \caption{Heat map of correlated $\chi^2_{{\rm dof},ij}$, highlighting the relative impact of each energy level on the total fit quality.}
    \label{fig:partial-chi2_M2S2L}
\end{minipage}
\end{figure*}

\begin{figure*}[h]
    \includegraphics[height=4cm]{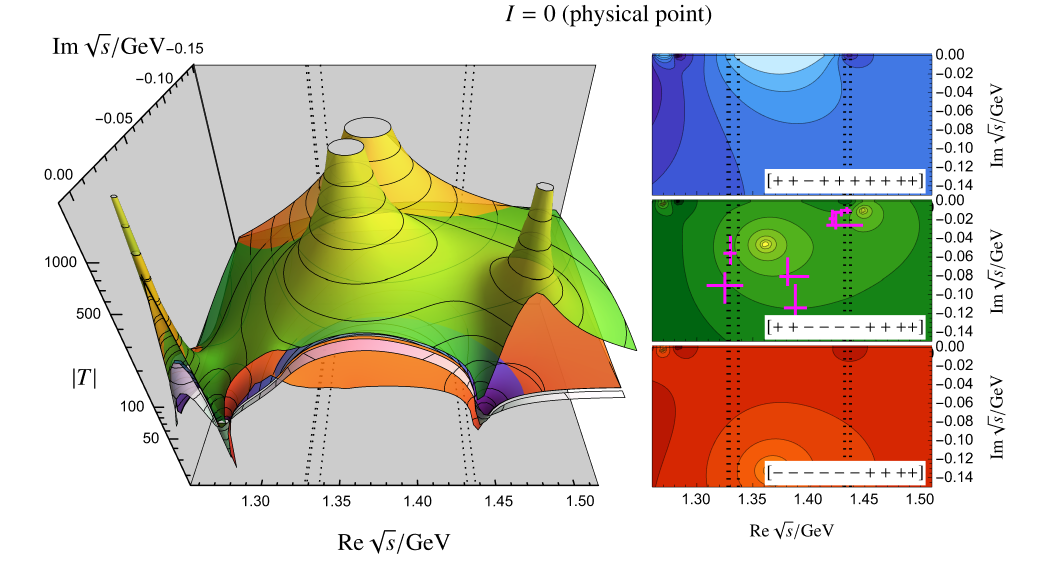}
    \includegraphics[height=4cm]{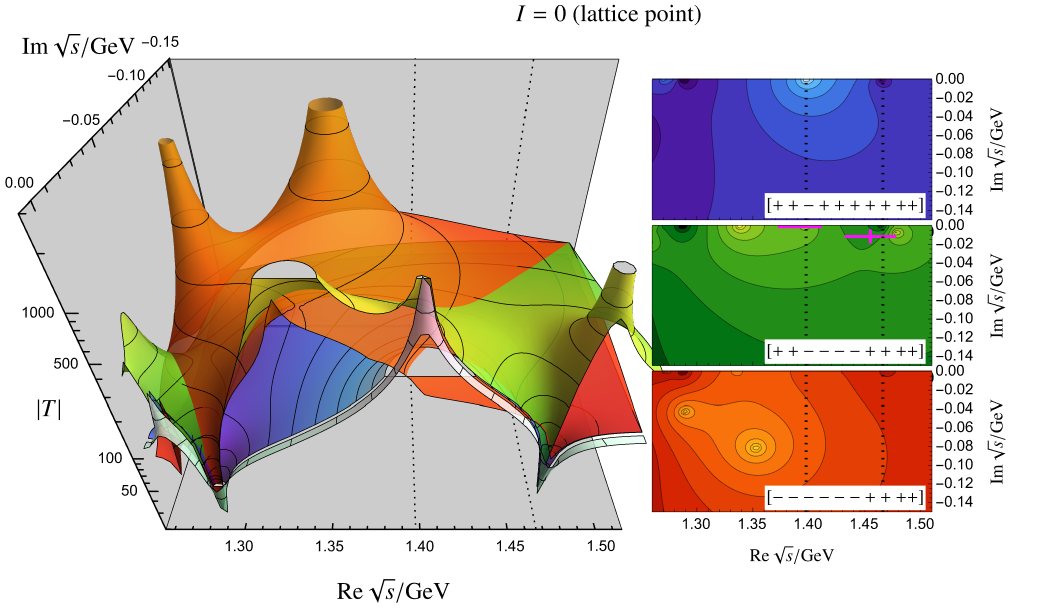}\\
    \includegraphics[height=4cm]{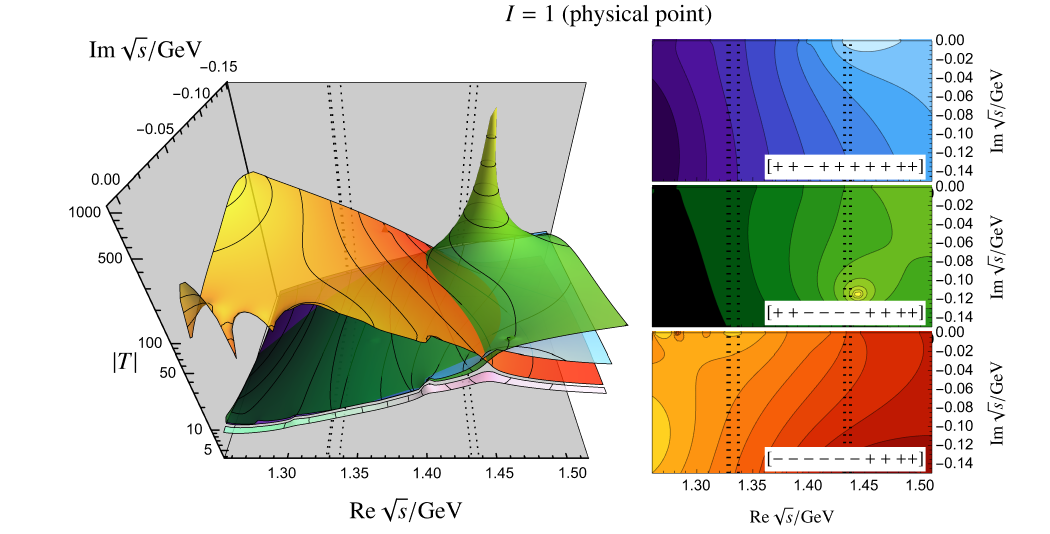}
    \includegraphics[height=4cm]{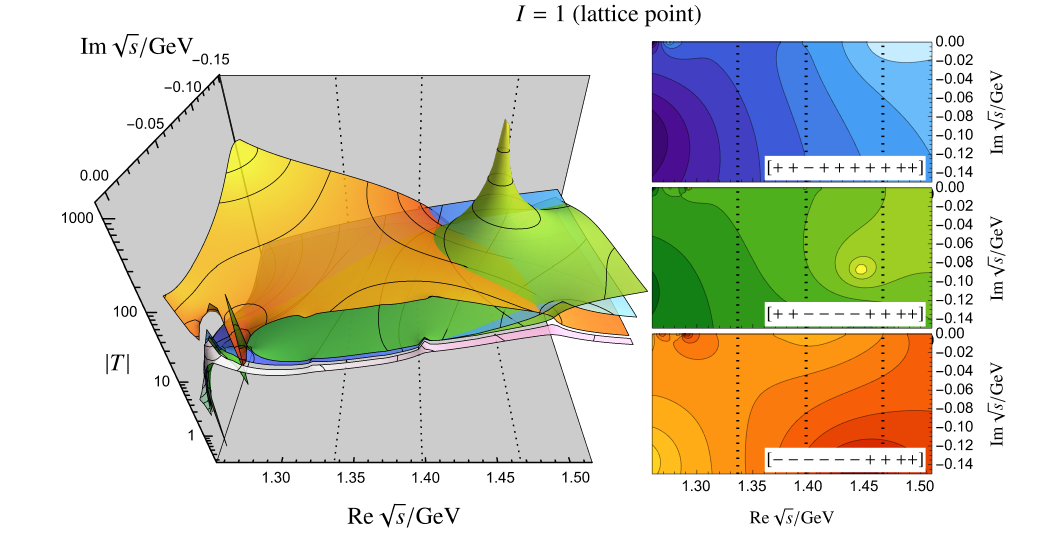}
    \caption{Isoscalar and isovector projected absolute value of the $\pi\Sigma\to \pi\Sigma$ scattering amplitude on the unphysical second Riemann sheets. Nomenclature as in the main text. Magenta crosses represent literature values from Refs.~\cite{BaryonScatteringBaSc:2023ori, BaryonScatteringBaSc:2023zvt} for the lattice point and Refs.~\cite{Ikeda:2012au,Guo:2012vv,Mai:2014xna} for the physical point.}
    \label{fig:app-25}
\end{figure*}

\clearpage
\subsection{M2S3L ($F_{25}$)}

\begin{figure*}[h]
\begin{minipage}[t!]{0.23\textwidth}
    \vspace{0pt} 
\centering
\caption{Total $\chi^2_\mathrm{dof}$ and the $\Lambda$ parameter
for M2S3L ($F_{18}$).}
\label{tab:details_25}
\begin{tabular}{|c|c|}
\hline
$\chi^2_{\rm dof}$ & 3.54 \\
$\Lambda[\mathrm{GeV}]$ & 0.7054119 \\
\hline
\end{tabular}
\label{tab:fit_summary-25}
\end{minipage}
\hfill
\begin{minipage}[t!]{0.7\textwidth}
    \vspace{-6pt} 
    \includegraphics[width=\linewidth]{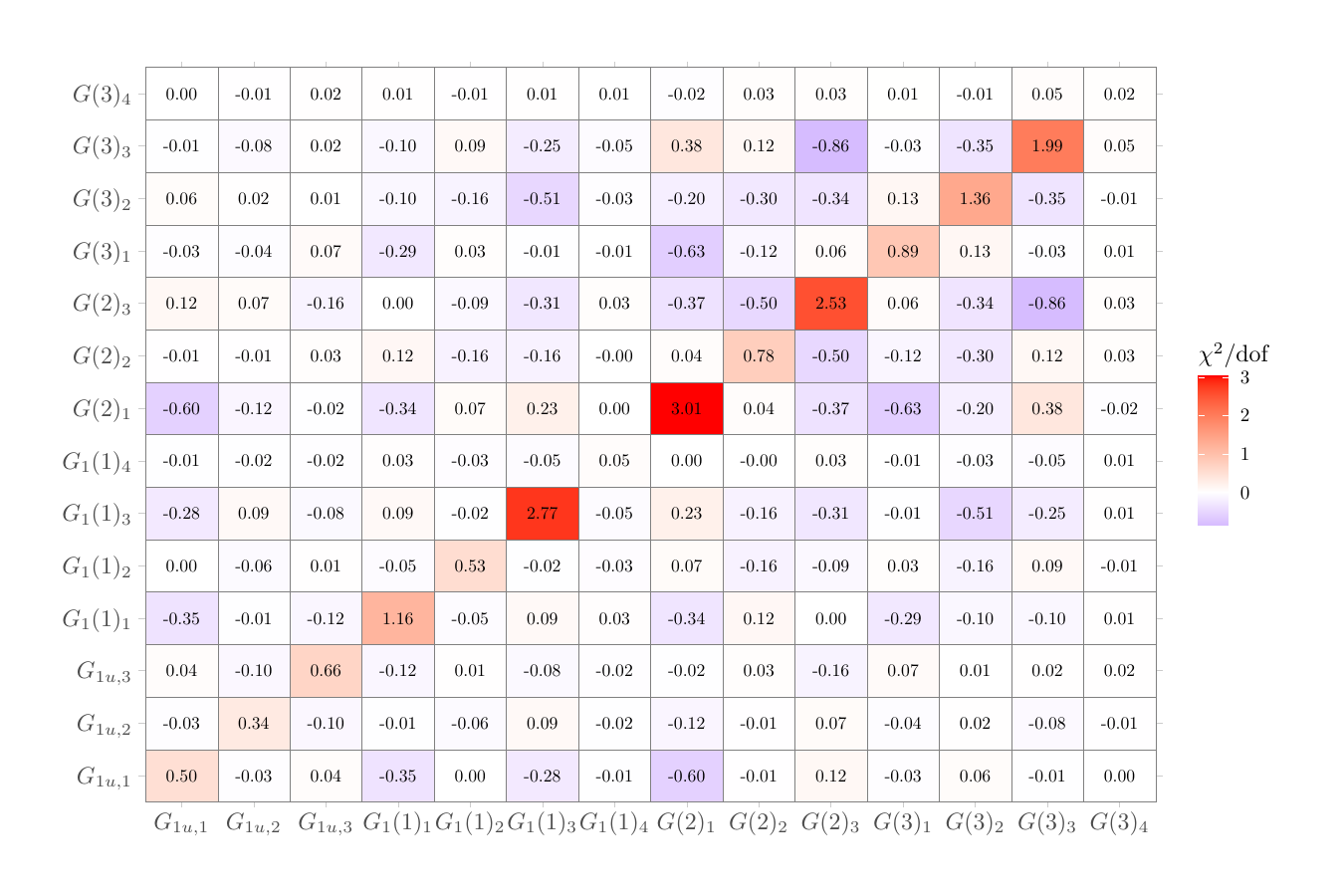}
    \vspace{-26pt}
    \caption{Heat map of correlated $\chi^2_{{\rm dof},ij}$, highlighting the relative impact of each energy level on the total fit quality.}
    \label{fig:partial-chi2_M2S3L}
\end{minipage}
\end{figure*}

\begin{figure*}[h]
    \includegraphics[height=4cm]{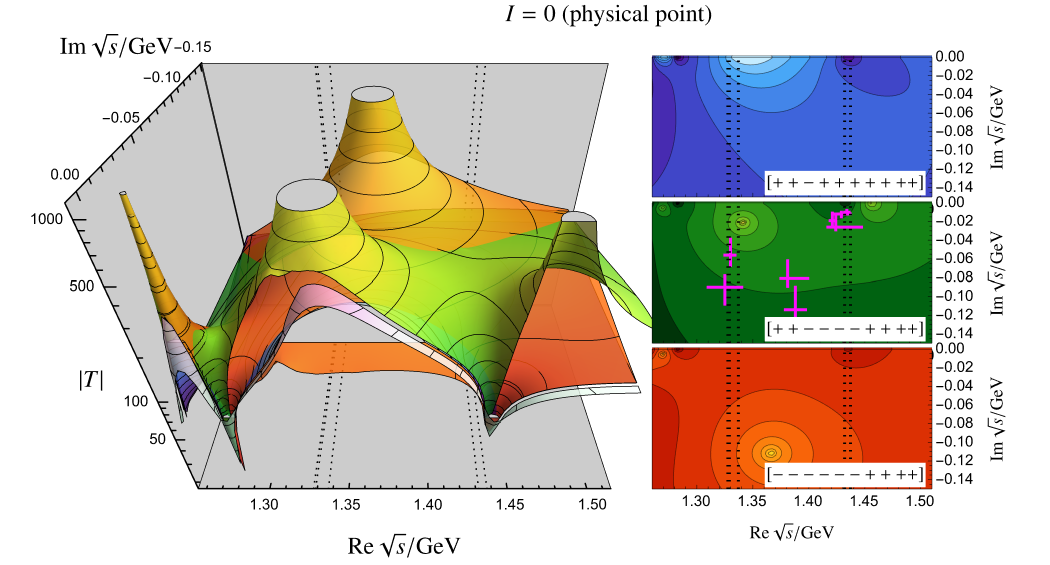}
    \includegraphics[height=4cm]{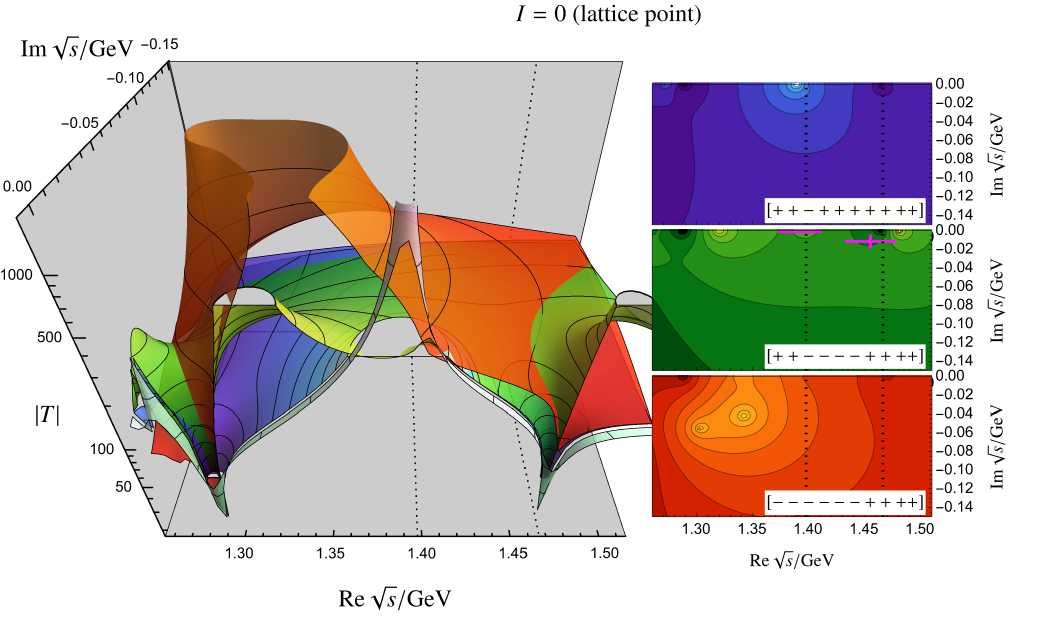}\\
    \includegraphics[height=4cm]{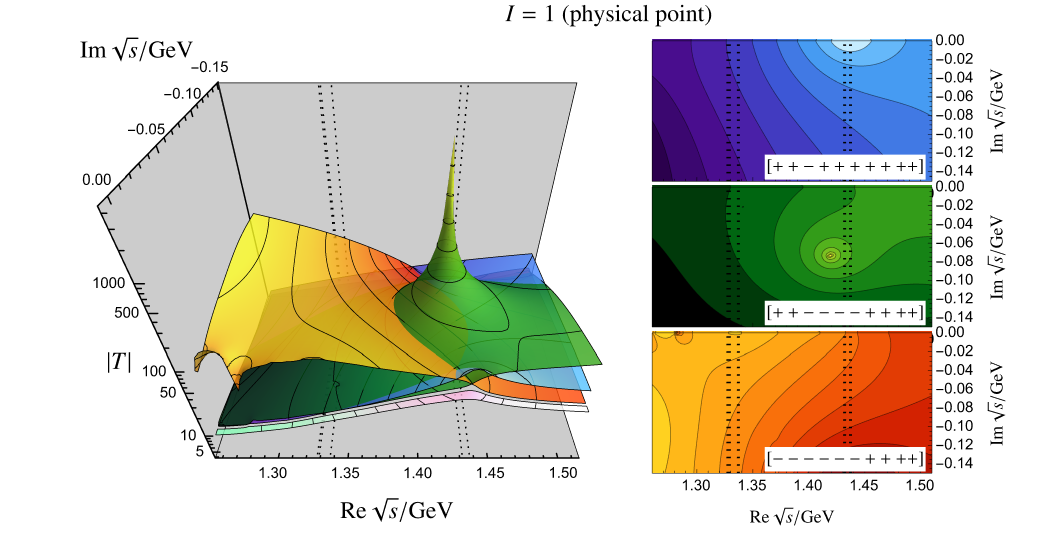}
    \includegraphics[height=4cm]{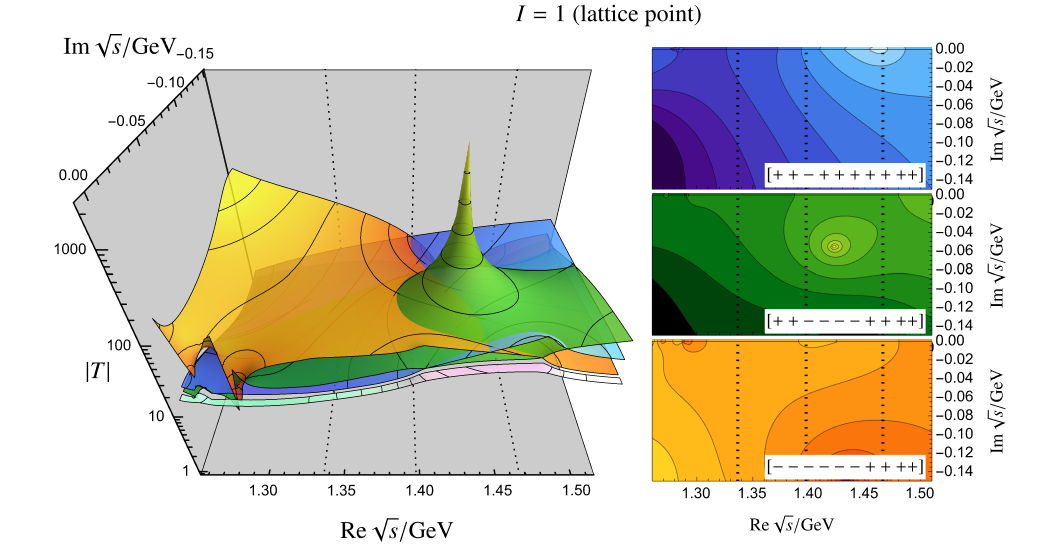}
    \caption{Isoscalar and isovector projected absolute value of the $\pi\Sigma\to \pi\Sigma$ scattering amplitude on the unphysical second Riemann sheets. Nomenclature as in the main text. Magenta crosses represent literature values from Refs.~\cite{BaryonScatteringBaSc:2023ori, BaryonScatteringBaSc:2023zvt} for the lattice point and Refs.~\cite{Ikeda:2012au,Guo:2012vv,Mai:2014xna} for the physical point.}
    \label{fig:app-}
\end{figure*}

\clearpage
\subsection{M3S1L  ($F_{01}$) }

\begin{figure*}[h]
\begin{minipage}[t!]{0.23\textwidth}
\centering
\caption{The total $\chi^2_\mathrm{dof}$, subtraction constants at the lattice point and LEC for M3S1L ($F_{01}$).}
\label{tab:details_01}
\begin{tabular}{|c|c|}
\hline
$\chi^2_{\rm dof}$ & 0.96 \\
\hline
$a_{\bar{K}N}$ & -4.088304e-03 \\
$a_{\pi\Lambda}$ & -1.079700e-01 \\
$a_{\pi\Sigma}$ & -4.088304e-03 \\
$a_{\eta\Lambda}$ & 3.560274e-03 \\
$a_{\eta\Sigma}$ & 2.163700e-01 \\
$a_{K\Xi}$ & 3.948000e-02 \\
\hline
$b_0[1/\mathrm{GeV}]$ & -4.730918e-01 \\
$b_D[1/\mathrm{GeV}]$ & 8.116358e-02 \\
$b_F[1/\mathrm{GeV}]$ & -3.145407e-01 \\
$d_1[1/\mathrm{GeV}]$ & 3.526983e-01 \\
$d_2[1/\mathrm{GeV}]$ & -7.041922e-02 \\
$d_3[1/\mathrm{GeV}]$ & -2.002697e-01 \\
$d_4[1/\mathrm{GeV}]$ & -5.572205e-01 \\
\hline
\end{tabular}
\end{minipage}
\hfill
\begin{minipage}[t!]{0.7\textwidth}
    \vspace{-6pt} 
    \includegraphics[width=\linewidth]{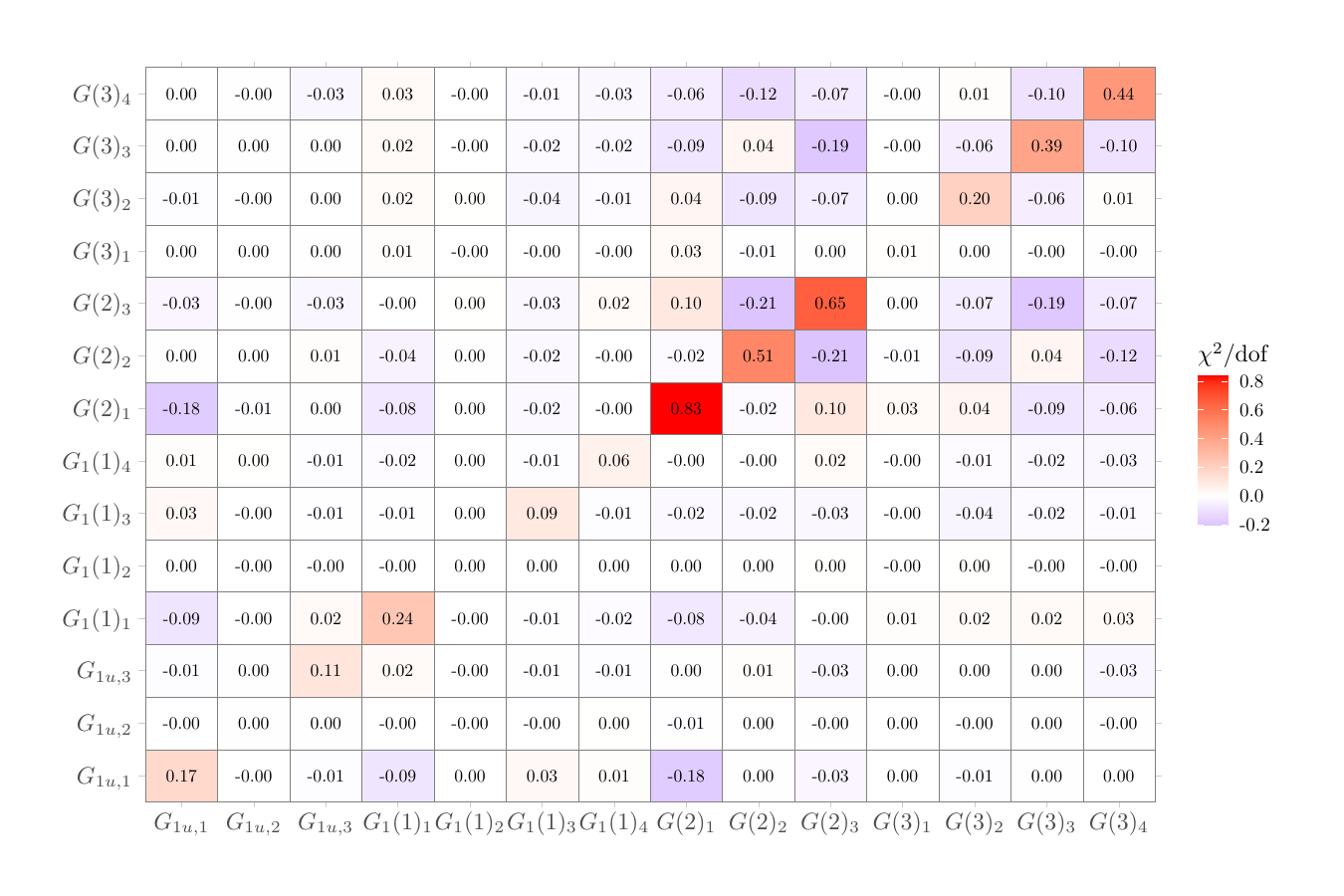}
    \vspace{-26pt}
    \caption{Heat map of correlated $\chi^2_{{\rm dof},ij}$, highlighting the relative impact of each energy level on the total fit quality.}
    \label{fig:partial-chi2_M3S1L}
\end{minipage}
\end{figure*}

\begin{figure*}[h]
    \includegraphics[height=4cm]{plots/3D-poles/fit-1-exp0.pdf}
    \includegraphics[height=4cm]{plots/3D-poles/fit-1-LAT0.pdf}\\
    \includegraphics[height=4cm]{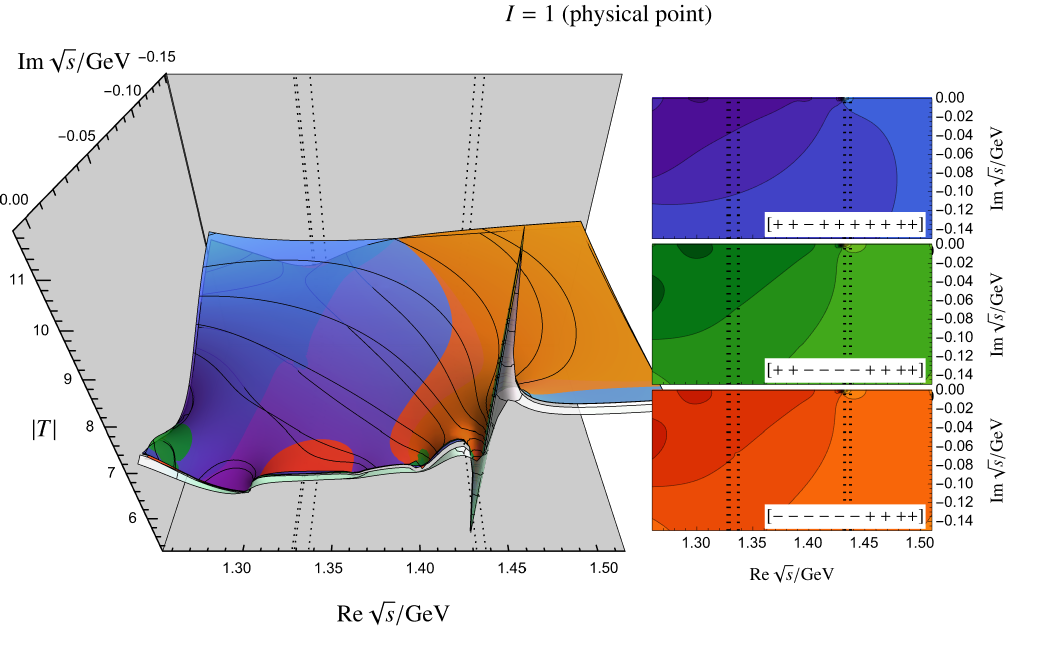}
    \includegraphics[height=4cm]{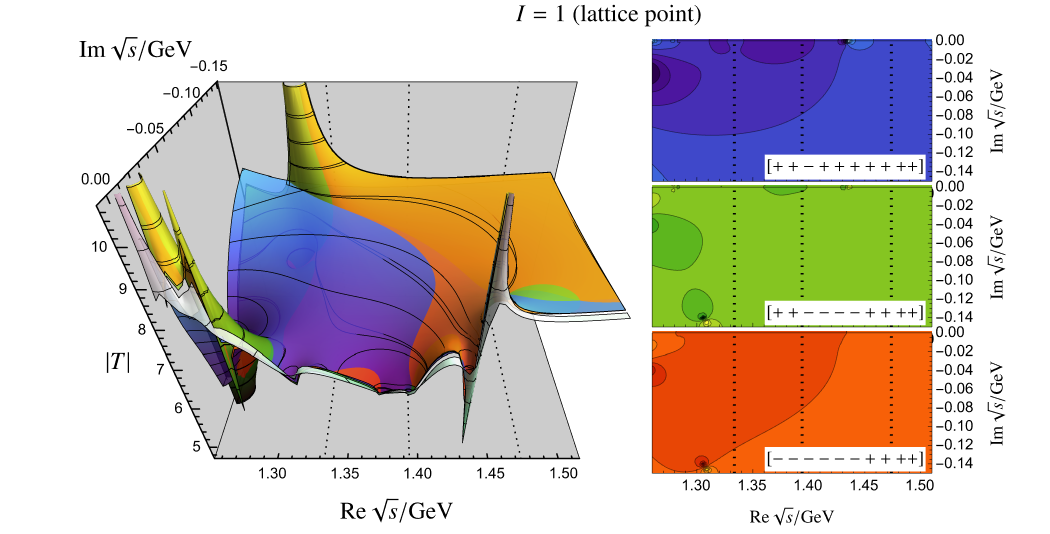}
    \caption{Isoscalar and isovector projected absolute value of the $\pi\Sigma\to \pi\Sigma$ scattering amplitude on the unphysical second Riemann sheets. Nomenclature as in the main text. Magenta crosses represent literature values from Refs.~\cite{BaryonScatteringBaSc:2023ori, BaryonScatteringBaSc:2023zvt} for the lattice point and Refs.~\cite{Ikeda:2012au,Guo:2012vv,Mai:2014xna} for the physical point.}
    \label{fig:3D-01}
\end{figure*}

\clearpage
\subsection{M3S2L ($F_{15}$)}

\begin{figure*}[h]
\begin{minipage}[t!]{0.23\textwidth}
    \vspace{0pt} 
\centering
\caption{The total $\chi^2_\mathrm{dof}$ and LECs for M3S1L ($F_{15}$).}
\label{tab:details_15}
\begin{tabular}{|c|c|}
\hline
$\chi^2_{\rm dof}$ & 0.90 \\
\hline
\centering
$b_0[1/\mathrm{GeV}]$ & -5.349500e-01 \\
$b_D[1/\mathrm{GeV}]$ & 9.599595e-02 \\
$b_F[1/\mathrm{GeV}]$ & -3.256764e-01 \\
$d_1[1/\mathrm{GeV}]$ & -8.386487e-01 \\
$d_2[1/\mathrm{GeV}]$ & 1.518967e-01 \\
$d_3[1/\mathrm{GeV}]$ & -4.546126e-01 \\
$d_4[1/\mathrm{GeV}]$ & 7.207285e-03 \\
\hline
\end{tabular}
\end{minipage}
\hfill
\begin{minipage}[t!]{0.7\textwidth}
    \vspace{-6pt} 
    \includegraphics[width=\linewidth]{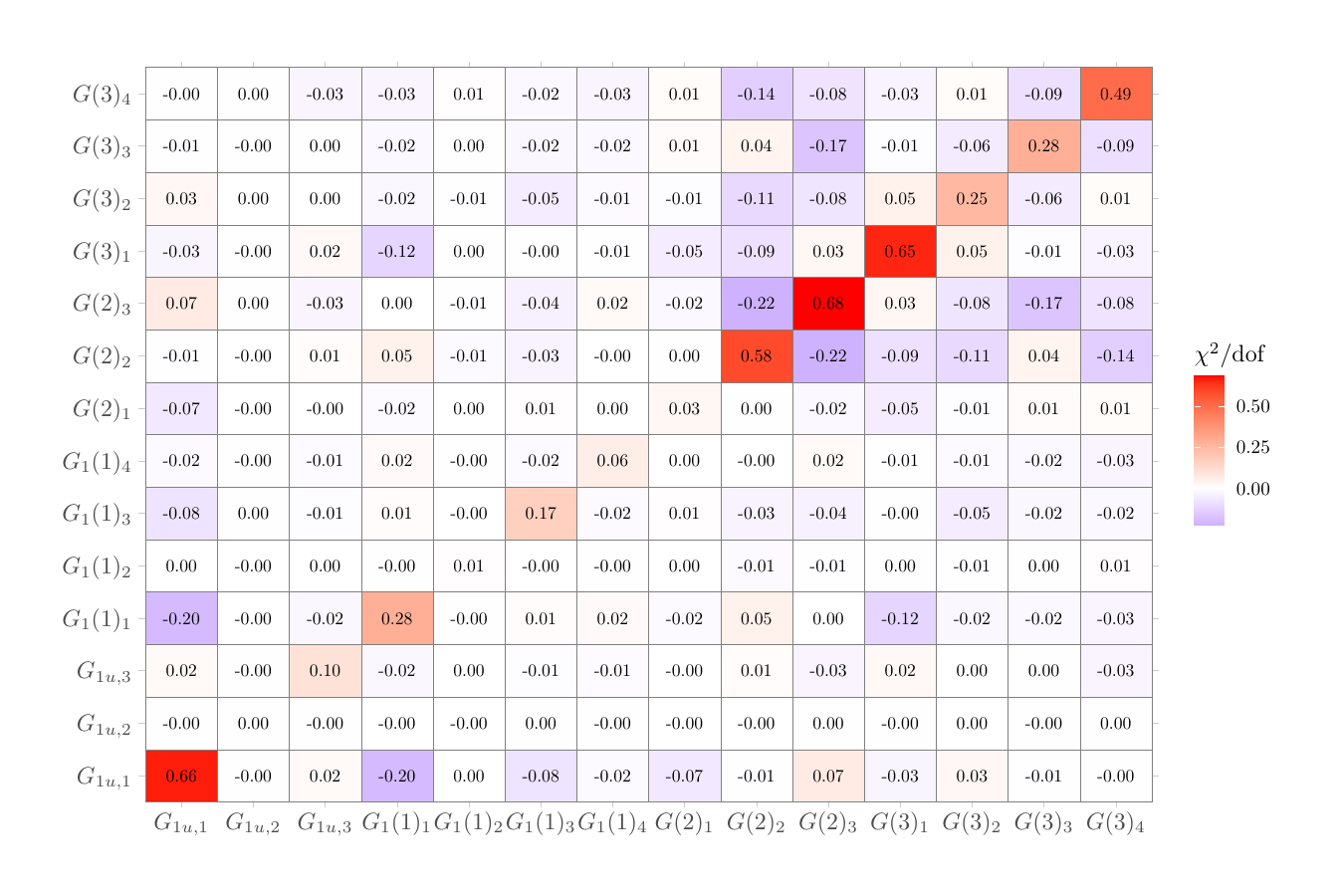}
    \vspace{-26pt}
    \caption{Heat map of correlated $\chi^2_{{\rm dof},ij}$, highlighting the relative impact of each energy level on the total fit quality.}
    \label{fig:partial-chi2_M3S2L}
\end{minipage}
\end{figure*}

\begin{figure*}[h]
    \includegraphics[height=4cm]{plots/3D-poles/fit-15-exp0.pdf}
    \includegraphics[height=4cm]{plots/3D-poles/fit-15-LAT0.pdf}\\
    \includegraphics[height=4cm]{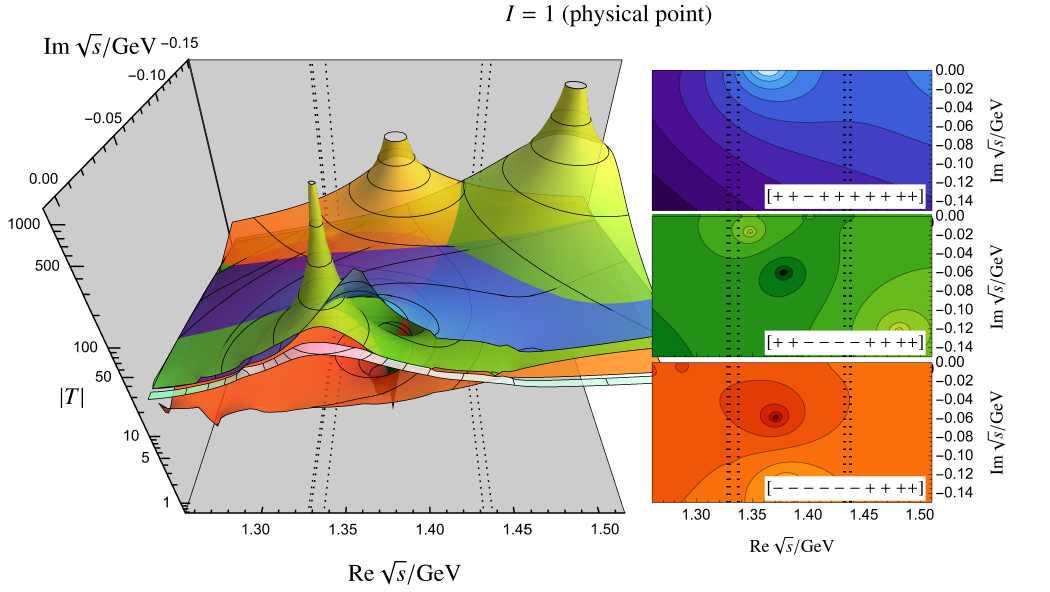}
    \includegraphics[height=4cm]{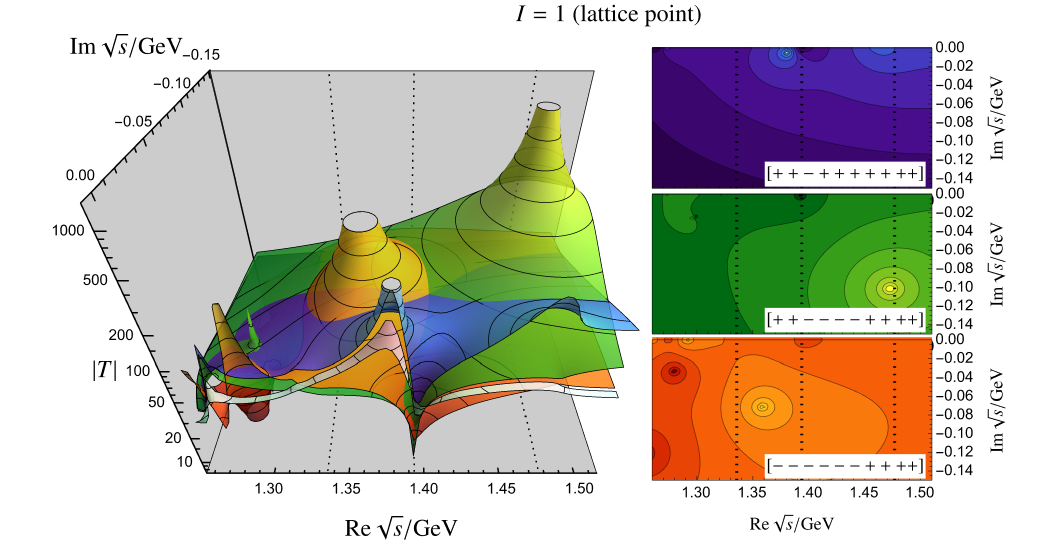}
    \caption{Isoscalar and isovector projected absolute value of the $\pi\Sigma\to \pi\Sigma$ scattering amplitude on the unphysical second Riemann sheets. Nomenclature as in the main text. Magenta crosses represent literature values from Refs.~\cite{BaryonScatteringBaSc:2023ori, BaryonScatteringBaSc:2023zvt} for the lattice point and Refs.~\cite{Ikeda:2012au,Guo:2012vv,Mai:2014xna} for the physical point.}
    \label{fig:app-15}
\end{figure*}

\clearpage
\subsection{M3S3L ($F_{10}$)}

\begin{figure*}[h]
\begin{minipage}[t!]{0.23\textwidth}
    \vspace{0pt} 
\centering
\caption{The total $\chi^2_\mathrm{dof}$, $\Lambda$ parameter and LECs for M3S3L ($F_{10}$).}
\label{tab:details_10}
\begin{tabular}{|c|c|}
\hline
$\chi^2_{\rm dof}$ & 0.92   \\
$\Lambda[\mathrm{GeV}]$ & 0.63786229 \\
\hline
$b_0[1/\mathrm{GeV}]$ & -5.493596e-01 \\
$b_D[1/\mathrm{GeV}]$ & 8.601840e-02 \\
$b_F[1/\mathrm{GeV}]$ & -3.226791e-01 \\
$d_1[1/\mathrm{GeV}]$ & 2.079492e+00 \\
$d_2[1/\mathrm{GeV}]$ & -6.344862e-02 \\
$d_3[1/\mathrm{GeV}]$ & 3.538223e-01 \\
$d_4[1/\mathrm{GeV}]$ & -2.262594e+00 \\
\hline
\end{tabular}
\end{minipage}
\hfill
\begin{minipage}[t!]{0.7\textwidth}
    \vspace{-6pt} 
    \includegraphics[width=\linewidth]{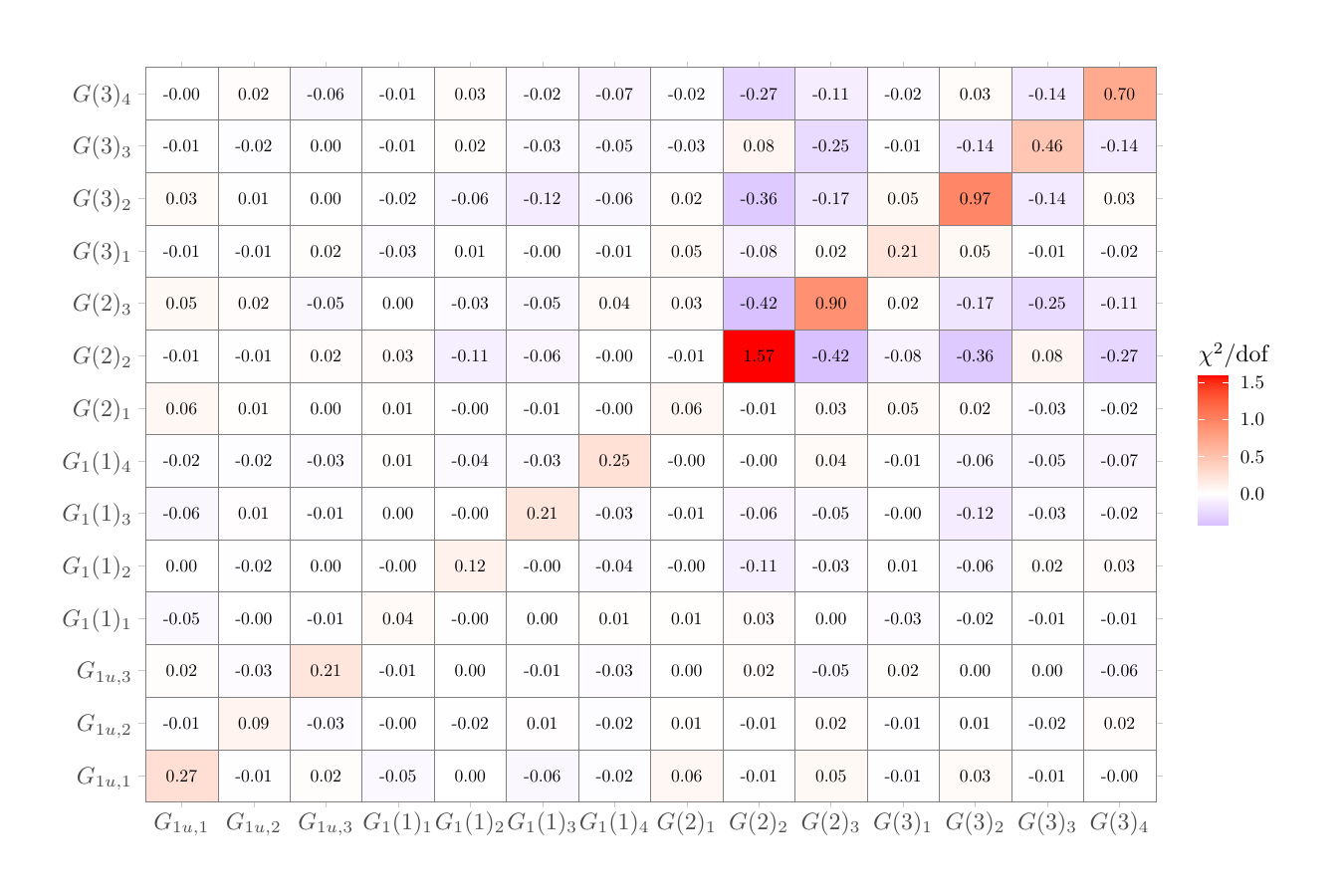}
    \vspace{-26pt}
    \caption{Heat map of correlated $\chi^2_{{\rm dof},ij}$, highlighting the relative impact of each energy level on the total fit quality.}
    \label{fig:partial-chi2_M3S3L}
\end{minipage}
\end{figure*}

\begin{figure*}[h]
    \includegraphics[height=4cm]{plots/3D-poles/fit-10-exp0.pdf}
    \includegraphics[height=4cm]{plots/3D-poles/fit-10-LAT0.pdf}\\
    \includegraphics[height=4cm]{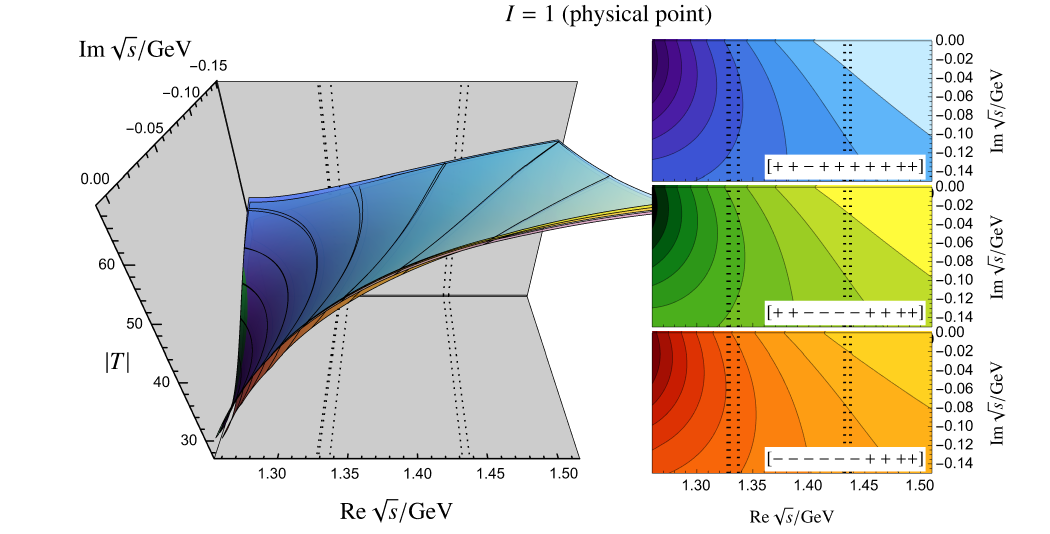}
    \includegraphics[height=4cm]{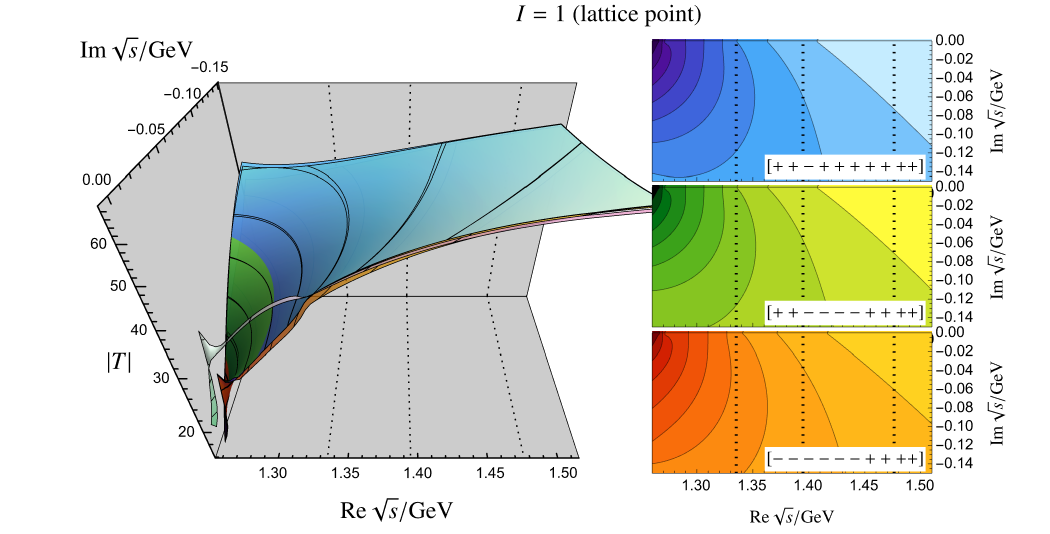}
    \caption{Isoscalar and isovector projected absolute value of the $\pi\Sigma\to \pi\Sigma$ scattering amplitude on the unphysical second Riemann sheets. Nomenclature as in the main text. Magenta crosses represent literature values from Refs.~\cite{BaryonScatteringBaSc:2023ori, BaryonScatteringBaSc:2023zvt} for the lattice point and Refs.~\cite{Ikeda:2012au,Guo:2012vv,Mai:2014xna} for the physical point.}
    \label{fig:app-10}
\end{figure*}

\clearpage
\section{Appendix B: Detailed fit results. Fits to the experimental data.}
\addcontentsline{toc}{section}{Appendix B: Detailed fit results. Fits to the experimental data.}
\label{app:fit-results-exp}

\subsection{M1S1P ($F_{21}$)}

\begin{figure*}[h]
\begin{minipage}[t!]{0.23\textwidth}
    \vspace{0pt} 
\centering
 \caption{The total $\chi^2_\mathrm{dof}$ as defined in \cref{eq:chi2}  and subtraction constants for M1S1P ($F_{21}$).}
    \label{tab:details_21}
\begin{tabular}{|c|c|}
\hline
$\chi^2_{\rm dof}$ & 4.23 \\
\hline
$a_{\bar{K}N}$ & -2.348949e-03 \\
$a_{\bar{K}N}$ & -2.348949e-03 \\
$a_{\pi\Lambda}$ & 3.881548e-01 \\
$a_{\pi\Sigma}$ & 8.779599e-04 \\
$a_{\pi\Sigma}$ & 8.779599e-04 \\
$a_{\pi\Sigma}$ & 8.779599e-04 \\
$a_{\eta\Lambda}$ & 1.527930e-03 \\
$a_{\eta\Sigma}$ & -7.466124e-01 \\
$a_{K\Xi}$ & -8.510607e-03 \\
$a_{K\Xi}$ & -8.510607e-03 \\
\hline
\end{tabular}
\end{minipage}
\hfill
\begin{minipage}[t!]{0.7\textwidth}
    \vspace{-6pt} 
    \includegraphics[width=\linewidth]{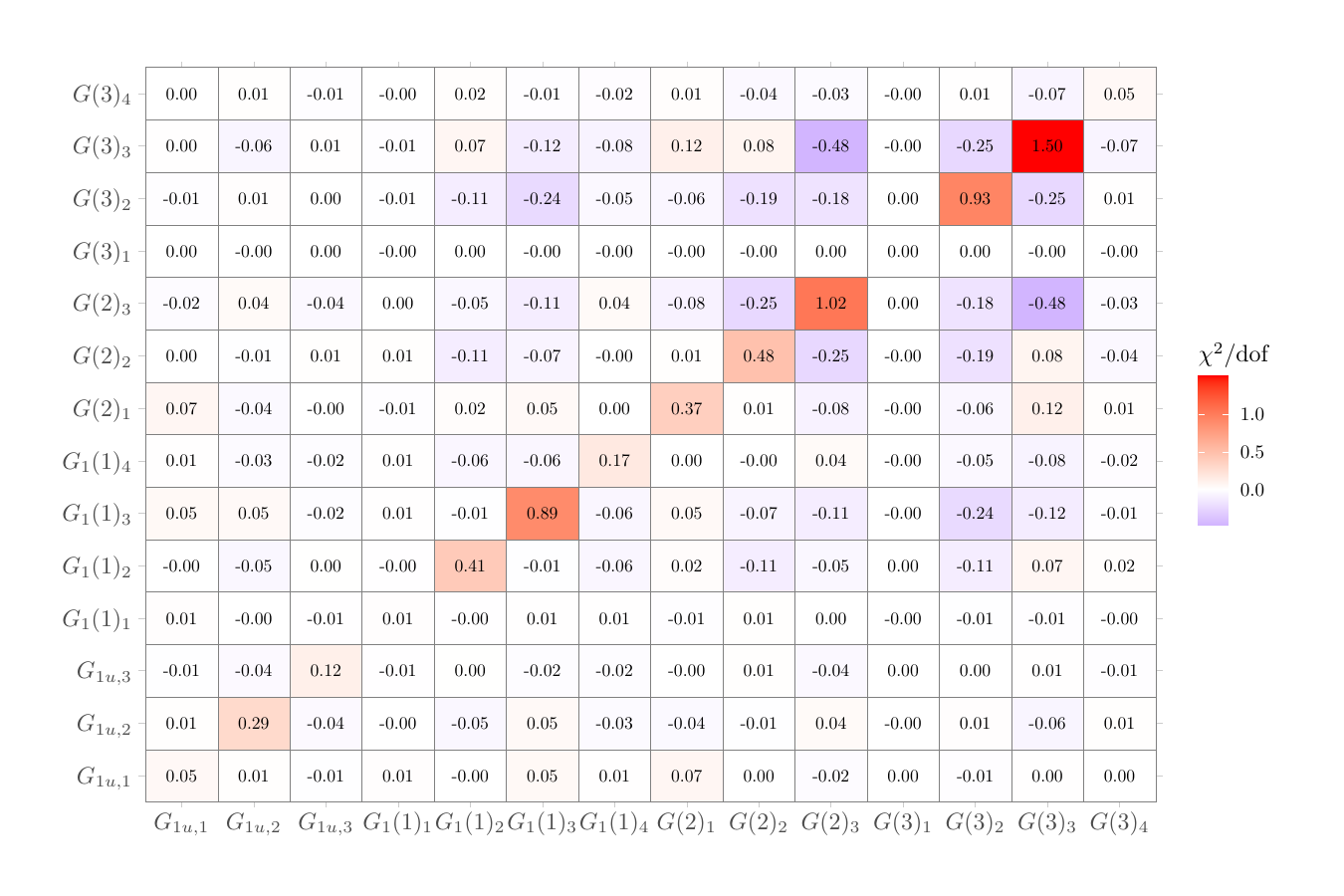}
    \vspace{-26pt}
    \caption{Heat map of correlated ${\chi^2}_{ij}/14$, highlighting the relative impact of each energy level on the total fit quality.}
    \label{fig:partial-chi2_M1S1P}
\end{minipage}
\end{figure*}

\begin{figure*}[h]
    \includegraphics[height=4cm]{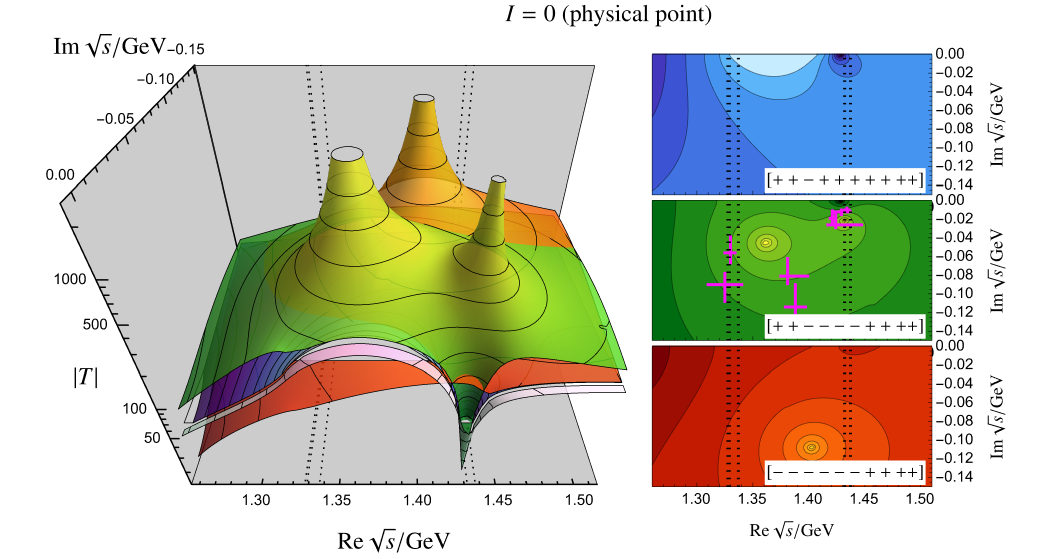}
    \includegraphics[height=4cm]{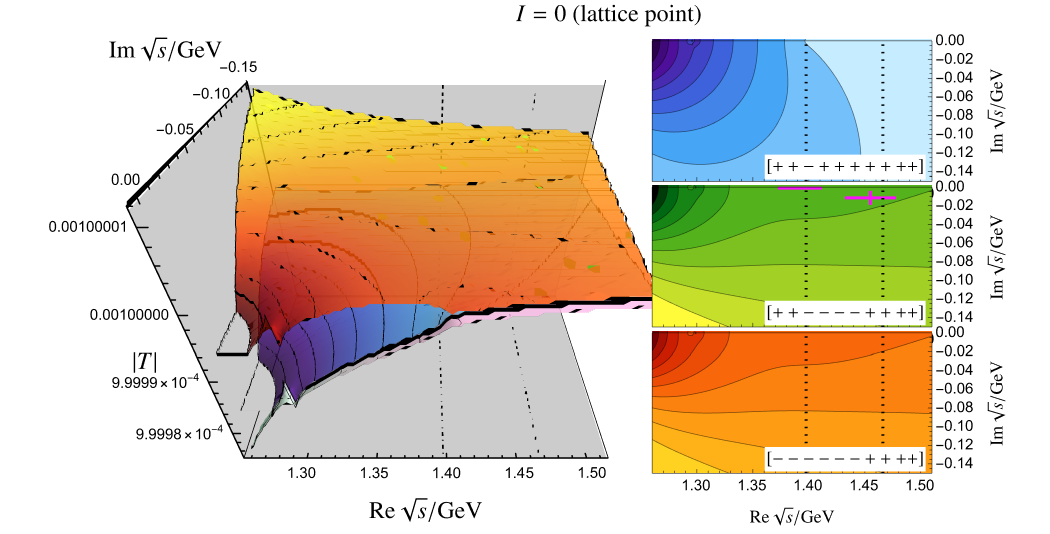}\\
    \includegraphics[height=4cm]{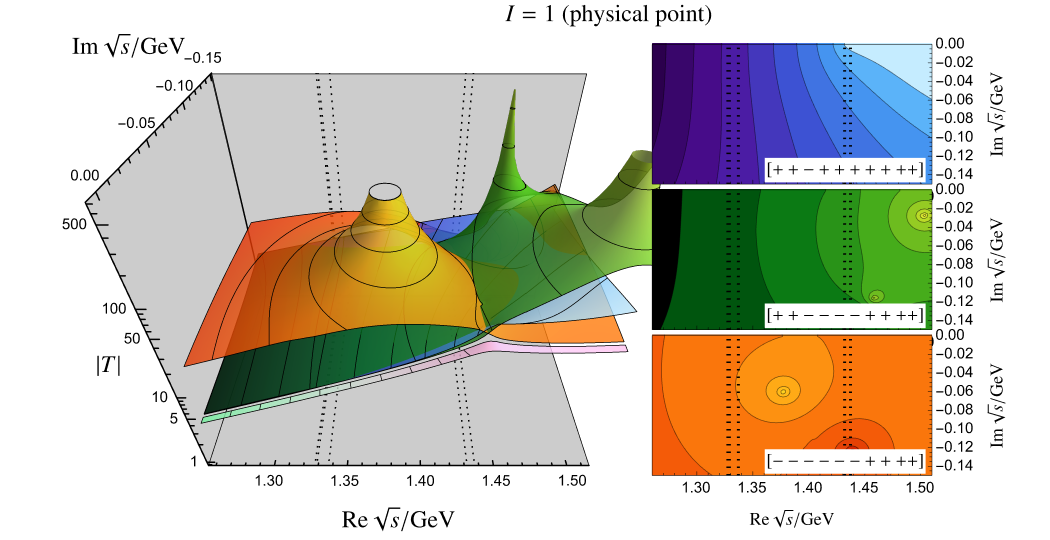}
    \includegraphics[height=4cm]{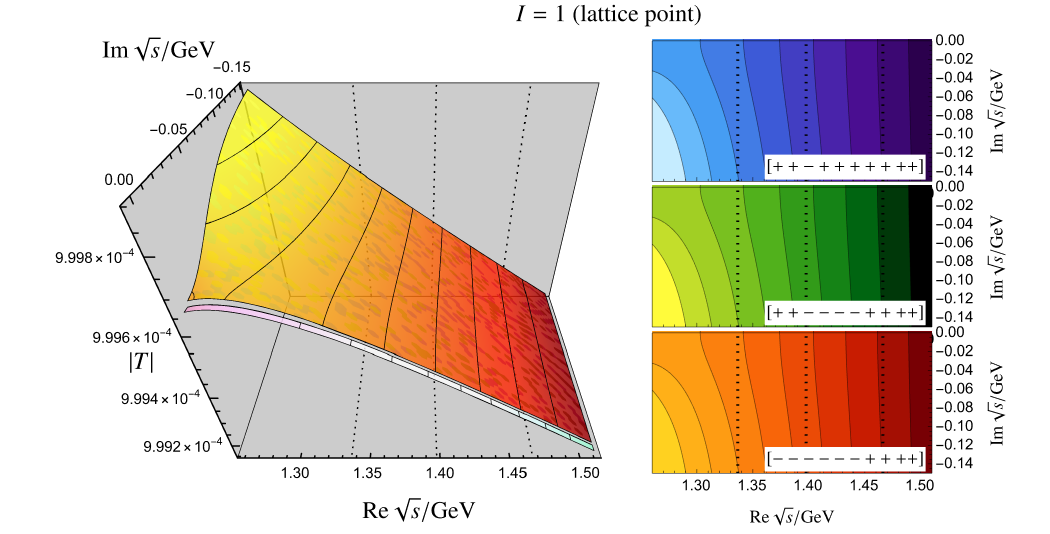}
    \caption{Isoscalar and isovector projected absolute value of the $\pi\Sigma\to \pi\Sigma$ scattering amplitude on the unphysical second Riemann sheets. Nomenclature as in the main text. Magenta crosses represent literature values from Refs.~\cite{BaryonScatteringBaSc:2023ori, BaryonScatteringBaSc:2023zvt} for the lattice point and Refs.~\cite{Ikeda:2012au,Guo:2012vv,Mai:2014xna} for the physical point.}
    \label{fig:app-21}
\end{figure*}

\clearpage
\subsection{M1S2P ($F_{28}$)}

\begin{figure*}[h]
\begin{minipage}[t!]{0.23\textwidth}
    \vspace{0pt} 
\centering
\caption{The total $\chi^2_\mathrm{dof}$ as defined in \cref{eq:chi2} for the parameter-free fit M1S2P ($F_{28}$).}
    \label{tab:details_28}
\begin{tabular}{|c|c|}
\hline
$\chi^2_{\rm dof}$ & 25.57 \\
\hline
\end{tabular}
\end{minipage}
\end{figure*}

\begin{figure*}[h]
    \includegraphics[height=4cm]{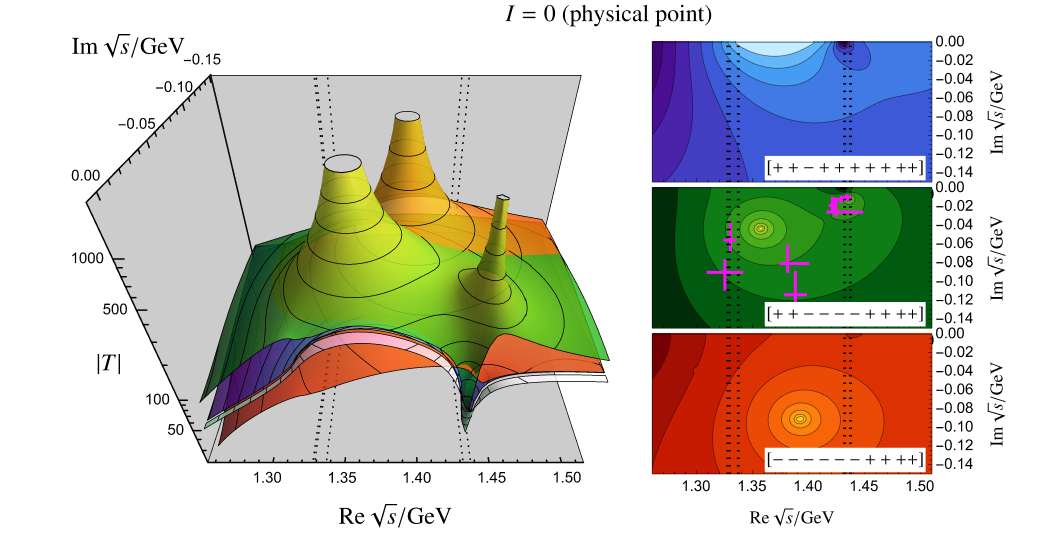}
    \includegraphics[height=4cm]{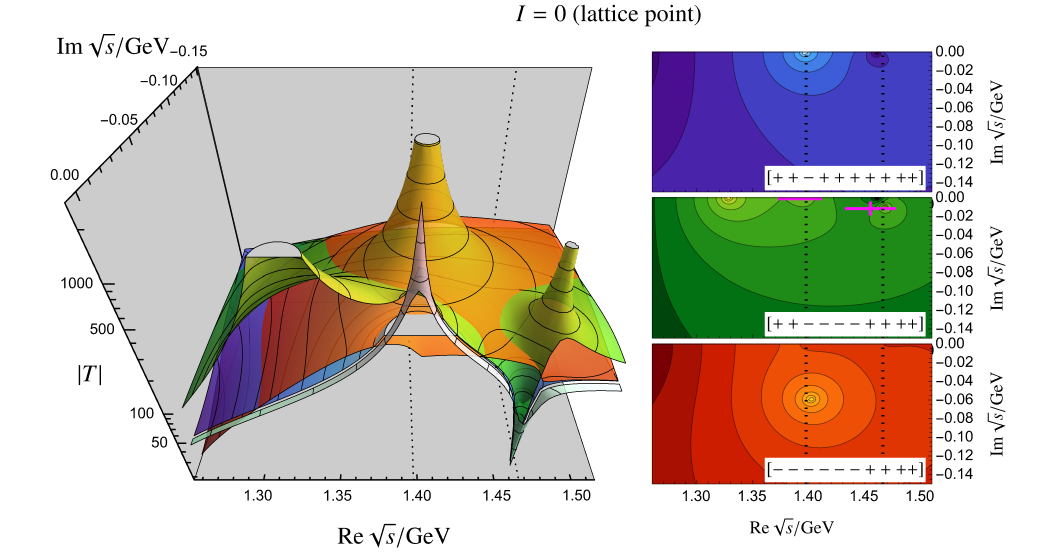}\\
    \includegraphics[height=4cm]{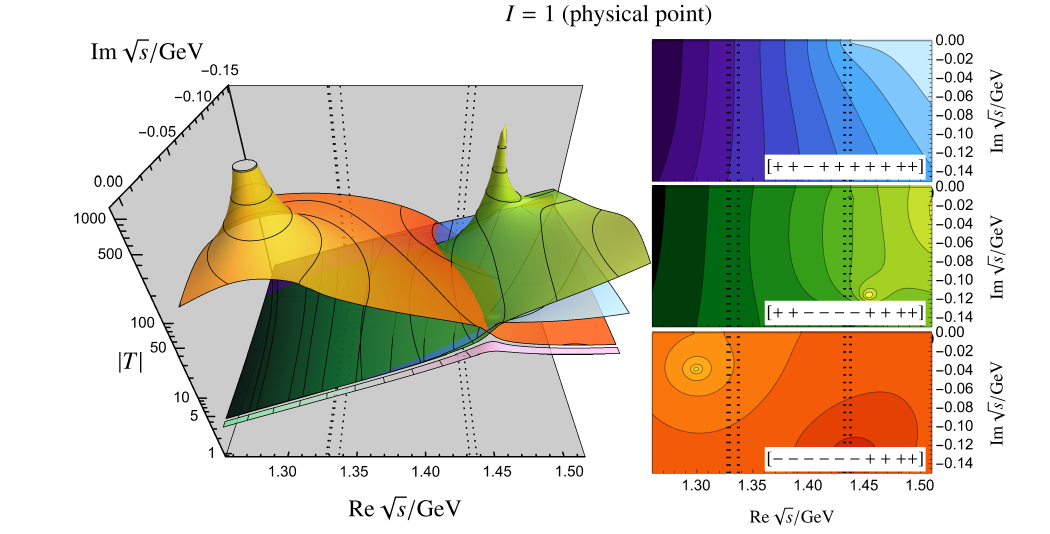}
    \includegraphics[height=4cm]{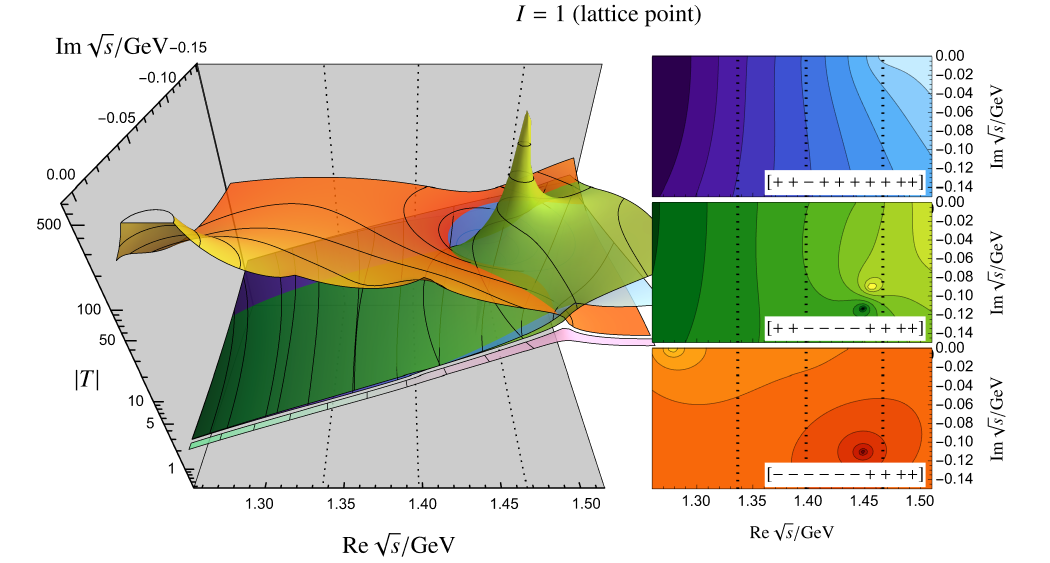}
    \caption{Isoscalar and isovector projected absolute value of the $\pi\Sigma\to \pi\Sigma$ scattering amplitude on the unphysical second Riemann sheets. Nomenclature as in the main text. Magenta crosses represent literature values from Refs.~\cite{BaryonScatteringBaSc:2023ori, BaryonScatteringBaSc:2023zvt} for the lattice point and Refs.~\cite{Ikeda:2012au,Guo:2012vv,Mai:2014xna} for the physical point.}
    \label{fig:app-28}
\end{figure*}

\clearpage
\subsection{M1S3P ($F_{27}$)}

\begin{figure*}[h]
\begin{minipage}[t!]{0.23\textwidth}
    \vspace{0pt} 
\centering
\caption{The total $\chi^2_\mathrm{dof}$ as defined in \cref{eq:chi2}  and $\Lambda$ parameter for M1S3P ($F_{27}$).}
    \label{tab:details_27}
\begin{tabular}{|c|c|}
\hline
$\chi^2_{\rm dof}$ & 30.28266 \\
\hline
$\Lambda[\mathrm{GeV}]$ & 0.8111258 \\
\hline
\end{tabular}
\label{tab:fit_summaryM1S3P}
\end{minipage}
\hfill
\begin{minipage}[t!]{0.7\textwidth}
    \vspace{-6pt} 
    \includegraphics[width=\linewidth]{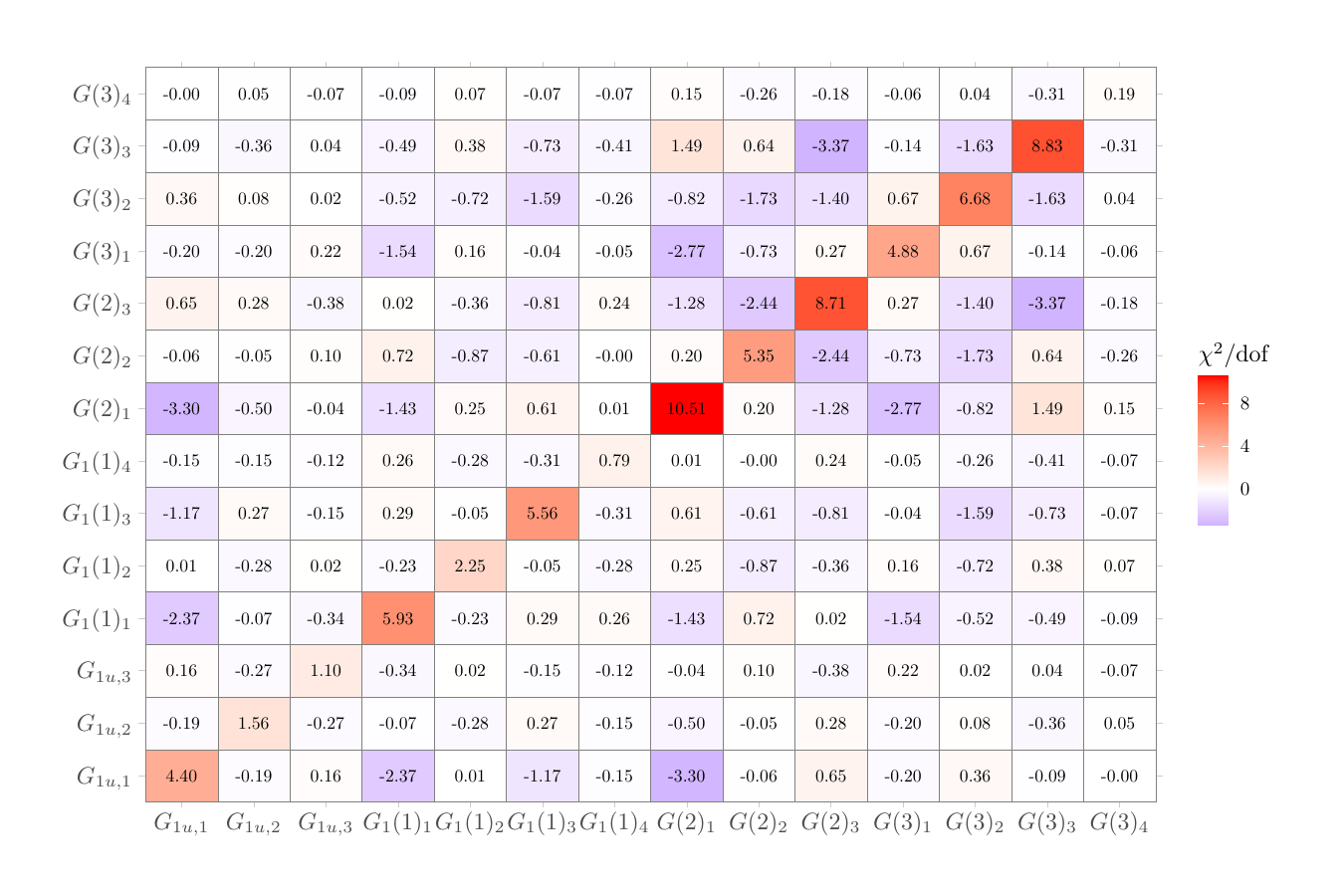}
    \vspace{-26pt}
    \caption{Heat map of correlated ${\chi^2}_{ij}/14$, highlighting the relative impact of each energy level on the total fit quality.}
    \label{fig:partial-chi2_M1S3P}
\end{minipage}
\end{figure*}

\begin{figure*}[h]
    \includegraphics[height=4cm]{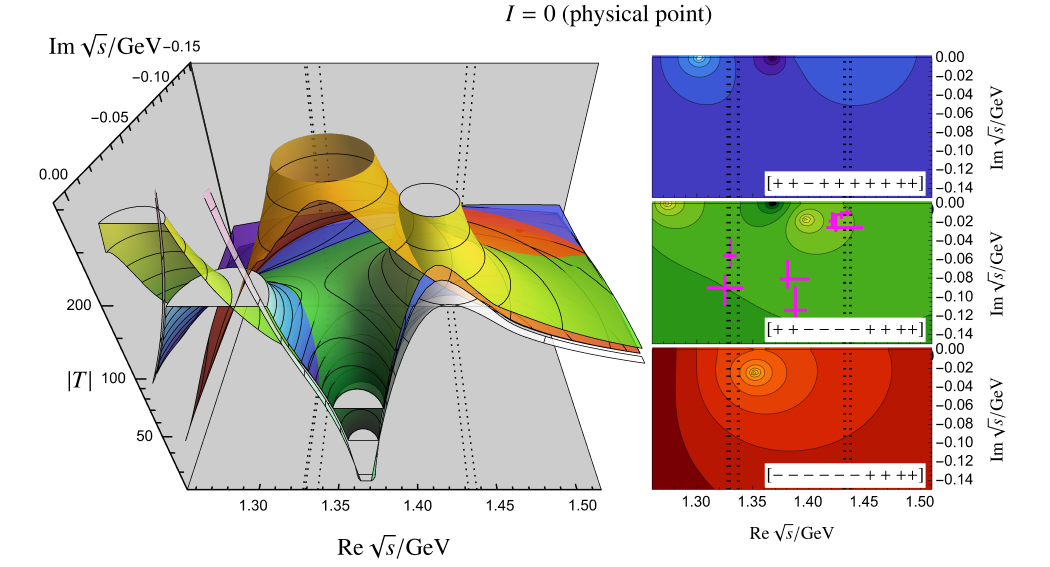}
    \includegraphics[height=4cm]{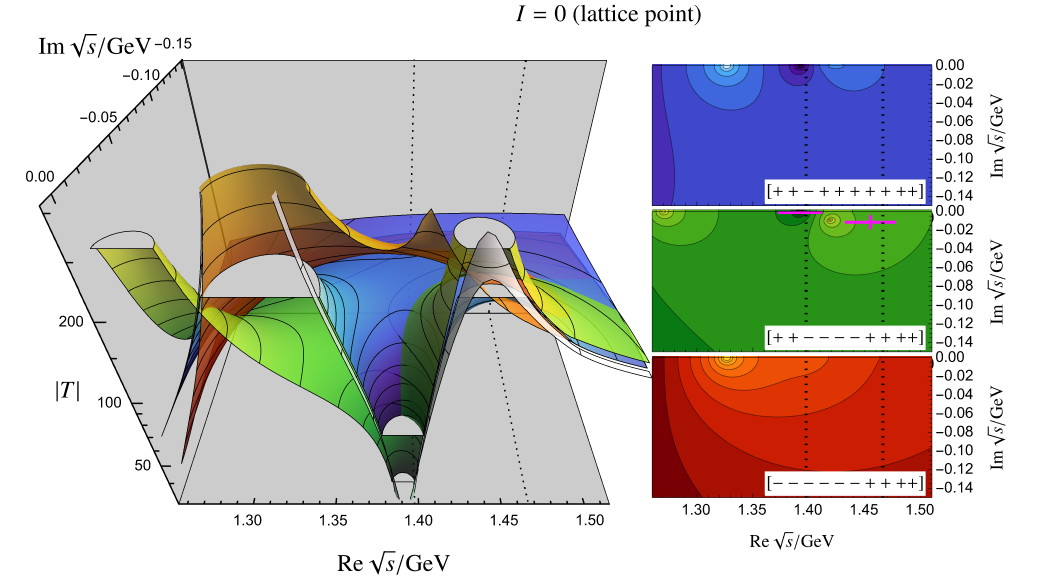}\\
    \includegraphics[height=4cm]{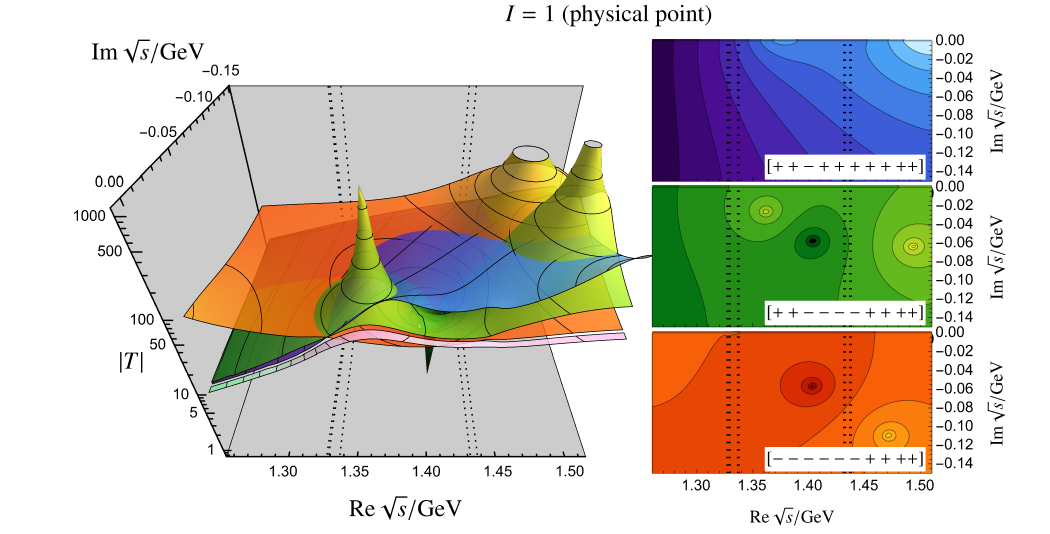}
    \includegraphics[height=4cm]{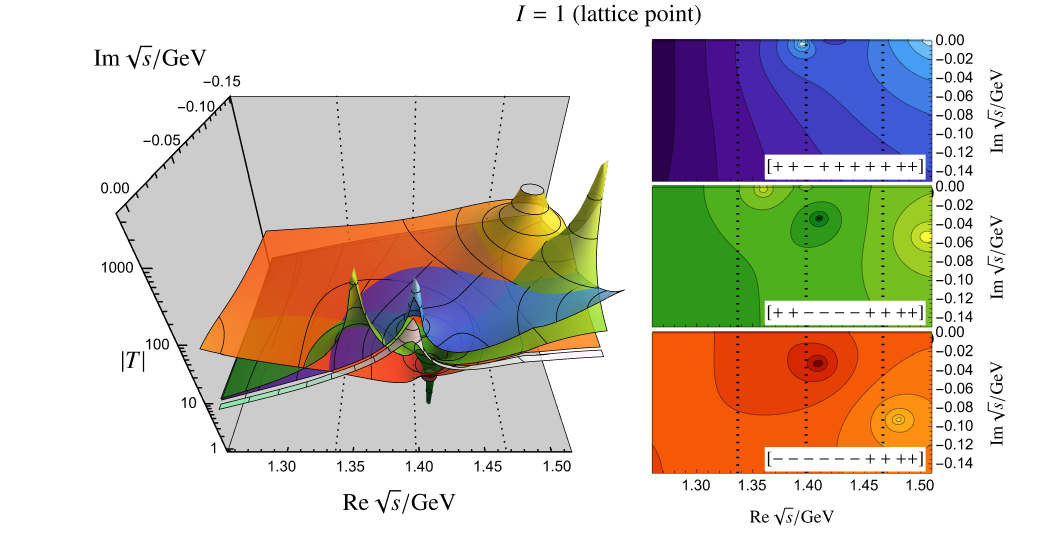}
    \caption{Isoscalar and isovector projected absolute value of the $\pi\Sigma\to \pi\Sigma$ scattering amplitude on the unphysical second Riemann sheets. Nomenclature as in the main text. Magenta crosses represent literature values from Refs.~\cite{BaryonScatteringBaSc:2023ori, BaryonScatteringBaSc:2023zvt} for the lattice point and Refs.~\cite{Ikeda:2012au,Guo:2012vv,Mai:2014xna} for the physical point.}
    \label{fig:app-27}
\end{figure*}

\clearpage
\subsection{M2S1P ($F_{22}$)}

\begin{figure*}[h]
\begin{minipage}[t!]{0.23\textwidth}
    \vspace{0pt} 
\centering
 \caption{The total $\chi^2_\mathrm{dof}$ as defined in \cref{eq:chi2}  and subtraction constants for M2S1P ($F_{22}$).}
     \label{tab:details_22}
\begin{tabular}{|c|c|}
\hline
$\chi^2_{\rm dof}$ & 8.86 \\
\hline
$a_{\bar{K}N}$ & -1.670707e-03 \\
$a_{\pi\Lambda}$ & 2.465099e-02 \\
$a_{\pi\Sigma}$ & -3.528238e-03 \\
$a_{\eta\Lambda}$ & -5.376287e-03 \\
$a_{\eta\Sigma}$ & -1.070200e-02 \\
$a_{K\Xi}$ & -1.305735e-02 \\
\hline
\end{tabular}
\end{minipage}
\hfill
\begin{minipage}[t!]{0.7\textwidth}
    \vspace{-6pt} 
    \includegraphics[width=\linewidth]{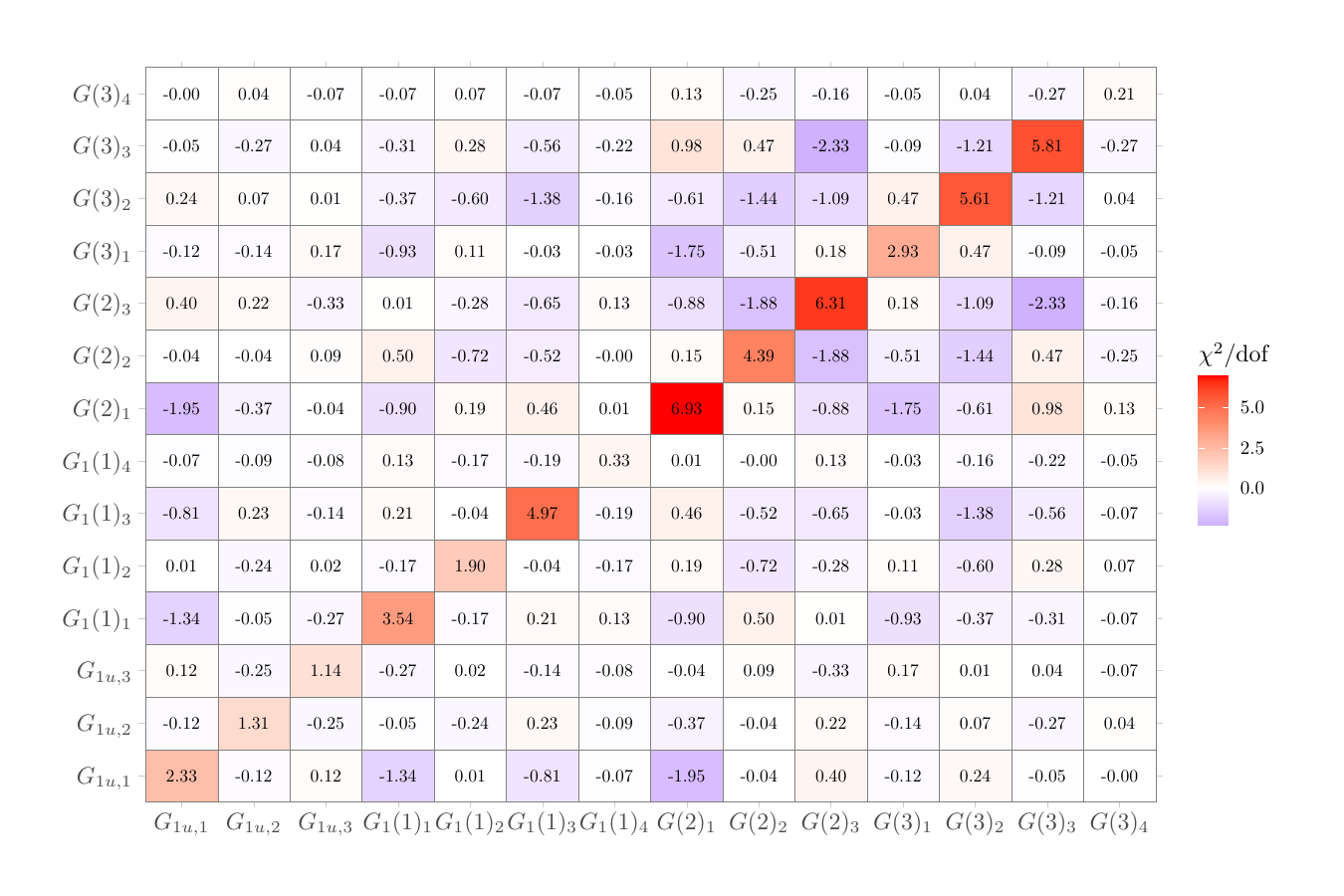}
    \vspace{-26pt}
    \caption{Heat map of correlated ${\chi^2}_{ij}/14$, highlighting the relative impact of each energy level on the total fit quality.}
    \label{fig::partial-chi2_M2S1P}
\end{minipage}
\end{figure*}

\begin{figure*}[h]
    \includegraphics[height=4cm]{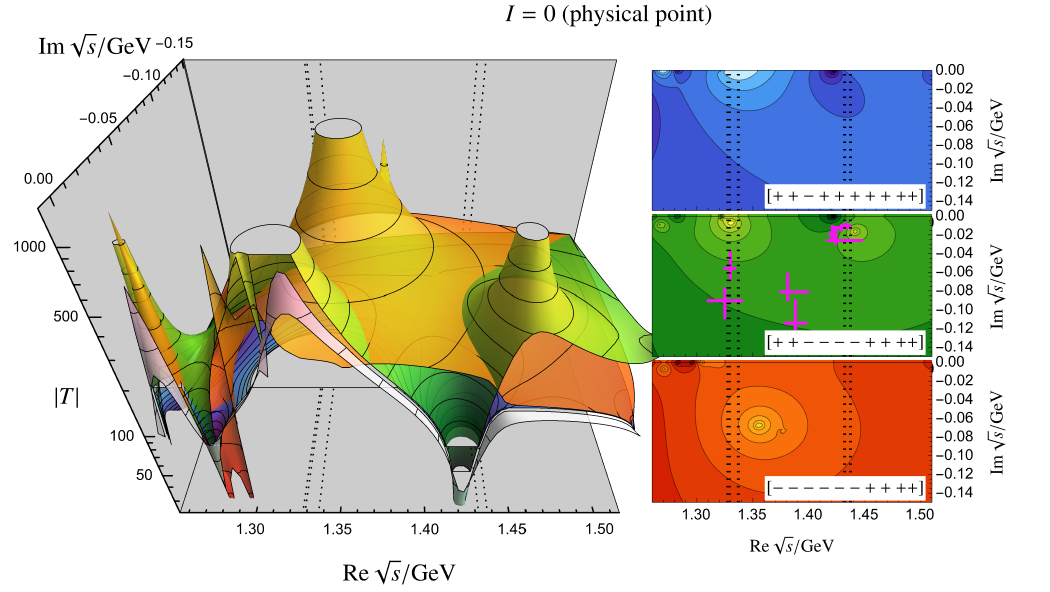}
    \includegraphics[height=4cm]{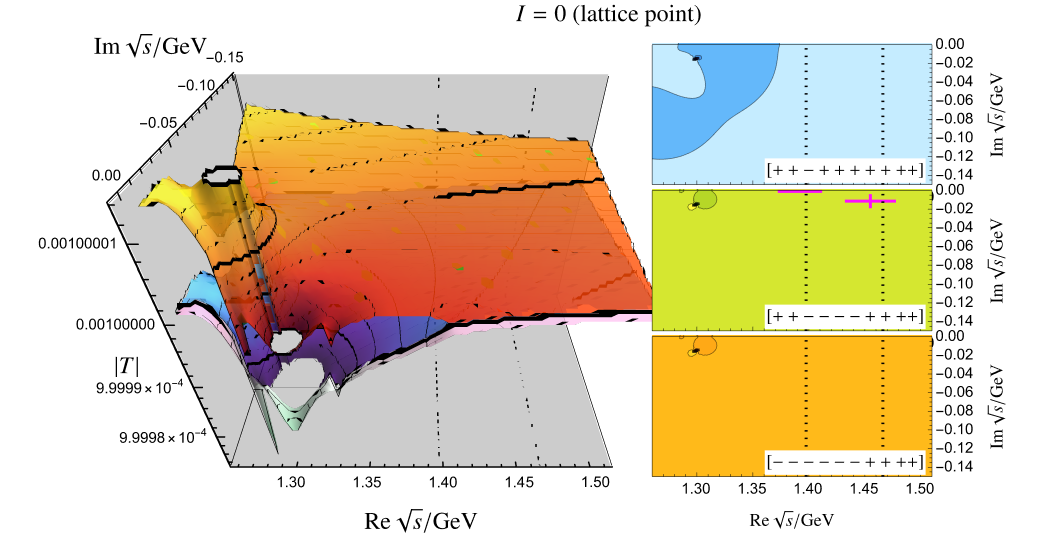}\\
    \includegraphics[height=4cm]{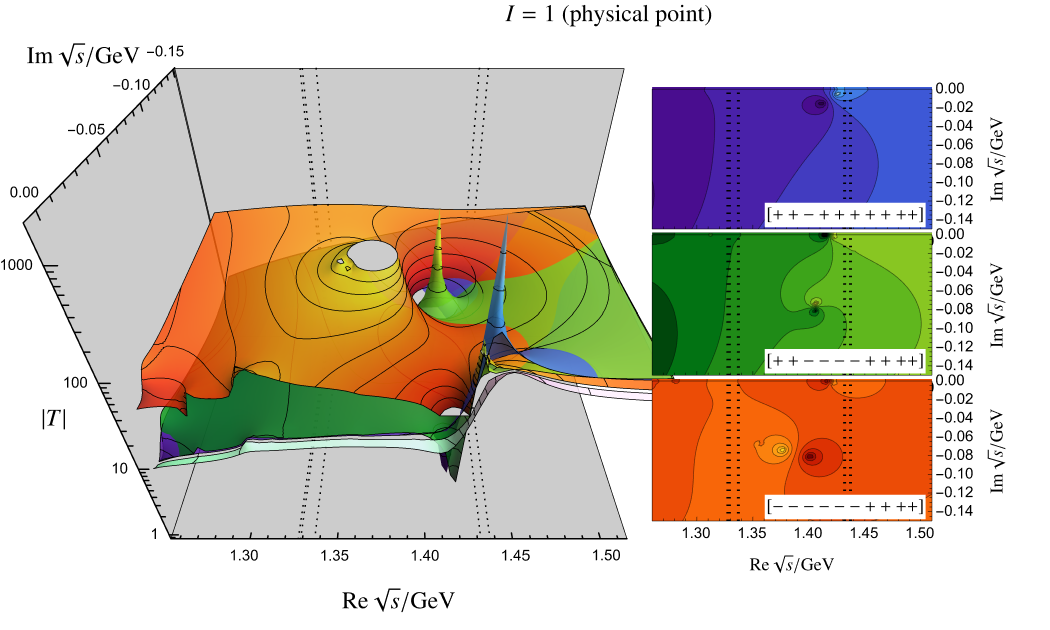}
    \includegraphics[height=4cm]{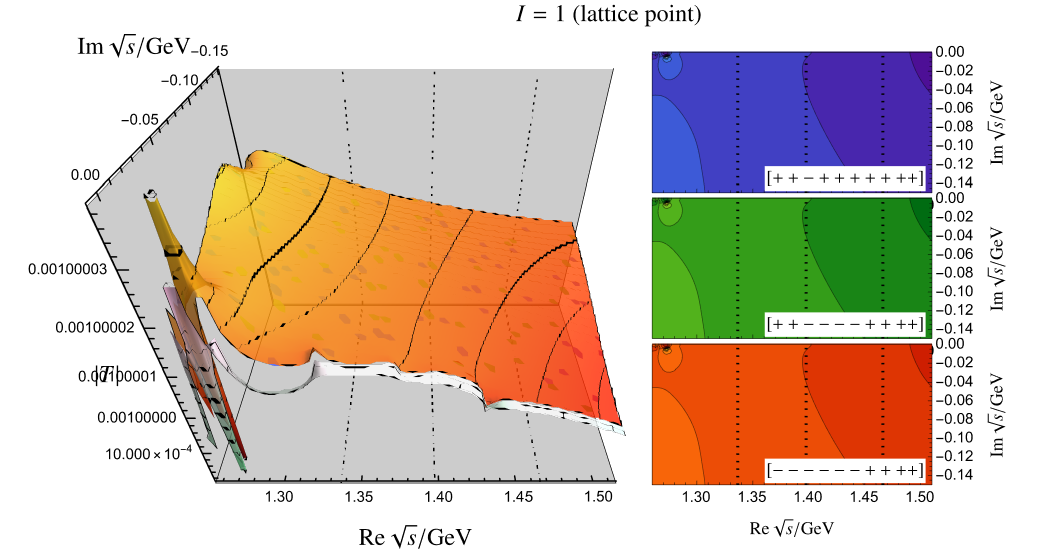}
    \caption{Isoscalar and isovector projected absolute value of the $\pi\Sigma\to \pi\Sigma$ scattering amplitude on the unphysical second Riemann sheets. Nomenclature as in the main text. Magenta crosses represent literature values from Refs.~\cite{BaryonScatteringBaSc:2023ori, BaryonScatteringBaSc:2023zvt} for the lattice point and Refs.~\cite{Ikeda:2012au,Guo:2012vv,Mai:2014xna} for the physical point.}
    \label{fig:app-22}
\end{figure*}

\clearpage
\subsection{M2S2P ($F_{29}$)}

\begin{figure*}[h]
\begin{minipage}[t!]{0.23\textwidth}
    \vspace{0pt} 
\centering
\caption{The total $\chi^2_\mathrm{dof}$ as defined in \cref{eq:chi2}  for the parameter-free fit, M2S2P ($F_{29}$).}
    \label{tab:details_29}
\begin{tabular}{|c|c|}
\hline
$\chi^2_{\rm dof}$ & 48.15 \\
\hline
\end{tabular}
\end{minipage}
    \end{figure*}

\begin{figure*}[h]
    \includegraphics[height=4cm]{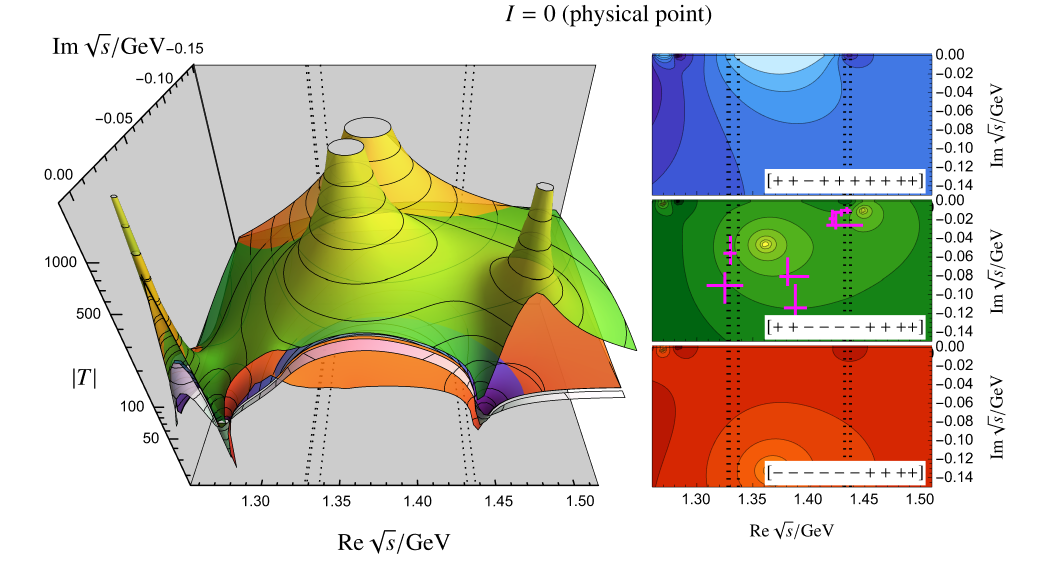}
    \includegraphics[height=4cm]{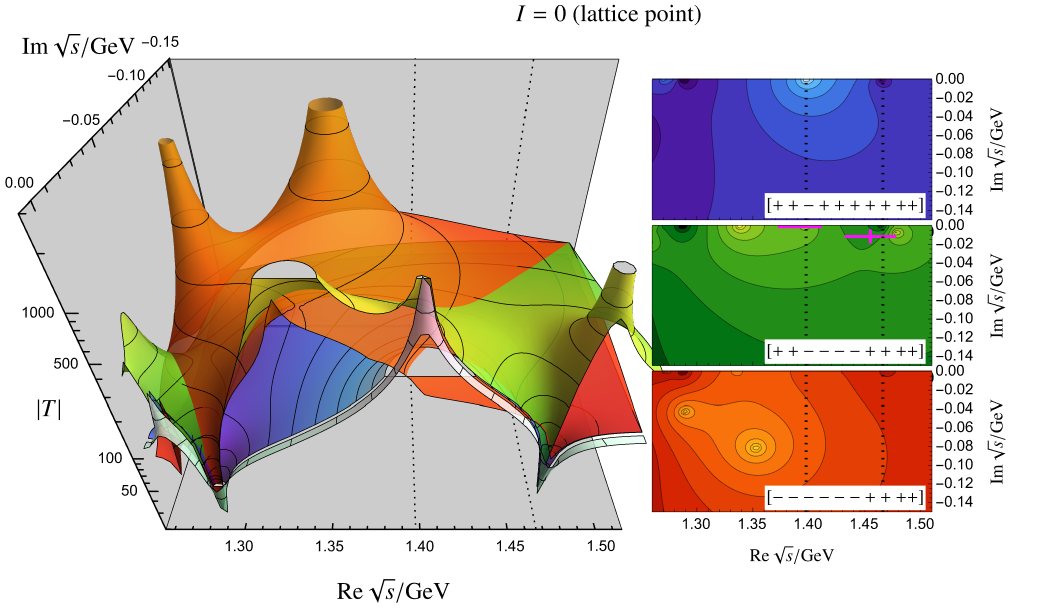}\\
    \includegraphics[height=4cm]{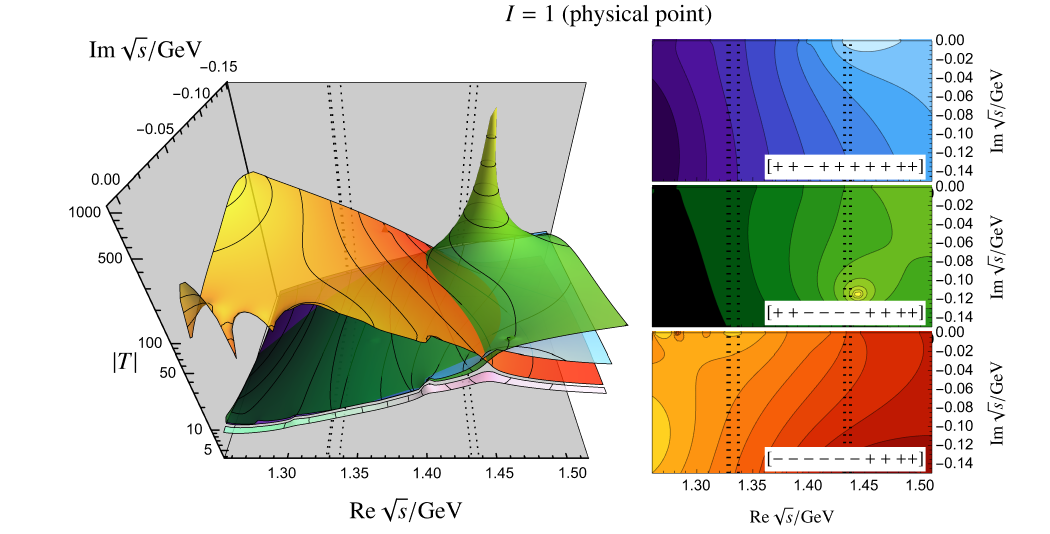}
    \includegraphics[height=4cm]{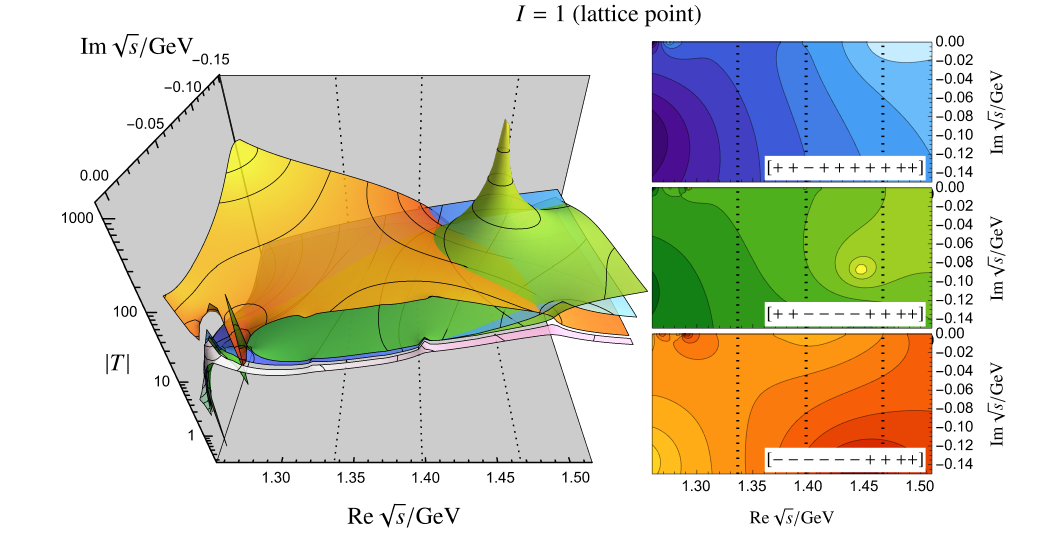}
    \caption{Isoscalar and isovector projected absolute value of the $\pi\Sigma\to \pi\Sigma$ scattering amplitude on the unphysical second Riemann sheets. Nomenclature as in the main text. Magenta crosses represent literature values from Refs.~\cite{BaryonScatteringBaSc:2023ori, BaryonScatteringBaSc:2023zvt} for the lattice point and Refs.~\cite{Ikeda:2012au,Guo:2012vv,Mai:2014xna} for the physical point.}
    \label{fig:app-29}
\end{figure*}

\clearpage
\subsection{M2S3P ($F_{26}$)}

\begin{figure*}[h]
\begin{minipage}[t!]{0.23\textwidth}
    \vspace{0pt} 
\centering
\caption{The total $\chi^2_\mathrm{dof}$ as defined in \cref{eq:chi2}  and the $\Lambda$ 
parameter for M1S1P ($F_{21}$).}
    \label{tab:details_26}
\begin{tabular}{|c|c|}
\hline
$\chi^2_{\rm dof}$ & 18.69 \\
\hline
$\Lambda[\mathrm{GeV}]$ & 1.087866 \\
\hline
\end{tabular}
\label{tab:fit_summaryM2S3E}
\end{minipage}
\hfill
\begin{minipage}[t!]{0.7\textwidth}
    \vspace{-6pt} 
    \includegraphics[width=\linewidth]{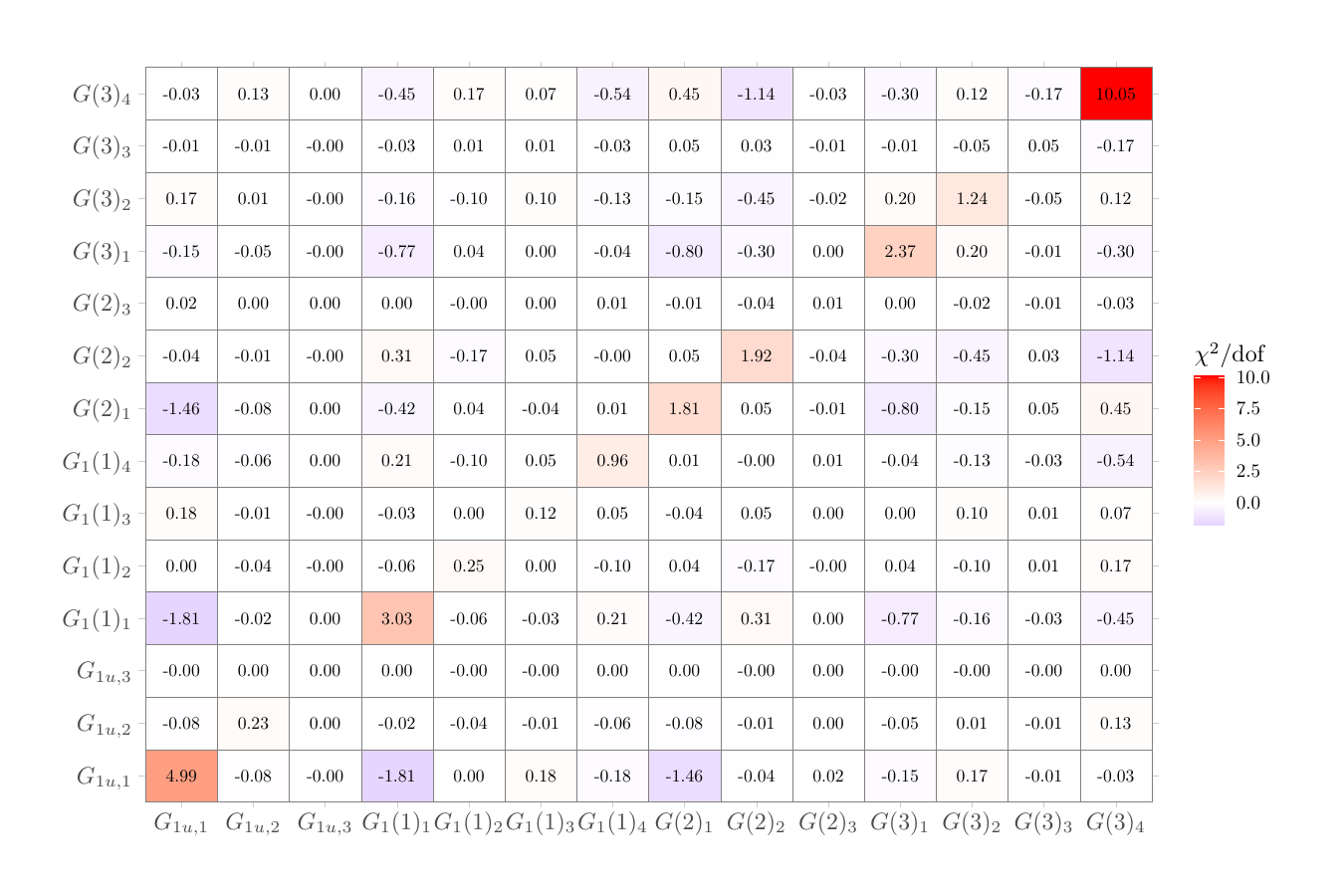}
    \vspace{-26pt}
    \caption{Heat map of correlated $\chi^2_{{\rm dof},ij}$, highlighting the relative impact of each energy level on the total fit quality.}
    \label{fig:partial-chi2_M2S3P}
\end{minipage}
\end{figure*}

\begin{figure*}[h]
    \includegraphics[height=4cm]{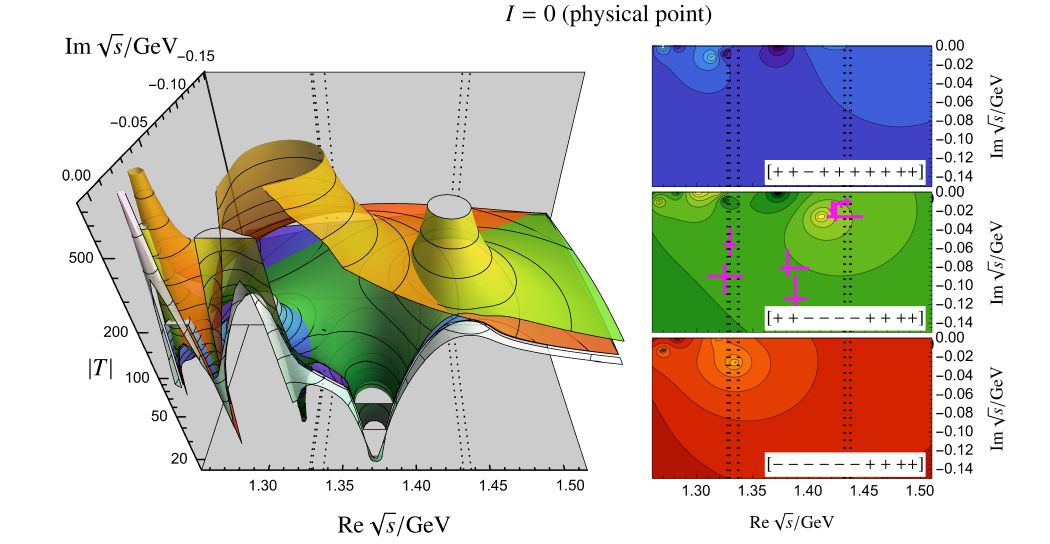}
    \includegraphics[height=4cm]{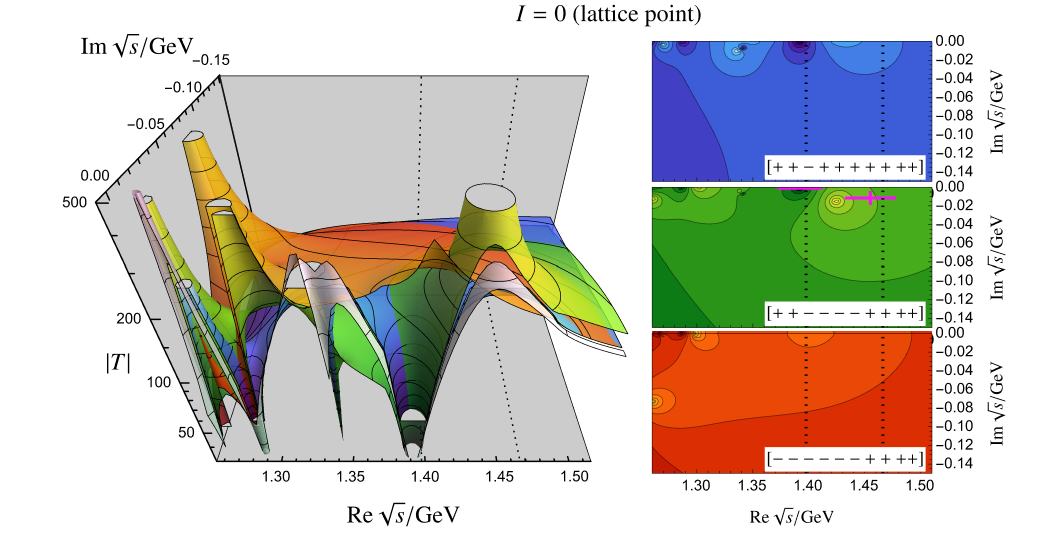}\\
    \includegraphics[height=4cm]{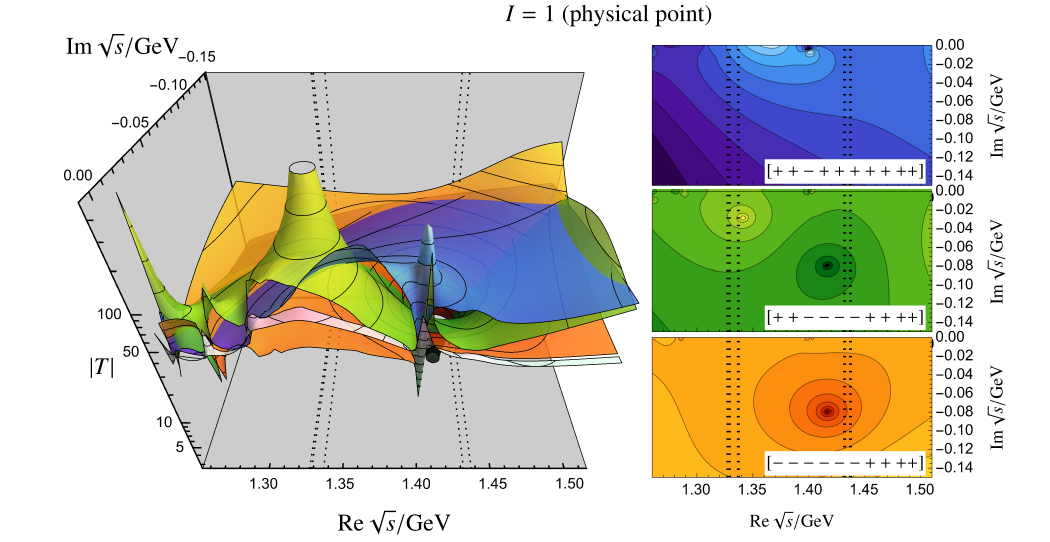}
    \includegraphics[height=4cm]{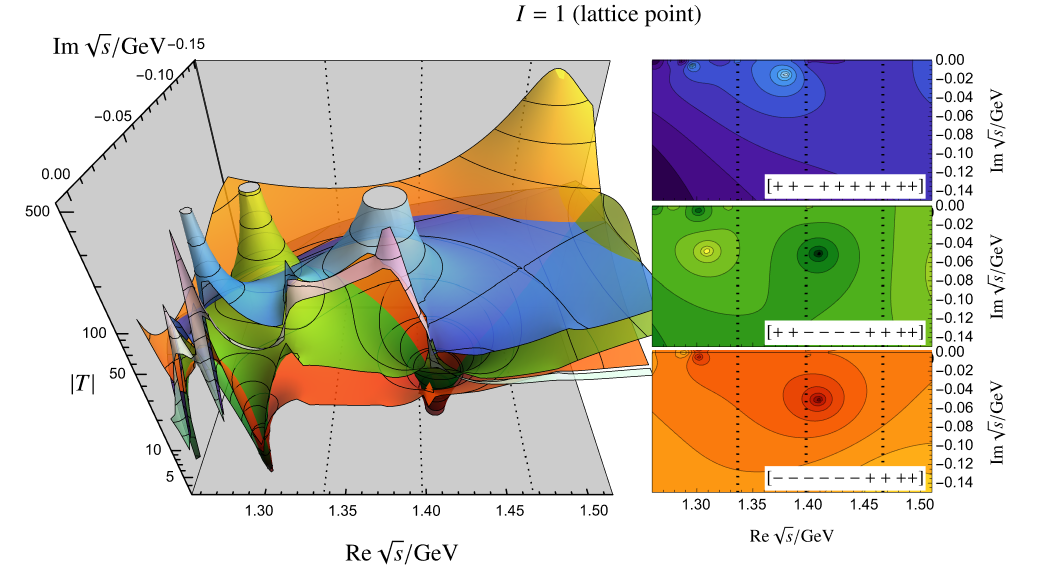}
    \caption{Isoscalar and isovector projected absolute value of the $\pi\Sigma\to \pi\Sigma$ scattering amplitude on the unphysical second Riemann sheets. Nomenclature as in the main text. Magenta crosses represent literature values from Refs.~\cite{BaryonScatteringBaSc:2023ori, BaryonScatteringBaSc:2023zvt} for the lattice point and Refs.~\cite{Ikeda:2012au,Guo:2012vv,Mai:2014xna} for the physical point.}
    \label{fig:app-26}
\end{figure*}
\clearpage
\subsection{M3S1P ($F_{30}$)}

\begin{figure*}[h]
\begin{minipage}[t!]{0.23\textwidth}
    \vspace{0pt} 
\centering
\caption{The total $\chi^2_\mathrm{dof}$ as defined in \cref{eq:chi2}, subtraction constants and LECs for M3S1P ($F_{30}$).}
    \label{tab:details_30}
\begin{tabular}{|c|c|}
\hline
$\chi^2_{\rm dof}$ & 1.51 \\
\hline
$a_{\bar{K}N}$ & 8.684939e-05 \\
$a_{\pi\Lambda}$ & 6.602418e-02 \\
$a_{\pi\Sigma}$ & -3.019298e-03 \\
$a_{\eta\Lambda}$ & 6.760390e-03 \\
$a_{\eta\Sigma}$ & -6.407026e-03 \\
$a_{K\Xi}$ & 3.279341e-03 \\
\hline
$b_0[1/\mathrm{GeV}]$ & -6.277105e-01 \\
$b_D[1/\mathrm{GeV}]$ & -3.489336e-01 \\
$b_F[1/\mathrm{GeV}]$ & -2.942295e-01 \\
$d_1[1/\mathrm{GeV}]$ & -1.759572e-01 \\
$d_2[1/\mathrm{GeV}]$ & -1.042583e-01 \\
$d_3[1/\mathrm{GeV}]$ & -4.012072e-01 \\
$d_4[1/\mathrm{GeV}]$ & -1.576405e-01 \\
\hline
\end{tabular}
\end{minipage}
\hfill
\begin{minipage}[t!]{0.7\textwidth}
    \vspace{-6pt} 
    \includegraphics[width=\linewidth]{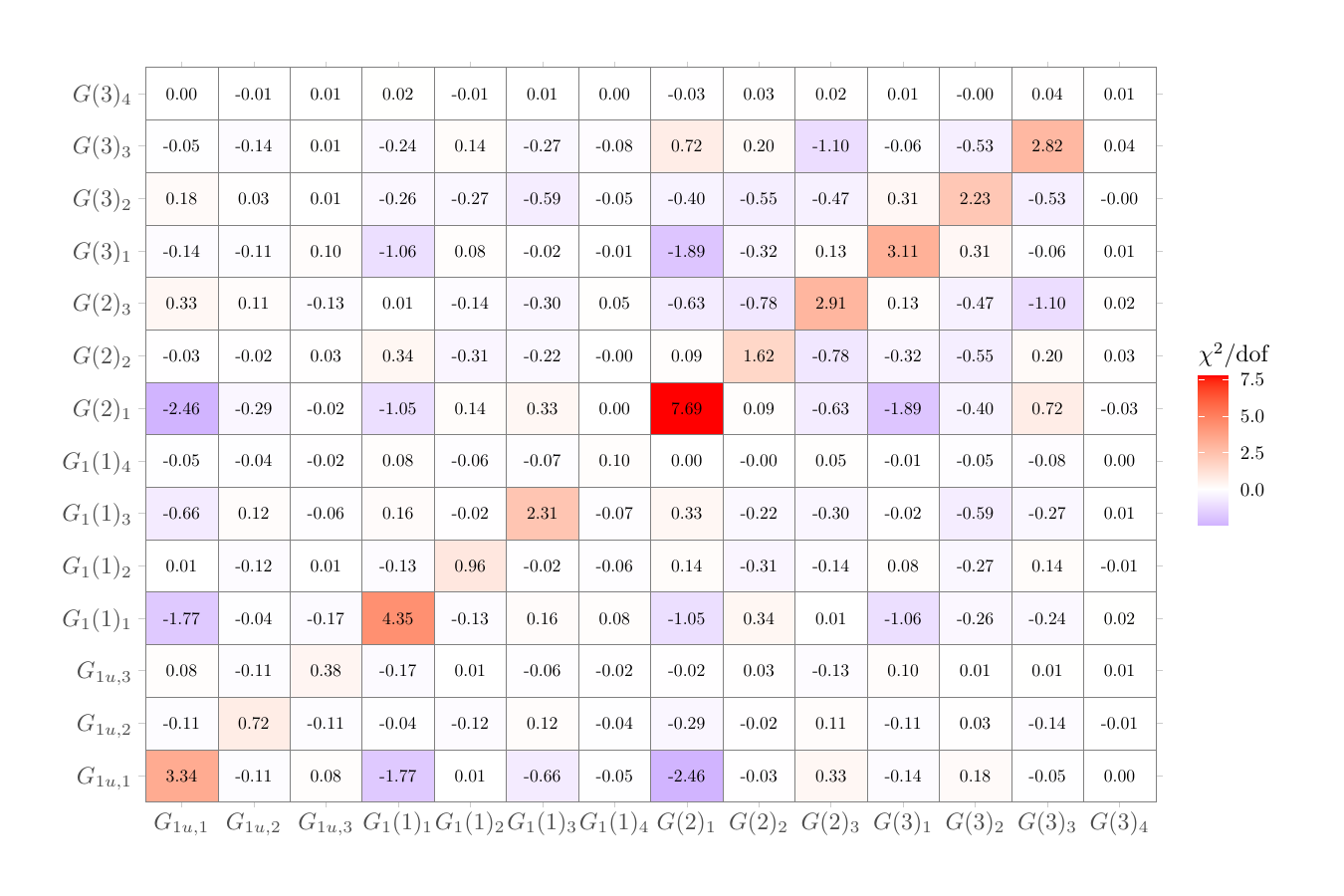}
    \vspace{-26pt}
    \caption{Heat map of correlated $\chi^2_{{\rm dof},ij}$, highlighting the relative impact of each energy level on the total fit quality.}
    \label{fig:partial-30}
\end{minipage}
\end{figure*}

\begin{figure*}[h]
    \includegraphics[height=4cm]{plots/3D-poles/fit-30-exp0.pdf}
    \includegraphics[height=4cm]{plots/3D-poles/fit-30-LAT0.pdf}\\
    \includegraphics[height=4cm]{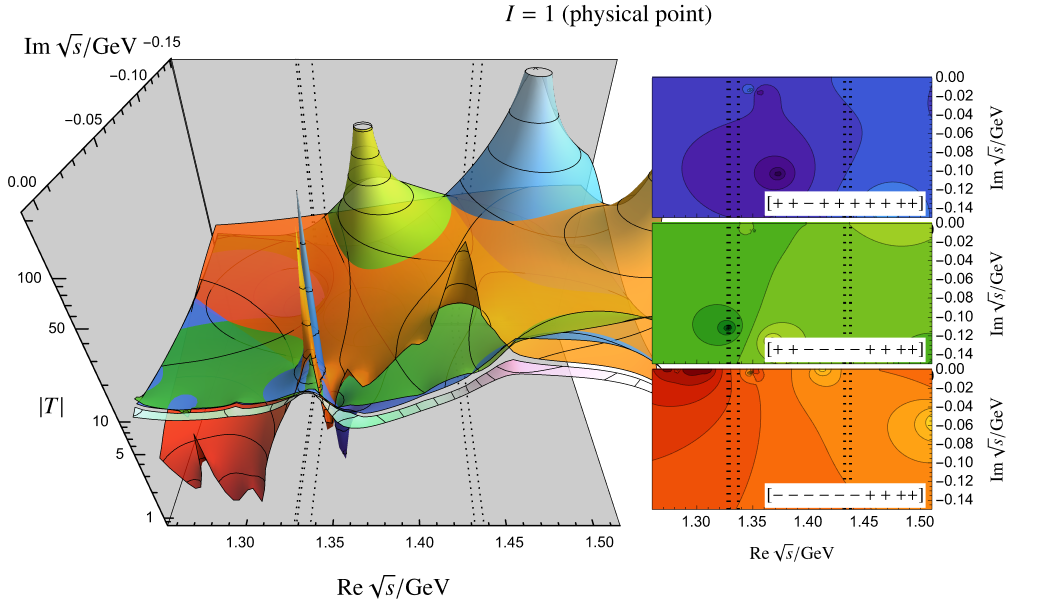}
    \includegraphics[height=4cm]{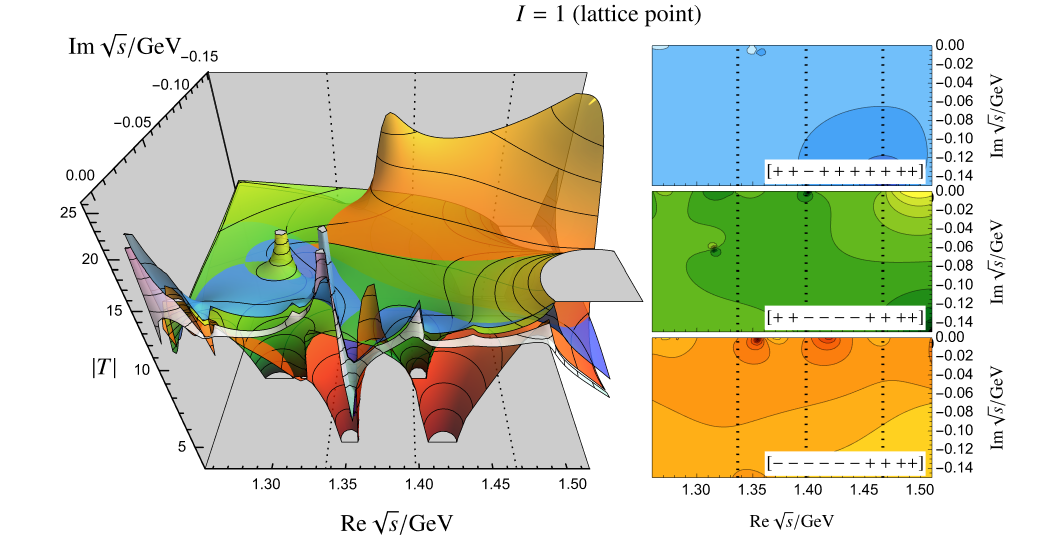}
    \caption{Isoscalar and isovector projected absolute value of the $\pi\Sigma\to \pi\Sigma$ scattering amplitude on the unphysical second Riemann sheets. Nomenclature as in the main text. Magenta crosses represent literature values from Refs.~\cite{BaryonScatteringBaSc:2023ori, BaryonScatteringBaSc:2023zvt} for the lattice point and Refs.~\cite{Ikeda:2012au,Guo:2012vv,Mai:2014xna} for the physical point.}
    \label{fig:app-30}
\end{figure*}

\clearpage
\subsection{M3S2P ($F_{13}$)}

\begin{figure*}[h]
\begin{minipage}[t!]{0.23\textwidth}
    \vspace{0pt} 
\centering
\caption{The total $\chi^2_\mathrm{dof}$ as defined in \cref{eq:chi2}  and LECs for Fit 13 (M3S2P).}
    \label{tab:details_13}
\begin{tabular}{|c|c|}
\hline
$\chi^2_{\rm dof}$ & 0.90 \\
\hline
\centering
$b_0[1/\mathrm{GeV}]$ & -5.349500e-01 \\
$b_D[1/\mathrm{GeV}]$ & 9.599595e-02 \\
$b_F[1/\mathrm{GeV}]$ & -3.256764e-01 \\
$d_1[1/\mathrm{GeV}]$ & -8.386487e-01 \\
$d_2[1/\mathrm{GeV}]$ & 1.518967e-01 \\
$d_3[1/\mathrm{GeV}]$ & -4.546126e-01 \\
$d_4[1/\mathrm{GeV}]$ & 7.207285e-03 \\
\hline
\end{tabular}
\end{minipage}
\hfill
\begin{minipage}[t!]{0.7\textwidth}
    \vspace{-6pt} 
    \includegraphics[width=\linewidth]{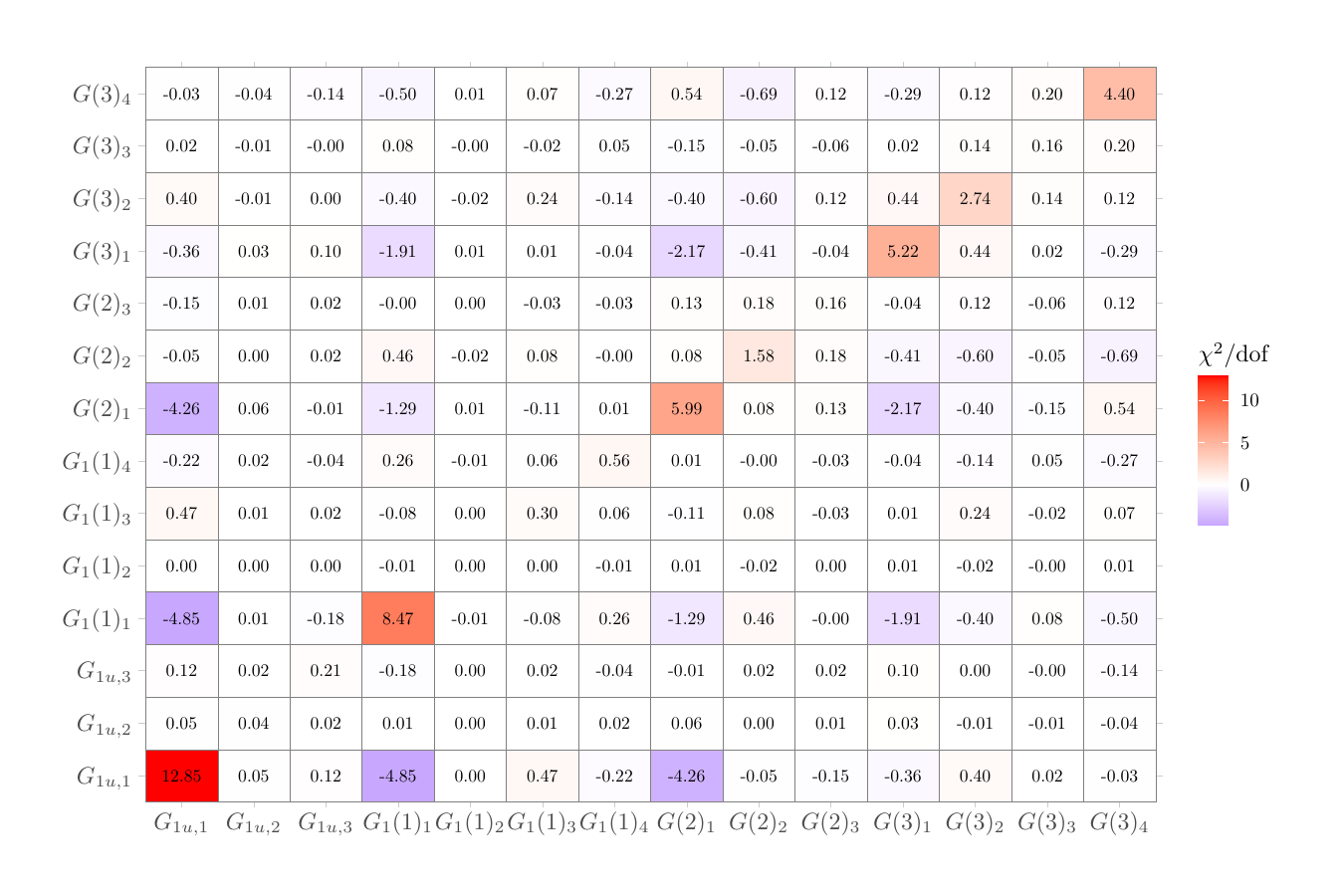}
    \vspace{-26pt}
    \caption{Heat map of correlated $\chi^2_{{\rm dof},ij}$, highlighting the relative impact of each energy level on the total fit quality.}
    \label{fig:partial-13P}
\end{minipage}
\end{figure*}

\begin{figure*}[h]
    \includegraphics[height=4cm]{plots/3D-poles/fit-13-exp0.pdf}
    \includegraphics[height=4cm]{plots/3D-poles/fit-13-LAT0.pdf}\\
    \includegraphics[height=4cm]{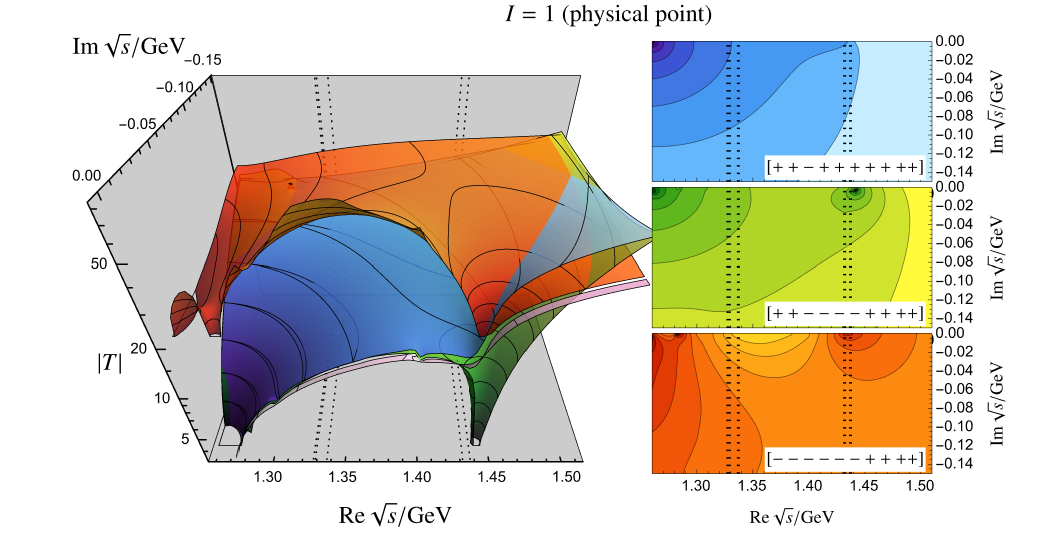}
    \includegraphics[height=4cm]{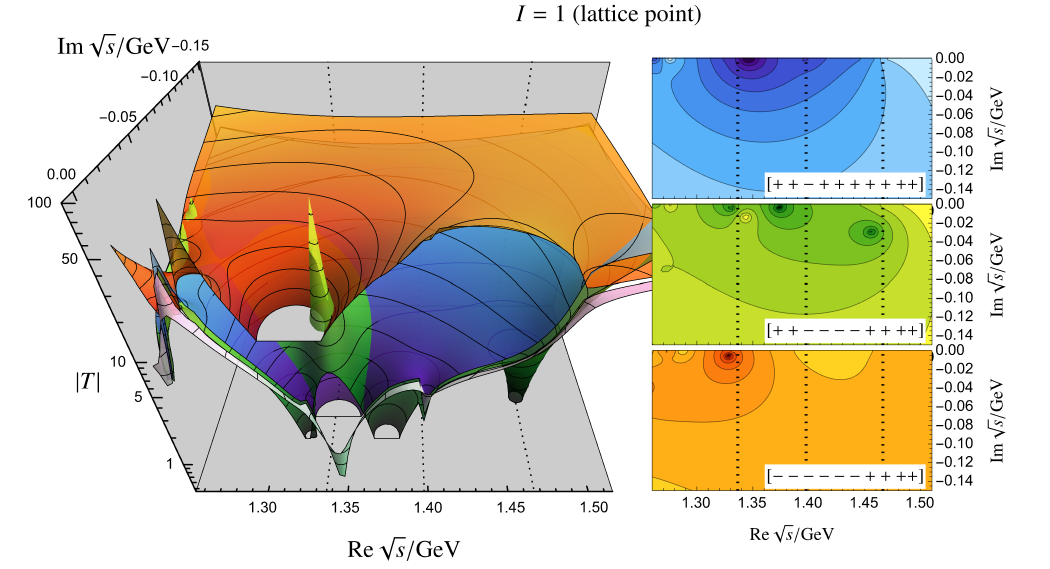}
    \caption{Isoscalar and isovector projected absolute value of the $\pi\Sigma\to \pi\Sigma$ scattering amplitude on the unphysical second Riemann sheets. Nomenclature as in the main text. Magenta crosses represent literature values from Refs.~\cite{BaryonScatteringBaSc:2023ori, BaryonScatteringBaSc:2023zvt} for the lattice point and Refs.~\cite{Ikeda:2012au,Guo:2012vv,Mai:2014xna} for the physical point.}
    \label{fig:app-13}
\end{figure*}

\clearpage
\subsection{M3S3P ($F_{11}$)}

\begin{figure*}[h]
\begin{minipage}[t!]{0.23\textwidth}
    \vspace{0pt} 
\centering
\caption{The total $\chi^2_\mathrm{dof}$ as defined in \cref{eq:chi2}, $\Lambda$ parameter  and LECs for M3S3P ($F_{11}$).}
    \label{tab:details_11}
\begin{tabular}{|c|c|}
\hline
$\chi^2_{\rm dof}$ & 1.46   \\
$\Lambda[\mathrm{GeV}]$ & 0.6802625 \\
\hline
$b_0[1/\mathrm{GeV}]$ & -5.440114e-01 \\
$b_D[1/\mathrm{GeV}]$ & -8.008565e-01 \\
$b_F[1/\mathrm{GeV}]$ & -4.127402e-01 \\
$d_1[1/\mathrm{GeV}]$ & -4.572182e-01 \\
$d_2[1/\mathrm{GeV}]$ & -2.716873e-01 \\
$d_3[1/\mathrm{GeV}]$ & -6.182698e-01 \\
$d_4[1/\mathrm{GeV}]$ & -9.100890e-01 \\
\hline
\end{tabular}
\end{minipage}
\hfill
\begin{minipage}[t!]{0.7\textwidth}
    \vspace{-6pt} 
    \includegraphics[width=\linewidth]{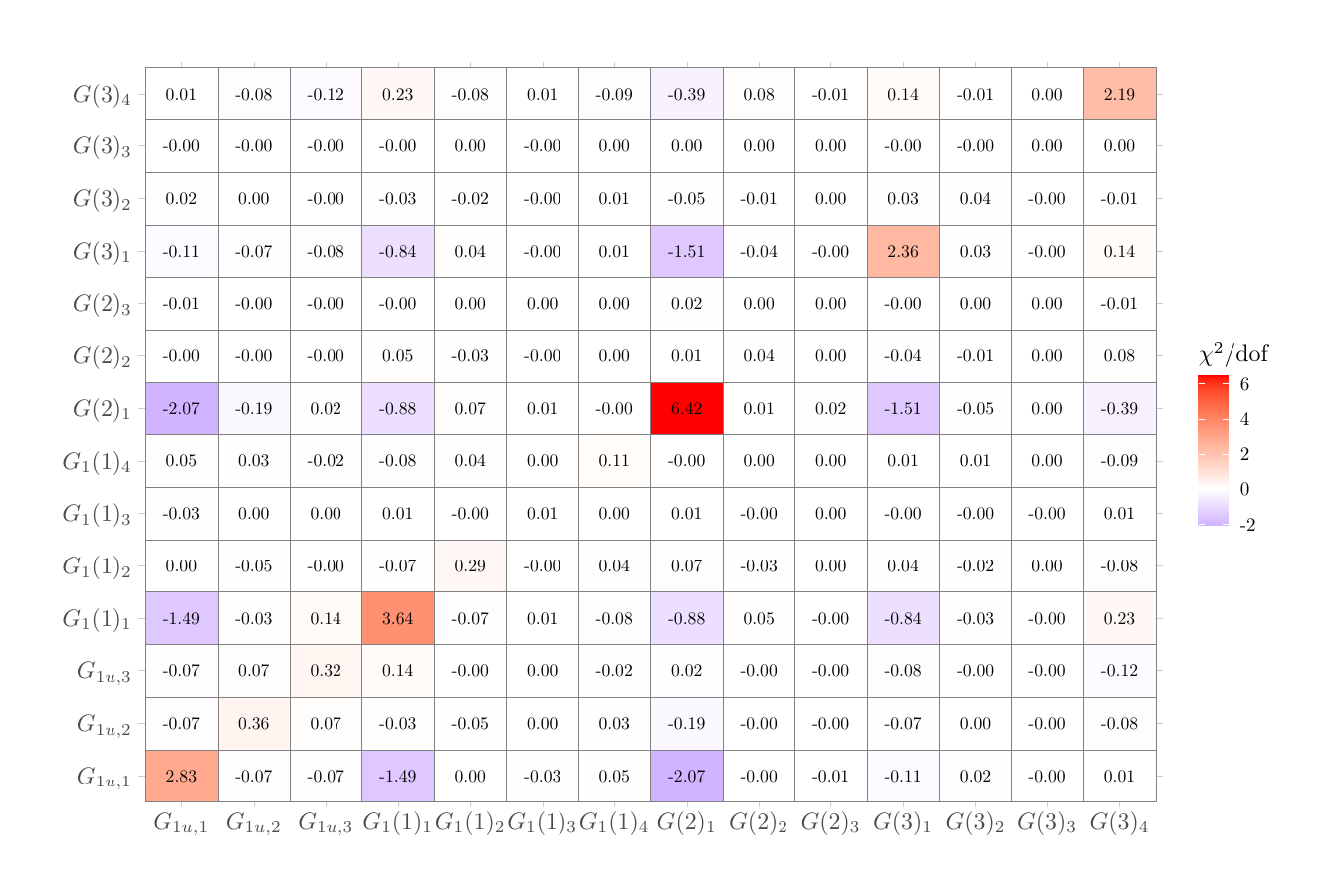}
    \vspace{-26pt}
    \caption{Heat map of correlated $\chi^2_{{\rm dof},ij}$, highlighting the relative impact of each energy level on the total fit quality.}
    \label{fig:partial-chi2_M3S3E}
\end{minipage}
\end{figure*}

\begin{figure*}[h]
    \includegraphics[height=4cm]{plots/3D-poles/fit-11-exp0.pdf}
    \includegraphics[height=4cm]{plots/3D-poles/fit-11-LAT0.pdf}\\
    \includegraphics[height=4cm]{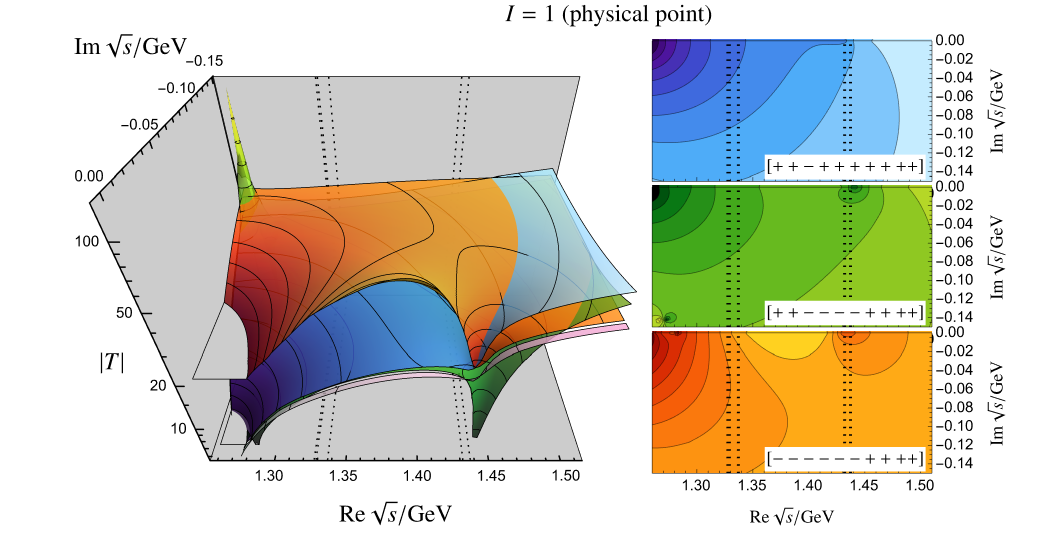}
    \includegraphics[height=4cm]{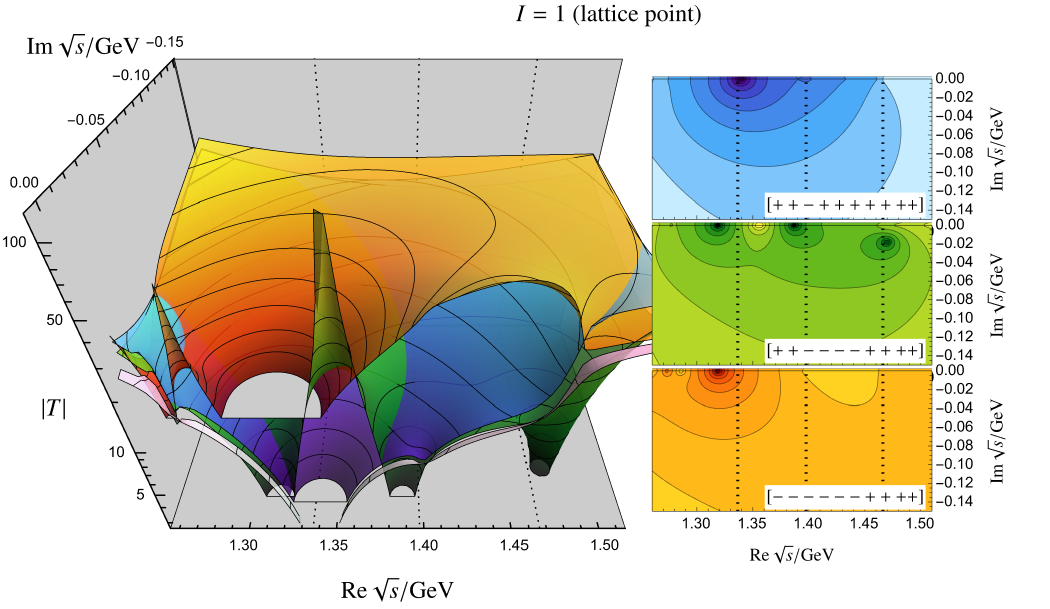}
    \caption{Isoscalar and isovector projected absolute value of the $\pi\Sigma\to \pi\Sigma$ scattering amplitude on the unphysical second Riemann sheets. Nomenclature as in the main text. Magenta crosses represent literature values from Refs.~\cite{BaryonScatteringBaSc:2023ori, BaryonScatteringBaSc:2023zvt} for the lattice point and Refs.~\cite{Ikeda:2012au,Guo:2012vv,Mai:2014xna} for the physical point.}
    \label{fig:app-11}
\end{figure*}

\clearpage
\section{Appendix C: Detailed fit results. Combined fits to the experimental data and lattice input.}
\addcontentsline{toc}{section}{Appendix C: Detailed fit results. Combined fits to the experimental data and lattice input.}
\label{app:fit-results-combined}

\subsection{M1S3PL ($F_{24}$)}

\begin{figure*}[h]
\begin{minipage}[t!]{0.23\textwidth}
    \vspace{0pt} 
\centering
\caption{The total $\chi^2_\mathrm{dof}$ as defined in \cref{eq:chi2}  and the $\Lambda$ parameter for M1S3PL ($F_{24}$).}
    \label{tab:details_24}
\begin{tabular}{|c|c|}
\hline
$\chi^2_{\rm dof}$ & 27.56447   \\
\hline
$\Lambda[\mathrm{GeV}]$ & 0.8106917 \\
\hline
\end{tabular}
\end{minipage}
\begin{minipage}[t!]{0.7\textwidth}
    \vspace{-6pt} 
    \includegraphics[width=\linewidth]{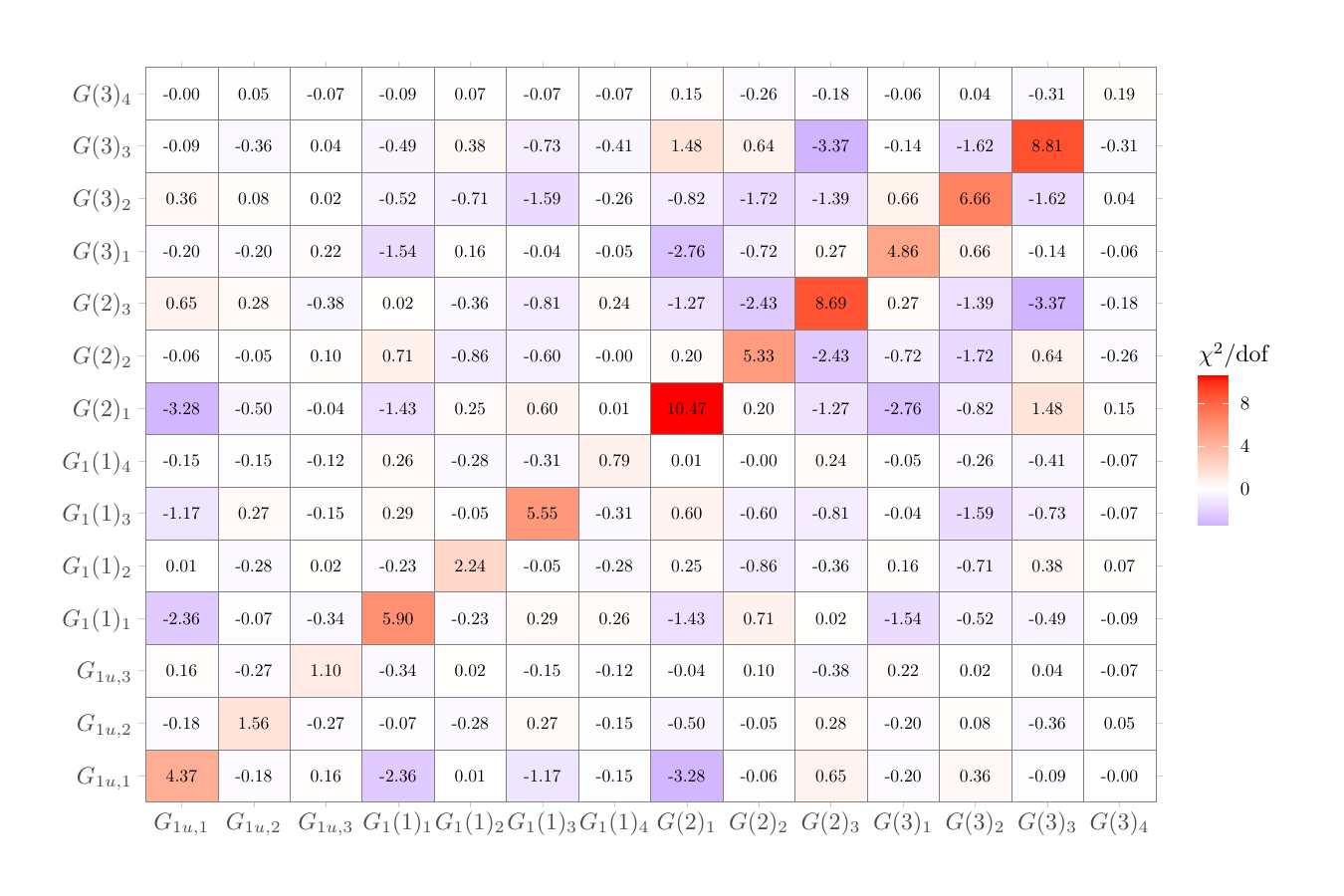}
    \vspace{-26pt}
    \caption{Heat map of correlated $\chi^2_{{\rm dof},ij}$, highlighting the relative impact of each energy level on the total fit quality.}
    \label{fig::partial-chi2_M1S3PL}
\end{minipage}

\end{figure*}

\begin{figure*}[h]
    \includegraphics[height=4cm]{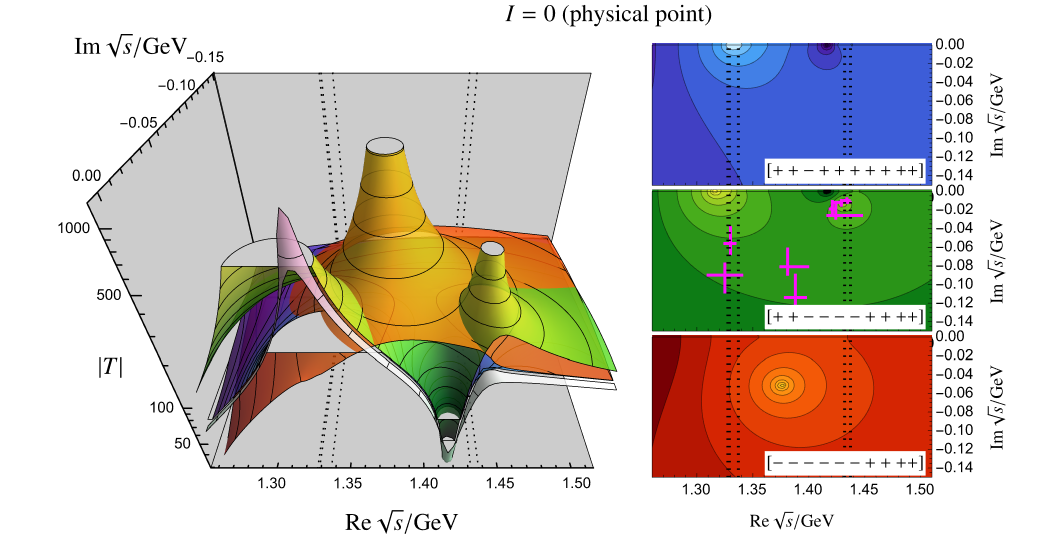}
    \includegraphics[height=4cm]{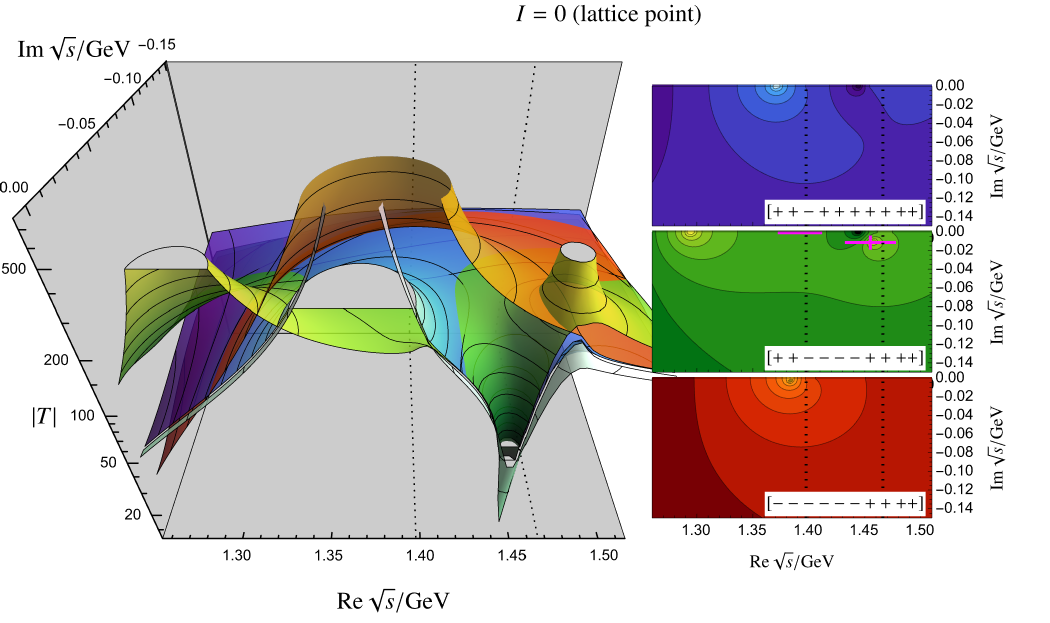}\\
    \includegraphics[height=4cm]{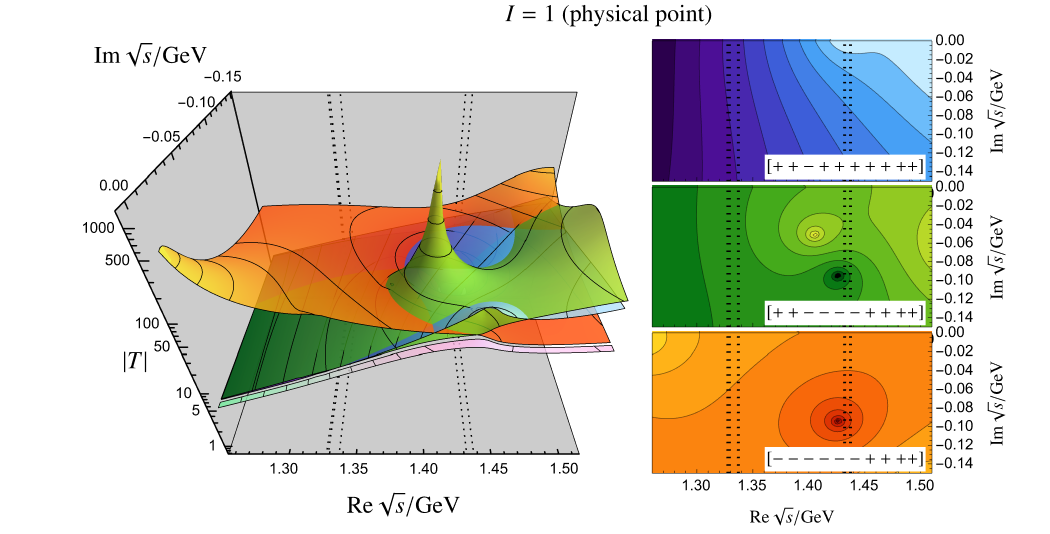}
    \includegraphics[height=4cm]{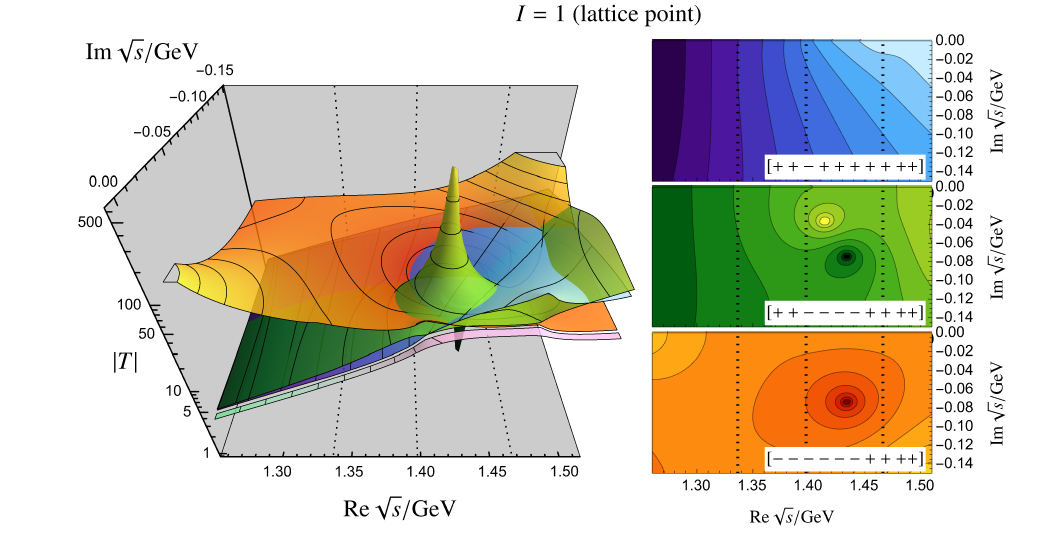}
    \caption{Isoscalar and isovector projected absolute value of the $\pi\Sigma\to \pi\Sigma$ scattering amplitude on the unphysical second Riemann sheets. Nomenclature as in the main text. Magenta crosses represent literature values from Refs.~\cite{BaryonScatteringBaSc:2023ori, BaryonScatteringBaSc:2023zvt} for the lattice point and Refs.~\cite{Ikeda:2012au, Guo:2012vv, Mai:2014xna} for the physical point.}
    \label{fig:app-24}
\end{figure*}

\clearpage
\subsection{M2S3PL ($F_{23}$)}

\begin{figure*}[h]
\begin{minipage}[t!]{0.23\textwidth}
    \vspace{0pt} 
\centering
\caption{The total $\chi^2_\mathrm{dof}$ as defined in \cref{eq:chi2}  and the $\Lambda$ parameter for M2S3PL ($F_{23}$).}
    \label{tab:details_23}
\begin{tabular}{|c|c|}
\hline
$\chi^2_{\rm dof}$ & 17.81 \\
\hline
$\Lambda[\mathrm{GeV}]$ & 1.088185 \\
\hline
\end{tabular}
\label{tab:fit_summaryM2S3PL}
\end{minipage}
\hfill
\begin{minipage}[t!]{0.7\textwidth}
    \vspace{-6pt} 
    \includegraphics[width=\linewidth]{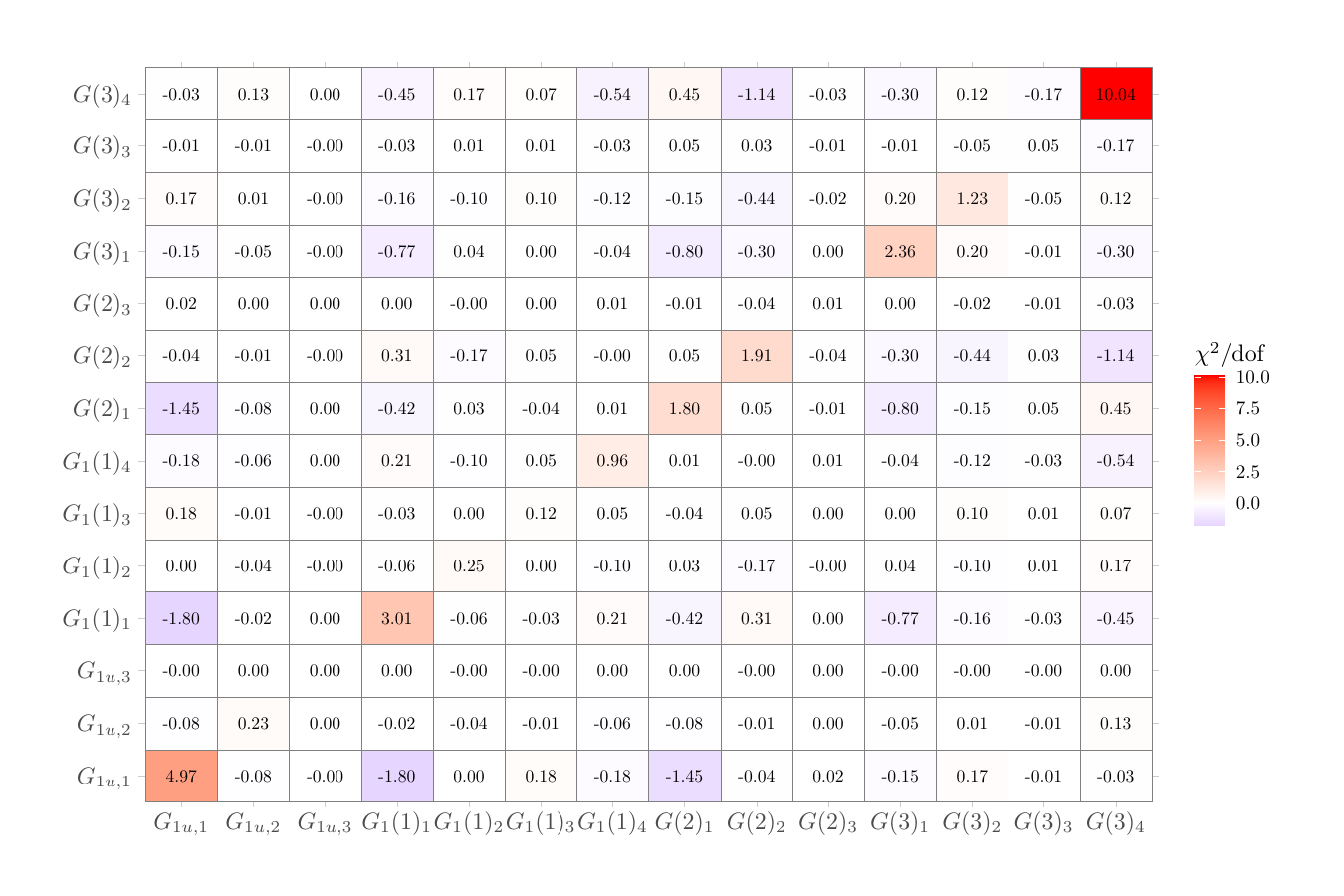}
    \vspace{-26pt}
    \caption{Heat map of correlated $\chi^2_{{\rm dof},ij}$, highlighting the relative impact of each energy level on the total fit quality.}
    \label{fig:partial-chi2_M2S3PL}
\end{minipage}
\end{figure*}

\begin{figure*}[h]
    \includegraphics[height=4cm]{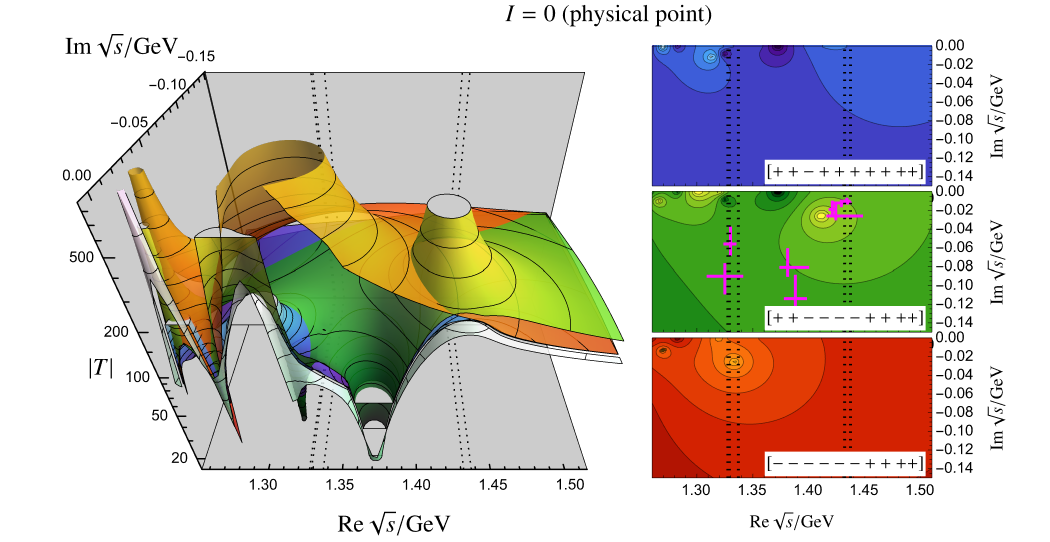}
    \includegraphics[height=4cm]{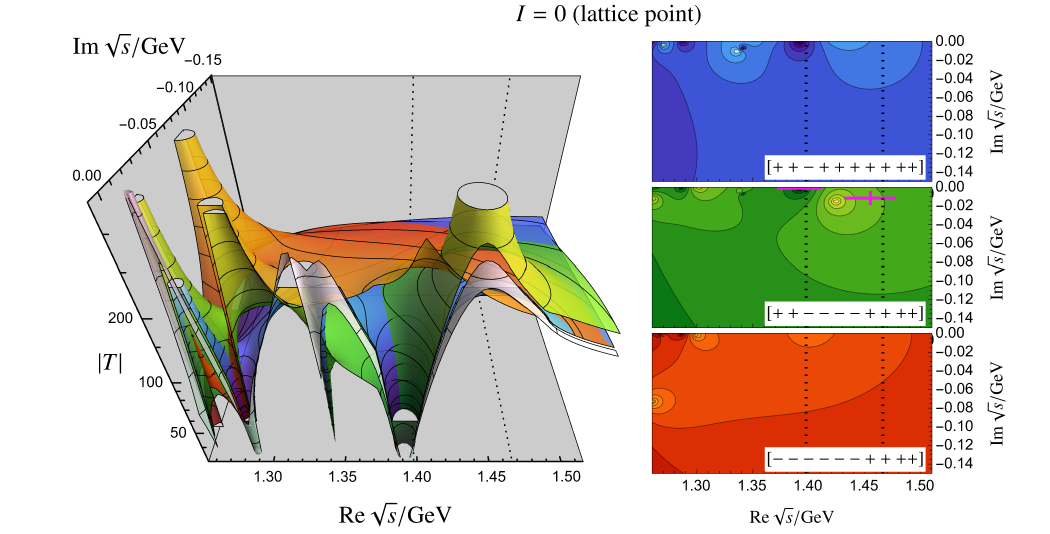}\\
    \includegraphics[height=4cm]{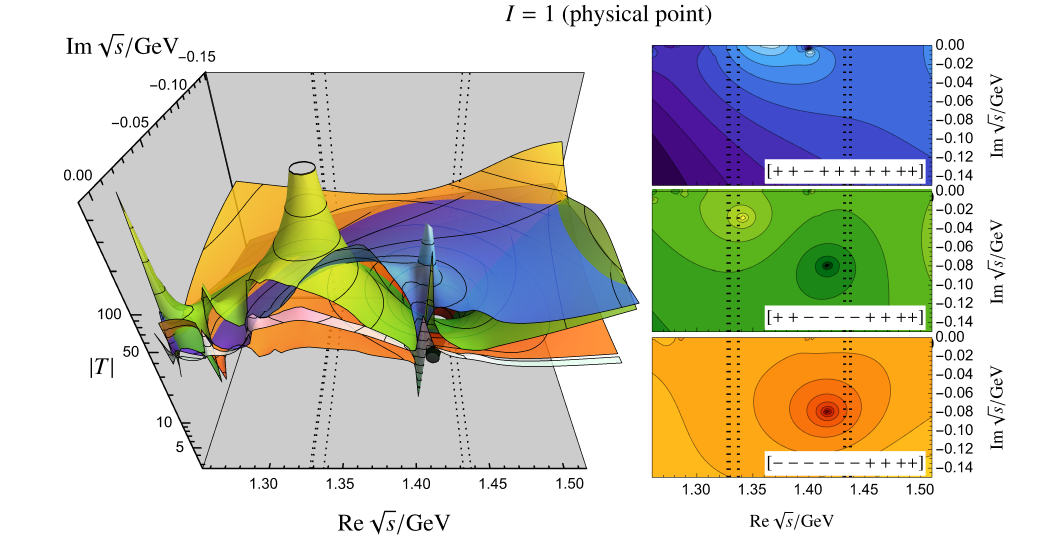}
    \includegraphics[height=4cm]{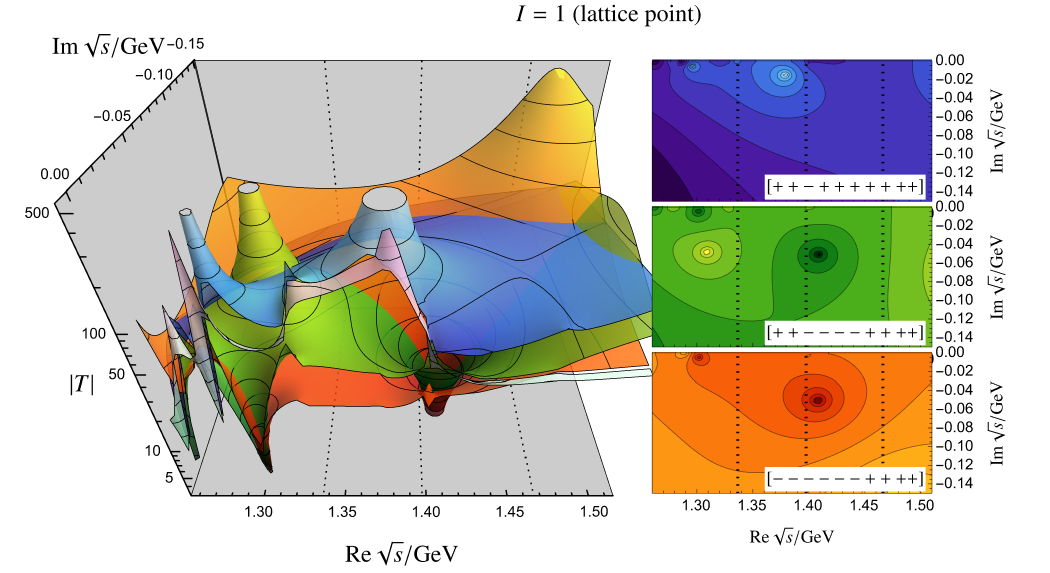}
    \caption{Isoscalar and isovector projected absolute value of the $\pi\Sigma\to \pi\Sigma$ scattering amplitude on the unphysical second Riemann sheets. Nomenclature as in the main text. Magenta crosses represent literature values from Refs.~\cite{BaryonScatteringBaSc:2023ori, BaryonScatteringBaSc:2023zvt} for the lattice point and Refs.~\cite{Ikeda:2012au, Guo:2012vv, Mai:2014xna} for the physical point.}
    \label{fig:app-23}
\end{figure*}

\clearpage
\subsection{M3S1PL ($F_{17}$)}

\begin{figure*}[h]
\hspace{-6pt}
\begin{minipage}[t!]{0.23\textwidth}
    \vspace{0pt} 
\centering
\caption{The total $\chi^2_\mathrm{dof}$ as defined in \cref{eq:chi2}, subtraction constants for both the lattice and the physical point and LECs for M3S1PL ($F_{17}$).}
    \label{tab:details_17}
\begin{tabular}{|c|c|c|}
\hline
$\chi^2_{\rm dof}$ & \multicolumn{2}{c|}{1.441434} \\
\hline
& Lattice & Experimental \\
\hline
$a_{\bar{K}N}$ & -1.564779e-03 & -1.546676e-03 \\
$a_{\pi\Lambda}$ & -1.079700e-01 & 8.517151e-02 \\
$a_{\pi\Sigma}$ & +3.721385e-03 & -2.728888e-03 \\
$a_{\eta\Lambda}$ & +1.376050e-02 & 9.982057e-03 \\
$a_{\eta\Sigma}$ & +2.163700e-01 & -5.741158e-03 \\
$a_{K\Xi}$ & +3.948000e-02 & 8.712392e-02 \\
\hline
$b_0[1/\mathrm{GeV}]$ & \multicolumn{2}{c|}{-6.569390e-01} \\
$b_D[1/\mathrm{GeV}]$ & \multicolumn{2}{c|}{6.740337e-02} \\
$b_F[1/\mathrm{GeV}]$ & \multicolumn{2}{c|}{-3.257189e-01} \\
$d_1[1/\mathrm{GeV}]$ & \multicolumn{2}{c|}{-2.520795e-01} \\
$d_2[1/\mathrm{GeV}]$ & \multicolumn{2}{c|}{3.095445e-02} \\
$d_3[1/\mathrm{GeV}]$ & \multicolumn{2}{c|}{-8.990016e-02} \\
$d_4[1/\mathrm{GeV}]$ & \multicolumn{2}{c|}{-5.497259e-02} \\
\hline
\end{tabular}
\end{minipage}
\hfill
\begin{minipage}[t!]{0.6\textwidth}
    \vspace{-6pt} 
    \includegraphics[width=\linewidth]{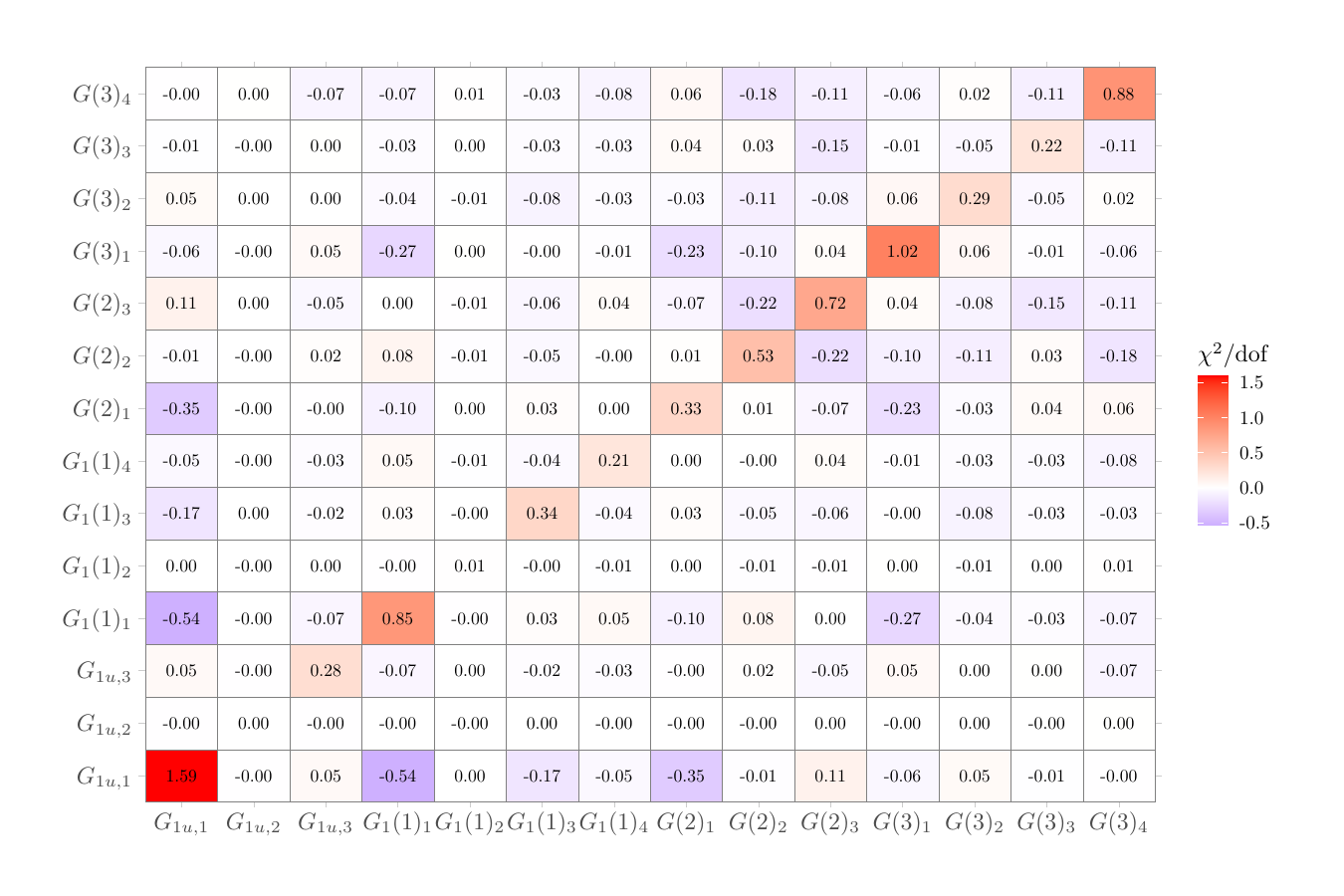}
    \vspace{-26pt}
    \caption{Heat map of correlated $\chi^2_{{\rm dof},ij}$, highlighting the relative impact of each energy level on the total fit quality.}
    \label{fig:partial-17}
\end{minipage}
\end{figure*}

\begin{figure*}[h]
    \includegraphics[height=4cm]{plots/3D-poles/fit-17-exp0.pdf}
    \includegraphics[height=4cm]{plots/3D-poles/fit-17-LAT0.pdf}\\
    \includegraphics[height=4cm]{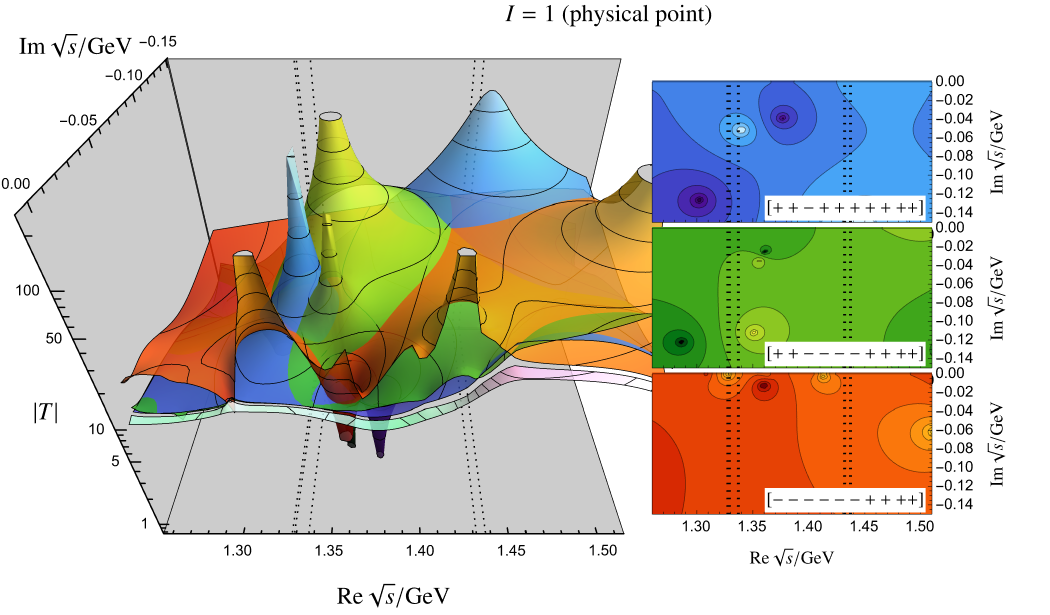}
    \includegraphics[height=4cm]{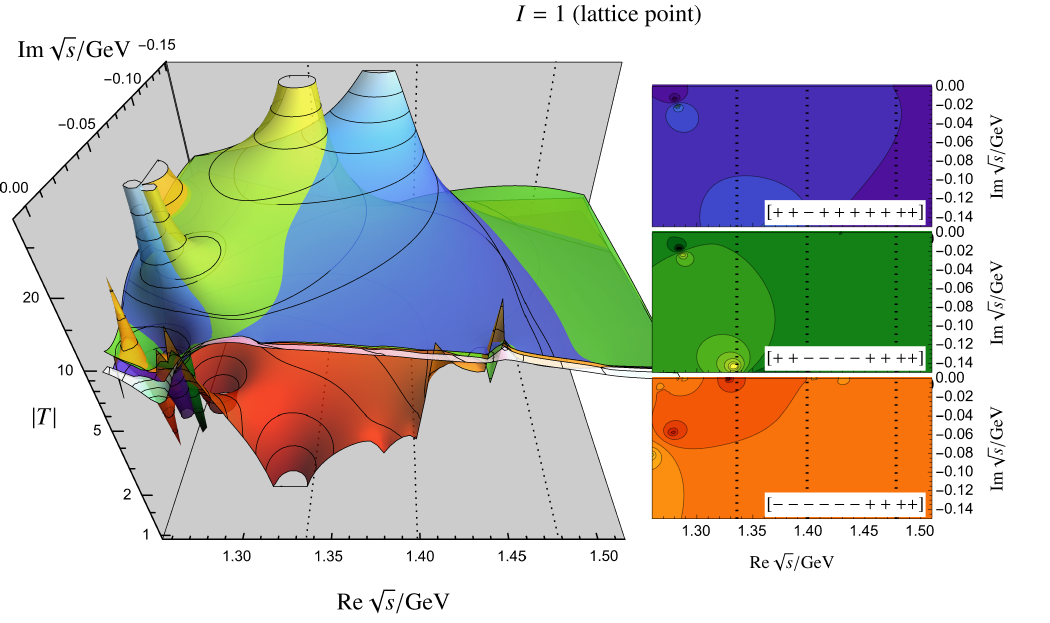}
    \caption{Isoscalar and isovector projected absolute value of the $\pi\Sigma\to \pi\Sigma$ scattering amplitude on the unphysical second Riemann sheets. Nomenclature as in the main text. Magenta crosses represent literature values from Refs.~\cite{BaryonScatteringBaSc:2023ori, BaryonScatteringBaSc:2023zvt} for the lattice point and Refs.~\cite{Ikeda:2012au, Guo:2012vv, Mai:2014xna} for the physical point.}
    \label{fig:app-17}
\end{figure*}

\clearpage
\subsection{M3S2PL ($F_{16}$)}

\begin{figure*}[h]
\begin{minipage}[t!]{0.23\textwidth}
    \vspace{0pt} 
\centering
\caption{The total $\chi^2_\mathrm{dof}$ as defined in \cref{eq:chi2}  and LECs for M3S2PL ($F_{16}$).}
    \label{tab:details_16}
\begin{tabular}{|c|c|}
\hline
$\chi^2_{\rm dof}$ & 2.1189 \\
\hline
\centering
$b_0[1/\mathrm{GeV}]$ & -3.414328e-01 \\
$b_D[1/\mathrm{GeV}]$ & 6.368574e-02 \\
$b_F[1/\mathrm{GeV}]$ & -3.021744e-01 \\
$d_1[1/\mathrm{GeV}]$ & -2.593481e-01 \\
$d_2[1/\mathrm{GeV}]$ & 4.433054e-02 \\
$d_3[1/\mathrm{GeV}]$ & 3.431286e-02 \\
$d_4[1/\mathrm{GeV}]$ & -3.704632e-01 \\
\hline
\end{tabular}
\end{minipage}
\hfill
\begin{minipage}[t!]{0.7\textwidth}
    \vspace{-6pt} 
    \includegraphics[width=\linewidth]{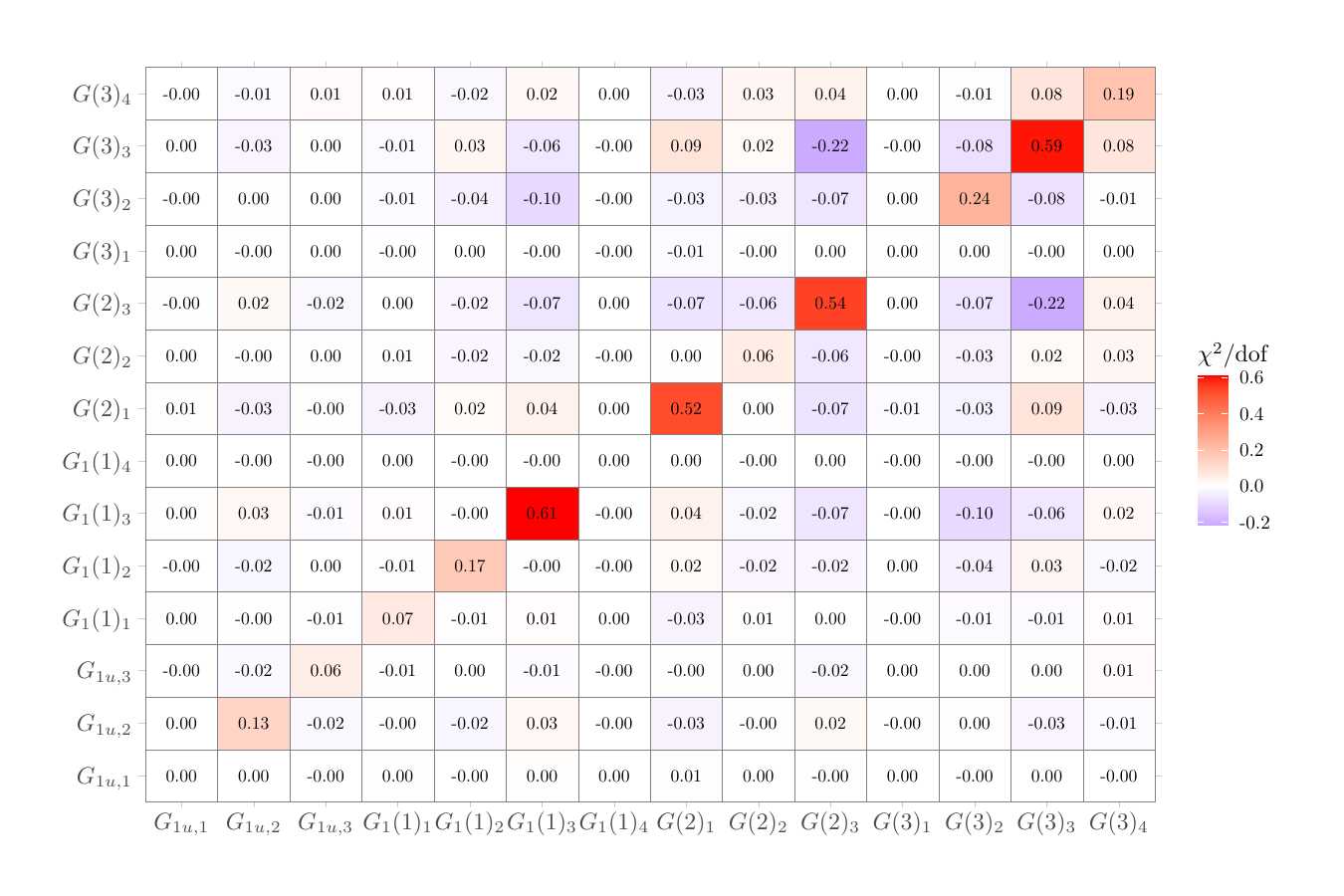}
    \vspace{-26pt}
    \caption{Heat map of correlated $\chi^2_{{\rm dof},ij}$, highlighting the relative impact of each energy level on the total fit quality.}
    \label{fig:partial-chi2_M3S2PL}
\end{minipage}
\end{figure*}

\begin{figure*}[h]
    \includegraphics[height=4cm]{plots/3D-poles/fit-16-exp0.pdf}
    \includegraphics[height=4cm]{plots/3D-poles/fit-16-LAT0.pdf}\\
    \includegraphics[height=4cm]{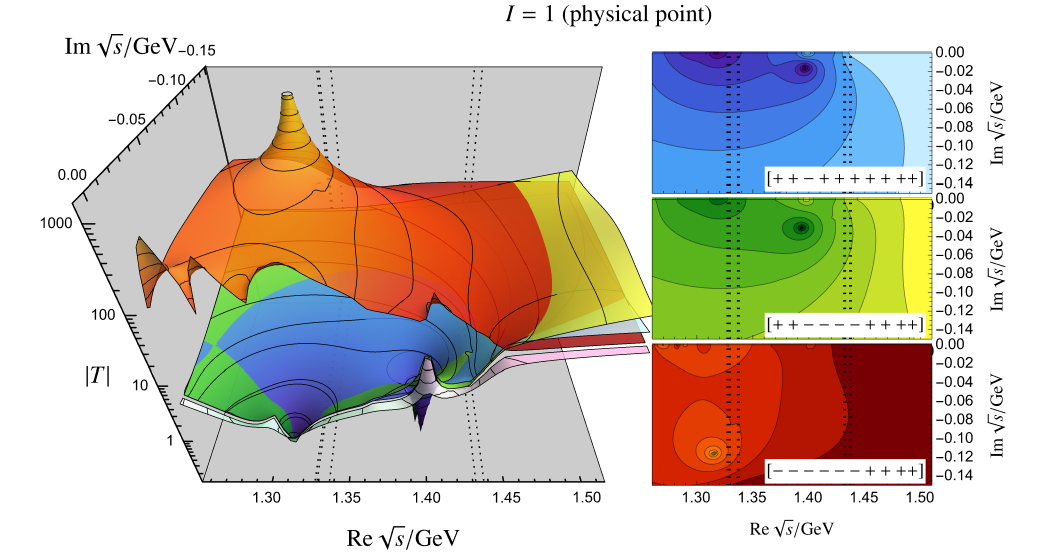}
    \includegraphics[height=4cm]{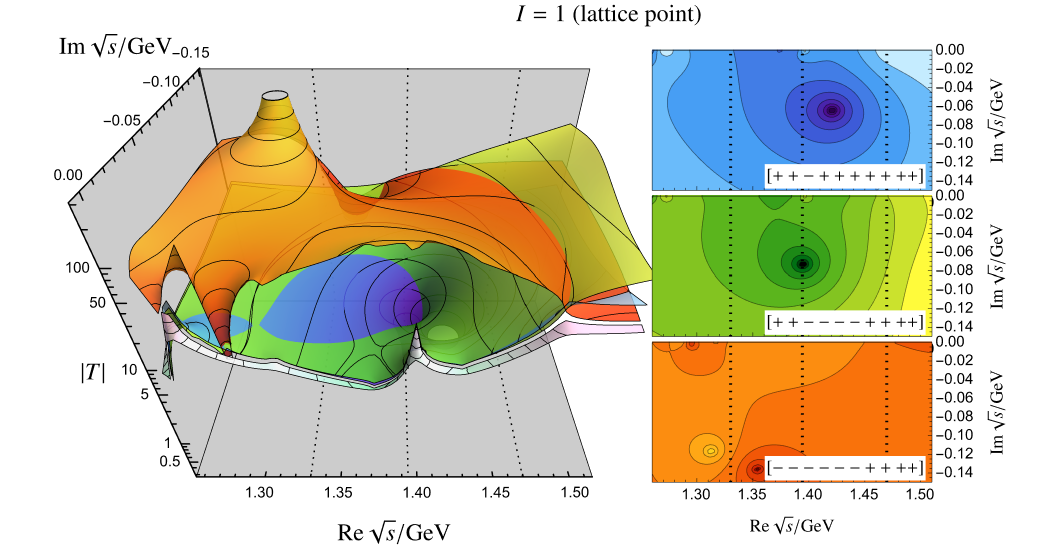}
    \caption{Isoscalar and isovector projected absolute value of the $\pi\Sigma\to \pi\Sigma$ scattering amplitude on the unphysical second Riemann sheets. Nomenclature as in the main text. Magenta crosses represent literature values from Refs.~\cite{BaryonScatteringBaSc:2023ori, BaryonScatteringBaSc:2023zvt} for the lattice point and Refs.~\cite{Ikeda:2012au,Guo:2012vv,Mai:2014xna} for the physical point.}
    \label{fig:app-16}
\end{figure*}

\clearpage
\subsection{M3S3PL ($F_{12}$)}

\begin{figure*}[h]
\begin{minipage}[t!]{0.23\textwidth}
    \vspace{0pt} 
\centering
\caption{The total $\chi^2_\mathrm{dof}$ as defined in \cref{eq:chi2}, $\Lambda$ parameter and LECs for M3S3PL ($F_{12}$).}
    \label{tab:details_12}
\begin{tabular}{|c|c|}
\hline
$\chi^2_{\rm dof}$ & 2.236   \\
\hline
$\Lambda[\mathrm{GeV}]$ & 0.4218104 \\
\hline
$b_0[1/\mathrm{GeV}]$ & -8.768647e-01 \\
$b_D[1/\mathrm{GeV}]$ & 5.246210e-02 \\
$b_F[1/\mathrm{GeV}]$ & -3.406325e-01 \\
$d_1[1/\mathrm{GeV}]$ & 1.201660e+00 \\
$d_2[1/\mathrm{GeV}]$ & -1.753693e-01 \\
$d_3[1/\mathrm{GeV}]$ & -4.544383e-01 \\
$d_4[1/\mathrm{GeV}]$ & 1.476738e-01 \\
\hline
\end{tabular}
\end{minipage}
\hfill
\begin{minipage}[t!]{0.7\textwidth}
    \vspace{-6pt} 
    \includegraphics[width=\linewidth]{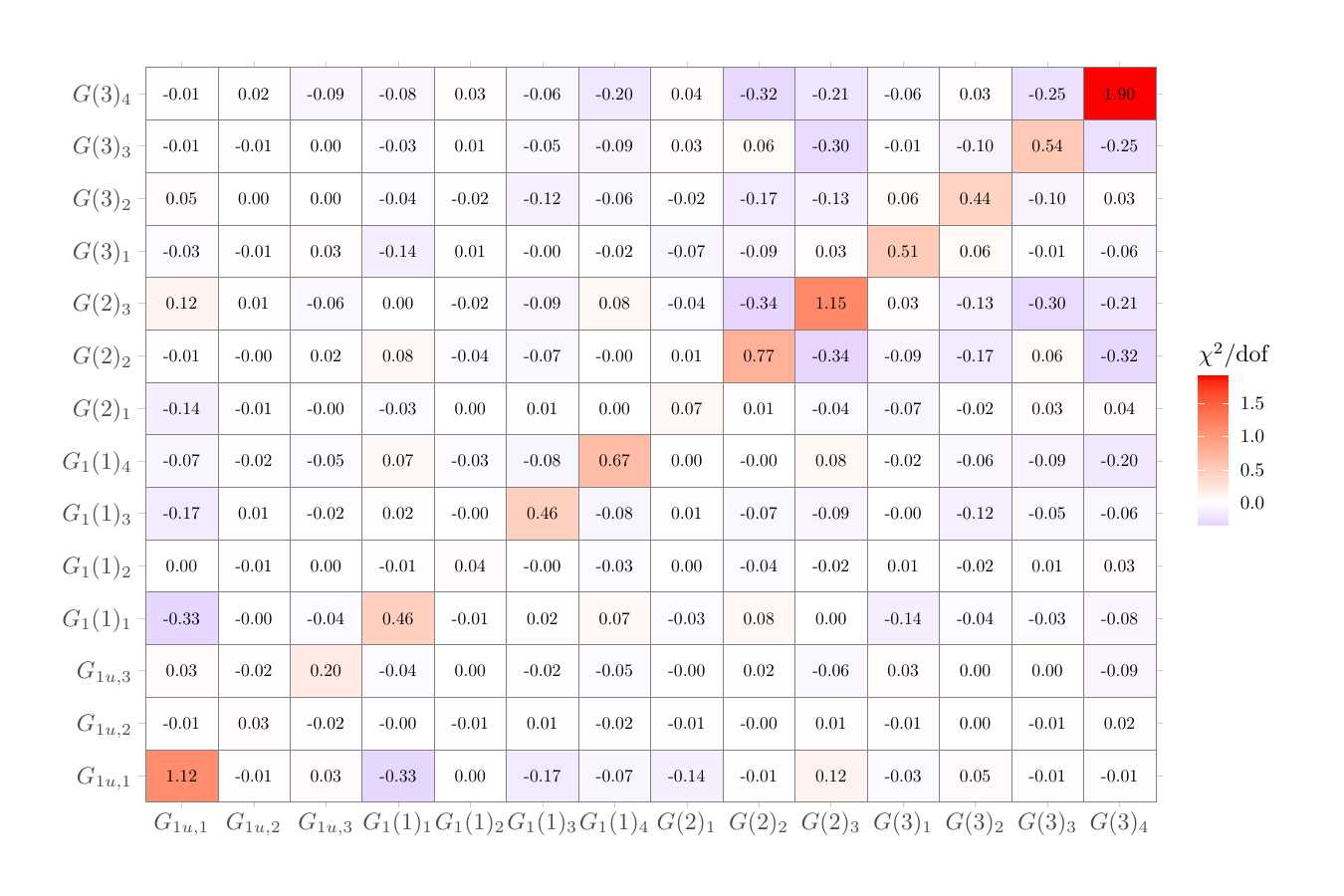}
    \vspace{-26pt}
    \caption{Heat map of correlated $\chi^2_{{\rm dof},ij}$, highlighting the relative impact of each energy level on the total fit quality.}
    \label{fig:partial-chi2_M3S3PL}
\end{minipage}
\end{figure*}

\begin{figure*}[h]
    \includegraphics[height=4cm]{plots/3D-poles/fit-12-exp0.pdf}
    \includegraphics[height=4cm]{plots/3D-poles/fit-12-LAT0.pdf}\\
    \includegraphics[height=4cm]{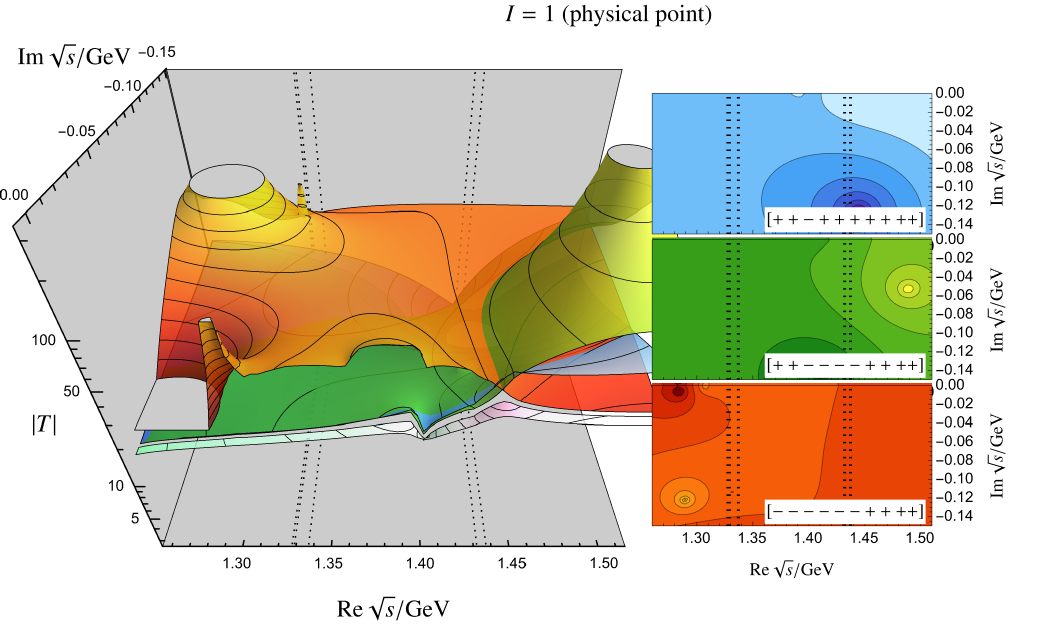}
    \includegraphics[height=4cm]{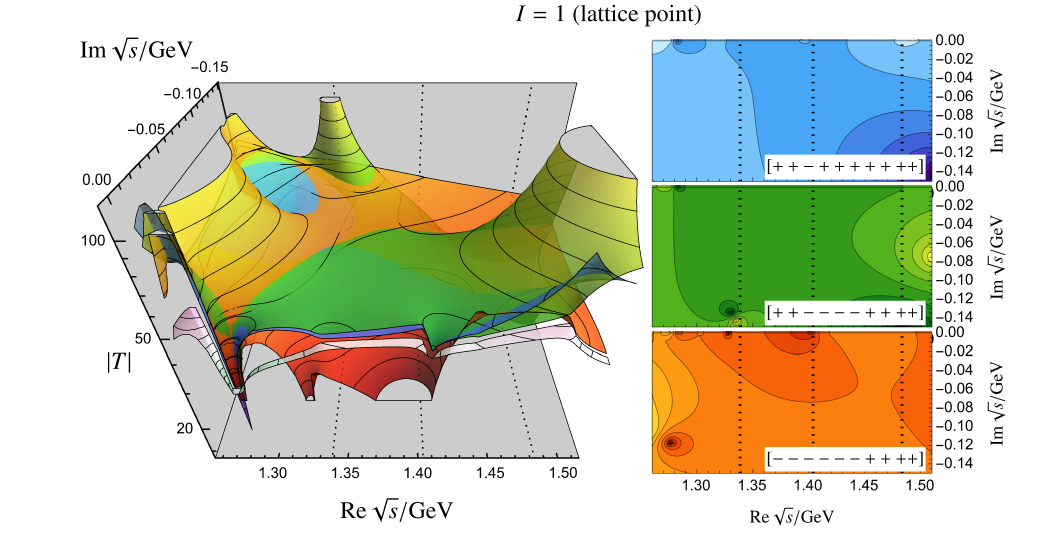}
    \caption{Isoscalar and isovector projected absolute value of the $\pi\Sigma\to \pi\Sigma$ scattering amplitude on the unphysical second Riemann sheets. Nomenclature as in the main text. Magenta crosses represent literature values from Refs.~\cite{BaryonScatteringBaSc:2023ori, BaryonScatteringBaSc:2023zvt} for the lattice point and Refs.~\cite{Ikeda:2012au,Guo:2012vv,Mai:2014xna} for the physical point.}
    \label{fig:app-12}
\end{figure*}

\end{document}